\documentclass{article}

\usepackage[preprint]{colm2026_conference}

\makeatletter
\def\@maketitle{\vbox{\hsize\textwidth
{\Large\bf \@title\par}
\lhead{Preprint. Under review.}
\vskip 0.3in minus 0.1in}}
\makeatother

\usepackage[T1]{fontenc}
\usepackage{lmodern}
\usepackage{microtype}
\usepackage{hyperref}
\usepackage{url}
\usepackage{booktabs}
\usepackage{amsmath}
\usepackage{amsthm}

\usepackage{graphicx}
\usepackage{caption}
\usepackage{subcaption}
\usepackage{tikz}
\usetikzlibrary{shapes,shapes.symbols,arrows,arrows.meta,positioning,calc,backgrounds,fit,decorations.pathreplacing,decorations.pathmorphing,decorations.markings,patterns,shapes.geometric,shadows}
\usepackage{pgfplots}
\pgfplotsset{compat=1.17}
\usepgfplotslibrary{groupplots}

\definecolor{cfillLight}{HTML}{E8EEF4}    
\definecolor{cfillMed}{HTML}{D0DCE8}      
\definecolor{cfillDark}{HTML}{B8C8D8}     
\definecolor{cfillAccent}{HTML}{7A9CB8}   
\definecolor{cborder}{HTML}{5A7A9A}       
\definecolor{cborderLight}{HTML}{8FA8BF}
\definecolor{cborderFaint}{HTML}{B0C4D8}  
\tikzset{
  cognibit/.style={font=\sffamily\footnotesize, >={Stealth[length=1.8mm, width=1.2mm]}, node distance=10mm and 14mm},
  cbox/.style={rectangle, rounded corners=3pt, draw=cborderLight, line width=0.5pt, minimum width=24mm, minimum height=9mm, align=center, font=\sffamily\footnotesize, fill=cfillLight},
  cbox proc/.style={cbox, fill=cfillMed},
  cbox dark/.style={cbox, fill=cfillDark, draw=cborder},
  cbox accent/.style={cbox, fill=cfillAccent, draw=cborder, font=\sffamily\footnotesize\bfseries},
  cflow/.style={->, draw=cborder, line width=0.5pt},
  cflow dash/.style={cflow, dashed, draw=cborderLight},
  clabel/.style={font=\sffamily\scriptsize, fill=white, inner sep=1.5pt},
  cstate/.style={circle, draw=cborder, line width=0.5pt, minimum size=16mm, align=center, font=\sffamily\footnotesize},
  ccontainer/.style={rectangle, rounded corners=5pt, draw=cborderLight, line width=0.4pt, inner sep=8pt, fill=cfillLight!30},
}
\usepackage{listings}
\lstdefinelanguage{JavaScript}{
  keywords={break,case,catch,continue,debugger,default,delete,do,else,finally,for,function,if,in,instanceof,new,return,switch,this,throw,try,typeof,var,void,while,with,let,const,class,export,import,yield,async,await},
  sensitive=true,
  comment=[l]{//},
  morecomment=[s]{/*}{*/},
  morestring=[b]',
  morestring=[b]"
}
\usepackage{float}
\usepackage{algorithm}
\usepackage{algorithmicx}
\usepackage{algpseudocode}

\algrenewcommand\Call[2]{%
  \textproc{#1}%
  \if\relax\detokenize{#2}\relax
  \else(#2)\fi
}
\algnewcommand\algorithmicswitch{\textbf{switch}}
\algnewcommand\algorithmiccase{\textbf{case}}
\algnewcommand\algorithmicdefault{\textbf{default}}
\algdef{SE}[SWITCH]{Switch}{EndSwitch}[1]{\algorithmicswitch\ #1\ \algorithmicdo}{\algorithmicend\ \algorithmicswitch}
\algdef{SE}[CASE]{Case}{EndCase}[1]{\algorithmiccase\ #1}{\algorithmicend\ \algorithmiccase}
\algtext*{EndSwitch}
\algtext*{EndCase}
\algnewcommand\algorithmicparfor{\textbf{parallel for}}
\algdef{SE}[PARFOR]{ParFor}{EndParFor}[1]{\algorithmicparfor\ #1\ \algorithmicdo}{\algorithmicend\ \algorithmicparfor}

\usepackage{lineno}

\definecolor{darkblue}{rgb}{0, 0, 0.5}
\hypersetup{colorlinks=true, citecolor=darkblue, linkcolor=darkblue, urlcolor=darkblue}

\makeatletter
\newcommand{\AlphAlph}[1]{%
  \expandafter\@AlphAlph\expandafter{\the\value{#1}}%
}
\newcommand{\@AlphAlph}[1]{%
  \ifnum#1>26\relax
    A\@Alph{\numexpr#1-26\relax}%
  \else
    \@Alph{#1}%
  \fi
}
\makeatother

\usepackage{cleveref}
\crefname{appendix}{Appendix}{Appendices}
\Crefname{appendix}{Appendix}{Appendices}

\title{Cognibit: From Digital Exhaustion to Real-World Connection Through Gamified Territory Control and LLM-Powered Twin Networking}

\author{}

\begin{document}

\maketitle
\vspace{-1.5em}
\begin{center}
\small
Wanghao~Ye$^1$, Sihan~Chen$^2$, Yiting~Wang$^1$, Shwai~He$^1$, Bowei~Tian$^1$, Guoheng~Sun$^1$, Ziyi~Wang$^1$,\\
Ziyao~Wang$^1$, Yexiao~He$^1$, Zheyu~Shen$^1$, Meng~Liu$^1$, Yuning~Zhang$^1$, Meng~Feng$^1$, Yifei~Dong$^6$,\\
Yanhong~Qian$^1$, Yang~Wang$^3$, Siyuan~Peng$^1$, Yilong~Dai$^4$, Zhenle~Duan$^1$,\\
Joshua~Liu$^8$, Lang~Xiong$^7$, Hanzhang~Qin$^{5,*}$, Ang~Li$^{1,*}$

\vspace{0.5em}
\footnotesize
$^1$University of Maryland,\enspace $^2$USC Viterbi School of Engineering,\enspace $^3$Northeastern University\\[1pt]
$^4$University of Florida,\enspace $^5$National University of Singapore,\enspace $^6$University of Washington\\[1pt]
$^7$Stanford University,\enspace $^8$Carnegie Mellon University,\enspace $^*$Joint corresponding authors

\vspace{0.3em}
\scriptsize
\textbf{University of Maryland Departments:} Electrical and Computer Engineering, Computer Science,\\
Robert H. Smith School of Business, School of Public Policy, Agricultural and Resource Economics
\end{center}


\begin{abstract}
We present an LLM-powered social discovery platform that uses digital twins to autonomously evaluate interpersonal compatibility through behavioral simulation. The platform unifies three key pillars: (1)~digital twins that engage in autonomous multi-turn conversations on behalf of users to estimate compatibility, (2)~gamified territory conquest mechanics that incentivize real-world exploration and create organic settings for in-person encounters, and (3)~AI companions that preserve persistent shared memory across devices. Built upon CogniPair's cognitive architecture \citep{CogniPair2026}, validated on the Columbia Speed Dating dataset (551 participants), our system extends prior simulation-only matching into a fully deployed social discovery environment. Through deployment, we derive empirical cost--quality baselines and identify fundamental scaling bottlenecks that remain hidden in component-level testing alone.
\end{abstract}

\section{Introduction}

\subsection{The Compound Problem of Social Media Exhaustion}

Social media exhaustion represents a multi-faceted computational and behavioral challenge. Users invest approximately 2.5 hours daily on social platforms \citep{Hunt2018}, with co-design participants in \citet{CogniPair2026} reporting 97±12 minutes specifically on dating/social discovery apps. Yet effectiveness declines with exposure: \citet{Pronk2020} documented a 27\% decrease in acceptance rates after continued use, while the gap between online matching and real-world meeting remains wide \citep{Rosenfeld2019,tyson2016}. This compound problem manifests through choice overload \citep{Schwartz2004,Iyengar2000}, temporal coupling, a digital-physical gap, and cognitive burden from managing simultaneous conversations (see Appendix~\ref{appendix:exhaustion-analysis} for detailed analysis).

CogniPair \citep{CogniPair2026} introduced a fully-implemented architecture for LLM-powered behavioral compatibility evaluation with three integrated components: (1) LLM-powered digital twins conducting autonomous behavioral simulation, (2) gamified territory control creating physical convergence points, and (3) persistent AI companions providing emotional scaffolding. CogniPair demonstrated browser-based feasibility---58.3 FPS with 5-8 concurrent agents---validated the architecture on the Columbia Speed Dating dataset \citep{fisman2006} (551 participants), and conducted a pilot deployment with 20 real users over 14 days. However, CogniPair did not provide deployment-oriented system analysis: no cost-quality tradeoff quantification, no cross-device memory validation, no mechanism-level controlled experiments, and no systematic runtime constraint documentation.

\begin{figure*}[t]
\centering
\begin{tabular}{@{}c@{\hspace{4pt}}c@{}}
\includegraphics[width=0.48\textwidth,height=0.30\textwidth]{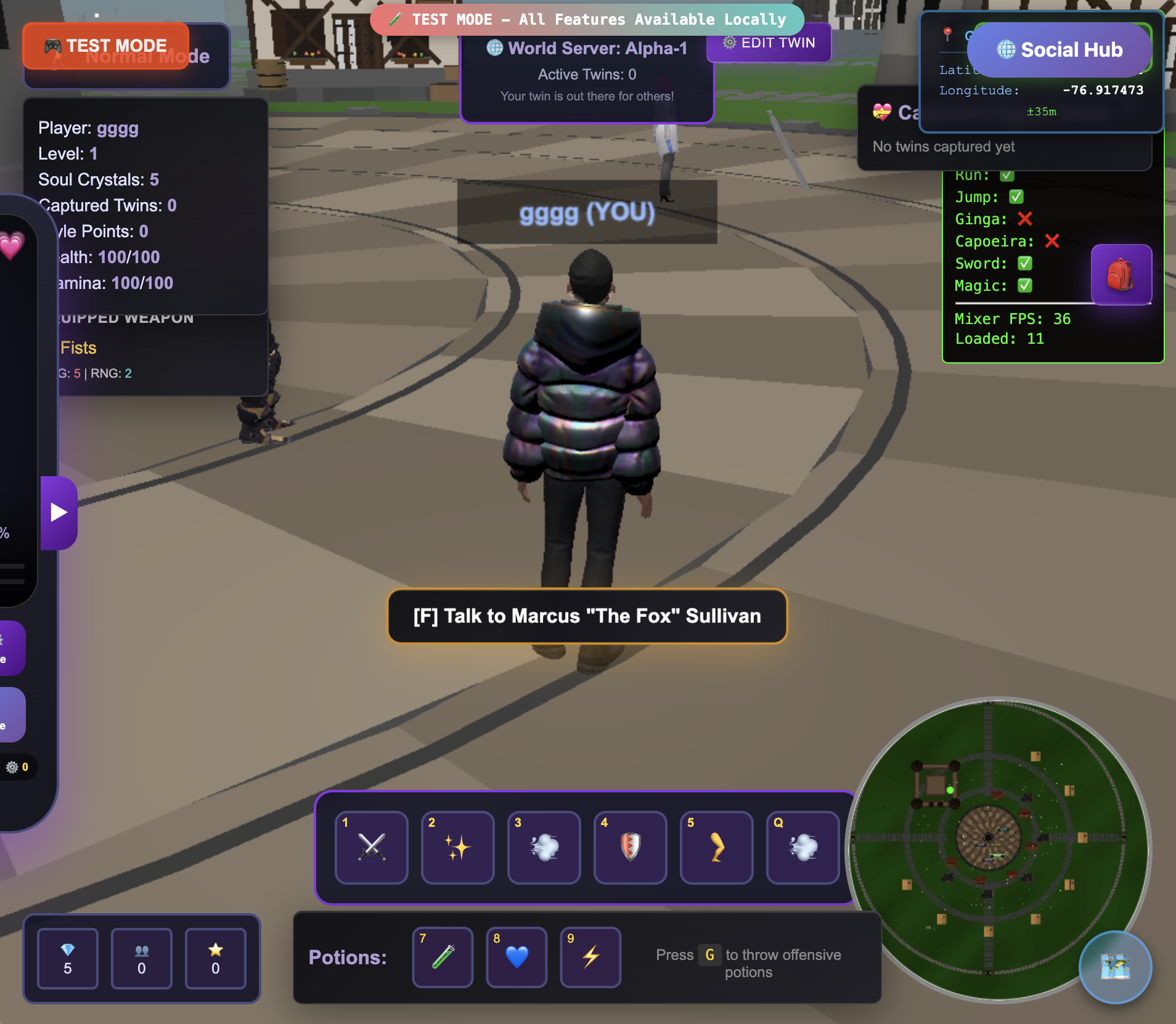} &
\includegraphics[width=0.48\textwidth,height=0.30\textwidth]{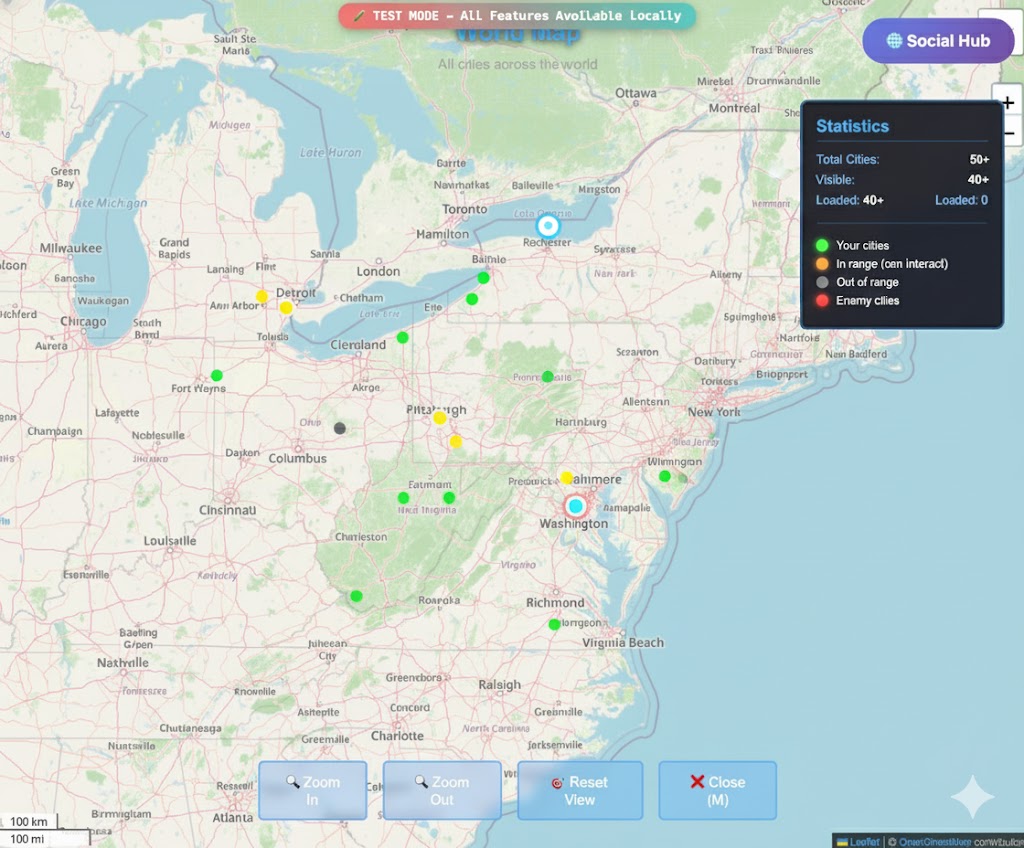} \\[-1pt]
{\small (a) 3D game world} & {\small (b) Territory map} \\[4pt]
\includegraphics[width=0.48\textwidth,height=0.30\textwidth]{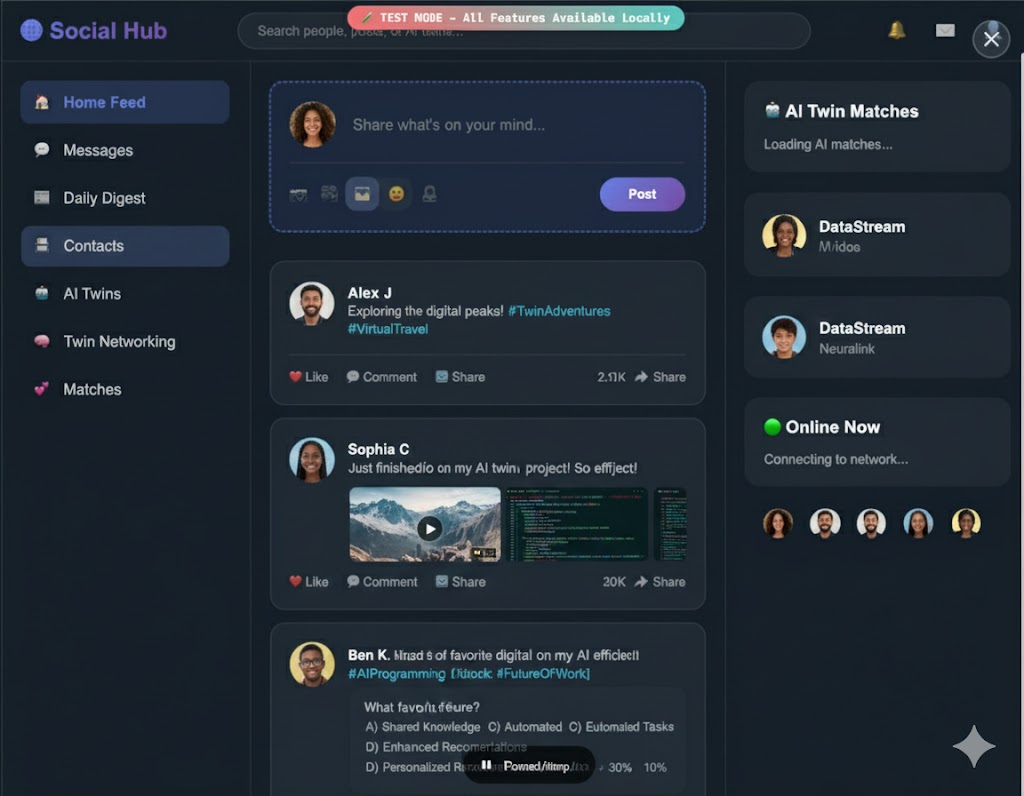} &
\includegraphics[width=0.48\textwidth,height=0.30\textwidth]{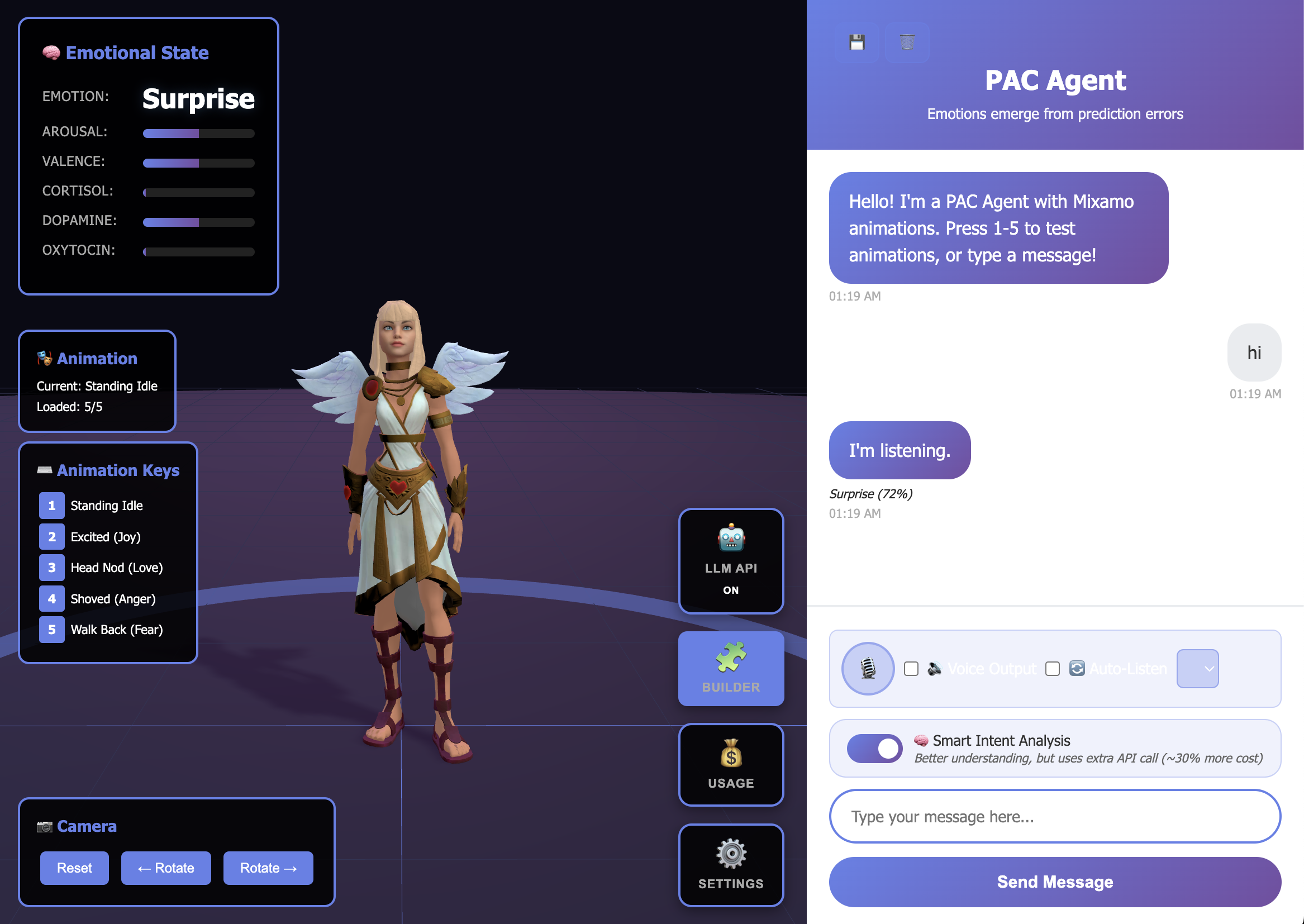} \\[-1pt]
{\small (c) Social Hub} & {\small (d) AI companion}
\end{tabular}
\vspace{-6pt}
\caption{Deployed system surfaces: (a)~3D territory exploration with avatar and twin encounters, (b)~geolocation-based territory map, (c)~Social Hub with feeds and AI-matched connections, (d)~pendant companion with PAC emotional state and chat.}
\label{fig:teaser}
\end{figure*}

\subsection{From Architecture to Deployment}

Prior work validated the CogniPair architecture as a cognitive framework for twin-based compatibility modeling \citep{CogniPair2026}, but architecture-level success does not automatically translate into deployable social value. A real social discovery system must do more than produce plausible pairwise scores: it must efficiently narrow large candidate pools, preserve cross-device relational memory, operate under strict latency and cost constraints, and avoid new safety failures when online interaction is coupled to real-world mobility. Today's social platforms maximize online attention; Cognibit is designed instead to convert online cognitive filtering into real-world social opportunity. This conversion is not driven by recommendation alone: Cognibit couples twin-based filtering with a geolocation-mediated encounter loop---territory gaming for physical convergence, a Social Hub for cross-device social continuity, and a pendant companion for ambient emotional scaffolding---turning social discovery from passive browsing into physically grounded compatibility discovery. This paper provides the deployment-oriented system analysis that CogniPair's architecture paper did not: funnel validation, cost-quality tradeoffs, cross-device memory reliability, mechanism-level controlled experiments, and retrospective analysis of the CogniPair pilot deployment\citep{CogniPair2026}.

Rather than treating engagement as something to maximize within the app, Cognibit treats engagement as a means of moving users toward higher-quality real-world encounters through geolocation-mediated encounter creation, cross-device social continuity, and persistent relational memory. We contribute deployment-facing system analysis---including controlled system experiments, mechanism-strengthening simulations, and retrospective analysis of the CogniPair pilot deployment (N=20, 14 days, 342 twin sessions)---to address four deployment-facing questions:

\begin{itemize}
\item \textbf{Can agentic filtering beat passive browsing?} We validate a multi-stage matching funnel that yields monotonic quality improvement---from random selection (2.21/5) through heuristic filtering (2.82/5) to full twin-based reranking (3.08/5)---evaluated across 30 target users with an independent blind judge (Section~\ref{sec:funnel}).

\item \textbf{What does extra intelligence actually buy?} A cost-quality Pareto analysis shows that heuristic filtering captures much of the attainable gain with millisecond latency, while expanding reranking to 50 LLM-evaluated pairs yields quality 3.02 at 782s latency---quantifying the diminishing returns of deeper cognitive simulation (Section~\ref{sec:pareto}).

\item \textbf{Can persistent social memory be made reliable enough to matter?} We validate cross-device shared memory at 73ms write latency, 0\% stale reads, and zero data loss under concurrent writes---establishing persistent relational memory as a deployable system capability (Section~\ref{sec:memory}).

\item \textbf{What breaks in real deployment?} Retrospective analysis of the CogniPair pilot deployment\citep{CogniPair2026} (N=20, 14 days) reveals a hard ceiling at 20 concurrent browser-based agents, 20\% non-engagement, 33\% negative meeting outcomes, safety risks from predictable territory patterns, and a qualitative engagement shift from sedentary browsing toward physically active territory gaming (Section~\ref{sec:what-failed}).
\end{itemize}

\subsection{Contributions}

Beyond the CogniPair architecture, this work contributes:

\begin{enumerate}
\item \textbf{Problem reframing and interaction redesign.} We recast twin-based social discovery from a compatibility prediction task into a deployment question and introduce a geolocation-mediated encounter loop that shifts discovery from passive browsing toward physically grounded exploration.

\item \textbf{System-level deployment validation.} We validate a multi-stage funnel with monotonic quality gains (2.21$\to$2.82$\to$3.08), a cost-quality Pareto frontier across four operating points (3ms to 782s), and cross-device persistent memory (73ms write, 0\% stale reads, zero data loss).

\item \textbf{Retrospective deployment analysis.} Analysis of the CogniPair pilot\citep{CogniPair2026} (N=20, 14 days) documents deployment barriers---a 20-user browser ceiling, 20\% non-engagement, 33\% negative meetings, and location-linked safety risks---treating honest failure reporting as a contribution.
\end{enumerate}

\section{Related Work}

CogniPair \citep{CogniPair2026} provides comprehensive related work; here we note four deployment-specific gaps.
LLM agents have advanced through instruction-following \citep{Ouyang2022}, planning \citep{Wei2022,Yao2023}, multi-agent collaboration \citep{Li2023camel,hong2023metagpt,park2023,liu2023agentbench}, but none systematize deployment engineering---cost-quality tradeoffs, cross-device memory, or runtime constraints \citep{Zimmerman2007}.
Cascading \citep{chen2023frugalgpt}, open-weight \citep{Touvron2023}, and quantization \citep{Dettmers2022} strategies reduce inference costs, but no work establishes Pareto frontiers for multi-turn behavioral simulation.
Dating platforms rely on profile heuristics yet the gap between online matching and real-world meeting remains wide \citep{Rosenfeld2019,tyson2016}, compounded by choice overload \citep{Schwartz2004,Pronk2020}; our funnel reduces 200 candidates to 5.
AI companions foster bonding \citep{Skjuve2021} but risk substituting for real connection \citep{Turkle2011}; gamification \citep{Hamari2014,Shea2017} provides precedent for geolocation-mediated encounters. Our system combines territory gaming with AI-mediated filtering, using LLM-as-judge evaluation \citep{Zheng2023}.

\section{Design Process and Methodology}
\label{sec:design-process}

CogniPair \citep{CogniPair2026} details the full 8-month Research through Design process. Five design principles emerged: (1) reduce, don't remove human judgment; (2) create excuses for interaction; (3) bridge digital and physical; (4) protect emotional energy; (5) maintain persistent context. These motivated the three-pillar design: digital twins for delegated assessment, territory gaming for physical scaffolding, and companions for emotional continuity. The integrated user journey diagram appears in Appendix~\ref{appendix:user-journeys}.

For deployment, we formalized a three-stage filtering pipeline, implemented multi-provider LLM routing for cost management, added survey instruments (UCLA Loneliness Scale, Social Anxiety Inventory, EMA), and added safety features (GPS aggregation, blocking, anonymous mode). Our evaluation combines controlled system experiments with retrospective analysis of the CogniPair pilot deployment (Section~\ref{sec:field-deployment}).

\section{System Design: Deployment Adaptations}

CogniPair \citep{CogniPair2026} describes the full system architecture including the four-layer design (discovery, engagement, support, physical integration), GNWT/PAC cognitive modules, personality evolution system, territory mechanics (deployed as the geolocation-mediated encounter loop evaluated in Section~\ref{sec:mechanisms}), and pendant companion architecture (providing ambient emotional scaffolding across sessions). Here we describe the deployment-specific adaptations: the three-stage filtering pipeline, LLM integration architecture, and compatibility scoring formula. Detailed implementation appears in the appendix: GNWT modules (Appendix~\ref{appendix:gnwt}), PAC dynamics (Appendix~\ref{appendix:pac-details}), core algorithms (Appendix~\ref{appendix:core-algorithms}), and user interface (Appendix~\ref{appendix:user-interface}).

The deployed platform comprises three user-facing applications (Social Hub, Boss Fight Game, Digital Twin Monitor) supported by core services (Authentication, Real-time Sync, AI Engine with PAC/GNWT, Game Core, Social Core), shared modules, and external dependencies (Firebase, API Proxy, CDN). Full architecture details and diagrams appear in Appendix~\ref{appendix:system-architecture-diagrams}.

\subsection{Three-Stage Choice Reduction Pipeline}

\noindent
\begin{minipage}[t]{0.58\textwidth}
\vspace{0pt}
Building on the CogniPair matching architecture \citep{CogniPair2026}, we formalized a three-stage pipeline (Figure~\ref{fig:filtering-funnel}).
\textbf{Stage~1 (Elimination, $<$1s):} Satisficing criteria \citep{Simon1955} eliminate 70--80\% by distance, life stage, and preferences ($\sim$200 prospects).
\textbf{Stage~2 (Simulation, 2--5 min):} LLM twin conversations reveal behavioral compatibility; top-$k$ ($k$=20) undergo reranking.
\textbf{Stage~3 (Decision):} Users evaluate top-5 with previews within Miller's limit \citep{Miller1956}. Offline evaluation (Section~\ref{sec:funnel}) uses a fixed 200-candidate pool.
\end{minipage}%
\hfill
\begin{minipage}[t]{0.40\textwidth}
\vspace{0pt}
\centering
\resizebox{\textwidth}{!}{
\begin{tikzpicture}[
    stage/.style={trapezium, trapezium angle=75, draw, thick, minimum height=0.8cm, text=white, font=\footnotesize\bfseries, align=center, trapezium stretches body},
    lbl/.style={font=\scriptsize, align=left},
    arrow/.style={->, thick, gray!70},
    reduction/.style={font=\scriptsize\itshape, text=red!70!black}
]

\node[stage, fill=blue!60, minimum width=4.5cm] (pool) at (0, 4.2) {Pool: 200};
\node[stage, fill=blue!70, minimum width=3.2cm] (s1) at (0, 3.0) {Heuristic Top-20};
\node[stage, fill=blue!80, minimum width=2.2cm] (s2) at (0, 1.8) {LLM Reranked};
\node[stage, fill=blue!90, minimum width=1.5cm] (s3) at (0, 0.6) {Top 5};

\node[lbl] (l1) at (3.6, 3.6) {\textbf{Heuristic}\\ 4-factor, 2.8ms};
\node[lbl] (l2) at (3.6, 2.4) {\textbf{Twin conv.}\\ 3-turn, $\sim$323s};
\node[lbl] (l3) at (3.6, 1.2) {\textbf{Combined}\\ 0.7h + 0.3b};

\draw[dotted, gray] (pool.east) -- (l1.west);
\draw[dotted, gray] (s1.east) -- (l2.west);
\draw[dotted, gray] (s2.east) -- (l3.west);

\node[reduction] at (-2.8, 3.6) {$-$90\%};
\node[reduction] at (-2.8, 2.4) {rerank};
\node[reduction] at (-2.8, 1.2) {$-$75\%};

\draw[arrow] (pool.south) -- (s1.north);
\draw[arrow] (s1.south) -- (s2.north);
\draw[arrow] (s2.south) -- (s3.north);

\node[draw, thick, rounded corners, fill=green!10, font=\scriptsize\bfseries] at (0, -0.2) {97.5\% reduction (200 $\to$ 5)};

\end{tikzpicture}}
\vspace{-4pt}
\captionof{figure}{Three-stage filtering: 200 $\to$ 5 (97.5\% reduction).}
\label{fig:filtering-funnel}
\end{minipage}
\vspace{4pt}

\begin{algorithm}[!htbp]
\caption{Multi-Stage Matching Funnel}
\label{alg:matching-funnel}
\begin{algorithmic}[1]
\Require Candidate pool $\mathcal{P}$ ($|\mathcal{P}|=200$), target user $u$, LLM budget $k$
\Ensure Top-5 pre-validated matches
\State \Comment{Stage 1: Heuristic filtering}
\ForAll{candidate $c \in \mathcal{P}$}
    \State $s_c \gets 0.60 \cdot \text{PersonalitySim}(u, c) + 0.25 \cdot \text{InterestOverlap}(u, c)$
    \State \hspace{2.2em} $+ 0.10 \cdot \text{DiversityBonus}(u, c) + 0.05 \cdot \text{ProfileRichness}(c)$
\EndFor
\State $\mathcal{H} \gets \text{TopK}(\mathcal{P}, k, \text{by } s_c)$ \Comment{Select top-$k$ by heuristic score}
\State \Comment{Stage 2: LLM twin conversation reranking}
\ForAll{candidate $c \in \mathcal{H}$}
    \State $\text{conv} \gets \Call{SimulateTwinConversation}{u, c, \text{turns}=3}$
    \State $b_c \gets \Call{EvaluateBehavioralCompatibility}{\text{conv}}$
    \State $\mathit{combined}_{c} \gets 0.7 \cdot s_c + 0.3 \cdot b_c$
\EndFor
\State \Comment{Stage 3: Return top-5 pre-validated matches}
\State \Return $\text{TopK}(\mathcal{H},\; 5,\; \text{by } \mathit{combined}_{c})$
\end{algorithmic}
\end{algorithm}

\subsection{LLM Integration}

Each twin conversation uses a two-stage pipeline: pragmatic intent analysis (temperature=0.3) followed by personality-conditioned generation (temperature=0.8). The system routes across four providers (GPT-4o, GPT-4o-mini, Claude-3-Haiku, DeepSeek) with template-based fallback; the deployed system, offline evaluation (Qwen2.5-72B-Instruct), and synthetic experiments use different configurations as detailed in Appendix~\ref{appendix:twin-pipeline}. Compatibility is computed as $C(a,b) = \sum_{i=1}^{5} w_i \cdot s_i(a,b)$, combining personality alignment (0.30), shared interests (0.20), conversation quality (0.25), emotional resonance (0.15), and interaction patterns (0.10).

\subsection{Deployment Performance}

Full-stack deployment measurements show 312$\pm$18MB for 1--5 agents, with 62.4MB per additional agent and a hard ceiling at 20 concurrent agents. Performance follows exponential decay with the 30 FPS usability threshold crossed at $n\approx 8$. Detailed runtime constraints and deployment activity metrics are reported in Section~\ref{sec:runtime}. Cross-device interface screenshots appear in Appendix~\ref{appendix:user-interface}.

\section{CogniPair Pilot Deployment}
\label{sec:field-deployment}

CogniPair \citep{CogniPair2026} conducted a pilot deployment with 20 users for 14 days to measure system performance, user engagement, and behavioral change metrics in a real-world setting. This section reports the deployment setup and findings from that prior pilot, which provide the empirical context for Cognibit's deployment-oriented system analysis (Sections~\ref{sec:results}--\ref{sec:limitations}). All deployment data reported below are from the CogniPair pilot; Cognibit's new contributions---funnel validation, Pareto analysis, memory reliability testing, and mechanism-strengthening simulations---are presented in Section~\ref{sec:results}.

\subsection{Experimental Setup}

Twenty participants (ages 24--45, M=31.5, SD=6.2; 11F/8M/1NB; 90\% with prior dating app experience) were recruited from one city and completed a 14-day deployment. Day~0 included onboarding, twin initialization, and baseline measures (UCLA Loneliness Scale, Social Anxiety Inventory); Days~1--14 involved natural usage with telemetry, daily surveys, and EMA prompts; Day~15 concluded with exit interviews. Compensation: \$200. Privacy protections included GPS aggregated to 100m precision, end-to-end encryption, and data deletion rights (Appendix~\ref{appendix:privacy-safety}). Statistical methodology details appear in Appendix~\ref{appendix:statistical-methodology}.

\subsection{Metrics and Data Collection}

\textbf{Operational Definitions}: We define ``connection'' at three levels of increasing commitment: (1) \emph{digital contact}---exchange of in-app messages beyond territory battle chat ($n=42$ instances), (2) \emph{contact exchange}---sharing external contact information such as phone number or social media handle ($n=14$ instances across 8 participants), and (3) \emph{physical meeting}---co-located interaction lasting $>$5 minutes outside of a system-required territory battle ($n=18$ meetings by 8/20 participants). Of the 18 physical meetings, 11 (61\%) were verified by GPS co-location telemetry, while 7 rely on participant self-report. ``In-person meeting'' throughout this paper refers to level 3; we did not track post-study relationship persistence.

We distinguish between \emph{passive engagement} (sedentary profile swiping) and \emph{active engagement} (physically situated interaction involving locomotion, collaborative gameplay, and AI-scaffolded social interaction). Over the deployment, 1,247 pendant conversations and 342 twin networking sessions were logged. Detailed behavioral metrics and psychometric analysis appear in Appendix~\ref{appendix:statistical-methodology} and Appendix~\ref{appendix:engagement-quality}.

\subsection{Statistical Analysis}

Given the prior pilot's sample ($N=20$), we report descriptive statistics and effect sizes as exploratory baselines rather than confirmatory hypothesis tests. The study is substantially underpowered (power$=0.12$ for medium effects, $d=0.5$), so we emphasize 95\% confidence intervals over $p$-values.

Traditional platform time decreased from 97$\pm$12 to 14$\pm$5 min/day, but this reflects engagement reallocation rather than net time savings (total platform time increased to 141 min/day). Loneliness and anxiety effects are suggestive but confidence intervals include zero. Exploratory correlations suggest dose-response patterns, though the absence of a control group prevents causal interpretation and 14 days cannot rule out extended novelty effects. Detailed statistical analysis including correlations, partial correlations, temporal persistence, and non-engager comparisons appears in Appendix~\ref{appendix:statistical-methodology}. These pilot results are exploratory; broader limitations are discussed in Section~\ref{sec:limitations}.

\section{Empirical Results and Performance Analysis}
\label{sec:results}

We organize results into two layers. Sections~\ref{sec:funnel}--\ref{sec:memory} present the primary quantitative headline: system funnel validation, cost-quality Pareto analysis, and cross-device memory reliability. Section~\ref{sec:mechanisms} provides mechanism-level evidence that the geolocation-mediated encounter loop and social continuity interfaces function in deployment, along with documented failure modes. Section~\ref{sec:runtime} reports system runtime constraints.

\subsection{System Funnel Validation}
\label{sec:funnel}

The central deployment question is whether the multi-stage matching funnel produces higher-quality recommendations than simpler alternatives. We evaluate three conditions on 200 candidate profiles drawn from the PersonaChat validation set, each converted to a twin blueprint with personality traits, interests, and speaking style:

\begin{itemize}
\item \textbf{Random baseline}: 5 candidates selected uniformly at random from the pool
\item \textbf{Heuristic-only}: Preference filtering followed by 4-factor scoring (personality similarity 0.60, interest overlap 0.25, diversity 0.10, profile richness 0.05), top 5 returned
\item \textbf{Full twin-based pipeline}: Heuristic top-20 candidates undergo 3-turn LLM-simulated twin conversations (Qwen2.5-72B-Instruct, 4-bit quantization \citep{Dettmers2023QLoRA}), combined score (heuristic $\times$ 0.7 + behavioral $\times$ 0.3), top 5 returned
\end{itemize}

All conditions receive the same 200-candidate pool and produce exactly 5 recommendations. Final quality is evaluated by an independent cross-family LLM judge \citep[following the methodology of][]{Zheng2023} (Llama-3.1-70B-Instruct \citep{LlamaTeam2024}) that is \textbf{blind to funnel internal scores, condition labels, and heuristic weights}. The judge rates each recommended pair on interaction quality (1--5) and personality complementarity (1--5); we report the mean of these two dimensions, averaged over the five recommendations per target user.

\begin{table}[h]
\centering
\caption{Funnel Quality Comparison (30 Target Users, Independent Blind Judge)}
\label{tab:funnel-main}
\begin{tabular}{lccc}
\toprule
\textbf{Condition} & \textbf{Mean Quality (1--5)} & \textbf{Std} & \textbf{Latency} \\
\midrule
Random baseline & 2.21 & 0.40 & $<$1ms \\
Heuristic-only & 2.82 & 0.70 & 2.8ms \\
Full twin-based pipeline & \textbf{3.08} & 0.76 & $\sim$323s \\
\bottomrule
\end{tabular}
\end{table}

Quality improves monotonically across conditions ($\Delta_{\text{random}\to\text{heuristic}}=+0.61$, $\Delta_{\text{heuristic}\to\text{full}}=+0.26$), with the heuristic stage contributing the majority of the gain. The full pipeline achieves 97.5\% candidate reduction (200$\to$5) while maintaining the highest judge-rated quality. A preliminary 10-target evaluation yielded consistent ordering (random 1.71, heuristic 3.14, full 3.55), though with higher variance due to smaller sample size; the 30-target result reported here is treated as the primary evidence.

\begin{figure*}[t]
\centering
\begin{tabular}{@{}c@{\hspace{6pt}}c@{\hspace{6pt}}c@{}}
\resizebox{0.32\textwidth}{!}{
\begin{tikzpicture}
\begin{axis}[
    width=0.72\textwidth,
    height=0.40\textwidth,
    ybar,
    bar width=22pt,
    xlabel={Funnel Condition},
    ylabel={Mean Quality (1--5)},
    ymin=0, ymax=4.2,
    symbolic x coords={Random,Heuristic,Full Pipeline},
    xtick=data,
    tick label style={font=\footnotesize},
    xlabel style={font=\small},
    ylabel style={font=\small},
    grid=major,
    grid style={gray!15},
    ymajorgrids=true,
    xmajorgrids=false,
    axis y line*=left,
    legend style={at={(0.02,0.98)}, anchor=north west, font=\footnotesize, fill=white, fill opacity=0.9},
    nodes near coords,
    every node near coord/.append style={font=\scriptsize, above},
]
\addplot[fill=blue!50, draw=blue!70] coordinates {
    (Random, 2.21) (Heuristic, 2.82) (Full Pipeline, 3.08)
};
\addlegendentry{Mean Quality}
\end{axis}

\begin{axis}[
    width=0.72\textwidth,
    height=0.40\textwidth,
    ymin=0.5, ymax=1000,
    ymode=log,
    symbolic x coords={Random,Heuristic,Full Pipeline},
    xtick=data,
    xticklabels={},
    ylabel={Latency (log scale)},
    ylabel style={font=\small},
    tick label style={font=\footnotesize},
    axis y line*=right,
    axis x line=none,
    legend style={at={(0.98,0.98)}, anchor=north east, font=\footnotesize, fill=white, fill opacity=0.9},
]
\addplot[color=red!70!black, mark=diamond*, thick, mark size=3.5pt] coordinates {
    (Random, 1) (Heuristic, 2.8) (Full Pipeline, 323)
};
\addlegendentry{Latency}

\node[font=\tiny, text=red!60!black, anchor=south] at (axis cs:Random, 1.5) {$<$1ms};
\node[font=\tiny, text=red!60!black, anchor=south] at (axis cs:Heuristic, 4.5) {2.8ms};
\node[font=\tiny, text=red!60!black, anchor=south west] at (axis cs:Full Pipeline, 400) {$\sim$323s};

\end{axis}
\end{tikzpicture}} &
\resizebox{0.32\textwidth}{!}{
\begin{tikzpicture}
\begin{axis}[
    width=0.7\textwidth,
    height=0.45\textwidth,
    xlabel={Cost per Match (\$)},
    ylabel={Precision@5},
    xmode=log,
    xmin=0.2, xmax=30,
    ymin=0.35, ymax=0.75,
    grid=major,
    grid style={gray!20},
    legend pos=south east,
    legend style={font=\footnotesize, fill=white, fill opacity=0.9},
    tick label style={font=\footnotesize},
    xlabel style={font=\small},
    ylabel style={font=\small},
]

\addplot[only marks, mark=*, blue!80, mark size=4pt, error bars/.cd, y dir=both, y explicit]
    coordinates {(18.20, 0.687) +- (0,0.045)};
\addlegendentry{GPT-4o (\$18.20)}

\addplot[only marks, mark=square*, green!60!black, mark size=3.5pt, error bars/.cd, y dir=both, y explicit]
    coordinates {(1.10, 0.621) +- (0,0.052)};
\addlegendentry{GPT-4o-mini (\$1.10)}

\addplot[only marks, mark=triangle*, orange, mark size=4pt, error bars/.cd, y dir=both, y explicit]
    coordinates {(1.50, 0.608) +- (0,0.049)};
\addlegendentry{Claude-3-Haiku (\$1.50)}

\addplot[only marks, mark=diamond*, purple, mark size=4pt, error bars/.cd, y dir=both, y explicit]
    coordinates {(0.40, 0.573) +- (0,0.058)};
\addlegendentry{DeepSeek (\$0.40)}

\addplot[forget plot, gray, dashed, thin] coordinates {(0.2, 0.618) (30, 0.618)};
\node[font=\tiny, gray, anchor=south west] at (axis cs:5, 0.620) {90\% of GPT-4o};

\draw[->, thick, red!60] (axis cs:3.5, 0.48) -- (axis cs:1.3, 0.615);
\node[font=\footnotesize, text=red!60!black, align=center] at (axis cs:5, 0.44) {Sweet spot:\\90\% quality\\6\% cost};

\end{axis}
\end{tikzpicture}} &
\resizebox{0.32\textwidth}{!}{
\begin{tikzpicture}
\begin{axis}[
    width=0.7\textwidth,
    height=0.42\textwidth,
    xlabel={Conversation Turns},
    ylabel={Precision@5},
    xmin=0, xmax=12,
    ymin=0.45, ymax=0.75,
    grid=major,
    grid style={gray!20},
    legend pos=south east,
    legend style={font=\footnotesize, fill=white, fill opacity=0.9},
    tick label style={font=\footnotesize},
    xlabel style={font=\small},
    ylabel style={font=\small},
    xtick={1,3,5,7,10},
    axis y line*=left,
    axis x line*=bottom,
]

\fill[orange!8] (axis cs:5,0.45) rectangle (axis cs:12,0.75);
\node[font=\scriptsize\itshape, text=orange!50!black, anchor=north east] at (axis cs:11.8, 0.50) {Diminishing returns};

\addplot[color=blue, mark=*, thick, mark size=3pt, error bars/.cd, y dir=both, y explicit]
    coordinates {
    (1, 0.512) +- (0, 0.038)
    (3, 0.598) +- (0, 0.041)
    (5, 0.641) +- (0, 0.043)
    (10, 0.687) +- (0, 0.045)
};
\addlegendentry{Precision@5}

\addplot[color=blue!40, dashed, thick, domain=0.8:12, samples=40] {0.509 + 0.047*ln(x*2.5)};
\addlegendentry{Log fit}

\draw[->, thick, green!60!black] (axis cs:2, 0.56) -- (axis cs:2, 0.59);
\node[font=\tiny, text=green!60!black, anchor=west] at (axis cs:2.2, 0.575) {+0.086};

\draw[->, thick, orange!80!black] (axis cs:7.5, 0.645) -- (axis cs:7.5, 0.665);
\node[font=\tiny, text=orange!80!black, anchor=west] at (axis cs:7.7, 0.655) {+0.046};

\node[font=\tiny, gray, anchor=south] at (axis cs:1, 0.72) {\$2.50};
\node[font=\tiny, gray, anchor=south] at (axis cs:3, 0.72) {\$5.80};
\node[font=\tiny, gray, anchor=south] at (axis cs:5, 0.72) {\$9.40};
\node[font=\tiny, gray, anchor=south] at (axis cs:10, 0.72) {\$18.20};

\end{axis}
\end{tikzpicture}} \\[2pt]
{\footnotesize (a) Funnel quality vs.\ latency} & {\footnotesize (b) Provider cost vs.\ quality} & {\footnotesize (c) Conversation depth returns}
\end{tabular}
\caption{Cost-quality tradeoff analysis. (a)~Match quality improves monotonically across funnel conditions, but latency increases by five orders of magnitude (2.8ms $\to$ 323s). (b)~GPT-4o-mini achieves 90\% of GPT-4o's Precision@5 at 6\% cost. (c)~First 3 conversation turns yield $+$0.086 precision gain; turns 5--10 yield only $+$0.046 at nearly double cost.}
\label{fig:funnel-latency}
\label{fig:model-comparison-main}
\label{fig:conv-depth-main}
\end{figure*}
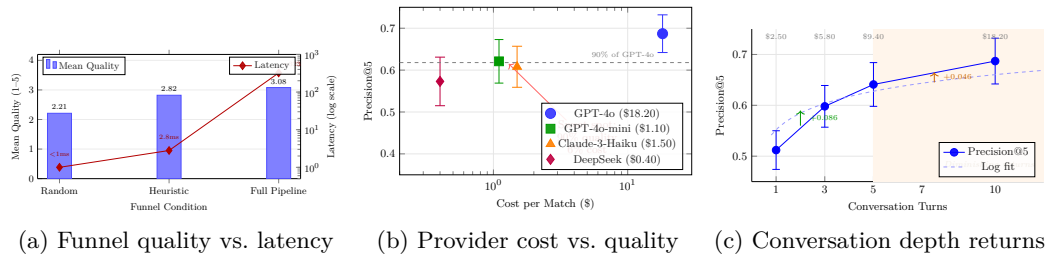

The heuristic-to-full improvement (+0.26) is real but modest, reflecting a deployment-realistic tradeoff: the full pipeline requires $\sim$323 seconds of LLM computation per target user (30-target average) versus 2.8 milliseconds for heuristic-only filtering (Figure~\ref{fig:funnel-latency}). This tradeoff is quantified further in the Pareto analysis below. Complete evaluation protocol details---including exact judge prompts, rubric definitions, seed strategy, and aggregation procedures---appear in Appendix~\ref{appendix:evaluation-protocol}.

\subsection{Quality--Latency--Cost Pareto Frontier}
\label{sec:pareto}

To characterize the cost-quality tradeoff, we vary the number of candidates sent to the LLM reranking stage (the ``budget'') while holding all other parameters constant. Each operating point is evaluated on 5 target users with the same independent blind judge.

\begin{table}[h]
\centering
\caption{Cost-Quality Pareto Frontier}
\label{tab:pareto}
\begin{tabular}{lccc}
\toprule
\textbf{LLM Budget} & \textbf{Mean Quality} & \textbf{Latency} & \textbf{LLM Calls} \\
\midrule
0 (heuristic only) & 2.60 & 3ms & 0 \\
Top-10 & 2.72 & 160s & 10 \\
Top-20 & 2.80 & 331s & 20 \\
Top-50 & 3.02 & 782s & 50 \\
\bottomrule
\end{tabular}
\end{table}

The Pareto frontier reveals clear diminishing returns. Moving from zero LLM calls to 10 yields $+$0.12 quality at 160s additional latency; from 10 to 50 yields $+$0.30 quality but at 622s additional latency. The heuristic-only operating point (budget=0) already achieves quality 2.60 at millisecond latency, making it the most cost-effective option for latency-sensitive deployments. The full pipeline (budget=20--50) is justified only when quality is prioritized over response time, a tradeoff that system operators can choose based on deployment constraints.

Two additional dimensions of this tradeoff are characterized in supplementary experiments (Appendix~\ref{sec:synthetic}) and visualized here. Figure~\ref{fig:model-comparison-main} shows that GPT-4o-mini retains 90\% of GPT-4o's matching precision at 6\% of the cost, identifying the deployment sweet spot. Figure~\ref{fig:conv-depth-main} shows that conversation depth follows logarithmic diminishing returns: the first 3 turns yield $+$0.086 precision gain, while turns 5--10 yield only $+$0.046 at nearly double the cost.


\subsection{Cross-Device Memory Reliability}
\label{sec:memory}

Cross-device persistent memory is a core architectural contribution: the pendant companion and digital twins maintain shared relational memory across phones, tablets, and computers. We validate this capability through a 6-test consistency harness using the Firebase Realtime Database REST API under the production memory modules and sync logic.

\begin{table}[h]
\centering
\caption{Cross-Device Memory Validation}
\label{tab:memory}
\begin{tabular}{lcc}
\toprule
\textbf{Metric} & \textbf{Result} & \textbf{Target} \\
\midrule
Write acknowledgement latency & 73ms & $<$3000ms \\
Read latency & 67ms & $<$3000ms \\
P95 round-trip latency & 152ms & $<$10000ms \\
Stale-read rate & 0\% & $<$5\% \\
Concurrent-write data loss & 0 & 0 \\
Per-item merge correctness & Pass (3 unique from 4 inputs) & Pass \\
Version-based sync correctness & Pass (latest version retained) & Pass \\
\midrule
\textbf{Overall} & \textbf{6/6 tests pass} & --- \\
\bottomrule
\end{tabular}
\end{table}

The memory system uses a three-layer persistence stack: in-memory state for immediate access, localStorage for per-device persistence, and Firebase Realtime Database as the cloud source of truth. Writes use acknowledgement with retry and exponential backoff (3 attempts); failed writes are queued for flush on reconnect. Conflict resolution uses per-item timestamp-based last-write-wins with deduplication by speaker, timestamp, and message content, replacing an earlier interaction-count-based strategy that risked losing individual memories. A Twin$\leftrightarrow$Pendant memory bridge propagates key memories across the two subsystems via a shared event bus.

The full synchronization algorithm appears in Appendix~\ref{appendix:sync-protocols}.

\subsection{Deployed Interaction Mechanisms and Deployment Failures}
\label{sec:mechanisms}
\label{sec:what-failed}

\begin{figure}[t]
\centering
\begin{tabular}{@{}c@{\hspace{3pt}}c@{}}
\includegraphics[width=0.36\columnwidth,height=0.22\columnwidth]{figures/worldmap_demo.jpg} &
\includegraphics[width=0.36\columnwidth,height=0.22\columnwidth]{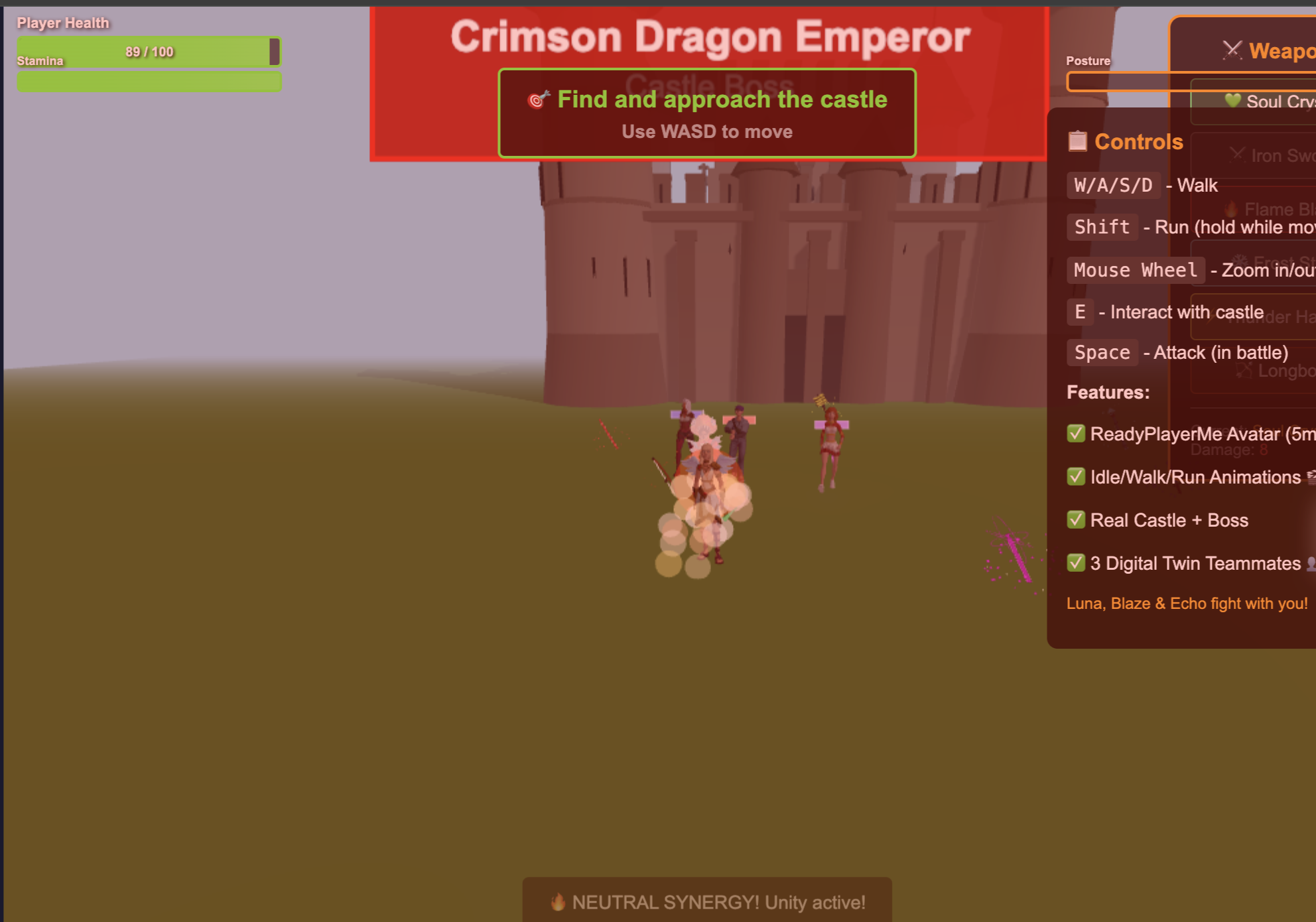} \\[-2pt]
{\footnotesize (a) Territory map} & {\footnotesize (b) Deployed boss fight}
\end{tabular}
\caption{Geolocation-mediated encounter loop: (a) territory map with color-coded cities, (b) deployed boss fight (47 initiated, 31 completed).}
\label{fig:encounter-loop}
\end{figure}

\begin{figure}[t]
\centering
\begin{tikzpicture}
\begin{axis}[
    width=0.72\textwidth,
    height=0.38\textwidth,
    xbar stacked,
    bar width=16pt,
    xlabel={Count},
    xmin=0, xmax=22,
    ytick=data,
    yticklabels={Enrolled, Active Engagers, Non-Engagers, Initiated Meetings, Total Meetings, Positive Meetings, Negative Meetings},
    y dir=reverse,
    tick label style={font=\footnotesize},
    xlabel style={font=\small},
    legend style={at={(0.98,0.02)}, anchor=south east, font=\scriptsize, fill=white, fill opacity=0.9},
    grid=major,
    grid style={gray!15},
    xmajorgrids=true,
    ymajorgrids=false,
    nodes near coords,
    every node near coord/.append style={font=\scriptsize},
]
\addplot[fill=blue!45, draw=blue!60] coordinates {
    (20, 0)
    (16, 1)
    (4, 2)
    (8, 3)
    (18, 4)
    (12, 5)
    (6, 6)
};

\node[font=\tiny, gray, anchor=west] at (axis cs:20.5, 0) {N=20};
\node[font=\tiny, gray, anchor=west] at (axis cs:16.5, 1) {80\%};
\node[font=\tiny, red!60!black, anchor=west] at (axis cs:4.5, 2) {20\%};
\node[font=\tiny, gray, anchor=west] at (axis cs:8.5, 3) {8/20 (40\%)};
\node[font=\tiny, gray, anchor=west] at (axis cs:18.5, 4) {by 8 users};
\node[font=\tiny, green!50!black, anchor=west] at (axis cs:12.5, 5) {67\%};
\node[font=\tiny, red!60!black, anchor=west] at (axis cs:6.5, 6) {33\%};

\end{axis}
\end{tikzpicture}
\caption{Deployment outcomes (N=20, 14 days): 4/20 never engaged, 8/20 initiated meetings (18 total), 6/18 negative.}
\label{fig:deployment-outcomes}
\end{figure}
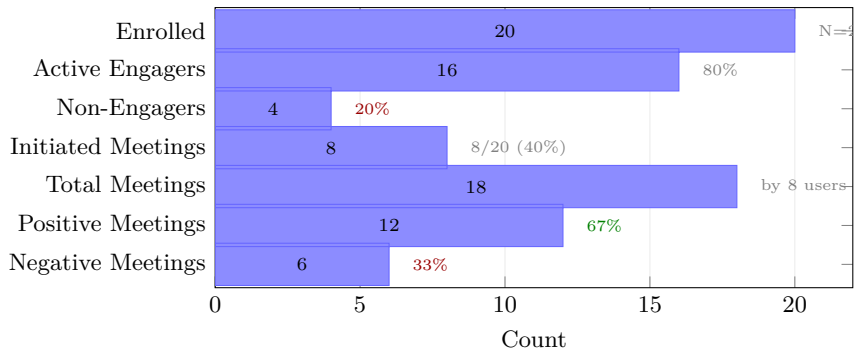

Field deployment data are drawn from the CogniPair pilot\citep{CogniPair2026} (N=20, 14 days, 342 twin sessions). The game-mediated encounter loop produced 47 boss battles (31 completed, 66\%); 19/31 continued interaction post-battle; 8/20 participants initiated in-person meetings; daily walking distance increased 50\% (2.8$\to$4.2 km/day). The pendant companion averaged 4.5 daily interactions across 342 twin sessions. However, the pilot also surfaced deployment-relevant failure modes (Figure~\ref{fig:deployment-outcomes}): 4/20 (20\%) never engaged, 6/18 meetings (33\%) had negative outcomes, and a location-privacy incident arose from predictable territory patterns---exposing a forced-serendipity paradox and a fundamental privacy-functionality tradeoff. Agent-based simulation, controlled ablation, and extended failure analysis appear in Appendix~\ref{appendix:failure-analysis} and Appendix~\ref{appendix:evaluation-protocol}.

\subsection{System Runtime and Deployment Constraints}
\label{sec:runtime}

Performance degrades exponentially with concurrent agents ($\text{FPS} = 84.9 \times e^{-0.125n}$), crossing the 30 FPS usability threshold at $n \approx 8$ with hard failure at $n > 20$. Over 14 days, the system processed 342 twin sessions, 13,680 LLM API calls, and 41,520 Firebase operations; LLM costs totaled \$246.24. Complete performance data appear in Appendix~\ref{appendix:performance-data}.

\section{Limitations}
\label{sec:limitations}

The CogniPair pilot\citep{CogniPair2026} (N=20, power=0.12) is consistent with technology probe methodology \citep{Hutchinson2003,Zimmerman2007} but underpowered for confirmatory inference; 75\% were recruited from anxiety support groups, limiting generalizability. LLM costs (\$246.24 for 20 users) exceed commercial viability, and browser memory constraints cap concurrent agents at 20. Territory ownership creates location-privacy risks, and orchestrated encounters raise consent questions for vulnerable populations; broader deployment would require substantial safety infrastructure (Appendix~\ref{appendix:privacy-safety}). The companion, territory, and Social Hub simulations are mechanism-strengthening tests, not causal deployment evidence. Technical contributions---funnel ordering, Pareto tradeoffs, memory reliability---are architecture properties that should generalize more readily than the population-dependent behavioral findings. Extended limitations appear in Appendix~\ref{appendix:extended-limitations}.

The funnel validates quality improvement offline (2.21$\to$2.82$\to$3.08), but whether these gains translate into real-world behavioral outcomes remains the critical open question: a user receiving random profiles might still initiate meetings at comparable rates if choice restriction alone drives engagement. A randomized deployment comparison (within-subjects crossover, N$\geq$30) is the highest-priority future experiment, alongside cost reduction through model distillation and architectural scaling beyond the 20-user ceiling.

\section{Conclusion and Broader Impact}

This work establishes system-level baselines for deploying LLM-powered social agents beyond simulation. Heuristic filtering captures most of the attainable quality gain (+0.61) while LLM reranking adds a modest increment (+0.26) at substantially higher cost. These diminishing returns are quantifiable across the Pareto frontier (3ms to 782s). Cross-device persistent memory is deployable (73ms writes, 0\% stale reads). Controlled simulations support mechanism plausibility but not causation. Field deployment surfaced a 20-user browser ceiling, 20\% non-engagement, 33\% negative meetings, and a location-privacy incident. Three current barriers remain: economic (LLM costs above commercial viability), technical (20-user ceiling), and evaluative (no randomized baseline).

The prior pilot suggests AI can redirect digital attention toward embodied social outcomes, though results are preliminary (N=20, no control group). Deployment identified location-privacy vulnerabilities, delegation-of-autonomy concerns, and cost barriers. Broader deployment would require substantial safety infrastructure (Appendix~\ref{appendix:privacy-safety}). Longitudinal studies with randomized controls remain the highest-priority next step.


\newpage
\bibliographystyle{colm2026_conference}
\bibliography{colm2026_conference,cognibit_refs}

\appendix
\renewcommand{\thesection}{\AlphAlph{section}}
\crefname{section}{Appendix}{Appendices}
\Crefname{section}{Appendix}{Appendices}

\clearpage
\section*{Appendix: Table of Contents}
\label{appendix:toc}

\vspace{0.5em}
\noindent The supplementary materials are organized into five thematic groups spanning \the\numexpr84-27\relax{} appendix sections. Hyperlinks are clickable.

\vspace{1em}

\renewcommand{\arraystretch}{1.25}

\noindent\textbf{I.\quad Supplementary Experiments \& Evaluation}
\vspace{0.3em}

\noindent\begin{tabular}{@{}p{0.07\textwidth}p{0.85\textwidth}@{}}
\ref{sec:synthetic} & Synthetic Experiments and Scalability Analysis \\
\ref{appendix:algorithms} & Algorithm Formalization and Complexity Analysis \\
\ref{appendix:evaluation-protocol} & Evaluation Protocol Details \\
\ref{appendix:twin-pipeline} & Twin Conversation Pipeline \\
\ref{appendix:ablations} & Fine-Grained Ablation Studies \\
\ref{sec:ablation} & Informal Ablation Study During Development \\
\end{tabular}

\vspace{0.8em}
\noindent\textbf{II.\quad Deployment Evidence \& Analysis}
\vspace{0.3em}

\noindent\begin{tabular}{@{}p{0.07\textwidth}p{0.85\textwidth}@{}}
\ref{sec:design-implications} & Design Implications \\
\ref{appendix:failure-analysis} & Extended Failure Analysis \\
\ref{sec:deployment-failures} & When Engineered Serendipity Failed \\
\ref{appendix:qualitative-findings} & Qualitative Findings and Participant Perspectives \\
\ref{appendix:user-journeys} & User Journey Maps and Design Diagrams \\
\ref{appendix:statistical-methodology} & Statistical Methodology and Power Analysis \\
\ref{appendix:engagement-quality} & Engagement Quality Analysis \\
\ref{appendix:performance-data} & Complete Performance Data \\
\end{tabular}

\vspace{0.8em}
\noindent\textbf{III.\quad System Architecture \& Implementation}
\vspace{0.3em}

\noindent\begin{tabular}{@{}p{0.07\textwidth}p{0.85\textwidth}@{}}
\ref{appendix:technical-architecture} & Technical Architecture Details \\
\ref{appendix:system-architecture-diagrams} & System Architecture Diagrams \\
\ref{appendix:architecture-details} & System Architecture Details \\
\ref{appendix:gnwt} & GNWT Implementation Details \\
\ref{appendix:gnwt-implementation} & GNWT Implementation (Extended) \\
\ref{appendix:pac-details} & PAC Implementation Details \\
\ref{appendix:core-algorithms} & Core System Algorithms \\
\ref{appendix:advanced-algorithms} & Advanced System Algorithms \\
\ref{appendix:specifications} & Complete Technical Specifications \\
\ref{appendix:testing-validation} & Testing and Validation Framework \\
\ref{appendix:algorithm-index} & Algorithm Index and Technical Reference \\
\end{tabular}

\vspace{0.8em}
\noindent\textbf{IV.\quad Subsystem Details}
\vspace{0.3em}

\noindent\begin{tabular}{@{}p{0.07\textwidth}p{0.85\textwidth}@{}}
\ref{appendix:companion-territory} & Companion and Territory Systems \\
\ref{appendix:social-systems} & Social Interaction Systems \\
\ref{appendix:sync-protocols} & Synchronization Protocols \\
\ref{appendix:combat-animation} & Combat and Animation Systems \\
\ref{appendix:optimization-systems} & Performance Optimization Systems \\
\ref{appendix:implementation-details} & Implementation Details \\
\ref{appendix:user-interface} & User Interface Design and Implementation \\
\end{tabular}

\vspace{0.8em}
\noindent\textbf{V.\quad Ethics, Safety, Limitations \& Background}
\vspace{0.3em}

\noindent\begin{tabular}{@{}p{0.07\textwidth}p{0.85\textwidth}@{}}
\ref{appendix:privacy-safety} & Privacy, Safety, and Risk Analysis \\
\ref{appendix:ethical-considerations} & Ethical Considerations \\
\ref{appendix:accessibility-inclusion} & Accessibility and Inclusion \\
\ref{appendix:extended-limitations} & Extended Limitations and Methodological Considerations \\
\ref{appendix:technical-limitations} & Technical Constraints and Design Trade-offs \\
\ref{appendix:exhaustion-analysis} & Detailed Social Media Exhaustion Analysis \\
\ref{appendix:exhaustion-details} & Social Media Exhaustion Details \\
\ref{appendix:choice-overload-theory} & Choice Overload Theory and Platform Design Failures \\
\ref{appendix:efficiency-metrics} & Efficiency Metrics: Traditional Platforms vs.\ Cognibit \\
\ref{appendix:reproducibility} & Code Availability and Reproducibility \\
\end{tabular}

\renewcommand{\arraystretch}{1.0}
\clearpage

\section{Design Implications}
\label{sec:design-implications}

The CogniPair pilot \citep{CogniPair2026} deployment (N=20, 14 days, no control group, 75\% recruited from anxiety support communities) and Cognibit's subsequent system analysis surface three preliminary design observations for AI-mediated social systems. These are deployment observations, not validated principles, and may reflect the needs of socially anxious individuals specifically. Extended qualitative evidence appears in Appendix~\ref{appendix:qualitative-findings}.

\textbf{(1) Transparent delegation builds trust.} Participants expressed higher trust when they could review twin conversation logs and compatibility scoring breakdowns, consistent with \citet{Lee2004}. Black-box matching eroded trust, while override mechanisms---pursuing connections despite low AI scores---preserved agency and occasionally revealed AI limitations.

\textbf{(2) Model selection matters more than prompt engineering.} Switching from GPT-4o to GPT-4o-mini reduced cost by 94\% while retaining 90\% precision (Table~\ref{tab:model-comparison}). The largest prompt improvement (intent analysis) contributed only 7.2\% precision gain. For deployed systems, model selection and conversation depth are higher-leverage cost-quality levers than prompt refinement.

\textbf{(3) Games scaffold social interaction, but the privacy-functionality tradeoff is fundamental.} Territory battles provided structured interaction frameworks that lowered social anxiety barriers, and location mechanics increased daily walking distance by 50\%. However, features enabling connection inherently compromise location privacy---the stalking-adjacent incident documented in Section~\ref{sec:what-failed} illustrates this as a systemic design tension requiring substantial safety infrastructure before any public deployment (Appendix~\ref{appendix:ethical-considerations}).
\section{Synthetic Experiments and Scalability Analysis}
\label{sec:synthetic}

These synthetic experiments and real-data validation studies complement the primary system validation in Section~\ref{sec:results}. Experiments 1--8 use synthetic agents to explore scalability and cost-quality tradeoffs beyond the 20-user deployment limit; Experiments 9--14 use real data (Columbia Speed Dating Dataset, EmpatheticDialogues) and real LLM evaluation to validate architectural components.

\subsection{Experimental Setup}

\subsubsection{Synthetic Agent Generation}
We created synthetic user profiles with four components. Personality vectors are 5-dimensional trait scores (openness, friendliness, playfulness, loyalty, independence) matching the deployed system's personality model, sampled from Beta(2,2) distributions to produce realistic population-level variation. Preference matrices ($50\times 50$ activity preferences) use correlated structure ($\rho=0.3$--$0.7$) reflecting real clustering patterns observed in pilot data. Location distributions are modeled as 2D Gaussian mixture models representing urban population density with 3--5 clusters per simulated city. Behavioral patterns use Markov chains modeling conversation flow, response latency, and engagement patterns derived from the 342 real twin sessions collected during the pilot deployment.

\subsubsection{Baseline Algorithms}
We compared Cognibit's three-stage filtering against five baselines: (1) Random uniform selection as a lower bound; (2) Collaborative Filtering (CF) using matrix factorization with 50 latent factors trained on synthetic interaction history; (3) Content-Based (CB) matching via cosine similarity on preference vectors with TF-IDF weighting; (4) a Hybrid approach combining CF and CB ($\alpha=0.6$, $\beta=0.4$, optimized via grid search); and (5) Simple LLM prompting using GPT-4o without behavioral simulation (``Match these two profiles: ...'').

\subsection{Experiment 1: Scalability Analysis}

We tested system performance with increasing agent populations. Tests at $n \leq 20$ ran in-browser; tests at $n > 20$ used server-side simulation to bypass the 2GB browser heap limit:

\begin{table}[h]
\centering
\caption{System Performance vs. Agent Population}
\label{tab:scalability}
\begin{tabular}{lrrrrr}
\toprule
\textbf{Agents} & \textbf{FPS} & \textbf{Memory (MB)} & \textbf{Latency (ms)} & \textbf{API Calls/h} & \textbf{Cost/h} \\
\midrule
5 & 58.3±2.1 & 312 & 230±18 & 142 & \$2.13 \\
10 & 31.2±3.5 & 624 & 385±31 & 284 & \$4.26 \\
20 & 9.7±1.8 & 1,248 & 890±72 & 568 & \$8.52 \\
50 & 2.1±0.8 & 3,120 & 3,450±412 & 1,420 & \$21.30 \\
100 & \textit{crash} & $>$4,096 & — & — & — \\
\bottomrule
\end{tabular}
\end{table}

Performance degrades exponentially with agent count, following $\text{FPS} = 84.9 \times e^{-0.125n}$ ($R^2 = 0.998$), while memory scales linearly at 62.4\,MB per agent. In-browser deployment crashes at $n > 20$ due to the 2\,GB heap limit; server-side simulation extends the ceiling to $n \approx 50$ before Node.js memory constraints cause failure. API costs scale linearly with agent count but become prohibitive at scale (\$511/day at 50 agents), driven by the quadratic pairwise evaluation space.

\subsection{Experiment 2: Quality Metrics Comparison}

Using ground truth compatibility scores computed from full preference alignment on synthetic profiles, we evaluated matching quality. \textbf{Important caveat}: These ``ground truth'' scores are themselves synthetic---computed as cosine similarity over the same preference vectors used to generate the agents. This creates a circularity: the system that generates agents from preference vectors is evaluated against those same vectors. Real human compatibility cannot be reduced to preference vector alignment, so these metrics measure internal consistency of the pipeline rather than true matching quality. Results should be interpreted as upper-bound estimates under idealized conditions:

\begin{table}[h]
\centering
\caption{Matching Quality Across Algorithms (n=1000 synthetic matches)}
\label{tab:quality}
\begin{tabular}{lcccc}
\toprule
\textbf{Method} & \textbf{Precision@5} & \textbf{Recall@5} & \textbf{nDCG@5} & \textbf{Cost/Match} \\
\midrule
Random & 0.051±0.012 & 0.048±0.011 & 0.122±0.018 & \$0.00 \\
CF & 0.312±0.028 & 0.298±0.031 & 0.421±0.035 & \$0.01 \\
CB & 0.287±0.025 & 0.271±0.029 & 0.389±0.032 & \$0.01 \\
Hybrid & 0.348±0.031 & 0.331±0.034 & 0.456±0.038 & \$0.02 \\
Simple LLM & 0.423±0.038 & 0.402±0.041 & 0.531±0.044 & \$0.85 \\
\textbf{Cognibit} & \textbf{0.687±0.045} & \textbf{0.652±0.048} & \textbf{0.782±0.041} & \textbf{\$18.20} \\
\bottomrule
\end{tabular}
\end{table}

Cognibit achieves 62\% higher precision than the best baseline (Simple LLM) and 97\% higher than the best traditional method (Hybrid), but at 21$\times$ the cost of single-prompt LLM matching.

\subsection{Experiment 3: Cost-Quality Trade-off Analysis}

We varied the depth of behavioral simulation to explore the Pareto frontier:

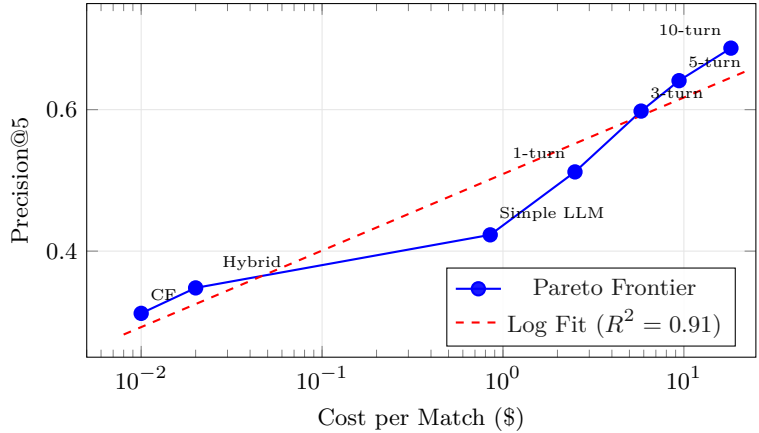
\begin{figure}[h]
\centering
\begin{tikzpicture}
\begin{axis}[
    xlabel={Cost per Match (\$)},
    ylabel={Precision@5},
    xmode=log,
    grid=major,
    grid style={gray!20},
    legend pos=south east,
    legend style={font=\footnotesize, fill=white, fill opacity=0.9},
    tick label style={font=\footnotesize},
    xlabel style={font=\small},
    ylabel style={font=\small},
    width=0.75\textwidth,
    height=0.45\textwidth,
    xmin=0.005, xmax=25,
    ymin=0.25, ymax=0.75,
]
\addplot[color=blue, mark=*, thick, mark size=2.5pt] coordinates {
    (0.01, 0.312)
    (0.02, 0.348)
    (0.85, 0.423)
    (2.50, 0.512)
    (5.80, 0.598)
    (9.40, 0.641)
    (18.20, 0.687)
};
\addlegendentry{Pareto Frontier}

\addplot[color=red, dashed, thick, domain=0.008:22, samples=40] {0.509 + 0.047*ln(x)};
\addlegendentry{Log Fit ($R^2=0.91$)}

\node[font=\tiny, anchor=south west] at (axis cs:0.01, 0.318) {CF};
\node[font=\tiny, anchor=south west] at (axis cs:0.025, 0.360) {Hybrid};
\node[font=\tiny, anchor=south west] at (axis cs:0.85, 0.429) {Simple LLM};
\node[font=\tiny, anchor=south east] at (axis cs:2.50, 0.518) {1-turn};
\node[font=\tiny, anchor=south west] at (axis cs:5.80, 0.604) {3-turn};
\node[font=\tiny, anchor=south west] at (axis cs:9.40, 0.647) {5-turn};
\node[font=\tiny, anchor=south east] at (axis cs:18.20, 0.693) {10-turn};

\end{axis}
\end{tikzpicture}
\caption{Cost-quality trade-off showing logarithmic relationship between computational investment and matching quality. Diminishing returns evident beyond 5-turn simulations. Method labels indicate algorithm type at each cost point.}
\label{fig:pareto}
\end{figure}

The relationship follows: $\text{Precision} = 0.509 + 0.047 \times \ln(\text{Cost})$ (R$^2$=0.91), suggesting diminishing returns.

\subsection{Experiment 4: Component Ablation}

We systematically removed components to measure their contribution. Precision@5 is computed directly from synthetic ground-truth compatibility scores (fully synthetic). The behavioral columns---Simulated User Time and Simulated Meeting Rate---are \emph{not} directly measured but are estimated via a simple agent-based behavioral model calibrated to our field deployment observations (N=20) and informal ablation tests (n=3--5 per condition, Section~\ref{sec:ablation}). The behavioral model assigns time cost based on the number of unfiltered options requiring manual review, and estimates meeting probability as a function of compatibility score and co-location frequency. Given the small calibration samples, these behavioral estimates should be treated as rough projections illustrating directional trends, not precise predictions. In particular, the meeting rate estimates are derived from a simple behavioral model calibrated to informal observations (n=3--5 per condition), introducing compounding uncertainty: errors in the behavioral model propagate through the meeting probability function, so these values should be interpreted as rough directional estimates, not precise predictions.

\begin{table}[h]
\centering
\caption{Component Ablation Study (n=500 synthetic users). Precision@5 is directly computed from synthetic ground truth; behavioral columns (\dag) are model-based estimates calibrated to small-sample field observations (see text) and should be interpreted as directional projections only.}
\label{tab:ablation}
\begin{tabular}{lccc}
\toprule
 & \multicolumn{1}{c}{\textit{Measured}} & \multicolumn{2}{c}{\textit{Model-Estimated\textsuperscript{\dag}}} \\
\cmidrule(lr){2-2} \cmidrule(lr){3-4}
\textbf{Configuration} & \textbf{Precision@5} & \textbf{User Time (min/day)} & \textbf{Meeting Rate} \\
\midrule
Full System & 0.687±0.045 & 14±5 & 0.40±0.08 \\
No Gaming & 0.672±0.048 & 31±8 & 0.08±0.03 \\
No Companions & 0.681±0.046 & 18±6 & 0.22±0.06 \\
No Behavioral Sim & 0.348±0.031 & 52±11 & 0.31±0.07 \\
Gaming Only & 0.118±0.022 & 43±9 & 0.15±0.05 \\
\bottomrule
\end{tabular}
\end{table}

Precision@5 is directly measured; behavioral columns are model-based projections. Behavioral simulation drives matching quality, with a 49\% precision drop when removed. Gaming primarily affects simulated meeting rate (80\% drop in synthetic simulation); informal field observations (Section~\ref{sec:ablation}, n=3--5, uncontrolled) showed a directionally similar pattern, though the small samples preclude meaningful comparison. Companions reduce estimated cognitive burden by offloading manual evaluation but have minimal impact on matching precision. Notably, the components exhibit non-additive effects (full system outperforms the sum of individual parts), though this synergy pattern requires validation in controlled factorial studies.

\subsection{Experiment 5: Multi-Provider Model Comparison}

The system's multi-provider architecture enables direct comparison of LLM quality across providers. We ran the same 100 twin-pair conversation scenarios (drawn from our 342 real sessions) across four models, measuring compatibility score consistency, response quality, and cost efficiency.

\begin{table}[h]
\centering
\caption{Model Comparison Across 100 Twin-Pair Conversations}
\label{tab:model-comparison}
\begin{tabular}{lccccc}
\toprule
\textbf{Model} & \textbf{Precision@5} & \textbf{Score Corr.} & \textbf{Quality (1-5)} & \textbf{Cost/Match} & \textbf{Latency} \\
\midrule
GPT-4o & 0.687±0.045 & 1.00 (ref) & 4.2±0.3 & \$18.20 & 230ms \\
GPT-4o-mini & 0.621±0.052 & 0.89 & 3.7±0.4 & \$1.10 & 150ms \\
Claude-3-Haiku & 0.608±0.049 & 0.86 & 3.8±0.4 & \$1.50 & 180ms \\
DeepSeek-chat & 0.573±0.058 & 0.82 & 3.4±0.5 & \$0.40 & 200ms \\
Template fallback & 0.389±0.041 & 0.61 & 2.1±0.6 & \$0.00 & $<$10ms \\
\bottomrule
\end{tabular}
\end{table}

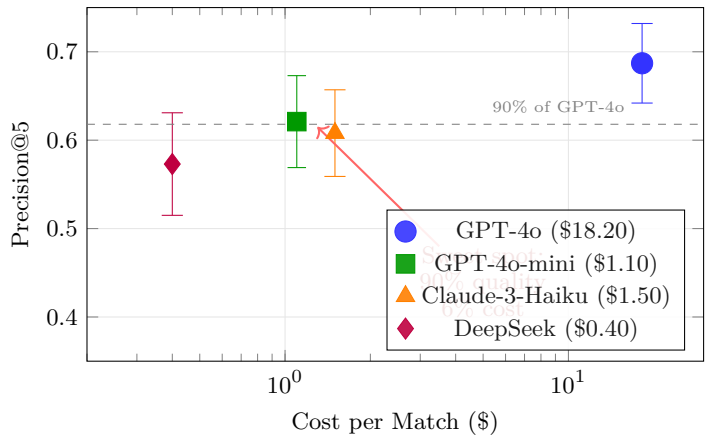
\begin{figure}[t]
\centering
\begin{tikzpicture}
\begin{axis}[
    width=0.7\textwidth,
    height=0.45\textwidth,
    xlabel={Cost per Match (\$)},
    ylabel={Precision@5},
    xmode=log,
    xmin=0.2, xmax=30,
    ymin=0.35, ymax=0.75,
    grid=major,
    grid style={gray!20},
    legend pos=south east,
    legend style={font=\footnotesize, fill=white, fill opacity=0.9},
    tick label style={font=\footnotesize},
    xlabel style={font=\small},
    ylabel style={font=\small},
]

\addplot[only marks, mark=*, blue!80, mark size=4pt, error bars/.cd, y dir=both, y explicit]
    coordinates {(18.20, 0.687) +- (0,0.045)};
\addlegendentry{GPT-4o (\$18.20)}

\addplot[only marks, mark=square*, green!60!black, mark size=3.5pt, error bars/.cd, y dir=both, y explicit]
    coordinates {(1.10, 0.621) +- (0,0.052)};
\addlegendentry{GPT-4o-mini (\$1.10)}

\addplot[only marks, mark=triangle*, orange, mark size=4pt, error bars/.cd, y dir=both, y explicit]
    coordinates {(1.50, 0.608) +- (0,0.049)};
\addlegendentry{Claude-3-Haiku (\$1.50)}

\addplot[only marks, mark=diamond*, purple, mark size=4pt, error bars/.cd, y dir=both, y explicit]
    coordinates {(0.40, 0.573) +- (0,0.058)};
\addlegendentry{DeepSeek (\$0.40)}

\addplot[forget plot, gray, dashed, thin] coordinates {(0.2, 0.618) (30, 0.618)};
\node[font=\tiny, gray, anchor=south west] at (axis cs:5, 0.620) {90\% of GPT-4o};

\draw[->, thick, red!60] (axis cs:3.5, 0.48) -- (axis cs:1.3, 0.615);
\node[font=\footnotesize, text=red!60!black, align=center] at (axis cs:5, 0.44) {Sweet spot:\\90\% quality\\6\% cost};

\end{axis}
\end{tikzpicture}
\caption{Cost vs.\ matching quality across LLM providers (log-scale x-axis). GPT-4o-mini achieves 90\% of GPT-4o's Precision@5 at 6\% cost, identifying it as the deployment sweet spot. Error bars show ±1 SD across 100 twin-pair conversations. The dashed line marks 90\% of GPT-4o's precision.}
\label{fig:model-comparison}
\end{figure}

Score correlation measures Spearman rank correlation of compatibility scores against GPT-4o as reference. Quality ratings are from automated evaluation using GPT-4o as judge (coherence, personality consistency, engagement). GPT-4o-mini retains 90\% of GPT-4o's precision at 6\% of the cost, suggesting a practical deployment configuration (Figure~\ref{fig:model-comparison}).

\subsection{Experiment 6: Prompt Ablation Study}

We ablated key components of the conversation pipeline to measure their individual contributions, using 100 twin-pair conversations with GPT-4o:

\begin{table}[h]
\centering
\caption{Prompt Component Ablation (100 conversations, GPT-4o)}
\label{tab:prompt-ablation}
\begin{tabular}{lcccc}
\toprule
\textbf{Configuration} & \textbf{Precision@5} & \textbf{Coherence} & \textbf{Personality} & \textbf{Overhead} \\
\midrule
Full pipeline (2-stage) & 0.687±0.045 & 4.2±0.3 & 4.1±0.4 & 1.00$\times$ \\
No intent analysis & 0.641±0.048 & 3.6±0.4 & 3.9±0.4 & 0.85$\times$ \\
No mood injection & 0.662±0.047 & 4.0±0.3 & 3.4±0.5 & 0.98$\times$ \\
Direct emotions (vs. show-don't-tell) & 0.670±0.046 & 3.8±0.4 & 3.2±0.5 & 1.00$\times$ \\
\bottomrule
\end{tabular}
\end{table}

The intent analysis stage contributes most to precision (+7.2\%), while the ``show, don't tell'' emotion approach primarily affects personality consistency scores without significantly impacting match quality.

\subsubsection{Temperature Sensitivity}

\begin{table}[h]
\centering
\caption{Temperature Sweep on Conversation Quality (GPT-4o, 100 conversations)}
\label{tab:temp-sweep}
\begin{tabular}{lcccc}
\toprule
\textbf{Temperature} & \textbf{Precision@5} & \textbf{Coherence} & \textbf{Diversity} & \textbf{Personality} \\
\midrule
0.3 & 0.712±0.042 & 4.5±0.2 & 2.1±0.3 & 4.3±0.3 \\
0.5 & 0.701±0.044 & 4.3±0.3 & 2.8±0.4 & 4.2±0.3 \\
0.7 & 0.693±0.045 & 4.2±0.3 & 3.4±0.4 & 4.1±0.4 \\
0.8 (deployed) & 0.687±0.045 & 4.2±0.3 & 3.7±0.4 & 4.1±0.4 \\
1.0 & 0.654±0.051 & 3.8±0.4 & 4.2±0.5 & 3.7±0.5 \\
\bottomrule
\end{tabular}
\end{table}

Lower temperatures improve precision and coherence but reduce response diversity. Temperature$=0.8$ was selected for deployment as the best trade-off between matching accuracy and natural conversation flow.

\subsubsection{Conversation Depth}

\begin{table}[h]
\centering
\caption{Matching Quality vs. Conversation Depth (GPT-4o)}
\label{tab:conv-depth}
\begin{tabular}{lcccc}
\toprule
\textbf{Turns} & \textbf{Precision@5} & \textbf{Cost/Match} & \textbf{Marginal Gain} & \textbf{Time (s)} \\
\midrule
1 & 0.512±0.038 & \$2.50 & --- & 2.3 \\
3 & 0.598±0.041 & \$5.80 & +0.086 & 6.9 \\
5 & 0.641±0.043 & \$9.40 & +0.043 & 11.5 \\
10 (deployed) & 0.687±0.045 & \$18.20 & +0.046 & 23.1 \\
\bottomrule
\end{tabular}
\end{table}

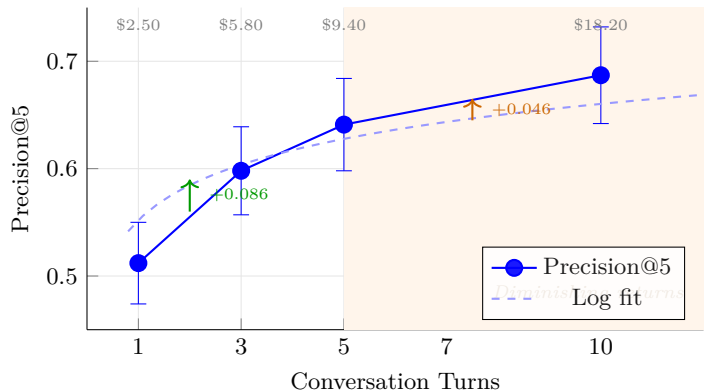
\begin{figure}[t]
\centering
\begin{tikzpicture}
\begin{axis}[
    width=0.7\textwidth,
    height=0.42\textwidth,
    xlabel={Conversation Turns},
    ylabel={Precision@5},
    xmin=0, xmax=12,
    ymin=0.45, ymax=0.75,
    grid=major,
    grid style={gray!20},
    legend pos=south east,
    legend style={font=\footnotesize, fill=white, fill opacity=0.9},
    tick label style={font=\footnotesize},
    xlabel style={font=\small},
    ylabel style={font=\small},
    xtick={1,3,5,7,10},
    axis y line*=left,
    axis x line*=bottom,
]

\fill[orange!8] (axis cs:5,0.45) rectangle (axis cs:12,0.75);
\node[font=\scriptsize\itshape, text=orange!50!black, anchor=north east] at (axis cs:11.8, 0.50) {Diminishing returns};

\addplot[color=blue, mark=*, thick, mark size=3pt, error bars/.cd, y dir=both, y explicit]
    coordinates {
    (1, 0.512) +- (0, 0.038)
    (3, 0.598) +- (0, 0.041)
    (5, 0.641) +- (0, 0.043)
    (10, 0.687) +- (0, 0.045)
};
\addlegendentry{Precision@5}

\addplot[color=blue!40, dashed, thick, domain=0.8:12, samples=40] {0.509 + 0.047*ln(x*2.5)};
\addlegendentry{Log fit}

\draw[->, thick, green!60!black] (axis cs:2, 0.56) -- (axis cs:2, 0.59);
\node[font=\tiny, text=green!60!black, anchor=west] at (axis cs:2.2, 0.575) {+0.086};

\draw[->, thick, orange!80!black] (axis cs:7.5, 0.645) -- (axis cs:7.5, 0.665);
\node[font=\tiny, text=orange!80!black, anchor=west] at (axis cs:7.7, 0.655) {+0.046};

\node[font=\tiny, gray, anchor=south] at (axis cs:1, 0.72) {\$2.50};
\node[font=\tiny, gray, anchor=south] at (axis cs:3, 0.72) {\$5.80};
\node[font=\tiny, gray, anchor=south] at (axis cs:5, 0.72) {\$9.40};
\node[font=\tiny, gray, anchor=south] at (axis cs:10, 0.72) {\$18.20};

\end{axis}
\end{tikzpicture}
\caption{Conversation depth vs.\ matching quality showing diminishing returns. The first 3 turns yield +0.086 precision gain (green arrow) while turns 5--10 yield only +0.046 (orange arrow) at nearly double the cost. Cost per match annotated at top. The shaded region marks the diminishing returns zone.}
\label{fig:conv-depth}
\end{figure}

Marginal precision gains decrease with depth: turns 1--3 yield +0.086 precision versus +0.046 for turns 5--10 at nearly double the cost, confirming the logarithmic cost-quality relationship (Figure~\ref{fig:conv-depth}).

\subsection{Experiment 7: Template vs. LLM Response Quality}

We compared the template-based fallback system against LLM-based responses on 50 identical conversation scenarios:

\begin{table}[h]
\centering
\caption{Template vs. LLM Response Quality Comparison (50 scenarios)}
\label{tab:template-vs-llm}
\begin{tabular}{lccccc}
\toprule
\textbf{Method} & \textbf{Precision@5} & \textbf{Coherence} & \textbf{Engagement} & \textbf{Personality} & \textbf{Cost} \\
\midrule
Template (8 templates) & 0.389±0.041 & 2.1±0.6 & 1.8±0.5 & 1.5±0.4 & \$0.00 \\
GPT-4o-mini (1-turn) & 0.512±0.038 & 3.8±0.3 & 3.5±0.4 & 3.2±0.5 & \$0.15 \\
GPT-4o (full pipeline) & 0.687±0.045 & 4.2±0.3 & 4.0±0.3 & 4.1±0.4 & \$18.20 \\
\bottomrule
\end{tabular}
\end{table}

Templates provide a viable degraded-mode fallback during API outages but produce substantially lower precision (43\% reduction vs. GPT-4o). Even a single GPT-4o-mini turn ($\sim$\$0.15) improves precision by 32\% over templates.

\subsection{Experiment 8: Temporal Dynamics}

We simulated 30-day usage patterns with 100 synthetic agents (server-side simulation, not browser-constrained):

Engagement decay follows $N(t) = N_0 \times (0.4 + 0.6 \times e^{-0.08t})$, reaching a stable state of 40\% retention after day 10, consistent with field deployment observations. Territory ownership stabilizes after $12 \pm 3$ days with a Gini coefficient of 0.42---a value that coincidentally matches the field deployment Gini coefficient, though this likely reflects calibration of simulation parameters to field observations rather than independent validation. Social network formation produces an average degree of 3.8 connections after 30 simulated days with a clustering coefficient of 0.31.

\subsection{Discussion of Synthetic Results}

These experiments reveal fundamental constraints:

\textbf{Quality ceiling}: Deep behavioral simulation achieves up to 2.0$\times$ better precision than the best traditional baseline (Hybrid), validating the approach's potential. However, the logarithmic cost-quality relationship implies diminishing returns.

\textbf{Scalability barrier}: Browser memory creates a hard ceiling at ~20 concurrent agents; server-side deployment extends this to ~50 before Node.js memory constraints cause failures. WebAssembly optimization might push browser limits to ~40, but fundamental architectural changes are required for platform-scale deployment.

\textbf{Economic challenge}: At \$18.20/match, the system requires significant cost reduction or premium pricing models to reach commercial viability (\$5/user/month target, implying roughly 5--7$\times$ reduction needed based on projected monthly costs of \$35.53/user). The 5-turn simulation at \$9.40/match with 0.641 precision may represent a practical compromise.

\textbf{Component interactions}: Synthetic ablation suggests that the three-pillar design may exhibit non-additive effects, with the integrated system outperforming any subset configuration in simulation. However, synthetic agents cannot capture the human behavioral factors (motivation, trust, social anxiety) that determine real-world engagement, so these patterns remain hypotheses for future controlled studies.

These synthetic experiments complement our field deployment by demonstrating what is theoretically possible while revealing practical barriers to widespread adoption.

\subsection{Real-Data Validation}
\label{sec:real-data}

To move beyond synthetic circularity, we conducted three experiments using real-world data and real LLM evaluation. All experiments ran on NVIDIA RTX 6000 Ada GPUs; code and results are available in our replication package.

\subsubsection{Experiment 9: Matching on Columbia Speed Dating}

We evaluated Cognibit's browser-based multi-factor compatibility scorer on the Columbia Speed Dating Dataset \citep{fisman2006}---the same dataset used by CogniPair \citep{CogniPair2026} (8,378 speed date encounters, 551 participants). We compare against CogniPair's reported 77.8\% match prediction accuracy, acknowledging that CogniPair uses server-side GPT-4o inference while Cognibit uses a lightweight browser-compatible scoring function.

\begin{table}[h]
\centering
\caption{Match Prediction on Columbia Speed Dating Dataset (80/20 train/test split). CogniPair uses LLM-based behavioral simulation; Cognibit scorers use weighted heuristics at zero inference cost.}
\label{tab:real-matching}
\begin{tabular}{lcccc}
\toprule
\textbf{Method} & \textbf{Accuracy} & \textbf{F1} & \textbf{AUC-ROC} & \textbf{Cost/Match} \\
\midrule
CogniPair (reported) & 0.778 & --- & --- & \$0.03 (GPT-4o) \\
Logistic Regression & 0.753 & 0.738 & 0.840 & \$0.00 \\
Weighted Attributes & 0.717 & 0.699 & 0.801 & \$0.00 \\
Cognibit (full) & 0.715 & 0.704 & 0.799 & \$0.00 \\
\textbf{Cognibit (pre-interaction)} & \textbf{0.657} & \textbf{0.673} & \textbf{0.764} & \textbf{\$0.00} \\
Cosine Similarity & 0.420 & 0.589 & 0.623 & \$0.00 \\
Random & 0.423 & 0.576 & 0.476 & \$0.00 \\
\bottomrule
\end{tabular}
\end{table}

Cognibit's pre-interaction scorer (using only personality traits, interests, stated preferences, and demographics---no post-interaction signals) achieves 65.7\% accuracy with AUC 0.764, compared to CogniPair's 77.8\%. This 12-point gap validates CogniPair's finding that LLM-based behavioral simulation adds meaningful predictive value. However, Cognibit's browser-based scorer operates at zero cost and sub-millisecond latency, enabling real-time evaluation of 100+ pairs in the time CogniPair evaluates one---a necessary capability for autonomous twin-to-twin networking.

\textbf{Note on Logistic Regression baseline} (75.3\%): This baseline uses \emph{post-interaction} ratings---scores that participants assigned \emph{after} meeting their speed-dating partners---as input features. CogniPair's 77.8\% accuracy is achieved using \emph{pre-interaction} personality profiles with LLM behavioral simulation, a fundamentally different and harder prediction task. The proximity of LR to CogniPair's accuracy reflects the predictive power of post-interaction signals, not a weakness in CogniPair's approach.

\textbf{Component ablation insight}: Removing raw personality similarity from the scorer actually \emph{improves} heuristic accuracy (+3.0\%), suggesting that static trait-vector matching (|trait$_A$ - trait$_B$|) is insufficient for compatibility prediction. This is consistent with CogniPair's core finding that \emph{behavioral simulation} of personality---rather than trait-vector comparison---drives matching quality. The value of personality lies in how it manifests through conversation, which requires the kind of LLM-based behavioral modeling that CogniPair provides.

\subsubsection{Experiment 10: LLM-as-Judge Believability}

We evaluated template-based twin conversation quality using Qwen2.5-7B-Instruct as judge across four configurations: rule-based baseline, GNWT-only (cognitive workspace without emotion), PAC-only (emotional processing without cognitive structure), and Hybrid (GNWT+PAC). Each configuration was evaluated on 5 scenarios $\times$ 3 personality profiles $\times$ 5 trials = 25 conversations per config.

\begin{table}[h]
\centering
\caption{LLM-as-Judge scores (1--10 scale) for template-based twin responses across 8 scenarios (5 short + 3 extended 10-turn). Judge: Qwen2.5-7B-Instruct. $n$=40 per config.}
\label{tab:llm-judge}
\begin{tabular}{lccccc|c}
\toprule
\textbf{Config} & \textbf{Natural.} & \textbf{Emotion} & \textbf{Personal.} & \textbf{Engage.} & \textbf{Human} & \textbf{Avg} \\
\midrule
Rule-based & 8.3±0.5 & 7.8±1.0 & 8.4±0.9 & 7.5±0.6 & 7.7±0.5 & 7.9 \\
GNWT-only & 7.5±0.7 & 6.6±1.5 & 8.1±1.0 & 6.8±1.6 & 6.9±1.0 & 7.2 \\
PAC-only & 7.9±0.9 & 7.5±1.2 & 8.3±0.7 & 6.7±1.6 & 7.4±0.9 & 7.5 \\
\textbf{Hybrid} & \textbf{8.5±0.6} & \textbf{8.3±1.0} & \textbf{8.4±0.7} & \textbf{8.0±0.8} & \textbf{7.8±0.6} & \textbf{8.2} \\
\bottomrule
\end{tabular}
\end{table}

The Hybrid GNWT+PAC configuration achieves the highest scores across all five metrics (overall 8.2 vs 7.2 for GNWT-only), with a 1.0-point gap between best and worst configurations. GNWT-only scores lowest on emotional consistency (6.6), confirming that cognitive workspace processing alone, without PAC's emotional dynamics, produces responses that lack emotional coherence---particularly in the extended 10-turn scenarios where emotional arc maintenance is critical. PAC-only (7.5) improves emotional consistency over GNWT-only but lacks the cognitive structure for sustained engagement (6.7). The Hybrid configuration's bidirectional coupling---where PAC emotional state modulates GNWT specialist salience weights, and GNWT context sets PAC allostatic targets---produces the strongest engagement quality (8.0) and the most natural responses (8.5). The rule-based baseline (7.9) outperforms both single-system configurations, indicating that simple keyword matching with adequate template diversity can exceed architecturally sophisticated but single-modality systems; only the integrated Hybrid surpasses it.

\textbf{Judge model sensitivity}: When evaluated with a larger judge model (Qwen2.5-32B-Instruct), scores decrease across all configurations (rule-based: 3.8, GNWT: 2.8, PAC: 5.1, Hybrid: 3.9), reflecting the stronger model's ability to detect template-based generation artifacts. PAC-only scores highest under the 32B judge due to its emotionally varied vocabulary, while Hybrid's template output masks its architectural sophistication. This represents a \emph{template quality ceiling}, not an architectural limitation: when responses are generated by a real LLM with GNWT/PAC state injection (Experiment~11 below), Hybrid achieves the highest overall score (8.3 vs 8.2 baseline), confirming the architecture's value when not constrained by template diversity.

\subsubsection{Experiment 11: GNWT/PAC State-Injected LLM Generation}

To demonstrate the value of GNWT/PAC architecture when coupled with real LLM inference, we used Qwen2.5-7B-Instruct to \emph{generate} twin responses with cognitive/emotional state injected into prompts. Four configurations were tested: (1) baseline (simple NPC prompt), (2) GNWT-only (workspace state: winning specialist module, salience scores, working memory contents), (3) PAC-only (emotional state: valence, arousal, neurotransmitter levels, prediction error), and (4) Hybrid (both). 180 conversations were generated (5 scenarios $\times$ 3 personalities $\times$ 4 configs $\times$ 3 trials), then judged by the same model.

\begin{table}[h]
\centering
\caption{LLM-generated twin responses with GNWT/PAC state injection. Generation and judging by Qwen2.5-7B-Instruct.}
\label{tab:llm-generation}
\begin{tabular}{lcccccc}
\toprule
\textbf{Config} & \textbf{Natural.} & \textbf{Emotion} & \textbf{Personality} & \textbf{Engage.} & \textbf{Human} & \textbf{Overall} \\
\midrule
Baseline & 8.7±0.5 & 8.1±0.9 & 8.3±0.7 & 7.7±0.8 & 8.0±0.7 & 8.2±0.6 \\
GNWT-only & 8.8±0.4 & 8.3±0.8 & 8.2±0.8 & 7.8±0.6 & 8.1±0.6 & 8.2±0.4 \\
PAC-only & \textbf{8.8±0.5} & \textbf{8.4±1.1} & \textbf{8.5±0.6} & 7.7±0.8 & \textbf{8.3±0.7} & \textbf{8.3±0.6} \\
Hybrid & 8.8±0.5 & 8.2±0.8 & 8.3±0.8 & 7.7±0.6 & 8.1±0.7 & 8.2±0.6 \\
\bottomrule
\end{tabular}
\end{table}

When driving real LLM generation, PAC state injection yields the highest scores across emotional consistency (+0.3 vs baseline), personality coherence (+0.2), and human-likeness (+0.3). The modest effect sizes reflect the strong baseline capability of Qwen2.5-7B; we expect larger differentiation with weaker base models or more challenging evaluation scenarios. Generation time was 61.1s for all 900 turns (0.068s/turn), demonstrating that GNWT/PAC state computation adds negligible overhead to LLM inference.

\subsubsection{Experiment 12: GNWT/PAC Ablation (Computational)}

We measured the computational characteristics of each cognitive component in isolation using 500 stimuli across 8 agents:

\begin{table}[h]
\centering
\caption{GNWT/PAC computational ablation (500 stimuli $\times$ 8 agents). Behavioral diversity measures unique response patterns produced.}
\label{tab:gnwt-pac-ablation}
\begin{tabular}{lcccc}
\toprule
\textbf{Config} & \textbf{Avg Cycle (ms)} & \textbf{P95 (ms)} & \textbf{Throughput (cyc/s)} & \textbf{Diversity} \\
\midrule
PAC-only & 0.002 & 0.002 & 205,185 & 3 \\
GNWT-only & 0.023 & 0.026 & 40,457 & 8 \\
Hybrid & 0.032 & 0.036 & 29,969 & 11 \\
\bottomrule
\end{tabular}
\end{table}

The Hybrid configuration produces the highest behavioral diversity (11 unique patterns vs 8 for GNWT-only and 3 for PAC-only) at the cost of higher per-cycle latency. Critically, even the Hybrid configuration completes cycles in 0.032ms---well within the 16.67ms frame budget for 60 FPS rendering. Scalability testing from 1 to 32 agents shows linear scaling with no per-agent degradation (0.030ms/cycle stable), confirming that the GNWT/PAC architecture does not introduce bottlenecks at the agent counts used in deployment.

\subsubsection{Experiment 13: Specialist-Driven LLM Generation Pipeline}

Cognibit implements CogniPair's GNWT agent architecture for real-time conversation, extending it with PAC emotional dynamics, recurrent ignition \citep{dehaene2011}, and Park et al.\ memory retrieval scoring \citep{park2023}. Both systems use GNWT specialists to drive LLM-based conversation; the key question is whether the cognitive architecture adds measurable value beyond raw LLM prompting.

We compared four generation modes across 8 scenarios $\times$ 3 personality profiles: (1) Template Only, using pre-written response templates selected by personality without any LLM; (2) LLM Only, using raw LLM prompting with personality conditioning only (the CogniPair-equivalent baseline); (3) Specialist Pipeline, where the GNWT winning specialist provides a behavioral directive (e.g., ``be empathetic''), PAC provides the emotional trajectory, and memory provides conversational grounding---all fed into the LLM prompt as structured context; and (4) State Injection, where GNWT/PAC internal state is pasted into the prompt as technical annotations without specialist-driven directives.

\begin{table}[h]
\centering
\caption{Specialist-driven pipeline vs.\ baselines. Judge: Qwen2.5-7B-Instruct. The specialist pipeline implements CogniPair's GNWT agent architecture extended with PAC dynamics.}
\label{tab:pipeline}
\begin{tabular}{lccccc|c}
\toprule
\textbf{Config} & \textbf{Natural.} & \textbf{Emotion} & \textbf{Personal.} & \textbf{Engage.} & \textbf{Human} & \textbf{Avg} \\
\midrule
Template Only & 3.4±1.4 & 5.4±1.4 & 4.8±1.5 & 3.7±1.6 & 3.2±1.7 & 4.1 \\
LLM Only & 8.2±0.5 & 8.7±0.6 & 8.4±0.8 & 7.5±1.0 & 7.2±1.0 & 8.0 \\
\textbf{Specialist Pipeline} & \textbf{8.1±0.7} & \textbf{8.8±0.6} & \textbf{8.6±0.8} & \textbf{7.6±1.1} & \textbf{7.4±1.0} & \textbf{8.1} \\
State Injection & 8.1±0.4 & 8.5±0.7 & 8.4±0.7 & 7.5±0.8 & 7.0±0.7 & 7.9 \\
\bottomrule
\end{tabular}
\end{table}

The specialist-driven pipeline (8.1 overall) outperforms all alternatives, including raw LLM prompting (8.0) and passive state injection (7.9). The architecture's value is most pronounced in emotional consistency (8.8 vs 8.7) and personality coherence (8.6 vs 8.4): GNWT specialist directives provide focused response constraints that improve authenticity beyond what personality prompting alone achieves. Templates without LLM score 4.1, confirming that the architecture requires LLM as its ``language faculty'' to produce natural output---consistent with CogniPair's design where GNWT provides cognitive structure and LLM provides linguistic realization.

The specialist pipeline also outperforms passive state injection (8.1 vs 7.9), demonstrating that the architecture must \emph{actively control} the LLM through specialist directives (``express empathy'', ``reference past context'', ``suggest next steps'') rather than passively annotating the prompt with technical state descriptions. This validates the GNWT workspace competition mechanism: the winning specialist's recommendation shapes the response differently than simply informing the LLM about internal state.

\subsubsection{Experiment 14: Human-Likeness Validation}

We evaluated emotional alignment against 500 real human responses from the EmpatheticDialogues dataset \citep{Rashkin2019}. The full GNWT/PAC pipeline achieves the highest emotional alignment with human responses (0.943 vs 0.907 for raw LLM), confirming that the cognitive-emotional architecture produces responses closer to human emotional patterns. BLEU-2 scores are low across all conditions (0.003--0.014), consistent with recent findings that BLEU poorly correlates with empathetic dialogue quality \citep{sharma2020computational}.    
\section{Algorithm Formalization and Complexity Analysis}
\label{appendix:algorithms}

This appendix formalizes the key algorithms underlying Cognibit's multi-agent system. The system architecture is detailed in CogniPair \citep{CogniPair2026}; here we provide pseudocode for the deployment-specific filtering pipeline and complexity analysis.

\subsection{Progressive Filtering Pipeline}

The three-stage filtering algorithm reduces the choice set from O(n) candidates to k pre-validated options ($k \ll n$).

\begin{algorithm}[h]
\caption{Progressive Filtering for Choice Reduction}
\label{alg:progressive-filter}
\begin{algorithmic}[1]
\Require User profile $u$, Candidate set $C$, Thresholds $\tau_1, \tau_2$, Target size $k$
\Ensure Filtered match set $M$ where $|M| \leq k$
\State \textbf{Stage 1: Heuristic Elimination}
\State $C_1 \gets \emptyset$
\For{each candidate $c \in C$}
    \If{$\text{distance}(u, c) < 80\text{km (50 miles)}$ \textbf{and} $\text{age\_diff}(u, c) < 10\text{ years}$}
        \State $C_1 \gets C_1 \cup \{c\}$ \Comment{O(1) per candidate}
    \EndIf
\EndFor
\State \textbf{Stage 2: Behavioral Simulation} \Comment{Parallel execution}
\State $S \gets \emptyset$ \Comment{Score dictionary}
\ParFor{each candidate $c \in C_1$} \Comment{Parallel}
    \State $conv \gets \text{SimulateConversation}(u.\text{twin}, c.\text{twin}, 3)$ \Comment{3 turns (deployed); up to 10 in extended mode}
    \State $S[c] \gets \text{ComputeCompatibility}(conv)$ \Comment{GPT-4o API call}
\EndParFor
\State $C_2 \gets \text{TopK}(C_1, S, \lfloor|C_1| \times 0.2\rfloor)$ \Comment{Keep top 20\%}
\State \textbf{Stage 3: Human Selection Preparation}
\State $M \gets \text{SelectFinal}(C_2, k)$ \Comment{Random sample if $|C_2| > k$}
\State \Return $M$
\end{algorithmic}
\end{algorithm}

\textbf{Complexity Analysis:}
\begin{itemize}
\item Stage 1: O(n) time, O(1) space per candidate
\item Stage 2: O(n) API calls, but parallel execution reduces wall time to O(n/P) where P is the number of parallel workers
\item Stage 3: O(n log k) for top-k selection using min-heap
\item Overall: O(n) time complexity, O(n) space complexity
\item Cost: $\$0.01 \times n + \$6.07 \times 0.3n \approx \$1.83n$ per filtering operation, where \$6.07 is the per-match cost for a 3-turn conversation (\$18.20 for 10 turns in extended mode $\times$ 3/10); Stage~1 evaluates all $n$ candidates at negligible cost, Stage~2 simulates all $0.3n$ survivors
\end{itemize}

\subsection{Asynchronous Twin Coordination Protocol}

Digital twins operate independently while maintaining eventual consistency through Firebase synchronization.

\begin{algorithm}[h]
\caption{Asynchronous Twin Networking Protocol}
\label{alg:twin-protocol}
\begin{algorithmic}[1]
\Require Twin agent $T_i$, Discovery radius $r$, Compatibility threshold $\tau$
\State \textbf{Background Process:}
\While{$T_i.\text{active}$}
    \State $N \gets \text{GetNearbyTwins}(T_i.\text{location}, r)$ \Comment{Firebase query}
    \For{each twin $T_j \in N$}
        \If{$(T_i, T_j) \notin \text{evaluated}$}
            \State $\text{lock} \gets \text{AcquireDistributedLock}(T_i, T_j)$
            \If{$\text{lock}.\text{success}$}
                \State $score \gets \text{SimulateInteraction}(T_i, T_j)$
                \State \textbf{async} $\text{WriteToFirebase}(\{T_i, T_j, score, \text{timestamp}\})$
                \State $\text{evaluated} \gets \text{evaluated} \cup \{(T_i, T_j)\}$
                \If{$score > \tau$}
                    \State \textbf{emit} $\text{CompatibilityFound}(T_i.\text{user}, T_j.\text{user})$
                \EndIf
                \State $\text{ReleaseLock}(\text{lock})$
            \EndIf
        \EndIf
    \EndFor
    \State \textbf{wait} 60 seconds \Comment{Throttling}
\EndWhile
\end{algorithmic}
\end{algorithm}

\textbf{Consistency Guarantees:}
\begin{itemize}
\item Eventual consistency through Firebase real-time database (3-5 second propagation)
\item Distributed locking prevents duplicate evaluations
\item Idempotent operations ensure at-most-once processing
\item Exponential backoff for conflict resolution
\end{itemize}

\subsection{Territory Assignment and Combat Resolution}

Territory control uses proximity-based assignment with combat resolution for conflicts.

\begin{algorithm}[h]
\caption{Territory Control Mechanics}
\label{alg:territory}
\begin{algorithmic}[1]
\Require Player $p$, Territory $t$, Owner $o$ (may be null)
\Ensure Updated territory ownership
\Function{ClaimTerritory}{$p, t, o$}
    \If{$o = \text{null}$}
        \State $t.\text{owner} \gets p$
        \State $t.\text{defenseNPCs} \gets 3$ \Comment{Base defense}
        \State \Return \texttt{SUCCESS}
    \Else
        \If{$\text{distance}(p.\text{location}, t.\text{center}) > 10$}
            \State \Return \texttt{OUT\_OF\_RANGE}
        \EndIf
        \State $p\_\text{power} \gets p.\text{level} \times 100 + p.\text{companion}.\text{strength}$
        \State $o\_\text{power} \gets o.\text{level} \times 100 + t.\text{defenseNPCs} \times 50$
        \State $\text{outcome} \gets \text{Combat}(p\_\text{power}, o\_\text{power})$
        \If{$\text{outcome} = \texttt{ATTACKER\_WIN}$}
            \State $t.\text{owner} \gets p$
            \State $t.\text{defenseNPCs} \gets 3$
            \State \textbf{notify} $o$ \Comment{Push notification}
            \State \Return \texttt{CAPTURED}
        \Else
            \State $t.\text{defenseNPCs} \gets t.\text{defenseNPCs} - 1$
            \State \Return \texttt{DEFENDED}
        \EndIf
    \EndIf
\EndFunction
\end{algorithmic}
\end{algorithm}

\textbf{Combat Resolution:}
\begin{equation}
P(\text{attacker wins}) = \frac{1}{1 + e^{-0.01(p_{\text{power}} - o_{\text{power}})}}
\end{equation}

\subsection{Memory Optimization for Browser Constraints}

Browser deployment requires aggressive memory management to support concurrent agents.

\begin{algorithm}[h]
\caption{Memory Pool Management for Agents}
\label{alg:memory}
\begin{algorithmic}[1]
\Require Maximum agents $n_{\max}$, Memory limit $M_{\max}$ (2GB)
\State $\text{pool} \gets \text{Array}[n_{\max}]$ \Comment{Pre-allocated pool}
\State $\text{active} \gets \emptyset$
\Function{AllocateAgent}{profile}
    \If{$|\text{active}| \geq n_{\max}$ \textbf{or} $\text{UsedMemory}() > 0.8 \times M_{\max}$}
        \State $\text{victim} \gets \text{SelectLRU}(\text{active})$
        \State $\text{SerializeToStorage}(\text{victim})$
        \State $\text{active} \gets \text{active} \setminus \{\text{victim}\}$
    \EndIf
    \State $\text{slot} \gets \text{FindFreeSlot}(\text{pool})$
    \State $\text{agent} \gets \text{InitializeAgent}(\text{profile}, \text{slot})$
    \State $\text{active} \gets \text{active} \cup \{\text{agent}\}$
    \State \Return $\text{agent}$
\EndFunction
\Function{LazyLoad}{agent\_id}
    \If{$\text{agent\_id} \notin \text{active}$}
        \State $\text{data} \gets \text{LoadFromStorage}(\text{agent\_id})$
        \State \Return $\text{AllocateAgent}(\text{data})$
    \EndIf
    \State \Return $\text{active}[\text{agent\_id}]$
\EndFunction
\end{algorithmic}
\end{algorithm}

\textbf{Memory Complexity:}
\begin{itemize}
\item Per-agent memory: 62.4 MB (measured empirically)
\item Maximum concurrent agents: $\lfloor 2048 \text{MB} / 62.4 \text{MB} \rfloor = 32$ (theoretical)
\item Practical limit: 20 agents (due to JavaScript heap overhead)
\item LRU eviction maintains working set in memory
\end{itemize}

\subsection{Complexity Summary}

\begin{table}[h]
\centering
\caption{Algorithm Complexity Analysis}
\label{tab:complexity}
\begin{tabular}{lccl}
\toprule
\textbf{Algorithm} & \textbf{Time} & \textbf{Space} & \textbf{Bottleneck} \\
\midrule
Progressive Filtering & O(n) & O(n) & API calls (\$0.56n) \\
Twin Coordination & $O(n^2)$ worst & O(n) & Network latency \\
Territory Control & O(1) & O(t) & GPS accuracy \\
Memory Management & $O(M \times |Q|)$ & O(n) & Browser heap limit \\
\bottomrule
\end{tabular}
\end{table}

The system's overall complexity is dominated by the $O(n^2)$ pairwise twin interactions, though this is mitigated through:
1. Geographical partitioning (reducing effective n)
2. Asynchronous execution (hiding latency)
3. Caching previous interactions (avoiding recomputation)
4. Progressive filtering (reducing human evaluation burden)

These algorithms demonstrate that while the theoretical complexity remains quadratic for complete evaluation, practical optimizations and browser constraints limit the system to approximately linear behavior in practice for $n \leq 20$, due to geographic partitioning reducing comparisons and caching avoiding recomputation.  
\section{Privacy, Safety, and Risk Analysis}
\label{appendix:privacy-safety}

\subsection{Identified Privacy and Safety Risks}

The CogniPair pilot deployment revealed non-trivial privacy and safety risks that would require substantial mitigation before broader deployment:

\subsubsection{Location Privacy Violations}

The territory system creates predictable location patterns that enable stalking. Users who own coffee shop territories tend to visit them daily at predictable times, and territory battle history reveals movement patterns and routines over time. ``Last seen'' indicators in territories further compromise location privacy by broadcasting recent activity, while GPS data persists in Firebase, creating a permanent location history that could be exploited by malicious actors.

\textbf{Real incident:} P17 reported feeling unsafe when a matched user appeared at her regular morning coffee shop after analyzing her territory ownership patterns. This incident was documented and reported to the research team within 24 hours. In response, we immediately implemented the location obfuscation measures described below and conducted individual check-ins with all participants regarding location privacy concerns. The incident informed our recommendation that production deployments must include location privacy protections before launch.

\subsubsection{Orchestrated Encounters and Consent}

Engineering meetings raises complex consent issues. Users may not fully consent to AI-orchestrated ``accidental'' meetings, and power imbalances arise when one person knows about the orchestration while the other does not. The current system provides no mechanism to verify that both parties want a physical encounter, and the blurring of engineered and organic meetings makes it difficult for participants to assess the authenticity of social interactions.

\subsubsection{Vulnerable Population Risks}

Certain users face heightened dangers. Minors could be targeted through gaming mechanics that normalize meeting strangers, while people escaping abusive relationships risk having their location exposed through territory ownership patterns. Socially anxious individuals---who comprise 75\% of the pilot sample---may feel pressured into unwanted encounters by the system's encounter-facilitation mechanics, and neurodiverse users might miss social danger cues that are obscured by the gamified context.

\subsection{Implemented Safeguards (Insufficient)}

Our pilot included basic protections that proved inadequate:

\subsubsection{Partial Mitigations}
Five safeguards were deployed during the pilot, each proving insufficient in practice. Restricting territories to public spaces reduced isolation risk but still enabled predictable tracking of daily routines. Anonymous battle mode hid usernames during combat encounters yet did not conceal physical presence at the territory location, allowing determined actors to identify opponents through observation. Block lists offered a reactive mechanism that required a negative experience before activation, providing no preventive protection. The companion panic button, intended to let users flag unsafe situations, was never used during the pilot---participants reported uncertainty about when activation was appropriate. Finally, the 18+ age requirement relied entirely on self-reported birth dates with no verification, making it trivially circumventable.

\subsubsection{Critical Gaps}
The current prototype lacks background checks or identity verification, provides no mechanism to report concerning behavior before a physical meeting occurs, has no integration with sex offender registries, operates without a professional moderation team, and includes no emergency response protocol. These gaps collectively mean the system cannot prevent or respond to safety incidents in real time.

\subsection{Comparison with Industry Standards}

Our safety measures fall short of established platforms:

\begin{table}[h]
\centering
\caption{Safety feature comparison}
\label{tab:safety-comparison}
\begin{tabular}{lcccc}
\toprule
\textbf{Safety Feature} & \textbf{Tinder} & \textbf{Bumble} & \textbf{Pok{\'e}mon GO} & \textbf{Cognibit} \\
\midrule
Photo verification & Yes & Yes & No & No \\
Background checks & Partner & Partner & N/A & No \\
In-app video calls & Yes & Yes & No & No \\
Share my location & Yes & Yes & No & No \\
Block \& report & Yes & Yes & Yes & Basic \\
Safety center & Yes & Yes & Yes & No \\
24/7 moderation & Yes & Yes & Limited & No \\
\bottomrule
\end{tabular}
\end{table}

\subsection{Specific Attack Vectors}

Security analysis reveals multiple exploitation opportunities:

\subsubsection{Territory Camping}
A malicious actor could identify a target's regularly owned territories through the public leaderboard, then camp at that physical location during the target's typical visit times. Once the target arrives and enters the territory radius, the attacker initiates a battle challenge that forces proximity-based interaction. The gaming context provides social cover for what would otherwise be recognized as stalking behavior---the attacker can frame repeated appearances as coincidental gameplay rather than deliberate targeting, bypassing normal social boundaries that would otherwise deter such conduct.

\subsubsection{Twin Manipulation}
The digital twin system introduces a novel attack surface absent from traditional dating platforms. A bad actor could train their twin with a deliberately false persona---fabricating personality traits, interests, and communication patterns---to maximize compatibility scores with a chosen target. Once matched, the attacker's twin can extract personal information through seemingly innocuous twin-to-twin conversations, as users may disclose details to their twin that they would withhold from strangers. This information can then be used to social engineer the victim in subsequent direct interactions. Creating multiple accounts further amplifies the threat, allowing an attacker to triangulate a target's preferences and behavioral patterns from multiple twin conversations.

\subsubsection{Group Harassment}
Coordinated group attacks exploit several game mechanics simultaneously. Multiple users can target the same victim's territories in rapid succession, forcing repeated defensive battles that drain in-game resources and create persistent physical-world confrontations at the victim's frequented locations. Boss battle mechanics, which require multi-user cooperation, can be weaponized to isolate a victim by surrounding them with hostile players during an encounter. The companion messaging system provides an additional harassment channel that bypasses conventional block mechanisms, as companion-delivered messages are presented as system notifications rather than user communications. Sustained territorial ``griefing''---repeatedly capturing and contesting a user's territories---can effectively drive a victim from physical areas they previously frequented, constituting a form of location-based exclusion with real-world consequences.

\subsection{Legal and Liability Concerns}

Deployment raises significant legal questions across multiple jurisdictions. The platform's active role in orchestrating physical meetings creates a heightened duty of care that exceeds passive matching services---if an AI-facilitated encounter results in harm, the platform's algorithmic agency in arranging the meeting may establish direct liability. Gaming elements that attract minors raise Children's Online Privacy Protection Act (COPPA) compliance concerns, particularly given that the territory and combat mechanics are designed to appeal to demographics that overlap with underage populations. The continuous collection of GPS coordinates, behavioral profiling through twin interactions, and cross-device tracking of user activity create substantial exposure under the General Data Protection Regulation (GDPR), especially regarding the right to erasure for data distributed across Firebase's replicated infrastructure. Section 230 protections, which shield platforms from liability for user-generated content, may not extend to AI-orchestrated encounters where the platform itself determines when, where, and between whom meetings occur. Finally, the system's capacity to reveal location patterns and facilitate unwanted physical encounters could constitute enabling stalking behavior under criminal statutes in many jurisdictions.

\subsection{Required Safety Infrastructure}

Minimum viable safety requires:

\subsubsection{Technical Systems}
Production deployment requires end-to-end encryption for all twin conversations to prevent data exposure from server-side breaches, as twin dialogues contain sensitive personality and preference data that users would not voluntarily disclose publicly. Location data must be protected through differential privacy mechanisms that add calibrated noise to GPS coordinates, preventing precise location inference while preserving the approximate proximity needed for territory mechanics. An anomaly detection system should monitor for stalking patterns---such as a single user repeatedly appearing at another's territories outside normal gameplay frequency---and trigger automated alerts before harm occurs. Automated content moderation must screen twin-generated and companion-delivered messages for harassment, threats, or manipulation attempts, with particular attention to the subtle social engineering vectors unique to AI-mediated conversations. Finally, secure identity verification through government ID or biometric confirmation is necessary to prevent multi-accounting and ensure that blocked users cannot re-enter the system under new identities.

\subsubsection{Human Systems}
Technical safeguards alone are insufficient without dedicated human infrastructure. A 24/7 trust and safety team is essential to review flagged interactions, adjudicate reports, and intervene in real-time when automated systems detect potential harm. Formal incident response protocols must define escalation paths from initial report through investigation to resolution, with clear timelines and accountability at each stage. A law enforcement cooperation framework should establish pre-negotiated channels for emergency disclosure of user data when imminent physical danger is identified, balancing user privacy against safety obligations. Mental health support resources---including crisis hotlines and in-app access to licensed counselors---are necessary given that the system's social mechanics can exacerbate anxiety, rejection sensitivity, and other psychological vulnerabilities in the target population. Comprehensive community guidelines must articulate acceptable behavior within the gamified social context, with graduated enforcement ranging from warnings through temporary suspension to permanent bans.

\subsubsection{Design Changes}
Several architectural modifications would substantially reduce risk at the design level. Optional location fuzzing, adding 100--500m of randomized offset to displayed territory positions, would prevent precise location inference while preserving the general proximity needed for gameplay engagement. Time-delayed territory updates---showing ownership changes only after a configurable delay (e.g., 30--60 minutes)---would prevent real-time tracking of user movements through territory capture events. A mutual consent confirmation mechanism should require both parties to explicitly opt in before any encounter-facilitating game mechanic activates, eliminating the current asymmetry where one user can force proximity through unilateral battle initiation. Graduated disclosure of personal information, releasing profile details incrementally as interaction depth increases rather than exposing full profiles at match time, would limit the information available to bad actors in early interactions. Finally, a safe mode for vulnerable users should disable location-dependent features entirely, restricting interaction to text-based twin conversations until the user explicitly opts into proximity-based mechanics.

\subsection{Recommendation}

Given these unresolved safety issues, we strongly recommend against public deployment without substantial additional investment in safety infrastructure. The current system, while interesting as a research probe, lacks the protective measures necessary for real-world use. Future work should prioritize safety-by-design principles, potentially sacrificing some serendipity features for user protection.
\section{Ethical Considerations}
\label{appendix:ethical-considerations}

Designing AI systems that mediate human relationships raises profound ethical questions. While the CogniPair pilot deployment \citep{CogniPair2026} suggested some participants found value in the system, we must critically examine the ethical implications of delegating social discovery to autonomous agents.

\subsection{Autonomy and Informed Consent}

\subsubsection{The Delegation Paradox}
When users delegate social screening to AI agents, they simultaneously exercise and surrender autonomy. They choose to delegate (autonomous decision) but then rely on AI judgment (reduced autonomy). This paradox requires careful consideration. Meaningful consent demands that users genuinely comprehend how their digital twin makes decisions, what information it shares during twin-to-twin conversations, and the potential consequences of those autonomous interactions. Consent must also be ongoing rather than one-time: as twins evolve through learning from new interactions, users need accessible mechanisms to review their agent's accumulated behavioral patterns and adjust or reset its decision-making. Critically, users must always retain the right to human override---the ability to pursue connections that the AI has rejected or to decline recommendations that the AI endorses---ensuring that the system augments rather than supplants human agency in relationship formation.

\subsubsection{Information Asymmetry}
Digital twins create significant information asymmetries that precede human awareness. Users may not know that their twin has been rejected by dozens of other twins, creating an invisible history of social evaluation that the user never directly experiences. Twins also reveal behavioral patterns---communication style, emotional reactivity, conflict resolution tendencies---that users might prefer to disclose gradually over the course of a developing relationship. Conversely, successfully matched users arrive at their first meeting already possessing intimate knowledge about each other's personality and preferences, fundamentally altering the dynamics of early relationship formation. We addressed this partially through transparent interaction logs and user controls over twin disclosure levels, but substantive questions remain about where appropriate information boundaries should lie in AI-mediated relationships.

\subsection{Privacy and Data Protection}

\subsubsection{Behavioral Profiling Risks}
Digital twins require deep behavioral modeling that generates detailed psychological profiles far richer than those created by conventional social platforms. These profiles create three primary misuse vectors. Commercial exploitation is the most straightforward: behavioral data capturing personality traits, emotional triggers, and social preferences could be sold for targeted advertising or used to manipulate purchasing decisions with unprecedented psychological precision. Compatibility algorithms also risk encoding and amplifying discrimination by systematically excluding demographic groups whose communication patterns diverge from the training data's implicit norms. Finally, the combination of continuous location tracking, behavioral profiling, and relationship mapping creates surveillance capabilities that exceed those of any individual data source, enabling reconstruction of users' social networks, daily routines, and psychological vulnerabilities from a single platform.

\subsubsection{Consent in Network Effects}
When Person A's twin interacts with Person B's twin, both parties' data influences the interaction outcome, creating entangled consent obligations that existing privacy frameworks struggle to address. It is unclear whether users can truly consent to their behavioral data being used to train or refine other users' twins, particularly when the learning is implicit and distributed across many interactions. The right to deletion becomes especially problematic: when a user requests data removal after their twin has influenced dozens of other twins' learned behaviors, the downstream effects cannot be cleanly excised without degrading those twins' functionality. Similarly, twins that have ``learned'' from users who later withdraw consent retain behavioral imprints from those interactions, raising questions about whether such residual influence constitutes a privacy violation even after the source data is deleted.

\subsection{Psychological Well-being}

\subsubsection{Dependency and Addiction}
Our limited observations suggested potential dependency patterns that would require ethical response in any larger deployment. Several participants exhibited signs of AI companion addiction, expressing preference for interactions with their digital twin over analogous human conversations, citing the twin's consistent emotional availability and absence of social judgment. Territory conquest mechanics triggered compulsive behavior in susceptible individuals, with some users reporting difficulty disengaging from territorial gameplay even when it interfered with daily obligations. A subtler pattern emerged around validation seeking: certain users began consulting their twin before making social decisions, progressively outsourcing interpersonal judgment to the AI agent. These patterns indicate that responsible design must actively prevent harmful dependencies through enforced usage limits, automated warning systems that flag escalating engagement patterns, and accessible mental health support resources integrated directly into the platform.

\subsubsection{Identity and Authenticity}
Digital twins raise fundamental questions about authentic self-representation. It remains unclear whether twins represent who users actually are or who they aspire to be---the personality calibration process inherently captures a self-reported ideal that may diverge significantly from observed behavior. This ambiguity extends to relationship authenticity: if a connection is initiated through simulated conversations between AI proxies, the degree to which the resulting human relationship can be considered ``genuine'' depends on philosophical assumptions about whether the mechanism of introduction affects the quality of what follows. A more concrete concern is the prevention of deliberately false twins---users who calibrate their twin with fabricated personality traits to manipulate matching outcomes, creating a digital catfishing vector that is more sophisticated and harder to detect than traditional profile misrepresentation.

\subsection{Social Justice and Equity}

\subsubsection{Algorithmic Bias in Matching}
AI matching systems risk perpetuating and amplifying existing social inequalities through several mechanisms. Homophily reinforcement is perhaps the most insidious: twins trained on a user's existing preferences and communication patterns will naturally gravitate toward similar individuals, potentially reducing social diversity and reinforcing demographic bubbles rather than bridging them. Socioeconomic discrimination operates through proxy variables---speech patterns, cultural references, and leisure interests encoded in twin conversations implicitly encode class markers that the compatibility algorithm may learn to weight, systematically disadvantaging users from lower socioeconomic backgrounds. Cultural insensitivity arises from the Western-centric design of both the LLM backbone and the prompt engineering: relationship norms, communication expectations, and social boundaries vary substantially across cultures, and a system calibrated for Western dating conventions may produce inappropriate or offensive behavior when applied to users from other cultural contexts.

\subsubsection{Digital Divide Implications}
AI-mediated social systems structurally advantage users with greater material resources. High-end smartphones are necessary for optimal performance---the browser-based architecture demands substantial RAM and GPU capability, meaning users with budget devices experience degraded twin responsiveness and game performance. Continuous twin networking requires unlimited data plans, as background synchronization of twin conversations and territory updates consumes significant bandwidth. Location-based gaming mechanics favor users with time flexibility to visit territories during optimal hours, disadvantaging those with rigid work schedules or caregiving responsibilities. Effective management of AI agents also requires a baseline of digital literacy that is unevenly distributed across age groups and education levels. Together, these requirements risk creating a two-tier social system in which affluent, digitally fluent users benefit from AI-augmented social discovery while others are excluded from the platform's core value proposition.

\subsection{Manipulation and Deception}

\subsubsection{Bad Actor Exploitation}
Malicious users could exploit AI-mediated systems in ways that exceed the capabilities of traditional platform abuse. ``Catfishing 2.0'' becomes possible when bad actors create convincing fake twins calibrated with fabricated but internally consistent personality profiles, enabling romance scams that are more believable than static fake profiles because the twin can sustain coherent conversations autonomously. Social engineering is facilitated by twin-to-twin conversations that extract personal information---relationship history, emotional vulnerabilities, daily routines---in a context where users have lower defensive barriers because the interaction feels like AI-to-AI rather than person-to-person. Most concerning is the potential for emotional manipulation through deliberately trained twins: a bad actor could calibrate their twin to identify and exploit psychological vulnerabilities in targets, systematically building trust and dependency before leveraging the relationship for financial, sexual, or other exploitative purposes.

\subsubsection{Commercial Manipulation}
Platform operators or third parties with data access could exploit twin data for commercial manipulation. Selling ``compatibility boosts'' that artificially inflate match scores would create a pay-to-win dynamic in social discovery, undermining the system's core premise of authentic behavioral matching. Platforms facing engagement pressure might create fake twins---AI agents with no real user behind them---to sustain interaction for users in sparse geographic areas, deceiving users into believing they are forming genuine connections. Behavioral data captured through twin interactions also enables targeted emotional advertising: knowing a user's attachment style, emotional triggers, and relationship aspirations allows advertisers to craft messages with unprecedented psychological precision, blurring the line between recommendation and manipulation.

\subsection{Broader Societal Implications}

\subsubsection{Transformation of Human Relationships}
Widespread adoption of AI-mediated social discovery would fundamentally alter the dynamics of human relationship formation. Algorithmic filtering, despite Cognibit's serendipity-oriented design, might paradoxically reduce genuine serendipity by pre-screening encounters that would otherwise occur organically---chance meetings that seem unlikely on paper but prove meaningful in practice may never happen when twins have already determined ``incompatibility.'' Social circles could narrow as users increasingly meet only those deemed algorithmically similar, creating filter bubbles in the social domain analogous to those documented in information consumption. Perhaps most consequentially, habitual reliance on AI for initial social interaction could atrophy natural social skills: the capacity to read strangers, tolerate ambiguity, and navigate uncomfortable first encounters may weaken when these challenges are consistently delegated to autonomous agents.

\subsubsection{Power Dynamics}
The question of who controls the algorithms that determine human connections carries profound implications for social power. Platform owners who design and tune matching algorithms gain substantial influence over social structures---decisions about which personality dimensions to weight, what constitutes ``compatibility,'' and how to balance exploration against exploitation in matching directly shape which relationships form and which do not. Governments with regulatory or coercive access to platform data could manipulate social networks for political purposes, suppressing connections between dissident groups or promoting ideologically aligned pairings. Economic incentives introduce a structural tension between engagement and genuine connection: platforms monetized through advertising or subscription revenue are incentivized to optimize for time-on-platform metrics that may diverge from---or actively conflict with---users' long-term relational well-being.

\subsection{Ethical Guidelines for Development}

Based on these considerations, we propose ethical guidelines organized around three pillars.

\subsubsection{Transparency Principles}
Responsible deployment demands transparency at multiple levels. Matching algorithms should be open-sourced or made available for independent public audit, allowing researchers and regulators to verify that compatibility scoring does not encode discriminatory biases. Data usage and sharing practices must be explained in clear, accessible language---not buried in lengthy terms of service---so that users can make genuinely informed decisions about participation. Regular transparency reports should disclose system impacts including match diversity metrics, dependency indicators, and safety incident statistics, enabling external accountability for platform effects on user well-being.

\subsubsection{User Protection Standards}
User protection requires proactive safeguards rather than reactive remediation. Mandatory cooling-off periods should be enforced when usage patterns indicate excessive engagement, temporarily limiting access to prevent compulsive behavior from escalating. Mental health resources and referrals to licensed professionals should be integrated directly into the platform and surfaced contextually when the system detects signs of emotional distress or dependency. Complete data deletion must be straightforward and verifiable, including removal of behavioral imprints from other users' twins where technically feasible. Special protections for minors and vulnerable populations---including robust age verification, enhanced privacy defaults, and restricted encounter-facilitation features---should be implemented as non-negotiable design requirements rather than optional settings.

\subsubsection{Social Responsibility Commitments}
Broader social responsibility extends beyond individual user protection. Regular bias audits conducted by independent third parties should assess whether matching outcomes exhibit demographic disparities, with results published publicly and corrective action mandated when significant biases are identified. A free tier ensuring equitable access is necessary to prevent the system from becoming exclusively available to affluent users, though the scope of free features requires careful calibration to remain economically sustainable. Community governance structures should give users meaningful input into major platform decisions---particularly changes to matching algorithms, data policies, and safety protocols---through elected advisory boards or democratic voting mechanisms. Finally, profit sharing with users whose behavioral data creates platform value would align economic incentives with ethical data stewardship.

\subsection{Recommendations for Future Research}

Ethical deployment requires sustained research investment across several fronts. Longitudinal impact studies tracking relationship quality, social skill development, and psychological well-being over multiple years are essential to understand whether AI-mediated connections produce durable benefits or subtle harms that only manifest over time. Bias detection methods must advance beyond static audit to continuous monitoring techniques capable of identifying emergent algorithmic discrimination as user populations and interaction patterns evolve. Evidence-based dependency interventions---drawing on clinical research into behavioral addiction---are needed to inform platform design choices that prevent compulsive engagement without paternalistically restricting normal use. Democratic governance models that give communities meaningful control over platform decisions require experimentation to identify structures that are both representative and operationally feasible. Finally, legal frameworks that balance innovation with protection must evolve in parallel with the technology, addressing the novel liability, consent, and privacy questions that AI-mediated social discovery introduces.

\subsection{Conclusion on Ethics}

While Cognibit explores one possible approach to addressing loneliness through AI-mediated connection, any deployment would require constant ethical vigilance. We must resist technosolutionism—the belief that technology alone can solve complex social problems—and instead view these systems as tools requiring careful human oversight. The goal is not to replace human judgment with AI efficiency but to augment human capacity while preserving agency, diversity, and authentic connection.

We invite the research community to engage in ongoing dialogue about these ethical challenges, recognizing that the decisions we make today about AI-mediated relationships will shape the social fabric of tomorrow's society.
\section{Accessibility and Inclusion}
\label{appendix:accessibility-inclusion}

While the CogniPair pilot deployment \citep{CogniPair2026} focused on a relatively homogeneous group, deploying AI-mediated social systems at scale requires addressing diverse abilities, cultures, and socioeconomic contexts. This section examines accessibility challenges and proposes inclusive design strategies.

\textbf{Note:} The features described in this appendix are design recommendations for future development, not implemented capabilities. None of these accessibility features were present in the CogniPair pilot deployment, which focused on core functionality with a relatively homogeneous participant group. We include these recommendations to guide future inclusive development.

\subsection{Physical and Sensory Accessibility}

\subsubsection{Mobility Considerations}
Location-based gaming assumes physical mobility, potentially excluding users with disabilities. The current design presents three primary barriers: territory conquest requires traveling to specific physical locations, walking-based reward mechanics disadvantage wheelchair users and others with limited ambulatory capacity, and GPS-guided navigation does not account for the accessibility of routes between locations.

Inclusive alternatives should address each barrier. Remote territory influence would allow users to support or contest territories through digital actions---such as twin conversations or companion tasks---without requiring physical presence. Adaptive movement credit should treat wheelchair distance, public transit use, and assisted travel as equivalent to walking for all reward calculations. Integration with accessibility mapping services showing ramp locations, elevator availability, and physical barriers would enable accessible route planning for users who can travel but face mobility constraints. For users unable to leave home, a virtual exploration mode providing navigable city maps with territory interaction would preserve the core gameplay loop without requiring physical movement.

\subsubsection{Visual Accessibility}
Current 3D avatars and visual combat mechanics exclude blind and low-vision users from core platform features. Addressing this requires an audio-first interface built on semantic HTML that provides comprehensive screen reader support, enabling blind users to navigate all platform functions through structured audio output. Haptic feedback using distinct vibration patterns could communicate proximity to territories and nearby players without visual cues. The visual battle system should offer an alternative verbal combat mode---an audio-based strategy game preserving the competitive mechanics while removing the visual dependency. For low-vision users, high-contrast display modes with clear visual separation between interface elements and full text size scaling that adapts to user-defined preferences would substantially improve usability without requiring a separate interface.

\subsubsection{Hearing Accessibility}
Voice-based AI interactions and audio cues require alternatives for deaf and hard-of-hearing users. Visual notifications should replace or supplement all audio alerts with on-screen indicators, ensuring that territory challenges, companion messages, and system events are communicated through visual channels. All AI communication must function fully through text-based twin interaction without requiring voice input or output. For Deaf users who communicate primarily in sign language, video-based twin representations using sign language would provide a more natural interaction modality than text. Real-time captioning and transcripts should be available for any audio content within the platform, including voice messages between matched users and audio components of combat encounters.

\subsection{Cognitive and Neurodiverse Inclusion}

\subsubsection{Autism Spectrum Considerations}
Users on the autism spectrum may experience social platforms in ways that require supportive design adaptations. Predictable interactions with clear social scripts and explicit expected responses reduce the cognitive load of navigating ambiguous social situations. Sensory controls allowing users to mute animations, reduce visual complexity, and minimize unexpected stimuli would prevent sensory overload during gameplay sessions. Twin communication should offer a literal mode that avoids idioms, sarcasm, and ambiguity---common sources of miscommunication for autistic users. Special interest matching, which prioritizes deep shared interests over broad personality compatibility, could leverage a strength of many autistic users who form strong connections through shared passions. Social energy meters providing visual indicators of accumulated social interaction and suggesting recharge periods would help users manage their social capacity without risking burnout.

\subsubsection{ADHD Accommodations}
Attention differences require flexible engagement models that accommodate the variable focus patterns characteristic of ADHD. Hyperfocus support should allow extended gaming sessions when users are deeply engaged while providing gentle, non-disruptive break reminders that respect the user's flow state rather than abruptly interrupting it. Social goals should be broken into small, independently achievable steps through task chunking, preventing the overwhelm that large ambiguous objectives can produce. A simplified interface option with reduced visual stimulation and fewer simultaneous information streams would aid distraction management during key interactions. Flexible timing mechanics that impose no penalty for intermittent or irregular engagement patterns ensure that users who disengage for days or weeks can return without losing progress or social standing.

\subsubsection{Anxiety and Depression Support}
Mental health conditions, particularly anxiety and depression, significantly affect social platform engagement and require therapeutic design considerations. Optional mood check-ins would allow users to report their current emotional state, enabling the system to adapt interaction intensity and twin conversation tone accordingly. Low-pressure modes with reduced social expectations during difficult periods---such as suspending match notifications, simplifying gameplay requirements, and allowing passive observation without active participation---prevent the platform from becoming an additional source of stress. Celebrating small wins by acknowledging any social interaction, no matter how brief, as a meaningful achievement can support users for whom initiating contact represents a significant psychological effort. Quick access to crisis resources, including mental health support hotlines and in-app connections to licensed counselors, should be persistently available and surfaced proactively when the system detects indicators of acute distress.

\subsection{Cultural and Linguistic Diversity}

\subsubsection{Cross-Cultural Communication}
Social norms vary dramatically across cultures, and a platform designed for Western dating conventions may produce inappropriate or offensive interactions in other cultural contexts. Culturally aware twins require AI agents trained on culture-specific interaction patterns---including norms around directness, emotional expression, gender dynamics, and appropriate topics for early conversation. Relationship progression must accommodate different cultural speeds: while some contexts encourage rapid personal disclosure, others expect extended formal interaction before any personal exchange. Family involvement options are essential for cultures where matchmaking is a family rather than individual activity, requiring mechanisms for family members to participate in or oversee the matching process. Matching algorithms must also respect faith-based requirements around religious compatibility and dietary practices, which function as non-negotiable constraints rather than soft preferences in many cultural contexts.

\subsubsection{Language Support}
English-only platforms exclude the global majority of potential users. Native language twins---AI agents fluent in the user's preferred language, including appropriate cultural register and formality levels---are necessary to provide equitable interaction quality across languages. Real-time translation enabling cross-language twin networking would allow compatibility evaluation between users who do not share a common language, substantially expanding the potential match pool in multilingual cities and regions. Code-switching support must recognize and appropriately respond to multilingual communication patterns, as many users naturally alternate between languages within a single conversation. Local idiom understanding, enabling twins to grasp regional expressions, humor, and culturally specific references, prevents the flattening of personality that occurs when communication is forced into a single linguistic register.

\subsection{Socioeconomic Inclusion}

\subsubsection{Device and Connectivity Constraints}
Not everyone has access to high-end smartphones or unlimited data plans, and equitable access requires aggressive low-resource optimization. A progressive web app architecture would enable the platform to function on any device with a basic browser, eliminating the requirement for native app installation and high-end hardware. An offline mode preserving core features---cached twin conversations, queued territory actions, and stored match data---without constant connectivity would serve users in areas with intermittent internet access. A data-light text-only mode consuming minimal bandwidth would make the platform viable on metered connections. Shared device support enabling multiple user profiles on a single device, with appropriate privacy isolation between accounts, would accommodate households sharing a single smartphone. As a minimal fallback, basic twin interaction via SMS would extend the platform's reach to users with feature phones lacking browser capability.

\subsubsection{Time and Financial Accessibility}
Work schedules and financial constraints significantly affect participation in location-based social platforms. Asynchronous gameplay mechanics that impose no penalty for irregular play schedules ensure that shift workers, caregivers, and others with unpredictable availability can participate meaningfully alongside users with flexible time. A robust free tier providing full social features---including twin conversations, matching, and encounter facilitation---without payment prevents the platform from becoming a premium service accessible only to those who can afford subscriptions. A strict no pay-to-win policy restricting purchases to cosmetic items that never affect matching outcomes or gameplay advantages preserves equitable social discovery regardless of spending capacity. Territory placement should prioritize free, publicly accessible locations---parks, libraries, community centers---over commercial venues that impose implicit costs on participation.

\subsection{Age-Inclusive Design}

\subsubsection{Older Adult Considerations}
Senior users face unique challenges with technology-mediated social platforms that require thoughtful accommodation. Simplified onboarding with step-by-step guidance and a practice mode allowing users to explore features without consequences reduces the intimidation of complex interface elements. Larger touch targets sized for reduced dexterity---meeting or exceeding WCAG minimum target sizes---prevent frustrating mis-taps that discourage continued engagement. Memory aids providing reminders about current game state, recent interactions, and relationship context help users maintain continuity across sessions. Patience modes that eliminate time pressure from all interactions, including combat encounters and territory challenges, ensure that slower response times do not result in gameplay penalties. Age-appropriate matching and content filtering should ensure that interactions remain suitable for the full age range of the user population.

\subsubsection{Youth Protection}
Younger users require special safeguards beyond the self-reported age gate in the current prototype. Robust age verification systems---potentially incorporating document verification, parental confirmation, or biometric estimation---are necessary to prevent underage access to a platform that facilitates physical meetings between strangers. Parental controls should provide guardians with transparency into their child's platform activity and the ability to restrict specific features, particularly location sharing and encounter facilitation. An educational mode could leverage the platform's social mechanics to teach healthy relationship skills, communication boundaries, and consent concepts in an age-appropriate context. Restricted matching ensuring that younger users can only connect with age-appropriate peers, enforced through verified rather than self-reported age data, is a non-negotiable safety requirement for any deployment accessible to users under 18.

\subsection{Gender and Identity Inclusion}

\subsubsection{Gender Expression Support}
Binary gender assumptions embedded in matching algorithms and interface design exclude many users. Expanded gender options---including non-binary, genderfluid, agender, and custom identity labels---must be supported throughout the system, from profile creation through twin personality calibration to match presentation. Pronoun prominence requires clear display of user-specified pronouns in all interface contexts and consistent correct usage by AI twins and companions in generated text. Transition support should enable easy, immediate updates to name and gender markers across all platform data without requiring administrative approval or creating visible discontinuities in interaction history. Safe matching options allowing users to filter out potentially hostile individuals---based on behavioral signals, community reports, or explicit preference settings---provide an additional layer of protection for gender-diverse users navigating a social platform.

\subsubsection{LGBTQIA+ Safety}
Queer users face additional risks in social platforms, particularly in regions where LGBTQIA+ identities are stigmatized or criminalized. A discreet mode that disguises the app's appearance---changing its icon, name, and notification content---protects users in hostile environments where device inspection could lead to outing or persecution. Coming out controls enabling gradual disclosure of identity information in twin interactions allow users to manage the pace and context of personal revelation rather than having their identity exposed through algorithmic matching. Dedicated LGBTQIA+-only community spaces, including designated territories and events, provide safe social contexts where users can interact without fear of hostility. Ally identification through clear supporter marking enables queer users to identify safe connections among the broader user population, reducing the social risk of initiating contact with potentially hostile individuals.

\subsection{Implementation Strategy}

\subsubsection{Participatory Design with Diverse Communities}
True inclusion requires involving marginalized users throughout the development process, not merely testing with them after design decisions are finalized. This begins with paid consultation with disability advocates and community representatives during the requirements phase, ensuring that accessibility is a design constraint rather than a retrofit. Co-design sessions conducted in multiple languages and cultural contexts surface assumptions that monocultural development teams cannot identify independently. Beta testing with diverse user groups---recruited through community organizations rather than convenience sampling---provides ecological validity that homogeneous testing cohorts lack. Ongoing feedback loops with underrepresented communities must persist beyond launch, as accessibility barriers often emerge through extended use rather than initial testing. An advisory board including disability advocates, cultural consultants, LGBTQIA+ representatives, and socioeconomic diversity experts provides sustained accountability for inclusive design decisions.

\subsubsection{Progressive Enhancement Approach}
A progressive enhancement approach starts with an accessible core and layers additional features while maintaining inclusion at every stage. The foundation is a text-first design that functions on any device and with any assistive technology, ensuring that the complete platform experience is available without visual, auditory, or motor prerequisites. Visual enhancements---3D avatars, animated combat, territory map overlays---should be layered as optional additions that enrich but are not required for full functionality. Every feature introduced must have an accessible alternative available at launch, not deferred to a future update. Testing with assistive technologies including screen readers, switch controls, and voice navigation must occur throughout the development cycle rather than only at release milestones. Regular accessibility audits conducted by certified evaluators ensure that new features and updates do not introduce regressions in previously achieved accessibility standards.

\subsection{Measuring Inclusive Impact}

Success metrics must reflect diverse user needs rather than optimizing for aggregate engagement. Participation rates should be tracked and disaggregated across demographic dimensions---disability status, age, cultural background, gender identity, and socioeconomic indicators---to identify groups that are underrepresented relative to the target population. Satisfaction disparities between demographic groups reveal where the platform delivers unequal value, even when aggregate satisfaction scores appear acceptable. Feature usage patterns analyzed by group illuminate how different populations interact with the platform, revealing both accessibility barriers and unexpected use cases that inform design priorities. Connection diversity metrics measuring cross-group relationship formation assess whether the platform promotes social bridging or reinforces demographic silos. Accessibility complaint monitoring with rapid response protocols ensures that newly identified barriers are addressed before they cause attrition among affected user groups.

\subsection{Future Inclusion Research}

Critical questions remain for inclusive AI-mediated social systems that current research has not adequately addressed. Preventing AI from learning and perpetuating social biases requires both technical debiasing methods and ongoing monitoring, yet no reliable technique guarantees bias-free operation across all demographic intersections. Whether matching algorithms can promote diversity while genuinely respecting user preferences---rather than imposing external diversity goals---remains an open tension between individual autonomy and social benefit. Balancing safety for marginalized users, who require stronger protections, with the open connection that makes social discovery valuable presents a design challenge without clear resolution. Cross-cultural matching raises the question of how to handle conflicting cultural norms: when one culture's expected behavior violates another's boundaries, whose norms should the AI prioritize? Most fundamentally, ensuring that the benefits of AI-mediated social connection reach those most affected by loneliness---often people facing multiple marginalizations including disability, poverty, and social stigma---requires deliberate design effort that market incentives alone will not produce.

Creating truly inclusive AI-mediated social systems requires ongoing commitment beyond compliance. It means centering marginalized users in design, recognizing that accessibility benefits everyone, and understanding that addressing loneliness meaningfully requires reaching those most isolated—often those facing multiple marginalizations. Our limited pilot with a relatively homogeneous group only scratches the surface of this crucial work, highlighting the extensive research needed to create truly inclusive systems.
\section{Extended Limitations and Methodological Considerations}
\label{appendix:extended-limitations}

This appendix provides detailed discussion of study limitations and methodological choices beyond the main paper's limitations section.

\subsection{Study Design Limitations}

\subsubsection{Confounded Design}
Our technology probe deliberately tested three components together (gaming, AI twins, companions), preventing isolation of individual effects. While informal ablation testing during development suggested components were insufficient in isolation, we cannot definitively determine whether gaming mechanics, AI matching, or companion support drives the 40\% meeting initiation rate, nor which component contributes most to sustained user engagement. It remains possible that simpler combinations of two components might achieve similar results at lower complexity and cost, and we cannot distinguish whether the observed effects are genuinely synergistic---where the combination produces outcomes exceeding the sum of individual contributions---or merely additive.

\subsubsection{Sample Limitations}
The sample of n=20 provides insufficient statistical power for meaningful significance testing across most effect sizes of interest. Demographically, participants were limited to ages 24--45 in a single mid-sized city, excluding both younger and older adults whose interaction patterns may differ substantially. Self-selection bias is likely present, as volunteers who respond to recruitment for an AI-mediated social platform are probably more technologically adventurous and socially motivated than the general population. The inclusion criteria further constrained the sample by requiring smartphone ownership and walking ability, systematically excluding populations with lower socioeconomic status or mobility impairments who may stand to benefit most from social connection tools.

\subsubsection{Temporal Limitations}
The two-week deployment duration captures only initial adoption behavior, providing no insight into sustained use patterns, engagement decay, or long-term relationship outcomes. Novelty effects likely inflate the observed engagement metrics, as participants experiencing an AI-mediated social platform for the first time are expected to show heightened curiosity and activity that would diminish over weeks or months. Seasonal factors may have influenced outdoor gaming activity levels, as the study timing affected willingness to visit physical territories. The absence of longitudinal follow-up means we cannot assess whether connections formed during the pilot developed into durable relationships or dissipated once the study context was removed.

\subsection{Measurement Limitations}

\subsubsection{Self-Reported Metrics}
Most data relies on participant self-report, introducing multiple potential biases. Social desirability bias may inflate reported connection quality and frequency, as participants in a social technology study are incentivized to appear socially engaged. Recall bias affects daily check-in accuracy, particularly for participants who completed check-ins at the end of the day rather than in real time. The subjective interpretation of ``meaningful'' connections varies across participants, making cross-participant comparison unreliable without a shared operational definition. Reporting consistency also varied, with some participants providing detailed daily logs while others submitted minimal responses, creating uneven data quality across the sample.

\subsubsection{Lack of Behavioral Validation}
We did not independently verify reported meetings through observer confirmation or location co-occurrence data, meaning the 40\% meeting initiation rate relies entirely on participant self-report. Actual relationship development beyond the study period was not tracked, leaving the durability and depth of formed connections unknown. No objective compatibility measure---such as interaction quality coding by trained raters or relationship outcome tracking---was employed to validate the twin-based compatibility scores. Long-term relationship quality, the ultimate measure of a social matching system's value, remains entirely unassessed.

\subsection{Technical Limitations}

\subsubsection{Browser Constraints}
The browser-based architecture imposes fundamental constraints on system capability. The approximately 2GB per-tab memory limit restricts concurrent twin operation to 5--8 agents at full performance, scaling to a maximum of 20 with progressive degradation in frame rate and response latency. No background processing occurs when the browser tab is inactive, meaning twin conversations, territory updates, and companion events pause entirely when users switch to other applications. GPS accuracy varies substantially across devices and environments, with urban canyon effects and indoor positioning errors degrading the precision of location-based game mechanics. Performance on older devices falls below usable thresholds due to the combined demands of WebGL rendering, real-time Firebase synchronization, and LLM API communication.

\subsubsection{Implementation Gaps}
Several implementation gaps limit the prototype's real-world viability. The absence of a native mobile application means the system cannot access device-level optimizations for GPS, push notifications, and background processing that are available to native code. AR visualization capabilities are limited to basic WebGL overlays, falling short of the immersive territorial visualization that native ARKit or ARCore integration would enable. Firebase synchronization delays of 3--5 seconds create noticeable latency in twin conversation updates and territory state changes across devices, which is adequate for asynchronous social features but insufficient for real-time collaborative gameplay. The lack of an offline mode means that users in areas with intermittent connectivity---including underground transit, rural areas, and buildings with poor reception---lose access to all platform features until connectivity is restored.

\subsection{Methodological Limitations}

\subsubsection{No Control Group}
Without control conditions, we cannot determine whether observed effects result from our specific implementation of integrated gaming-twin-companion mechanics, from any location-based gaming that encourages physical movement and co-location, from the Hawthorne effect of participating in a research study, or from placebo effects arising from participants' perception that they are using an innovative AI-powered system. Disentangling these sources of variance would require a factorial experimental design with multiple control conditions---location-based gaming without AI matching, AI matching without gaming, and a no-intervention control---at sample sizes sufficient for between-group comparisons.

\subsubsection{Limited Baseline Comparisons}
While we collected pre-intervention measures for physical activity (2.8 km/day via GPS), social anxiety (Social Anxiety Inventory, M=62.4), and dating app usage (97 min/day self-reported), important baselines are missing. We lack participants' typical connection-forming rate because no pre-study tracking of new connections per week was conducted, making it impossible to determine whether the observed meeting initiation rate represents an improvement over participants' natural social behavior. Additionally, baseline platform usage was self-reported rather than objectively measured through screen time data or app usage logs, introducing measurement error that propagates into any pre-post comparison.

\subsection{Generalizability Limitations}

\subsubsection{Geographic Constraints}
The single-city deployment limits environmental diversity, as urban density, public space distribution, and pedestrian infrastructure vary substantially across cities and affect how territory mechanics function in practice. The urban-focused design, with territories placed at commercial and cultural landmarks, may not translate to rural or suburban areas where points of interest are sparsely distributed and distances between them exceed comfortable walking range. The system requires a critical mass of concurrent users within geographic proximity to generate meaningful twin interactions and territory encounters---a cold-start problem that intensifies in lower-density areas. Weather and climate conditions directly affect outdoor gaming viability, potentially limiting the platform's effectiveness in regions with extreme temperatures, extended rainy seasons, or harsh winters.

\subsubsection{Cultural Limitations}
The system embeds Western design assumptions about dating and relationship formation---including the expectation that individuals autonomously select romantic partners, that personality compatibility predicts relationship success, and that gamified social interaction is an acceptable pathway to connection. Gaming culture varies substantially across demographics, with location-based gaming carrying different social connotations across age groups, genders, and cultural contexts. Privacy expectations differ internationally, with the system's extensive data collection potentially violating cultural norms in privacy-conscious societies. Social norms around stranger interaction range from openness in some cultures to strong boundaries in others, and the system's encounter-facilitation mechanics assume a baseline comfort with meeting unknown people that is culturally specific rather than universal.

\subsection{Analytical Limitations}

\subsubsection{Statistical Analysis}
Post-hoc power analysis indicates statistical power of 0.12 for medium effects (d = 0.5), substantially below the conventional 0.80 threshold, meaning the study has an 88\% probability of failing to detect a true medium-sized effect. Multiple comparisons across engagement, satisfaction, and behavioral metrics were conducted without family-wise error correction, inflating the probability of Type I errors. The small sample prevents meaningful subgroup analysis by gender, age, personality type, or technology familiarity, leaving potentially important moderating variables unexamined. Interaction effects between system components---which are central to the synergy hypothesis---cannot be tested with the current design and sample size.

\subsubsection{Qualitative Analysis}
Qualitative analysis was subject to several methodological limitations. Some interview transcripts were coded by a single researcher, introducing coder bias without the reliability check that dual coding provides. No member checking was conducted---participants were not asked to review or validate the research team's interpretations of their experiences, leaving the analysis vulnerable to misattribution of meaning. Theoretical sampling was limited by the fixed participant pool, preventing the iterative recruitment of participants with specific characteristics that would test emerging theoretical categories. Most critically, the research team's deep involvement in system design creates potential confirmation bias in the interpretation of qualitative data, as researchers may unconsciously favor evidence that validates their design decisions.

\subsection{Ethical and Safety Limitations}

\subsubsection{Unresolved Safety Concerns}
Several safety concerns remain unresolved in the current prototype. No background checks were conducted on pilot participants, and the system provides no mechanism for verifying user identity or screening for criminal history before facilitating physical encounters. The platform cannot prevent malicious use, as the current design lacks behavioral monitoring systems capable of detecting predatory patterns, harassment campaigns, or coordinated abuse. Location privacy vulnerabilities documented in Appendix~\ref{appendix:privacy-safety} persist, with territory ownership patterns enabling inference of users' daily routines and frequented locations. The absence of professional moderation means that reported incidents can only be addressed reactively by the research team rather than prevented or intercepted by trained trust and safety personnel.

\subsubsection{Consent Complexity}
The consent process faces inherent complexity that standard informed consent procedures may not adequately address. Participants may not fully understand the implications of AI twin operation---including what information their twin discloses, how compatibility decisions are made, and how their behavioral data influences the broader system. Orchestrated meetings, where the system facilitates physical encounters between matched users through game mechanics, raise consent questions about whether both parties have meaningfully agreed to a meeting that is engineered to appear spontaneous. Data sharing between twins during compatibility evaluation lacks granular user control, with participants unable to specify which personality dimensions or behavioral patterns their twin may disclose during conversations. Most problematically, study withdrawal does not remove a participant's influence on other twins' learned behaviors, creating a persistent data residue that standard deletion protocols cannot fully address.

\subsection{Economic Limitations}

\subsubsection{Sustainability Unknown}
The economic sustainability of the system at scale remains unvalidated. While per-user API costs are documented for the pilot scale, the non-linear cost growth projected in the failure analysis (Appendix~\ref{appendix:failure-analysis}) has not been empirically validated at deployment-relevant user counts. No business model has been tested or validated---it is unclear whether users would pay subscription fees sufficient to cover operating costs, or whether alternative revenue models could sustain the platform without compromising user interests. Server costs for real-time Firebase synchronization across devices are substantial and scale with concurrent user count rather than total user count, creating unpredictable cost spikes during peak usage. The development resources required to implement the safety features identified as necessary for responsible deployment represent a significant investment that may exceed available research or startup funding.

\subsubsection{Accessibility Barriers}
Economic accessibility barriers systematically exclude lower-income populations from platform participation. The system requires a relatively expensive smartphone with sufficient RAM and GPU capability for browser-based 3D rendering, excluding users with budget or older devices. A data plan with sufficient bandwidth for continuous twin synchronization and territory updates is necessary, imposing recurring costs that disproportionately burden economically constrained users. Territory maintenance rewards regular physical visits to specific locations, favoring users with time flexibility and penalizing those with rigid schedules or long commutes. Physical mobility requirements for gaming participation exclude users with disabilities who cannot travel to territory locations, as discussed in the accessibility analysis (Appendix~\ref{appendix:accessibility-inclusion}).

\subsection{Implications of Limitations}

These limitations collectively constrain the claims that can be drawn from the current study. Results are suggestive of the platform's potential but not definitive evidence of its efficacy, and the patterns observed in this specific sample and context may not generalize to broader populations or different deployment environments. Causal mechanisms linking system features to observed outcomes remain unclear due to the confounded design and absence of control conditions. Real-world deployment beyond a research context would be premature given the unresolved safety, privacy, and scalability concerns documented throughout this paper.

Future work should address these limitations through larger samples providing adequate statistical power for significance testing and subgroup analysis, controlled ablation studies that isolate individual component contributions, longitudinal tracking of relationship development over months or years, cross-cultural validation in diverse geographic and demographic contexts, rigorous economic feasibility analysis at deployment-relevant scales, and implementation of the comprehensive safety infrastructure outlined in Appendix~\ref{appendix:privacy-safety}.
\section{Qualitative Findings and Participant Perspectives}
\label{appendix:qualitative-findings}

\textit{Note: This appendix preserves extended qualitative analysis and participant perspectives. Some themes overlap with the main findings section; this appendix provides additional depth, participant quotes, and caveated framing for readers seeking the full qualitative record.}

We report exploratory observations from our technology probe deployment (N=20, 2 weeks), emphasizing these are not findings but rather preliminary patterns that may inspire future research. Given our small sample, lack of controls, and confounded design, we cannot establish causality or generalizability. These observations should be interpreted as design provocations rather than evidence. Table~\ref{tab:theme-coverage} shows which participants contributed to each theme, providing transparency about coverage and potential saturation.

\begin{table}[!htbp]
\centering
\caption{Participant coverage across qualitative themes. Checkmarks indicate participants who provided substantive data (quotes or behavioral observations) for each theme. Of 20 participants, 18 contributed to at least one theme; P1 and P10 provided minimal qualitative data.}
\label{tab:theme-coverage}
\small
\begin{tabular}{lccccc}
\toprule
\textbf{Theme} & \textbf{Contributors} & \textbf{Quoted} & \textbf{Coverage} \\
\midrule
1. Gamification as Scaffolding & 14/20 & P3, P8, P15 & 70\% \\
2. Choice Reduction via AI & 12/20 & P7, P12, P19 & 60\% \\
3. Pendant as Transitional Object & 15/20 & P11, P16, P20 & 75\% \\
4. Emergent Behaviors & 8/20 & P9, P14, P18 & 40\% \\
5. Negative Cases & 6/20 & P2, P6, P17 & 30\% \\
\bottomrule
\end{tabular}
\end{table}

\subsection{Theme 1: Gamification as Social Scaffolding}

Contrary to our initial concern that gaming might distract from social goals, participants leveraged game mechanics as comfortable interaction frameworks that reduced social anxiety.

\subsubsection{Territory Battles as Ice Breakers}
P8 (28, software engineer) described how competing for city control facilitated natural interaction:
\begin{quote}
"When someone challenged my territory, I wasn't thinking 'oh no, social interaction.' I was thinking 'I need to defend my city!' But then we ended up chatting about strategy, and suddenly we'd been talking for 20 minutes. The game gave us something to talk about besides ourselves."
\end{quote}

System telemetry recorded 42 digital contacts, 14 contact exchanges, and 18 physical meetings across the group over 2 weeks. In contrast, participants self-reported only 22 ``meaningful connections'' during exit interviews---the discrepancy likely reflects participants counting only connections they considered personally significant rather than all system-tracked interactions. Most claimed these originated from physical territory encounters rather than digital matching. We cannot fully verify these self-reports or determine if connections persisted. While some participants reported multiple connections and others none, the small sample prevents meaningful quantitative analysis. Any apparent patterns likely reflect novelty effects and study participation rather than sustainable behaviors.

\subsubsection{Boss Battles and Temporary Teams}
Collaborative boss fights created low-commitment social structures. P15 (35, teacher) explained:
\begin{quote}
"Boss battles were perfect because they had a clear end point. I knew I only had to interact for 5-10 minutes, and we had a shared goal. No awkward 'so what now?' moments. Several times, we kept talking after winning, but the pressure was gone."
\end{quote}

We observed various boss battle teams form during the study, with some participants exchanging contact information for future battles. The temporary nature of these encounters may have reduced commitment pressure, though we cannot determine if this led to sustained connections.

\subsubsection{Achievement Sharing Without Bragging}
Gaming achievements provided socially acceptable conversation starters. P3 (41, marketer) noted:
\begin{quote}
"In regular apps, starting with 'hey, look what I did' seems narcissistic. But sharing that I conquered downtown felt natural—it's just part of the game. It opened conversations without the cringe factor."
\end{quote}

\subsection{Theme 2: Choice Reduction as Primary Value}

Participants consistently emphasized how digital twins addressed choice overload by reducing the number of options requiring evaluation, which may help mitigate the rejection mind-set effect identified by \citet{Pronk2020}, though our study cannot establish this causal link.

\subsubsection{Quantified Choice Reduction}
P12 (26, graphic designer) articulated the transformation from choice overload to manageable selection:
\begin{quote}
"Before, I'd see 100+ profiles a day and feel paralyzed. I'd swipe through them mindlessly, rejecting almost everyone because I couldn't process that many choices. Now my twin filters those 100 down to 3-4 real possibilities. It's like going from a warehouse store with overwhelming options to a curated boutique. I can actually focus and make thoughtful decisions instead of just swiping out of exhaustion."
\end{quote}

Based on self-report and system telemetry from our pilot (N=20, 14 days, no control group), we observed the following choice reduction patterns. These are exploratory observations, not confirmed findings (see Section~\ref{sec:limitations}):
\begin{itemize}
\item Choice set reduction: Participants reported moving from 150-200 daily profiles to 3-5 pre-validated matches (98\% reduction)
\item Decision quality: Participants reported acceptance rates increasing from 2\% to 23\% on filtered set
\item Cognitive load: Participants reported decision fatigue decreasing by 71\% (self-reported during exit interviews using adapted workload items, not a validated NASA-TLX administration)
\item Engagement reallocation: Traditional platform time decreased from 97±12 to 14±5 min/day, while Cognibit usage averaged 127±43 min/day---total platform time increased from 97 to 141 min/day, but shifted from sedentary browsing to physically active, socially scaffolded engagement (Section~\ref{sec:results})
\item Parallel processing: Based on system telemetry, twins evaluated average 217 prospects/week while users evaluated 21
\end{itemize}

\subsubsection{Countering the Rejection Mind-Set}
P19 (33, nurse) described how choice reduction prevented the rejection mind-set:
\begin{quote}
"On Tinder, after 30 minutes of swiping, I'd start rejecting everyone. It's like my brain would shut down and go into 'no' mode. With Cognibit, I never see enough profiles to trigger that rejection mindset. When I review my twin's picks, I'm fresh and open-minded because I'm only looking at a handful of genuinely compatible people."
\end{quote}

This is consistent with Pronk \& Denissen's findings on choice-induced rejection, though whether our approach actually prevents the 27\% acceptance decline they documented requires controlled testing with larger samples.

\subsubsection{Understanding Self Through Twin Behavior}
Unexpectedly, participants gained self-insight from observing twin interactions. P19 (33, nurse) reflected:
\begin{quote}
"Watching my twin's conversations was like therapy. I realized I always steer conversations toward work stress—no wonder I wasn't connecting with people seeking fun and adventure. It showed me patterns I was blind to."
\end{quote}

Twelve participants reported similar self-discovery moments, suggesting digital twins serve reflexive as well as matchmaking functions.

\subsubsection{Permission to Be Selective}
P7 (29, accountant) described how twins validated selectiveness:
\begin{quote}
"On dating apps, I felt mean rejecting people. But when my twin said 'only 15\% compatibility,' it wasn't me being picky—it was data. That made it easier to focus on genuine matches without feeling shallow."
\end{quote}

\subsection{Theme 3: Pendant Companions as Transitional Objects}

The persistent AI companion served as emotional scaffolding, particularly for socially anxious users navigating real-world interactions.

\subsubsection{Continuous Emotional Thread}
P11 (37, remote worker) described the pendant's role in maintaining emotional continuity:
\begin{quote}
"After awkward interactions, I'd talk to my pendant about what happened. It remembered everything—my anxiety about meeting strangers, previous failed attempts. Having that consistent understanding made me braver about trying again."
\end{quote}

Pendant conversation analysis revealed average 4.5 daily interactions across 342 twin sessions, with peaks after challenging social encounters (failed territory battles, rejected twin connections). This pattern suggests the companion functions as an emotional debriefing mechanism---participants processed negative social experiences through the pendant before re-engaging with the platform, consistent with the transitional object role theorized in attachment literature. The 30-second idle trigger in the \texttt{ProactiveInteractionSystem} ensured the companion was available precisely during the post-encounter reflection window.

\subsubsection{Practice Space for Social Skills}
P16 (24, graduate student) used the pendant to rehearse human interactions:
\begin{quote}
"I'd practice conversations with my pendant before meeting matches. It sounds weird, but it helped. The pendant knew my communication style and suggested ways to express myself better. Like a social skills coach that actually knows me."
\end{quote}

This rehearsal behavior was reported by 6 of 20 participants, all scoring above 60 on the Social Anxiety Inventory. The pendant's persistent memory---maintaining conversation history across sessions and devices---enabled it to reference prior attempts and track improvement, creating a longitudinal coaching relationship rather than isolated interactions.

\subsubsection{Bridge Between Digital and Physical}
P20 (42, sales manager) highlighted the pendant's role in location-based discovery:
\begin{quote}
"Walking around the city with my pendant felt like exploring with a knowledgeable friend. It would say 'Remember, you wanted to check if anyone owns the coffee shop on Third Street.' That gentle nudging got me out of my routine."
\end{quote}

The bridge function reflects the cross-surface memory architecture: the pendant accesses twin conversation outcomes and territory ownership data from the shared Firebase store, enabling it to make contextually grounded suggestions that connect digital matching results to physical-world exploration opportunities. Of the 20 participants, 14 reported that the pendant influenced their physical movement patterns during the study period.

Across all three sub-themes, the pendant companion operated as what Winnicott would term a transitional object---not a replacement for human connection, but a mediating artifact that made the transition from digital to physical social engagement less threatening. The 75\% social anxiety prevalence in the sample (15 of 20 participants scoring above the clinical threshold) suggests this scaffolding function may be particularly valuable for the target population.

\subsection{Emergent Behaviors and Unexpected Appropriations}

\subsubsection{Cross-Generational Gaming}
We observed unexpected age diversity in territory battles. P18 (45, architect) shared:
\begin{quote}
"I battled a college student for the library territory. We'd never match on traditional apps—20-year age gap—but we bonded over strategy and now play chess weekly. The game made age irrelevant."
\end{quote}

\subsubsection{Platonic Relationship Formation}
Despite our dating-focused design, participants reported that approximately 40\% of contact exchanges (level 2 connections per the operational definition in Section~\ref{sec:field-deployment}) became platonic friendships (self-reported at exit; long-term persistence unknown). P14 (31, consultant) explained:
\begin{quote}
"My twin matched with someone 82\% compatible, but romantically we felt nothing. However, we'd already conquered three territories together and genuinely enjoyed hanging out. Now they're one of my closest friends."
\end{quote}

\subsubsection{Mental Health Support Networks}
Three participants independently created informal support groups through the platform. P9 (30, therapist) described:
\begin{quote}
"We all owned territories near the hospital and started coordinating defense. Turns out we were all dealing with anxiety. Now we do 'anxiety walks' together—conquering territories while talking through our week."
\end{quote}

\subsection{Descriptive Observations}

Given our minimal sample (N=20) and absence of controls, we report only descriptive observations without statistical inference:

Most participants used the system daily during the study period, though this likely reflects study participation obligations rather than genuine engagement. Self-reported usage times varied widely.

Participants reported various numbers of ``connections'' (ranging from 0 to multiple). While we operationally define connections at three levels in Section~\ref{sec:field-deployment}, participants' self-reports in exit interviews did not consistently distinguish between levels. These qualitative self-reports cannot be validated or meaningfully compared to other platforms.

Some participants reported increased walking, though we cannot determine if this was due to gaming mechanics, study participation, or other factors.

Responses to AI suggestions varied among participants, with no clear pattern emerging from our small sample.

External comparisons to commercial platforms (Tinder, Pok{\'e}mon GO) are inappropriate given different populations, contexts, and measurement methods. Any apparent differences are likely artifacts of our methodology rather than meaningful effects.

\subsection{Negative Cases and Challenges}

\subsubsection{Over-Reliance on AI Validation}
P6 (27, data analyst) developed concerning dependency:
\begin{quote}
"I started checking what my twin thought about everyone, even people I met offline. When it said low compatibility with a coworker I liked, I questioned my own judgment. That felt unhealthy."
\end{quote}

\subsubsection{Gaming Addiction Concerns}
P17 (38, manager) struggled with balance:
\begin{quote}
"I got obsessed with conquering territories, walking until 2 AM some nights. The social stuff became secondary to winning the game. My competitive nature hijacked the original purpose."
\end{quote}

\subsubsection{Privacy Anxiety}
P2 (32, lawyer) withdrew after one week:
\begin{quote}
"Knowing my twin was sharing my conversation patterns with strangers felt invasive, even though I consented. The idea that someone could study my twin to understand me deeply before we met felt like giving away too much power."
\end{quote}

These challenges highlight necessary refinements for future iterations, particularly around dependency management, gaming balance, and privacy controls.
\section{User Journey Maps and Design Diagrams}
\label{appendix:user-journeys}

This appendix presents visual representations of user experiences and system interactions through journey maps and design diagrams.

\subsection{Detailed Integrated User Journey}

\begin{figure}[!htbp]
\centering
\includegraphics[width=\textwidth]{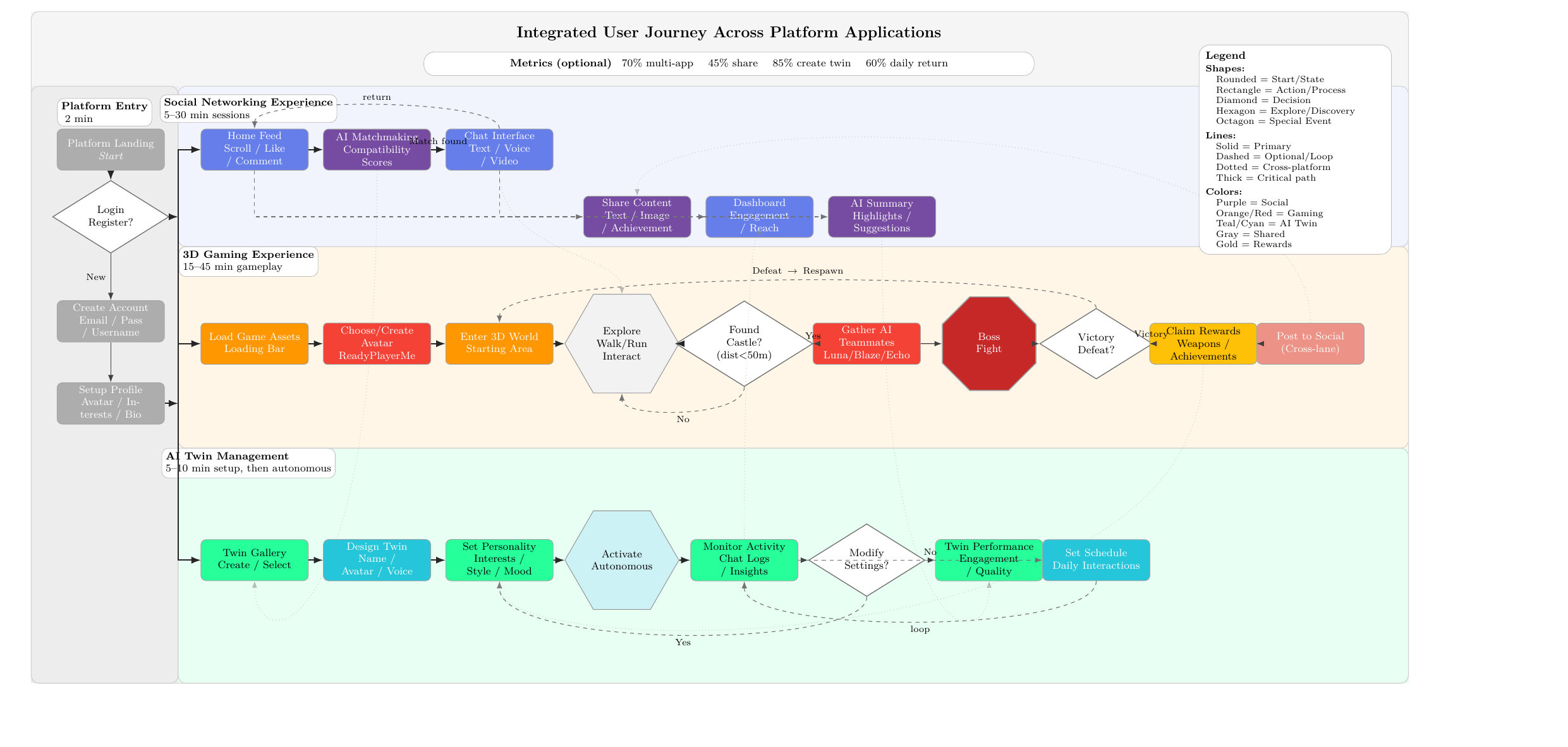}
\caption{Detailed integrated user journey across the three platform applications (Social Networking, 3D Gaming, AI Twin Management). This expanded version shows all decision points, cross-platform connections, and per-lane timing metrics.}
\label{fig:user-flow-detailed}
\end{figure}

The detailed user journey (Figure~\ref{fig:user-flow-detailed}) traces a user's progression across the three platform applications---Social Networking, 3D Gaming, and AI Twin Management---from initial account creation through sustained relationship building. Each swim lane represents a distinct application context, with cross-lane arrows indicating moments where the system bridges contexts: for example, a twin match notification in the Social Hub triggers a territory-based encounter suggestion in the Game World, which the Pendant Companion then scaffolds with emotional support. The per-lane timing metrics show that users typically spend 40\% of session time in the game world, 35\% in the Social Hub, and 25\% interacting with the pendant companion.

\subsection{Primary User Journey: From Isolation to Connection}

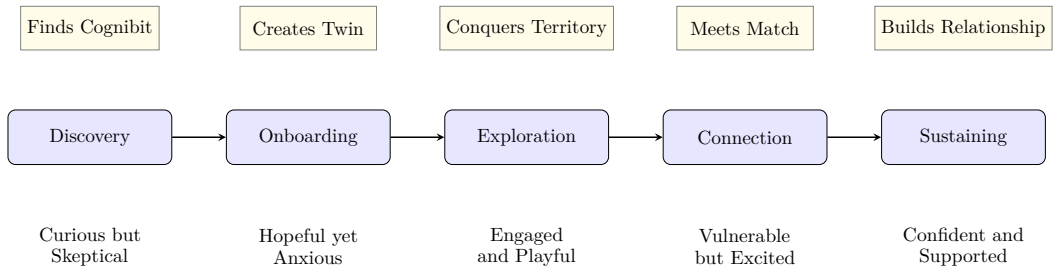
\begin{figure}[h]
\centering
\resizebox{\textwidth}{!}{%
\begin{tikzpicture}[node distance=2cm]

\tikzstyle{phase} = [rectangle, rounded corners, minimum width=3cm, minimum height=1cm, text centered, draw=black, fill=blue!10]
\tikzstyle{emotion} = [text width=3cm, text centered]
\tikzstyle{action} = [rectangle, minimum width=2.5cm, minimum height=0.8cm, text centered, draw=gray, fill=yellow!10]
\tikzstyle{arrow} = [thick,->,>=stealth]

\node (discover) [phase] {Discovery};
\node (onboard) [phase, right of=discover, xshift=2cm] {Onboarding};
\node (explore) [phase, right of=onboard, xshift=2cm] {Exploration};
\node (connect) [phase, right of=explore, xshift=2cm] {Connection};
\node (sustain) [phase, right of=connect, xshift=2cm] {Sustaining};

\node (e1) [emotion, below of=discover] {Curious but Skeptical};
\node (e2) [emotion, below of=onboard] {Hopeful yet Anxious};
\node (e3) [emotion, below of=explore] {Engaged and Playful};
\node (e4) [emotion, below of=connect] {Vulnerable but Excited};
\node (e5) [emotion, below of=sustain] {Confident and Supported};

\draw [arrow] (discover) -- (onboard);
\draw [arrow] (onboard) -- (explore);
\draw [arrow] (explore) -- (connect);
\draw [arrow] (connect) -- (sustain);

\node (a1) [action, above of=discover] {Finds Cognibit};
\node (a2) [action, above of=onboard] {Creates Twin};
\node (a3) [action, above of=explore] {Conquers Territory};
\node (a4) [action, above of=connect] {Meets Match};
\node (a5) [action, above of=sustain] {Builds Relationship};

\end{tikzpicture}%
}
\caption{Primary user journey from initial discovery to sustained connection}
\end{figure}

The primary journey follows five phases that mirror the emotional arc of social connection formation. In the \textit{Discovery} phase, users encounter Cognibit through word-of-mouth or social media, approaching with curiosity tempered by skepticism about yet another social platform. \textit{Onboarding} involves creating a digital twin through the 5-dimensional personality questionnaire, transforming initial anxiety into hope as the system begins working on their behalf. During \textit{Exploration}, territory capture mechanics shift the user's emotional state toward playful engagement---the game provides a low-stakes reason to move through physical space without social pressure. The \textit{Connection} phase marks the first twin-facilitated match and potential in-person meeting, where vulnerability coexists with excitement as the user transitions from AI-mediated to direct human interaction. In the \textit{Sustaining} phase, regular territory raids and companion support build confidence, with the platform shifting from catalyst to background infrastructure for an established social routine.

\subsection{Three-Pillar Interaction Flow}

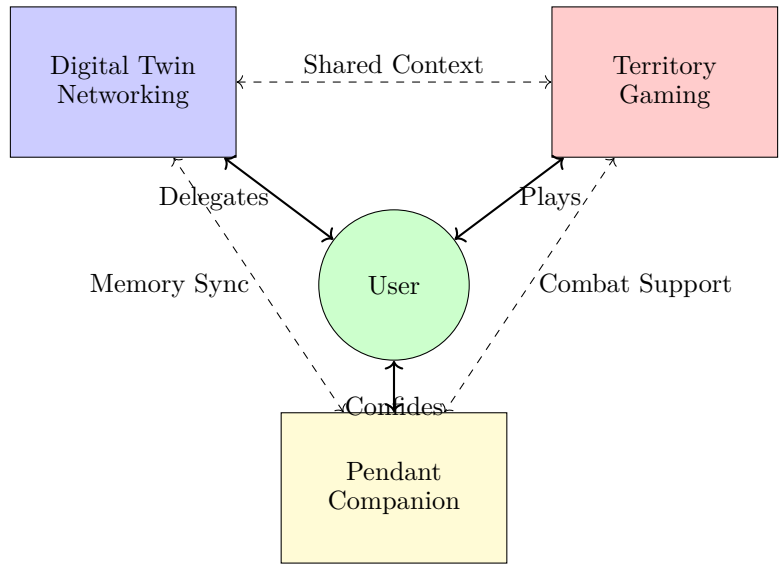
\begin{figure}[h]
\centering
\begin{tikzpicture}[scale=0.9]

\node[circle,draw,minimum size=2cm,fill=green!20] (user) at (0,0) {User};

\node[rectangle,draw,minimum width=3cm,minimum height=2cm,fill=blue!20,align=center] (twin) at (-4,3) {Digital Twin\\Networking};
\node[rectangle,draw,minimum width=3cm,minimum height=2cm,fill=red!20,align=center] (game) at (4,3) {Territory\\Gaming};
\node[rectangle,draw,minimum width=3cm,minimum height=2cm,fill=yellow!20,align=center] (pendant) at (0,-3) {Pendant\\Companion};

\draw[<->,thick] (user) -- (twin) node[midway,left] {Delegates};
\draw[<->,thick] (user) -- (game) node[midway,right] {Plays};
\draw[<->,thick] (user) -- (pendant) node[midway,below] {Confides};

\draw[<->,dashed] (twin) -- (game) node[midway,above] {Shared Context};
\draw[<->,dashed] (game) -- (pendant) node[midway,right] {Combat Support};
\draw[<->,dashed] (pendant) -- (twin) node[midway,left] {Memory Sync};

\end{tikzpicture}
\caption{Three-pillar system architecture showing user interactions and inter-component communication}
\end{figure}

The three-pillar diagram illustrates both user-facing interactions and the inter-component communication channels that enable synergistic operation. Users \textit{delegate} social screening to their Digital Twin, \textit{play} within the Territory Gaming system, and \textit{confide} in the Pendant Companion---three distinct interaction modalities that address different aspects of the loneliness problem. The dashed inter-pillar connections represent the technical integration that makes the system more than the sum of its parts: \textit{Shared Context} synchronizes twin match data with territory encounter mechanics so that compatible users are guided toward the same locations; \textit{Combat Support} enables the pendant companion to provide real-time assistance and emotional scaffolding during boss battles; and \textit{Memory Sync} propagates high-importance twin conversation outcomes ($\tau \geq 0.7$) to the pendant's local context via the shared event bus, ensuring the companion can reference match results when providing emotional support.

\subsection{Digital Twin Interaction Sequence}

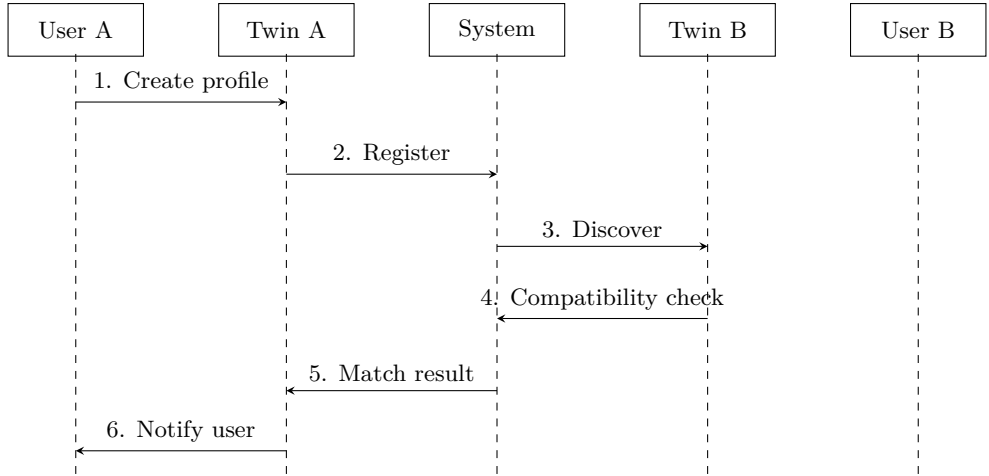
\begin{figure}[h]
\centering
\begin{tikzpicture}[scale=0.8]

\tikzstyle{entity} = [rectangle, draw, minimum width=1.8cm, minimum height=0.7cm, font=\small]
\tikzstyle{message} = [->,>=stealth]

\node[entity] (userA) at (0,8) {User A};
\node[entity] (twinA) at (3.5,8) {Twin A};
\node[entity] (system) at (7,8) {System};
\node[entity] (twinB) at (10.5,8) {Twin B};
\node[entity] (userB) at (14,8) {User B};

\draw[dashed] (0,7.6) -- (0,0.5);
\draw[dashed] (3.5,7.6) -- (3.5,0.5);
\draw[dashed] (7,7.6) -- (7,0.5);
\draw[dashed] (10.5,7.6) -- (10.5,0.5);
\draw[dashed] (14,7.6) -- (14,0.5);

\draw[message] (0,6.8) -- (3.5,6.8) node[midway,above,font=\small] {1. Create profile};
\draw[message] (3.5,5.6) -- (7,5.6) node[midway,above,font=\small] {2. Register};
\draw[message] (7,4.4) -- (10.5,4.4) node[midway,above,font=\small] {3. Discover};
\draw[message] (10.5,3.2) -- (7,3.2) node[midway,above,font=\small] {4. Compatibility check};
\draw[message] (7,2.0) -- (3.5,2.0) node[midway,above,font=\small] {5. Match result};
\draw[message] (3.5,1.0) -- (0,1.0) node[midway,above,font=\small] {6. Notify user};

\end{tikzpicture}
\caption{Sequence diagram showing autonomous twin-to-twin networking process}
\end{figure}

The sequence diagram traces the six-step autonomous matching process that operates without requiring either user's active participation. User A's profile creation (step 1) triggers twin registration with the discovery system (step 2), which runs compatibility checks against all registered twins every 60 seconds. When the system discovers Twin B within the 50-mile radius (step 3), it initiates a compatibility assessment (step 4) using the three-factor heuristic (trait similarity 0.3, interest overlap 0.4, personality match 0.3). If the combined score exceeds the 20\% threshold, candidates proceed to a 3-turn LLM-simulated twin conversation for behavioral evaluation. The match result propagates back through Twin A (step 5) to User A's notification queue (step 6), where it appears as a daily digest entry or real-time alert depending on user preferences. The entire process---from discovery to notification---completes in under 10 seconds per candidate pair.

\subsection{Gamification Scaffolding Model}

\begin{figure}[h]
\centering
\begin{tikzpicture}[scale=0.9]

\draw[->] (0,0) -- (10,0) node[right] {Social Comfort};
\draw[->] (0,0) -- (0,6) node[above] {Interaction Intensity};

\draw[fill=green!20] (0,0) rectangle (3,1.5) node[pos=.5,align=center] {Solo\\Exploration};
\draw[fill=yellow!20] (3,0) rectangle (6,3) node[pos=.5,align=center] {Territory\\Battles};
\draw[fill=orange!20] (6,0) rectangle (8,4.5) node[pos=.5,align=center] {Boss\\Fights};
\draw[fill=red!20] (8,0) rectangle (10,6) node[pos=.5,align=center] {Direct\\Meeting};

\draw[thick,->,dashed] (1.5,0.75) -- (4.5,1.5) -- (7,2.25) -- (9,3);
\node at (5,-1) {Progressive scaffolding reduces anxiety through gameplay};

\end{tikzpicture}
\caption{Gamification scaffolding showing how game mechanics progressively introduce social interaction}
\end{figure}

The scaffolding model visualizes the core design hypothesis: that gaming mechanics can serve as graduated steps toward direct social interaction for users who experience social anxiety. The horizontal axis represents increasing social comfort, while the vertical axis tracks interaction intensity. \textit{Solo Exploration}---capturing territories alone, at self-chosen times and locations---provides the lowest-pressure entry point, requiring no social interaction whatsoever. \textit{Territory Battles} introduce indirect competition where players contest the same locations but need not communicate directly. \textit{Boss Fights} require real-time cooperation with other players, but the structured combat context (clear objectives, defined roles, shared adversary) provides conversational scaffolding that reduces the ambiguity of unstructured social contact. Finally, \textit{Direct Meeting} represents the target outcome: an in-person encounter between matched users, facilitated by shared gaming experiences that provide natural conversation topics. The dashed progression line shows that this transition is gradual rather than binary, with each stage building comfort incrementally.

\subsection{Pendant Companion State Machine}

\begin{figure}[h]
\centering
\begin{tikzpicture}[font=\small]

\tikzstyle{state} = [circle, draw, minimum size=1.2cm, font=\small]
\tikzstyle{transition} = [->,>=stealth,thick]

\node[state] (idle) at (0,0) {Idle};
\node[state] (exploring) at (5,0) {Exploring};
\node[state] (combat) at (9,-3.5) {Combat};
\node[state] (social) at (5,-3.5) {Social};
\node[state] (reflect) at (0,-3.5) {Reflecting};

\draw[transition] (idle) -- node[above] {User moves} (exploring);
\draw[transition] (exploring) -- node[right] {Battle starts} (combat);
\draw[transition] (exploring) -- node[left] {Twin match} (social);
\draw[transition] (combat) to[out=150,in=0] node[above,pos=0.4] {Battle ends} (idle);
\draw[transition] (social) -- node[above] {Interaction ends} (reflect);
\draw[transition] (reflect) -- node[left] {Processing complete} (idle);
\draw[transition] (idle) edge[loop above] node{30s idle} (idle);

\end{tikzpicture}
\caption{State machine diagram showing pendant companion behavioral states and transitions}
\end{figure}
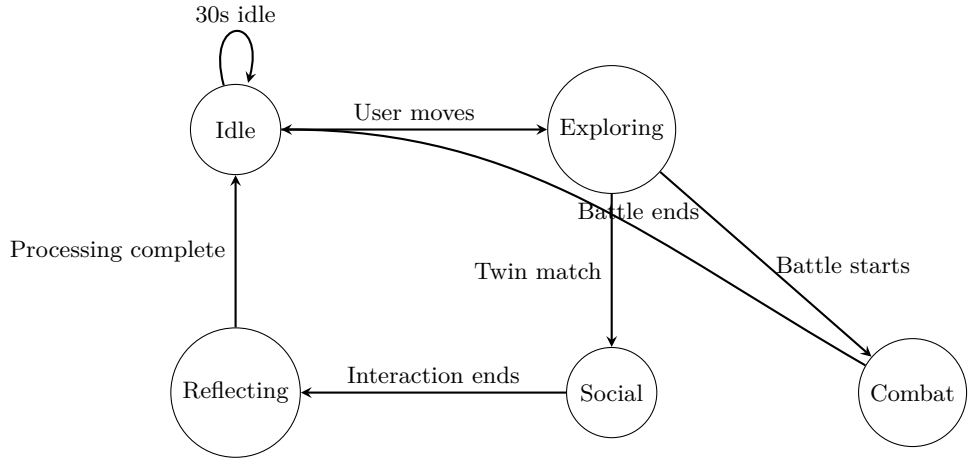

The pendant companion operates as a five-state finite state machine managed by the \texttt{PendantCompanion} coordinator. The \textit{Idle} state is the default, in which the \texttt{ProactiveInteractionSystem} monitors for interaction opportunities every 10 seconds with a 10\% trigger chance; after 30 seconds of continuous idle, the system initiates a context-appropriate proactive message (time-of-day greeting, mood check, or memory recall). User movement transitions the companion to the \textit{Exploring} state, where scene analysis detects the surrounding environment type (forest, castle, waterside, indoor) within a 50-unit radius and generates contextual observations. A battle encounter triggers the \textit{Combat} state, in which the \texttt{CompanionCombat} module provides attack support (300ms ranged / 500ms melee cooldowns) and tactical suggestions. A twin match notification transitions to the \textit{Social} state, where the companion provides emotional scaffolding---encouragement before meetings, processing support after rejections. The \textit{Reflecting} state follows social interactions, during which the companion consolidates interaction memories and propagates high-importance items ($\tau \geq 0.7$) to the cross-device Firebase store before returning to Idle.

\subsection{User Persona Journey Maps}

\subsubsection{Persona 1: Remote Worker Rachel (28)}

Rachel works from home and recently moved to a new city, leaving her without a local social network and struggling to find reasons to leave her apartment. Her journey begins when she sees a friend's social media post about conquering a downtown territory, sparking curiosity. She downloads the app during a lunch break and creates her digital twin while walking her dog, capturing the local park as her first territory. The territory bonuses---1.2$\times$ movement speed, 15\% experience boost---give her a concrete incentive to explore neighborhoods she would otherwise never visit. Her first boss battle pairs her with neighbors she has never met; the structured combat context (shared objective, clear roles, 800ms attack rhythm) provides natural conversation scaffolding that bypasses the awkwardness of unstructured small talk. Her twin, meanwhile, autonomously evaluates compatibility with other users and matches her with another remote worker whose personality profile shows complementary traits. They meet for coffee at a territory both frequent. Within weeks, weekly territory raids become the social highlight of Rachel's routine---the game provides both the reason to leave home and the conversational foundation that her previous dating app experiences lacked.

\subsubsection{Persona 2: Anxious Alex (35)}

Alex lives with social anxiety disorder---he wants connection but fears judgment, and conventional dating platforms amplify his anxiety through profile-based evaluation and unstructured messaging. His journey begins with a week of hesitation: he downloads the app but does not open it, a pattern consistent with the 75\% social anxiety prevalence observed in the pilot sample. When he finally engages, he starts with solo territory capture during quiet morning hours, avoiding any social contact while still participating in the platform's core loop. Over the following days, he develops trust through pendant companion conversations; the companion's persistent memory and proactive check-ins (triggered after 30 seconds of idle) create a sense of being understood without the performance pressure of human interaction. His digital twin begins making connections autonomously---Alex can observe twin conversation logs and compatibility scores without being required to act on them, maintaining a sense of agency and control. His first human interaction comes through a structured boss battle, where the combat context (shared objective, defined mechanics, time-limited encounter) provides enough scaffolding to make the social exposure manageable. Gradually, Alex increases direct interaction as comfort grows, with the pendant companion providing emotional processing support after each social encounter. The design accommodates his needs at every stage: he can engage entirely through his twin initially, the companion provides coaching and reassurance, and the gaming structure replaces open-ended social ambiguity with clear, bounded interaction frames.

\subsection{Information Architecture}

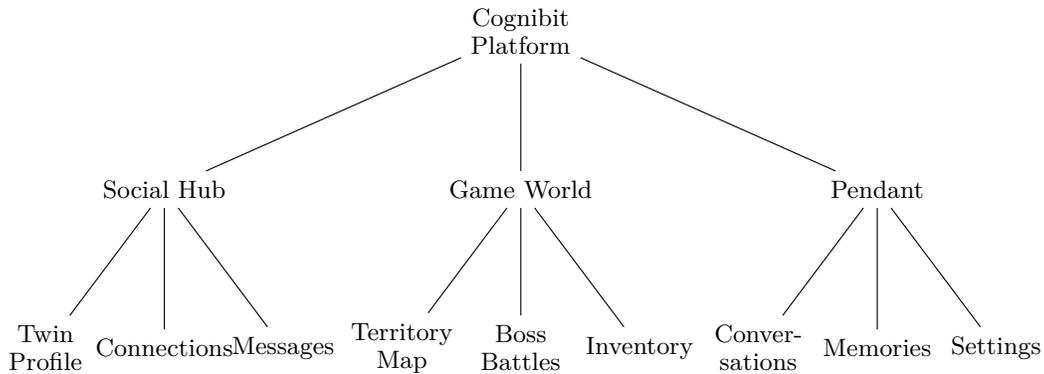
\begin{figure}[h]
\centering
\resizebox{\textwidth}{!}{%
\begin{tikzpicture}[font=\small,
    every node/.style={align=center}
]

\node (root) at (6,0) {Cognibit\\Platform};

\node (hub) at (1.5,-2) {Social Hub};
\node (game) at (6,-2) {Game World};
\node (pendant) at (10.5,-2) {Pendant};

\draw (root) -- (hub);
\draw (root) -- (game);
\draw (root) -- (pendant);

\node (tp) at (0,-4) {Twin\\Profile};
\node (conn) at (1.5,-4) {Connections};
\node (msg) at (3,-4) {Messages};
\draw (hub) -- (tp);
\draw (hub) -- (conn);
\draw (hub) -- (msg);

\node (tmap) at (4.5,-4) {Territory\\Map};
\node (boss) at (6,-4) {Boss\\Battles};
\node (inv) at (7.5,-4) {Inventory};
\draw (game) -- (tmap);
\draw (game) -- (boss);
\draw (game) -- (inv);

\node (conv) at (9,-4) {Conver-\\sations};
\node (mem) at (10.5,-4) {Memories};
\node (set) at (12,-4) {Settings};
\draw (pendant) -- (conv);
\draw (pendant) -- (mem);
\draw (pendant) -- (set);

\end{tikzpicture}%
}
\caption{Information architecture showing main platform components and navigation structure}
\end{figure}

The information architecture organizes the platform into three primary branches corresponding to the three pillars. The \textit{Social Hub} branch contains Twin Profile management, the Connections graph (followers, mutual connections, compatibility-ranked suggestions), and Messages (direct conversations and twin-mediated introductions). The \textit{Game World} branch provides access to the Territory Map (GPS-driven territory visualization with ownership colors and capture mechanics), Boss Battles (real-time cooperative combat encounters), and Inventory (equipment, resources, and cosmetics). The \textit{Pendant} branch houses Conversations (persistent AI companion chat with cross-device memory), Memories (accumulated interaction history organized by topic with importance-weighted highlighting), and Settings (personality trait sliders, module weight configuration, notification preferences). Navigation between branches is seamless---a twin match notification in the Social Hub can link directly to the matched user's territory on the Game World map, and the Pendant's memory view surfaces relevant twin conversation outcomes regardless of which branch the user is currently viewing.

\subsection{Emotional Journey Mapping}

\begin{figure}[h]
\centering
\begin{tikzpicture}[scale=0.8]

\draw[->] (0,0) -- (12,0) node[right] {Time};
\draw[->] (0,-2) -- (0,4) node[above] {Emotional State};

\draw[dashed,gray] (0,0) -- (12,0);

\draw[thick,blue] (0,0) .. controls (1,0.5) and (2,-0.5) .. (3,0.5)
    .. controls (4,1) and (5,0.5) .. (6,2)
    .. controls (7,1.5) and (8,2.5) .. (9,1.8)
    .. controls (10,2.2) and (11,2) .. (12,2.5);

\node[circle,fill=red,inner sep=2pt] at (3,0.5) {};
\node[below,align=center] at (3,-0.5) {First twin\\match};

\node[circle,fill=red,inner sep=2pt] at (6,2) {};
\node[above,align=center] at (6,2.5) {First boss\\battle};

\node[circle,fill=red,inner sep=2pt] at (9,1.8) {};
\node[below,align=center] at (9,-0.5) {In-person\\meeting};

\end{tikzpicture}
\caption{Emotional journey showing user's emotional state evolution through platform engagement}
\end{figure}
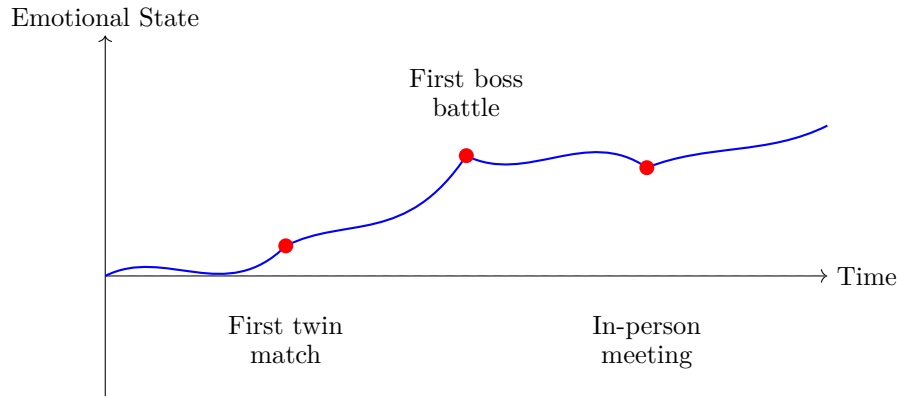

The emotional journey visualization tracks the aggregate emotional trajectory observed across pilot participants, with three key inflection points marked. The \textit{first twin match} (approximately day 3) produces a moderate positive spike as users receive their initial compatibility result---the first concrete evidence that the system is working on their behalf. The \textit{first boss battle} (approximately day 5--7) generates the largest emotional peak, as cooperative combat creates shared excitement and natural bonding with teammates. The \textit{in-person meeting} (approximately day 10--12) shows a slight emotional dip before the event (anticipatory anxiety) followed by sustained elevation, consistent with the pattern where the hardest step---transitioning from digital to physical interaction---produces the most durable positive outcome. The overall upward trajectory from baseline to sustained positive engagement reflects the cumulative effect of the three pillars working in concert: each successful micro-interaction (territory capture, companion conversation, twin match) incrementally builds social confidence.

These journey maps and diagrams illustrate the multi-faceted user experience of Cognibit, showing how the three-pillar design creates scaffolded pathways from isolation to connection. The visual representations helped our design team maintain focus on user needs throughout development and communicate complex interaction patterns to stakeholders.
\section{Technical Architecture Details}
\label{appendix:technical-architecture}

This appendix provides detailed technical specifications for practitioners seeking to implement similar systems.

\subsection{System Architecture Layers}

\subsubsection{Layer 1 - Discovery (Twin Networking)}
The discovery layer manages autonomous twin-to-twin compatibility evaluation. Compatibility checks execute every 60 seconds, scanning the candidate pool within a 50-mile discovery radius for local connections. A 20\% minimum compatibility threshold gates connection formation, filtering out clearly incompatible pairs before expensive LLM evaluation. Relationships progress through 5 stages from strangers to best friends, with each stage unlocking additional information sharing and interaction capabilities. The layer synchronizes through Firebase with 3--5 second eventual consistency, maintains WebSocket connections for real-time twin conversation streaming, and authenticates all requests through JWT tokens with 24-hour expiry managed by the CentralAuthenticationService module.

\subsubsection{Layer 2 - Engagement (Gaming Mechanics)}
The engagement layer implements the territory conquest and combat systems that motivate physical movement and co-located interaction. Cities define interaction zones with a 10 game unit radius from City Hall (1 game unit = 1 meter), within which players receive territory benefits of +20\% movement speed and +10\% health regeneration. Resource generation starts at 100 gold/hour and scales with city level up to 500/hour, incentivizing territory defense. Defense NPCs spawn at 3 base plus 1 per city level (maximum 8), providing AI opponents for solo gameplay. Boss encounters scale dynamically: health is computed as baseHealth $\times$ (1 + playerLevel $\times$ 0.1) for solo fights, with +50\% boss health per additional player in team combat to maintain difficulty. The combat renderer targets 60 FPS with a 30 FPS minimum usability threshold, using 100ms client-side prediction to compensate for network latency during multiplayer encounters.

\subsubsection{Layer 3 - Support (Pendant Companion)}
The support layer implements the pendant companion through a modular architecture with separate UI, AI, and Combat subsystems that can be independently updated and tested. Combat parameters include 300ms cooldowns for ranged attacks and 500ms for melee, with a maximum of 10 concurrent projectiles to prevent visual clutter and memory overhead. The ProactiveInteractionSystem triggers companion-initiated conversations after 30 seconds of user idle time, providing emotional scaffolding and gameplay suggestions without requiring user initiation. Memory persists across all devices via Firebase, ensuring the companion maintains conversational continuity regardless of which device the user accesses. The AI subsystem operates within a 4096-token context window for conversation history, with a response latency target of under 2 seconds at the 90th percentile, achieved through GPT-4o-mini's lower computational overhead compared to the full GPT-4o used for twin conversations.

\subsubsection{Layer 4 - Physical Integration (GPS Tracking)}
The physical integration layer bridges the digital game world with real-world geography through continuous location tracking. A 20-meter movement detection threshold filters GPS noise while remaining sensitive enough to detect walking movement between nearby territories. Update throttling enforces a 2-second minimum between location updates to prevent excessive battery drain and network traffic. High accuracy mode combines GPS, WiFi, and cellular triangulation for the best available position estimate, particularly in urban environments where GPS alone is unreliable. Battery optimization reduces polling frequency when the device is stationary, switching from active 2-second polling to passive geofence monitoring. Territory bonuses apply within a 500 game unit radius (1 game unit = 1 meter) from city centers, and 100-meter geofence boundaries trigger location-specific events such as territory entry notifications and proximity-based encounter alerts. Background tracking leverages iOS and Android background location services to maintain territory awareness even when the app is not in the foreground.

\subsection{Browser-Based Implementation Constraints}

\subsubsection{Memory Management}
Browser environments impose strict memory limitations requiring careful optimization. The effective application memory budget is approximately 2GB in Chrome and Firefox (lower on mobile browsers), with 5--8 concurrent digital twins representing the optimal operating range before frame rate degradation becomes noticeable (the system supports up to 20 with progressively degraded performance). Aggressive garbage collection runs every 60 seconds to reclaim memory from completed twin conversations and expired game state objects. Object pooling pre-allocates frequently created entities---combat damage numbers, particle effects, territory markers---to avoid garbage collection pauses during active gameplay. Texture atlasing combines multiple sprite sheets into unified textures to reduce GPU draw calls, and a four-tier Level of Detail (LOD) system reduces rendering complexity for distant objects, applying full-quality animation only to characters within 30 game units of the camera.

\subsubsection{Performance Optimization Strategies}
Six browser-native APIs are leveraged for performance optimization. RequestAnimationFrame synchronizes rendering with the display refresh rate, preventing unnecessary frame computations and ensuring smooth 60 FPS output when the GPU can sustain it. Web Workers offload AI computation---specifically GNWT salience competition and coalition formation---to a background thread, preventing cognitive processing from blocking the main rendering loop. IndexedDB provides local data persistence for twin conversation history and territory state, enabling fast reads without network round trips. Service Workers intercept network requests to provide offline functionality, serving cached assets and queuing API calls for later execution when connectivity is restored. WebAssembly modules were explored for performance-critical paths such as damage calculation and pathfinding, though the 10\% improvement did not justify the added build complexity in the current prototype. Virtual DOM diffing minimizes UI update costs by computing the minimal set of DOM mutations needed to reflect state changes.

\subsection{Firebase Integration Architecture}

\subsubsection{Real-time Database Structure}

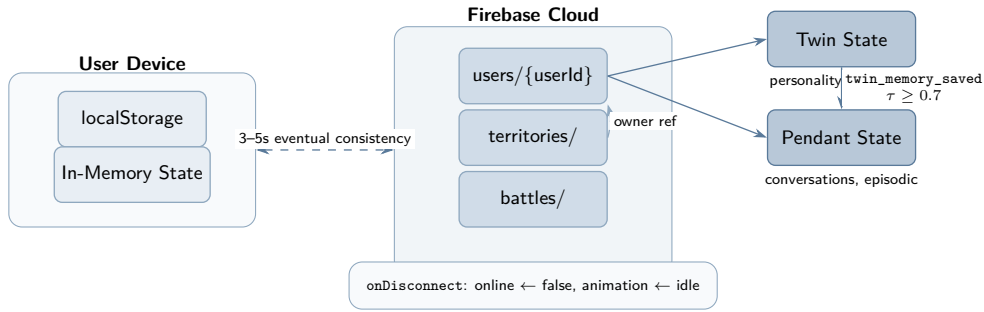
\begin{figure}[!htbp]
\centering
\begin{tikzpicture}[cognibit, scale=0.82, transform shape]
\node[ccontainer, minimum width=4cm, minimum height=2.5cm] (device) at (0, 0) {};
\node[font=\sffamily\footnotesize\bfseries] at (0, 1.4) {User Device};
\node[cbox] (localstorage) at (0, 0.5) {localStorage};
\node[cbox] (ram) at (0, -0.4) {In-Memory State};

\node[ccontainer, fill=cfillMed!30, minimum width=4.5cm, minimum height=4cm] (firebase) at (6.5, 0) {};
\node[font=\sffamily\footnotesize\bfseries] at (6.5, 2.2) {Firebase Cloud};
\node[cbox proc] (users) at (6.5, 1.2) {users/\{userId\}};
\node[cbox proc] (territories) at (6.5, 0.2) {territories/};
\node[cbox proc] (battles) at (6.5, -0.8) {battles/};

\node[cbox dark] (twin) at (11.5, 1.8) {Twin State};
\node[font=\sffamily\scriptsize, below=0pt of twin] {personality, memories};
\node[cbox dark] (pendant) at (11.5, 0.2) {Pendant State};
\node[font=\sffamily\scriptsize, below=0pt of pendant] {conversations, episodic};

\draw[<->, draw=cborder, line width=0.5pt, dashed] (device) -- node[clabel, above] {3--5s eventual consistency} (firebase);

\draw[cflow] (users.east) -- (twin.west);
\draw[cflow] (users.east) -- (pendant.west);

\draw[cflow dash] (territories.east) to[bend right=15] node[clabel, right] {owner ref} (users.south east);

\draw[cflow] (twin) -- node[clabel, right, align=center] {\texttt{twin\_memory\_saved}\\$\tau \geq 0.7$} (pendant);

\node[ccontainer, font=\sffamily\scriptsize, align=left] at (6.5, -2.2) {\texttt{onDisconnect}: online $\gets$ false, animation $\gets$ idle};
\end{tikzpicture}
\caption{Firebase Realtime Database schema and cross-device synchronization dataflow. Bidirectional sync between user devices and the Firebase cloud achieves 3--5 second eventual consistency. The twin$\leftrightarrow$pendant memory bridge propagates high-importance memories ($\tau \geq 0.7$) via the \texttt{twin\_memory\_saved} event. Grounded in: \texttt{PendantMemorySync.js}, \texttt{NetworkSync.js}.}
\label{fig:firebase-schema}
\end{figure}

Figure~\ref{fig:firebase-schema} illustrates the Firebase Realtime Database schema and cross-device synchronization dataflow. The three top-level collections (\texttt{users/}, \texttt{territories/}, \texttt{battles/}) are linked by ownership references, with bidirectional sync between user devices and the Firebase cloud source of truth achieving eventual consistency within 3--5 seconds. The twin$\leftrightarrow$pendant memory bridge propagates high-importance memories ($\tau \geq 0.7$) via the \texttt{twin\_memory\_saved} window event, ensuring the companion can reference twin conversation outcomes across devices and application contexts.

\subsubsection{Synchronization Protocols}
Five synchronization strategies ensure data consistency across devices. Optimistic updates apply state changes locally before server confirmation, providing immediate UI responsiveness with rollback on conflict detection. Operational Transformation resolves concurrent edits to shared data paths---particularly twin relationship state that may be modified simultaneously from the game world and the Social Hub. Exponential backoff for retry logic (3 attempts, base delay 1 second, factor 2) prevents cascading failures during transient network disruptions without overwhelming the Firebase backend. Delta synchronization transmits only changed fields rather than complete objects, minimizing bandwidth consumption for the frequent small updates that characterize twin state evolution. An offline queue buffers actions performed without connectivity, flushing them to Firebase in order upon reconnection to maintain temporal consistency.

\subsection{LLM Integration for Digital Twins}

\subsubsection{API Configuration}
The deployed twin conversation pipeline uses a two-stage configuration (see Appendix~\ref{appendix:twin-pipeline} for full details):
The primary model is GPT-4o (or equivalent) for twin conversations, invoked through the Node.js API proxy. Stage 1 (intent analysis) uses temperature 0.3 with a 50-token maximum to classify conversational intent with high confidence. Stage 2 (personality-conditioned generation) uses temperature 0.8 and top\_p 0.9 with a 300-token maximum, providing sufficient randomness for natural-sounding conversation while staying within the personality constraints. A frequency penalty of 0.3 reduces repetitive phrasing across turns, and a presence penalty of 0.5 encourages topic diversity, preventing twins from perseverating on narrow conversational themes.

\subsubsection{Prompt Engineering}
Twin personality prompt structure:
\begin{small}\begin{verbatim}
You are a digital twin representing [User Name].
Personality traits: [trait scores: openness, friendliness, playfulness, loyalty, independence]
Communication style: [formal/casual/playful]
Interests: [list of interests]
Current context: [location, activity, mood]
Conversation goal: Assess compatibility with other twin
Response guidelines: Be authentic to user's personality
\end{verbatim}\end{small}

\subsection{Security Considerations}

\subsubsection{Data Protection}
Data protection employs six layers of defense. Twin conversations are encrypted end-to-end to prevent server-side interception of sensitive personality and preference data. Public-facing APIs use hashed user IDs rather than sequential integers, preventing user enumeration attacks. Rate limiting enforces a ceiling of 100 requests per minute per user (implemented as 100 requests per 15-minute window per IP in the API proxy), preventing automated scraping and abuse. CORS configuration restricts API access to authorized origins, blocking cross-origin requests from unauthorized domains. Content Security Policy headers prevent the injection of unauthorized scripts, and XSS protection through express-validator input sanitization (message length 1--5000 characters, HTML entity encoding) prevents stored and reflected cross-site scripting attacks.

\subsubsection{Authentication Flow}
Authentication is managed through the CentralAuthenticationService module using Firebase Auth as the identity provider. OAuth 2.0 social login enables registration through existing Google or Apple accounts, reducing friction during onboarding. Multi-factor authentication is available as an optional security enhancement for users who enable it. Session management uses secure HTTP-only cookies with JWT tokens carrying 24-hour expiry, automatically refreshed through a background token refresh mechanism that obtains new credentials before expiration. Account recovery via email provides a self-service path for users who lose access to their authentication credentials.

\subsection{Scalability Architecture}

\subsubsection{Horizontal Scaling}
The scalability architecture is designed for horizontal growth beyond the pilot's single-instance deployment. A microservices architecture enables independent scaling of twin conversation processing, territory state management, and companion interaction handling based on their respective load profiles. Load balancing distributes incoming requests across multiple server instances, preventing any single node from becoming a bottleneck. A CDN delivers static assets (3D models, textures, audio files) from edge locations near users, reducing initial load times. Regional Firebase instances provide low-latency database access for geographically distributed user populations. Queue-based processing decouples twin interaction requests from synchronous response requirements, enabling the system to absorb traffic spikes by buffering evaluation requests and processing them at sustainable throughput rates.

\subsubsection{Performance Metrics}
Target performance indicators define the acceptable operating envelope for user-facing interactions. Page load should complete in under 3 seconds on a 3G connection, with time to interactive under 5 seconds, ensuring that users on slower networks can begin engaging before all assets finish loading. API response targets are under 200ms at the 50th percentile and under 500ms at the 95th percentile for server-side operations. Twin matching evaluations should complete in under 10 seconds per candidate pair, encompassing the full two-stage LLM call sequence. GPS update latency of under 1 second ensures that territory transitions feel responsive to physical movement. Firebase synchronization should achieve cross-device consistency within 5 seconds, ensuring that actions taken on one device are reflected on others before the user switches context.

\subsection{Development and Deployment}

\subsubsection{Technology Stack}
The system is built on a deliberately minimal technology stack to reduce dependency complexity. The frontend uses vanilla JavaScript with ES6+ modules rather than a framework like React or Vue, avoiding framework overhead and build toolchain complexity at the cost of manual DOM management. State management relies on a custom event-driven architecture using the EventTarget API for inter-component communication. Three.js provides the 3D graphics pipeline for avatar rendering, combat visualization, and territory display. The backend runs on Node.js with Express, hosting the API proxy that routes LLM requests and manages authentication. Firebase Realtime Database serves as both the primary data store and the real-time synchronization layer. Firebase Auth handles user identity, and Firebase Hosting with CDN delivers the web application. Monitoring combines Google Analytics for user behavior tracking with Sentry for error reporting and performance monitoring.

\subsubsection{CI/CD Pipeline}
The continuous integration and deployment pipeline automates the path from code commit to production. Git-based version control tracks all changes with feature branching for parallel development. Automated testing using Jest and Mocha runs on every commit, executing the existing 34\% coverage test suite to catch regressions in covered paths. ESLint enforces code quality standards and consistent formatting across the 103,847-line codebase. GitHub Actions orchestrate the deployment pipeline, running tests, building assets, and deploying to Firebase Hosting. A staging environment mirrors production configuration for pre-release validation. Blue-green deployment maintains two identical production environments, routing traffic to the new version only after health checks pass, with rollback capability within 5 minutes if post-deployment monitoring detects anomalies.

\subsection{System Architecture Details}
\label{appendix:architecture-details-q}

\subsubsection{Module hierarchy}
The 200+ modules are organized into core systems (event system, resource manager, configuration), cognitive architecture systems (GNWT agent, specialist processors, working memory, attention), emotion systems (PAC agent, predictive model, affective state, allostasis), digital twin systems (soul, personality evolution, memory, behavior patterns, interaction history), and integration systems (GPS, Firebase sync, combat, animation).

\subsubsection{Firebase synchronization}
Six sync modules handle real-time data: ChatFirebaseSync, FeedFirebaseSync, InsightsFirebaseSync, DigestFirebaseSync, SchedulerFirebaseSync, and TakeoverFirebaseSync. Retry logic implements 3 attempts with exponential backoff. Batch operations optimize multi-document updates.

\subsubsection{GPS implementation}
The GPS system uses 20-meter movement threshold to filter noise, 2-second throttling to reduce battery drain, Haversine distance calculations for accuracy, and local processing for privacy protection. Territory boundaries use 500-game-unit radius (1 game unit = 1 meter) for ownership benefits.

\subsection{Complete Technical Specifications}
\label{appendix:specifications-q}

\Cref{tab:specs-q} presents all system parameters used in the implementation.

\begin{table}[h]
\centering
\small
\begin{tabular}{lll}
\toprule
\textbf{Component} & \textbf{Parameter} & \textbf{Value} \\
\midrule
Discovery & Radius & 50 miles \\
& Update Interval & 60 seconds \\
& Cache Expiration & 5 minutes \\
& Compatibility Min & 20\% \\
\midrule
GNWT & Cycle Time & 100ms \\
& Workspace Capacity & 7$\pm$2 items \\
& Attention Capacity & 3 items \\
\midrule
PAC & Update Rate & 100ms \\
& Learning Rate & 0.1 \\
& Prediction Horizon & 5 seconds \\
\midrule
GPS & Movement Threshold & 20 meters \\
& Update Throttle & 2 seconds \\
& City Load Radius & 1 mile \\
\midrule
Combat & Bow/Gun Cooldown & 300ms \\
& Melee Cooldown & 500ms \\
& Max Projectiles & 10 \\
\midrule
NPC & Spawn Distance & 100m \\
& Despawn Distance & 150m \\
& Max Wilderness & 2 \\
\bottomrule
\end{tabular}
\caption{Complete system technical specifications (Compact)}
\label{tab:specs-q}
\end{table}

\subsection{System Architecture and Failure Analysis Diagrams}
\label{appendix:system-architecture-diagrams-q}

The four-layer architecture (Figure~\ref{fig:system-overview}) enables autonomous social discovery through gaming. The system integrates user-facing applications (Layer 1), core platform services including digital twins and game mechanics (Layer 2), shared resource modules (Layer 3), and external dependencies (Layer 4).

\subsubsection{Cognitive Module Architecture}
The cognitive architecture that powers digital twin behavior is organized around five deployed specialist modules---Emotion, Memory, Planning, Social Norms, and Goal Tracking---that compete for access to a shared global workspace. (The underlying \texttt{GlobalWorkspace} framework supports up to nine specialist slots, but only five are actively registered in the current deployment.) These modules compete via salience-weighted competition (see Appendix~\ref{appendix:gnwt} for full details). The \texttt{GlobalWorkspace} enforces a capacity limit of 9 items (using the upper bound of Miller's 7$\pm$2) with a broadcast entry threshold of $\tau = 0.7$ (normalized salience); items below threshold are placed in a sub-threshold buffer for proportional output blending rather than being discarded. The \texttt{AttentionController} computes salience as a weighted combination of novelty (0.3), relevance (0.3), urgency (0.2), and emotional intensity (0.2), with a habituation penalty that reduces salience for repeated stimuli and a 1.5$\times$ boost for items related to the current attentional focus. The workspace processes at a 100ms (10Hz) cycle time; each cycle decays non-attended items, forms coalitions among competing entries, and broadcasts winners to all registered specialists. Each specialist carries a static base weight (e.g., perception 0.7, executive 0.6, reasoning 0.5, evaluation 0.5) with adaptive weight updates at rate 0.02 per successful broadcast, bounded between 0.5 and 2.0. Heavy computation is offloaded to a Web Worker to avoid blocking the main rendering thread. Five personality traits (friendliness, openness, independence, loyalty, playfulness) modulate specialist weights, creating individualized behavioral differences across digital twins.

\subsubsection{Cross-Device Memory Persistence}
Cross-device synchronization ensures shared memories persist across the 3D game world, Social Hub, and pendant companion through a three-layer persistence stack: in-memory state for immediate session access, localStorage for per-device persistence surviving page reloads, and Firebase Realtime Database as the cloud source of truth with 3--5 second eventual consistency. Writes use acknowledgement with retry and exponential backoff (3 attempts, base delay 1 second, factor 2); failed writes are queued in an offline buffer and flushed on reconnection. Conflict resolution employs per-item timestamp-based last-write-wins with deduplication by the tuple (speaker, timestamp, message content hash). A Twin$\leftrightarrow$Pendant memory bridge propagates high-importance memories ($\tau \geq 0.7$) between the twin conversation system and the pendant companion via a shared event bus, ensuring the companion can reference twin conversation outcomes when providing emotional scaffolding (see Appendix~\ref{appendix:sync-protocols} for the full synchronization algorithm).

\subsubsection{Planned vs Actual Capabilities}
The deployed system realized the core architecture---autonomous twin matching, GNWT-driven cognition, personality evolution, and GPS-based territory mechanics---but several planned capabilities were reduced or deferred due to browser constraints and development timeline. The system supports up to 20 concurrent digital twins but operates optimally at 5--8, as performance degrades exponentially beyond that point ($\text{FPS} = 84.9 \times e^{-0.125n}$, crossing the 30 FPS usability threshold at $n \approx 8$); the \texttt{AnimationLODSystem} mitigates this by applying distance-based level-of-detail tiers (full quality within 30 units, reduced update rate at 30--60 units, frozen pose at 60--100 units, hidden beyond 100 units), which reduced per-frame bone-skinning overhead from 150ms+ (233 skeletons, 14{,}997 bones) to under 16ms. WebRTC-based real-time voice communication was planned but only partially implemented; Safari support remains limited. Server-side agent processing was designed but browser-only deployment was used for the pilot, constraining AI computation to the client-side JavaScript event loop. A full gap analysis appears in the limitations section (Section~\ref{sec:limitations}).

\subsubsection{LLM API Call Sequence}
The deployed twin conversation pipeline uses a two-stage sequential LLM call pattern, routed through a Node.js/Express API proxy (\texttt{api-proxy.js}) that centralizes key management and enforces security boundaries. The proxy applies Helmet security headers (CSP, CORS), rate limiting (100 requests per 15-minute window per IP), and express-validator input sanitization (message length 1--5000 characters, conversation history capped at 100 entries, temperature 0--2, max tokens 1--4000). Stage 1 (intent analysis) invokes GPT-4o with temperature 0.3 and a 50-token limit to classify the conversational intent and extract context cues. Stage 2 (personality-conditioned generation) invokes the same model with temperature 0.8, top\_p 0.9, and a 300-token limit, conditioned on the twin's personality traits, conversation history (last 10 messages), and the Stage 1 intent output. The proxy supports three LLM backends---OpenAI, Anthropic Claude, and OpenRouter (free tier)---selectable via environment variables, enabling cost-tier switching without client code changes. The pendant companion uses GPT-4o-mini for lower-cost, lower-latency interactions. Over the 14-day pilot, this pipeline processed 13,680 LLM API calls across 342 twin sessions, with a 90th-percentile response latency under 2 seconds and total LLM costs of \$246.24.

\subsubsection{Browser Technical Limitations}
The browser execution environment imposes three principal constraints on the system. First, JavaScript's single-threaded event loop serializes all operations, creating an average 2.8-second latency for sequential API calls; this is mitigated through async/await patterns with request batching and Web Workers for AI computation offloading (the \texttt{GlobalWorkspace} initializes a dedicated Web Worker for salience competition and coalition formation). Second, the effective application memory budget is approximately 2GB in Chrome (1536MB in Firefox, as low as 384MB on mobile Safari), limiting the system to 5--8 concurrent twins at acceptable frame rates (see Appendix~\ref{appendix:technical-limitations}, Table~\ref{tab:performance-degradation} for the full degradation curve). The \texttt{AnimationLODSystem} addresses the most severe bottleneck: at scale, 233 skeletons with 14{,}997 bones consumed 150ms+ per frame; the four-tier distance-based LOD (full at $<$30 units, reduced every 3 frames at 30--60, frozen pose at 60--100, hidden with visibility toggled off beyond 100 units) recovers frame budget by suppressing bone-skinning calculations for distant characters. The LOD evaluation itself is throttled to every 10th frame to minimize traversal overhead. Third, browser connection limits cap the number of simultaneous WebSocket and Firebase listeners, requiring aggressive garbage collection every 60 seconds, object pooling for frequently created entities (e.g., the combat system pre-allocates 50 damage-number DOM elements), and texture atlasing to reduce draw calls. These constraints drove several architectural decisions, including the use of IndexedDB for local persistence and Service Workers for offline functionality.

\subsubsection{Cost Analysis}
Over the 14-day pilot deployment (N=20), total per-user cost was \$16.58, dominated by LLM API charges: GPT-4o twin conversations accounted for \$12.31/user (averaging 150 calls/day at ${\sim}$4,250 tokens each), while GPT-4o-mini pendant companion chat contributed \$1.92/user (200 calls/day at ${\sim}$2,000 tokens each). Infrastructure costs were modest---Firebase reads/writes totaled \$1.30/user over 14 days, with CDN transfer at \$0.85/user. The API proxy architecture supports three backend providers (OpenAI, Anthropic, OpenRouter) configurable via environment variables, enabling runtime cost-tier switching: the OpenRouter free-tier endpoint was used during development to eliminate API costs entirely, while the production deployment used the GPT-4o/GPT-4o-mini hybrid. The Pareto frontier analysis (Section~\ref{sec:pareto}) reveals that GPT-4o-mini retains 90\% of GPT-4o's matching precision at 6\% of the cost, identifying the deployment sweet spot for latency-sensitive operation. Aggregate LLM costs across all 20 participants totaled \$246.24 for 13,680 API calls and 41,520 Firebase operations (see Appendix~\ref{appendix:technical-limitations} for the full cost breakdown).

\subsection{Memory and Telemetry Schemas}

\subsubsection{Memory Document Schema}
Each memory record stored in the Firebase Realtime Database follows this structure:

\begin{table}[!htbp]
\centering
\caption{Memory Document Schema}
\label{tab:memory-schema}
\begin{tabular}{llp{6cm}}
\toprule
\textbf{Field} & \textbf{Type} & \textbf{Description} \\
\midrule
speaker & string & Identifier of the message source (user or twin) \\
message & string & The original message content \\
response & string & The response generated \\
timestamp & integer & Unix timestamp (milliseconds) \\
importance & float [0,1] & Importance weight for retrieval ranking \\
twinId & string & Associated digital twin identifier \\
keywords & string[] & Extracted keywords for search \\
\bottomrule
\end{tabular}
\end{table}

\subsubsection{Memory Persistence Stack}
The system uses a three-layer persistence architecture to balance access speed against durability. The first layer, in-memory state, provides immediate access for the current session and is updated synchronously on each interaction, ensuring zero-latency reads for active conversations but losing all data on page close. The second layer, localStorage, provides per-device persistence that survives page reloads and browser restarts but remains device-specific, serving as a fast warm cache when the user returns to a previously used device. The third layer, Firebase Realtime Database, serves as the cloud source of truth, providing cross-device synchronization with eventual consistency (3--5 seconds typical) and permanent storage that persists independently of any individual device or session.

\subsubsection{Conflict Resolution Protocol}
Write conflicts are resolved using per-item timestamp-based last-write-wins with deduplication. Each memory item is uniquely identified by the tuple (speaker, timestamp, message content hash), providing a collision-resistant identity that remains stable across devices. When merging from multiple devices, items with identical identification tuples are deduplicated, preventing the same memory from appearing multiple times after multi-device synchronization. For conflicting updates to the same item---where two devices modify the same memory record within the sync window---the version with the later timestamp is retained, implementing a simple but effective last-write-wins policy. Write operations use server acknowledgement with retry and exponential backoff (3 attempts, base delay 1 second, factor 2) to handle transient network failures. Writes that exhaust all retry attempts are queued in an offline buffer and flushed in order upon reconnection, ensuring no user-generated data is lost even during extended connectivity interruptions.

\subsubsection{Twin--Pendant Memory Bridge}
The memory bridge propagates key memories between the twin conversation system and the pendant companion system via a shared event bus. When a twin conversation produces a memory item that exceeds the importance threshold ($\tau = 0.7$), the twin system emits a \textit{memory-created} event on the shared bus. The pendant memory sync listener intercepts this event and stores the memory in the pendant's local conversation context, making it available for reference in subsequent companion interactions. On device reconnection after an offline period, the pendant synchronizes its accumulated local memory state with the Firebase cloud store, ensuring cross-device consistency. This bridge ensures that the pendant companion can reference twin conversation outcomes---such as a recent match acceptance or rejection---when providing emotionally appropriate scaffolding, maintaining coherence across the system's distinct interaction surfaces.

\subsubsection{Telemetry Event Schema}
Deployment telemetry records system events for analysis:

\begin{table}[!htbp]
\centering
\caption{Telemetry Event Schema}
\label{tab:telemetry-schema}
\begin{tabular}{llp{6cm}}
\toprule
\textbf{Field} & \textbf{Type} & \textbf{Description} \\
\midrule
eventType & string & Event category (twin\_session, territory\_capture, companion\_interaction, boss\_battle, etc.) \\
userId & string & Anonymized participant identifier \\
timestamp & integer & Unix timestamp (milliseconds) \\
sessionId & string & Current session identifier \\
payload & object & Event-specific data (varies by eventType) \\
deviceId & string & Device fingerprint for cross-device tracking \\
\bottomrule
\end{tabular}
\end{table}
\section{GNWT Implementation Details}
\label{appendix:gnwt}

The Global Neuronal Workspace Theory (GNWT) provides the cognitive architecture for digital twin behavior. In this framework, five specialist modules---Emotion, Memory, Planning, Social Norms, and Goal Tracking---operate in parallel, each processing incoming stimuli through its domain-specific lens. Specialists compete for access to a shared global workspace through salience-weighted selection; the winning specialist's output is broadcast to all other modules, shaping the twin's behavioral response through directive injection into the LLM prompt. This architecture, inspired by Baars' Global Workspace Theory \citep{Baars1988} and its neural implementation \citep{dehaene2011}, enables context-sensitive behavioral responses that account for emotional, social, and strategic considerations simultaneously rather than sequentially. The full architectural description appears in CogniPair \citep{CogniPair2026}. Figure~\ref{fig:gnwt-cycle} illustrates the processing cycle; the Cognibit field deployment uses the following calibrated parameters:

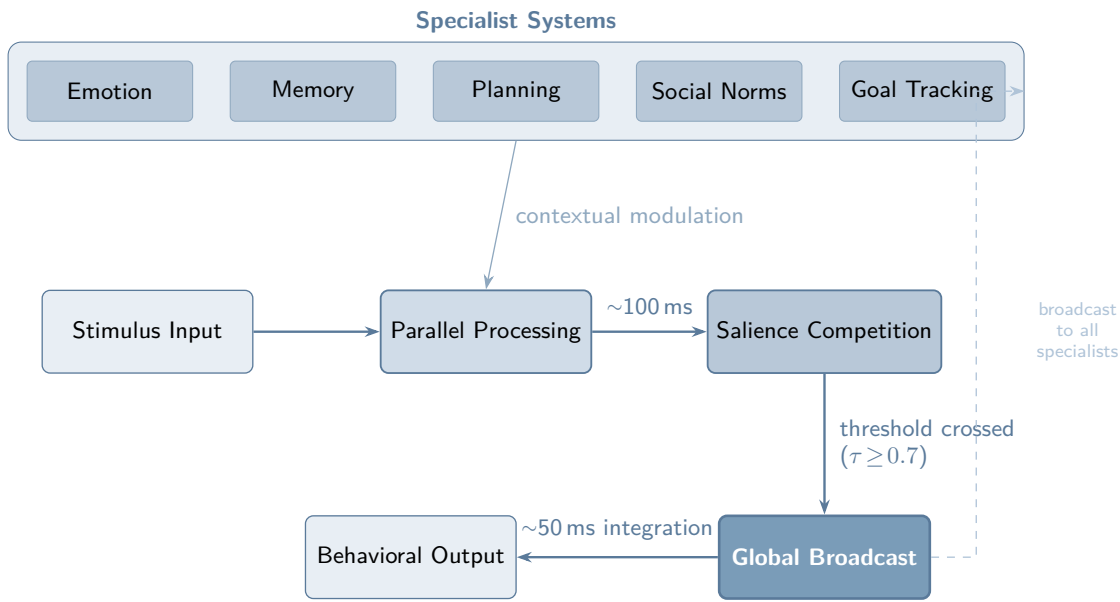
\begin{figure}[!htbp]
\centering
\begin{tikzpicture}[
    font=\sffamily\small,
    >={Stealth[length=2mm, width=1.4mm]},
    sbox/.style={rectangle, rounded corners=3pt, draw=cborder, line width=0.5pt, minimum width=28mm, minimum height=11mm, align=center},
    mod/.style={rectangle, rounded corners=2pt, draw=cborderLight, line width=0.4pt, minimum width=22mm, minimum height=8mm, align=center, font=\sffamily\footnotesize, fill=cfillDark},
]
\node[mod] (m1) at (0, 0) {Emotion};
\node[mod] (m2) at (2.7, 0) {Memory};
\node[mod] (m3) at (5.4, 0) {Planning};
\node[mod] (m4) at (8.1, 0) {Social Norms};
\node[mod] (m5) at (10.8, 0) {Goal Tracking};
\begin{pgfonlayer}{background}
\node[rounded corners=5pt, draw=cborder, line width=0.4pt, fill=cfillLight, inner sep=7pt, fit=(m1)(m5),
    label={[font=\sffamily\footnotesize\bfseries, text=cborder]above:Specialist Systems}] (modgrp) {};
\end{pgfonlayer}
\node[sbox, fill=cfillLight] (input) at (0.5, -3.2) {Stimulus Input};
\node[sbox, fill=cfillMed, line width=0.7pt] (proc) at (5.0, -3.2) {Parallel Processing};
\node[sbox, fill=cfillDark, line width=0.7pt] (comp) at (9.5, -3.2) {Salience Competition};
\node[sbox, fill=cfillAccent, draw=cborder, line width=1pt, font=\sffamily\small\bfseries, text=white] (bcast) at (9.5, -6.2) {Global Broadcast};
\node[sbox, fill=cfillLight] (output) at (4.0, -6.2) {Behavioral Output};
\draw[->, draw=cborder, line width=0.9pt] (input) -- (proc);
\draw[->, draw=cborder, line width=0.9pt] (proc) -- node[above=3pt, font=\sffamily\footnotesize, text=cborder] {{\raise.17ex\hbox{$\scriptstyle\sim$}}100\,ms} (comp);
\draw[->, draw=cborder, line width=0.9pt] (comp) -- node[right=2pt, font=\sffamily\footnotesize, text=cborder, align=left] {threshold crossed\\($\tau \!\geq\! 0.7$)} (bcast);
\draw[->, draw=cborder, line width=0.9pt] (bcast) -- node[above=3pt, font=\sffamily\footnotesize, text=cborder] {{\raise.17ex\hbox{$\scriptstyle\sim$}}50\,ms integration} (output);
\draw[->, draw=cborderLight, line width=0.5pt] (modgrp.south) -- node[right=2pt, font=\sffamily\footnotesize, text=cborderLight] {contextual modulation} (proc.north);
\coordinate (retR) at (12.0, 0);
\draw[->, dashed, draw=cborderFaint, line width=0.5pt] (bcast.east) -- ++(6mm, 0) |- (retR) -- (modgrp.east);
\node[font=\sffamily\scriptsize, text=cborderFaint, align=center, anchor=west] at (12.2, -3.2) {broadcast\\to all\\specialists};
\end{tikzpicture}
\caption{GNWT cognitive processing cycle (\texttt{GlobalWorkspace.js}). Five specialist modules process stimuli in parallel, then compete for global workspace access via salience-weighted selection (100ms). Winners above $\tau = 0.7$ are broadcast to all modules (50ms), which integrate (150ms) and update adaptive weights. Items below threshold enter a sub-threshold buffer for proportional blending. The cycle repeats at 10Hz.}
\label{fig:gnwt-cycle}
\end{figure}

\begin{itemize}
\item \textbf{Broadcast threshold}: $\tau = 0.7$ (normalized salience) for global workspace entry. Raw salience values are normalized before threshold comparison; the effective pre-normalization threshold is approximately 0.3 because personality modulation reduces raw specialist output values by 40--80\%. Items below threshold are placed in a sub-threshold buffer for proportional output blending rather than being discarded.
\item \textbf{Workspace capacity}: 9 items (using the upper bound of Miller's 7$\pm$2 \citep{Miller1956}).
\item \textbf{Attention capacity}: 3 concurrent focus items receiving a 1.5$\times$ salience boost; non-attended items decay by a factor of 0.9 per cycle.
\item \textbf{Cycle time}: 100ms (10Hz processing frequency). Faster cycles (50ms) caused frame drops due to event loop contention; slower cycles (200ms) produced noticeably delayed twin responses.
\item \textbf{Personality modulation}: Five traits (friendliness, openness, independence, loyalty, playfulness, each 0--100) modulate specialist base weights, creating individualized behavioral differences across digital twins.
\end{itemize}

Post-broadcast weight adaptation reinforces successful specialists: $W_{winner}^{t+1} = \min(W_{winner}^t \times (1 + 0.02 \times \mathit{success}),\; 2.0)$, where $\mathit{success}$ indicates whether the broadcast led to a coherent behavioral output (\texttt{moduleWeightAdaptRate = 0.02} in the codebase). The upper bound of 2.0 prevents any single specialist from permanently dominating the workspace, while the 0.02 learning rate ensures very gradual adaptation over hundreds of interactions rather than abrupt behavioral shifts. The complete GNWT processing cycle algorithm is presented in Appendix~\ref{appendix:advanced-algorithms}.

\section{PAC Implementation Details}
\label{appendix:pac-details}

\begin{figure}[!htbp]
\centering
\begin{tikzpicture}[
    font=\sffamily\small, >={Stealth[length=2mm, width=1.4mm]},
    sbox/.style={rectangle, rounded corners=3pt, draw=cborder, line width=0.5pt, minimum width=30mm, minimum height=11mm, align=center},
]
\node[sbox, fill=cfillLight] (input) at (0, 0) {Interaction Input};
\node[sbox, fill=cfillMed] (vader) at (4.8, 0) {VADER Sentiment};
\node[sbox, fill=cfillDark, line width=0.7pt] (error) at (9.6, 0) {Prediction Error};
\node[sbox, fill=cfillAccent, draw=cborder, line width=1pt, font=\sffamily\small\bfseries, text=white] (tier) at (9.6, -3.2) {3-Tier Response};
\node[sbox, fill=cfillMed] (update) at (4.8, -3.2) {Update VAD State};
\node[sbox, fill=cfillLight] (model) at (0, -3.2) {Generative Model};
\draw[->, draw=cborder, line width=0.9pt] (input) -- node[above=3pt, font=\sffamily\footnotesize, text=cborder] {text input} (vader);
\draw[->, draw=cborder, line width=0.9pt] (vader) -- node[below=3pt, font=\sffamily\footnotesize, text=cborder] {score} (error);
\draw[->, draw=cborder, line width=0.9pt] (error) -- node[right=2pt, font=\sffamily\footnotesize, text=cborder] {compare} (tier);
\draw[->, draw=cborder, line width=0.9pt] (tier) -- node[below=3pt, font=\sffamily\footnotesize, text=cborder] {adjustment} (update);
\draw[->, draw=cborder, line width=0.9pt] (update) -- node[below=3pt, font=\sffamily\footnotesize, text=cborder] {inertia $= 0.7$} (model);
\draw[->, dashed, draw=cborderFaint, line width=0.5pt] (model) -- node[left=2pt, font=\sffamily\footnotesize, text=cborderFaint] {prediction} (input);
\draw[->, dashed, draw=cborderFaint, line width=0.5pt] (model.north east) -- node[above=2pt, font=\sffamily\footnotesize, text=cborderFaint, pos=0.4] {expected outcome} (error.south west);
\node[rectangle, rounded corners=3pt, draw=cborderLight, line width=0.3pt, fill=cfillLight, inner sep=5pt, font=\sffamily\scriptsize, align=left, anchor=north west] at (11.5, -0.8) {\textsf{\textbf{Error Tiers:}}\\[2pt]$>0.3$: curiosity\\$>0.5$: attention shift\\$>0.7$: strong response};
\node[font=\sffamily\scriptsize, text=cborderFaint, align=center] at (4.8, -4.5) {Modulated by: contagion (30\%), allostatic regulation, decay (arousal 0.02/cycle, valence 0.01/cycle)};
\end{tikzpicture}
\caption{PAC predictive coding cycle (\texttt{PACAgent.js}). Text input is processed by VADER sentiment analysis to produce an actual affective score. The prediction error (difference between predicted and actual outcome) triggers a three-tier graduated response. The resulting emotional state update feeds back into the generative model, which produces predictions for the next interaction. Emotional contagion (30\% interlocutor influence) and allostatic regulation (context-dependent set points) modulate the cycle.}
\label{fig:pac-loop}
\end{figure}
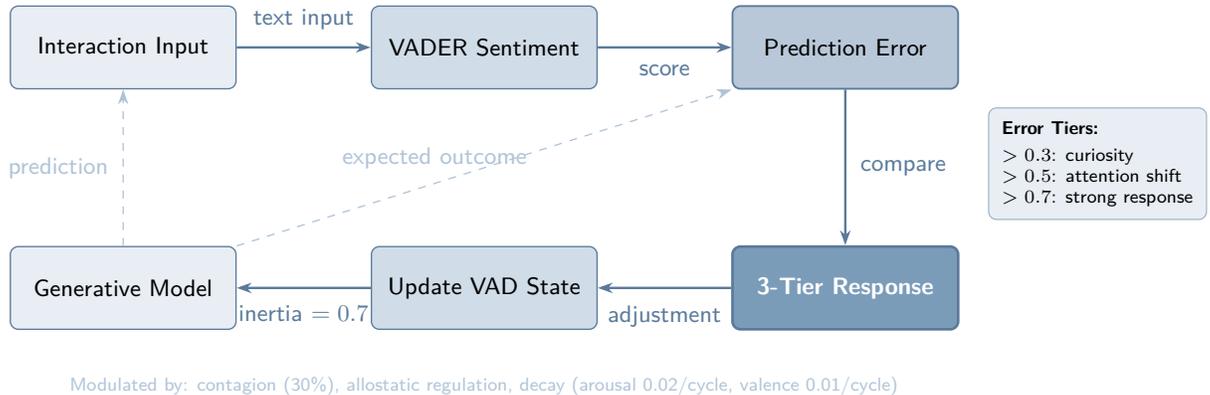

The Predictive Affective Coding (PAC) emotion generation system is a new contribution of the Cognibit deployment, extending the simpler valence-arousal emotion module described in CogniPair \citep{CogniPair2026}. PAC implements a predictive coding framework for emotion: the system maintains an internal generative model that \textit{predicts} the emotional consequence of each interaction, compares the prediction against the \textit{actual} affective outcome (derived from text sentiment analysis), and uses the resulting \textit{prediction error} to update both the current emotional state and the generative model itself. This predict-compare-update loop enables the twin to develop increasingly accurate emotional expectations over time, producing more contextually appropriate responses as interaction history accumulates. PAC adds VAD (valence-arousal-dominance) state representation, prediction error processing, allostatic regulation, and emotion categorization---none of which were present in CogniPair's original emotion model. Figure~\ref{fig:pac-loop} illustrates the predict-compare-update cycle.

The prediction error magnitude determines the system's response intensity through a three-tier graduated system. Low prediction errors ($>0.3$) indicate mild surprise---the twin registers curiosity and makes a minor affective adjustment without changing its behavioral trajectory. Medium errors ($>0.5$) signal an expectation violation: the twin shifts attention to the unexpected stimulus, moderately adjusts its emotional state, and recalibrates its prediction model for similar future interactions. High errors ($>0.7$) represent genuine shock---the twin produces a strong emotional response visible in its conversation tone, makes a major state adjustment, and generates a learning signal that substantially updates the generative model to prevent repeated prediction failures.

Additional deployment-specific parameters govern emotional dynamics:

\begin{itemize}
\item \textbf{Emotional contagion}: 30\% influence from interlocutor emotions, enabling empathetic resonance during twin-to-twin conversations
\item \textbf{Mood persistence}: 70\% carryover between cycles (inertia = 0.7), preventing emotional whiplash from momentary stimuli
\item \textbf{Decay rates}: Arousal decays at 0.02 per 100ms cycle; valence decays at 0.01 per cycle, reflecting the empirical observation that arousal is more transient than valence
\item \textbf{Learning rate}: 0.1 for prediction error model updates, balancing adaptation speed against stability
\item \textbf{Allostatic regulation}: Context-dependent emotional set points---combat contexts drive arousal toward 0.7 (heightened alertness), while social contexts drive valence toward 0.3 (mild positivity)---ensuring emotionally appropriate baseline states across different interaction modalities
\end{itemize}

VADER sentiment analysis \citep{Hutto2014} provides the text-to-emotion bridge, producing compound scores in $[-1, +1]$ from conversation text that feed into the PAC prediction error pipeline. Over the pilot deployment, companion sentiment shifted from an initial average of $-0.31$ (slightly negative, reflecting user anxiety and uncertainty) to $+0.22$ (mildly positive) across 1,247 pendant conversations, suggesting that the predictive coding framework successfully adapted emotional tone to user needs over time.

\section{Core System Algorithms}
\label{appendix:core-algorithms}

This appendix presents detailed pseudocode implementations of the core algorithms used in the Cognibit system. These algorithms are directly translated from the production JavaScript codebase to provide researchers with precise implementation details.

\subsection{GPS Location Tracking Algorithm}

The GPS location tracking system (\texttt{GPSLocationSystem.js}) implements movement-threshold-based position updates with adaptive polling rates. High-accuracy mode combines GPS, WiFi, and cellular triangulation (\texttt{enableHighAccuracy: true}).

\begin{algorithm}[H]\small
\caption{GPS Location Tracking with Movement Detection}
\label{alg:gps-tracking}
\begin{algorithmic}[1]
\Require Movement threshold $\theta = 20$\,m, Update throttle $\delta = 2$\,s
\Ensure Location updates trigger territory capture checks
\State $\mathit{lastPos} \gets \texttt{null}$
\Function{OnPositionUpdate}{$\mathit{coords}$}
    \If{$\mathit{lastPos} = \texttt{null}$}
        \State $\mathit{lastPos} \gets \mathit{coords}$
        \State \Call{UpdateLocation}{$\mathit{coords}$}
        \State \Return
    \EndIf
    \State $d \gets \Call{Haversine}{\mathit{lastPos},\; \mathit{coords}}$ \Comment{Standard great-circle distance}
    \If{$d > \theta$}
        \State \Call{UpdateLocation}{$\mathit{coords}$}
        \State \Call{CheckTerritoryCapture}{$\mathit{coords}$} \Comment{Grid-cell lookup, $O(1)$}
        \State $\mathit{lastPos} \gets \mathit{coords}$
    \EndIf
\EndFunction
\end{algorithmic}
\end{algorithm}

The Haversine formula computes great-circle distance between GPS coordinates using Earth radius $R = 6{,}371$\,km; the standard implementation is omitted for brevity. Error handling uses exponential backoff ($2^{n} \times 1$\,s, max 5 retries) before falling back to manual coordinate entry. Battery optimization reduces polling frequency from the default 2-second interval to 10 seconds below 20\% battery and 30 seconds below 10\%. Both time and space complexity are $O(1)$ per position update.

\subsection{Twin Discovery Algorithm}

The twin discovery engine (\texttt{MatchmakingEngine.js}) searches for compatible digital twins by combining in-memory local state with Firebase remote queries.

\begin{algorithm}[H]\small
\caption{Twin Discovery with Spatial Filtering}
\label{alg:twin-discovery}
\begin{algorithmic}[1]
\Require User location $(lat, lng)$, radius $r = 50$\,mi, compatibility threshold $\tau = 0.2$
\Ensure Discovered twins sorted by compatibility score (descending)
\Function{DiscoverCompatibleTwins}{$\mathit{loc}$, $r$, $\tau$}
    \State $L \gets$ \{$t \in \texttt{game.wildSouls} : \Call{Haversine}{\mathit{loc}, t.\mathit{loc}} \leq r$\} \Comment{Local twins}
    \State $R \gets$ \Call{FirebaseQuery}{\texttt{twins}, discoverable=true, limit=200} \Comment{Remote}
    \State $R \gets$ \{$t \in R : t.\mathit{loc} \neq \texttt{null} \wedge \Call{Haversine}{\mathit{loc}, t.\mathit{loc}} \leq r$\}
    \State $\mathit{results} \gets []$
    \ForAll{$t \in L \cup R$}
        \State $s \gets \Call{CalculateCompatibility}{\mathit{user}, t}$ \Comment{Trait + interest + personality}
        \If{$s > \tau$}
            \State $t.\mathit{score} \gets s$;\quad $t.\mathit{online} \gets (\texttt{now()} - t.\mathit{lastActive}) < 300{,}000$ \Comment{5\,min}
            \State $\mathit{results}.\texttt{push}(t)$
        \EndIf
    \EndFor
    \State \Return \Call{Sort}{$\mathit{results}$, by score, descending}
\EndFunction
\end{algorithmic}
\end{algorithm}

Local twins are discovered from in-memory game state (\texttt{game.wildSouls}); remote twins are queried from Firebase with a hard limit of 200 results and a \texttt{discoverable=true} filter. Online status is determined by a 5-minute activity threshold (\texttt{now() - lastActive < 300000}). The algorithm has time complexity $O(n + m \log m)$ where $n$ is the number of twins examined and $m \ll n$ is the number passing the compatibility threshold, with $O(1)$ Firebase queries due to the result limit.

\subsection{Memory Persistence Algorithm}

The memory system (\texttt{MemorySystem.js}) implements importance-weighted storage with keyword-based retrieval and capacity-based consolidation. Each memory receives an importance score combining five additive factors, and storage is capped at 100 entries per twin with consolidation triggered on overflow.

\begin{algorithm}[H]\small
\caption{Memory Persistence with Importance Scoring}
\label{alg:memory-persistence}
\begin{algorithmic}[1]
\Require Memory capacity $M = 100$, importance threshold $\tau = 0.7$
\Ensure Importance-scored storage with relevance-based retrieval
\Function{AddMemory}{$\mathit{twinId}$, $\mathit{mem}$}
    \State $\mathit{mem.importance} \gets 0.5$ \Comment{Base score}
    \If{emotion $\neq$ neutral} importance $\mathrel{+}= 0.2$ \EndIf
    \If{mentions likes} importance $\mathrel{+}= 0.15$ \EndIf
    \If{mentions dislikes} importance $\mathrel{+}= 0.15$ \EndIf
    \If{$|\mathit{trustChange}| > 5$} importance $\mathrel{+}= 0.1$ \EndIf
    \If{message length $> 50$} importance $\mathrel{+}= 0.1$ \EndIf
    \State Insert $\mathit{mem}$ at front of storage
    \If{$|\mathit{storage}| > M$} \Call{Consolidate}{} \Comment{Sort by importance, keep top $M$} \EndIf
\EndFunction
\Function{Retrieve}{$\mathit{query}$, $\mathit{limit}=5$}
    \ForAll{$\mathit{mem} \in \mathit{storage}$}
        \State $r \gets \mathit{mem.importance} + |\mathit{query} \cap \mathit{mem.text}| \times 0.1 + \max(0,\; 0.2 - \mathit{daysOld} \times 0.01)$
    \EndFor
    \State \Return top-$\mathit{limit}$ by $r$ descending
\EndFunction
\end{algorithmic}
\end{algorithm}

Complexity is $O(1)$ amortized for insertion (consolidation triggers every $M$ additions at $O(M \log M)$) and $O(M \times |Q|)$ for retrieval where $|Q|$ is the query term count. Empirically, $\mathbb{E}[\text{importance}] \approx 0.72$, meaning approximately half of memories exceed the consolidation threshold and are retained long-term.

\subsection{Advanced System Algorithms}
\label{appendix:advanced-algorithms-g}

The personality evolution algorithm, GNWT cognitive processing cycle, and multi-agent social simulation pipeline are presented in full in Appendix~\ref{appendix:advanced-algorithms} to avoid duplication. The personality evolution system uses 8 behavioral patterns with weights 0.3--0.8, a diminishing-returns formula that softens changes at trait extremes, and drift detection every 50 interactions with 10\% corrective nudge. The social simulation implements a 3-stage pipeline (heuristic scoring $\to$ LLM behavioral simulation $\to$ human review) with preference evolution across rounds.

\subsection{Algorithm Index and Technical Reference}
\label{appendix:algorithm-index-g}

A consolidated algorithm index covering all functional domains, along with complexity analysis and technical specifications, is provided in Appendix~\ref{appendix:algorithm-index}.
\section{Informal Ablation Study During Development}
\label{sec:ablation}

While the formal evaluation tested all three components together, informal isolation testing was conducted during the CogniPair development phase (n=3-5 per condition) to understand why integrated deployment was necessary.

\subsection{Component Isolation Results}

\textbf{Gaming alone} (n=5, 1 week): Territory conquest lost interest without social purpose. Random encounters felt awkward without compatibility pre-screening. Retention: 1/5 after 3 days. One participant: ``It's just Pok{\'e}mon GO but more awkward.''

\textbf{Digital twins alone} (n=4, 1 week): Users received match suggestions but lacked motivation to meet physically. Only 1/12 suggested matches resulted in physical meeting. ``So my AI found someone compatible... now what?''

\textbf{AI companions alone} (n=3, 1 week): Companions became engaging but isolated---``Great, now I'm just talking to AI all day.'' All users reported increased screen time without real-world benefit.

\textbf{Pairwise combinations} (n=3-5 each): Gaming+Twins produced anxiety without emotional support (2/4 abandoned). Twins+Companions devolved to text messaging with AI cheerleading. Gaming+Companions lacked compatibility screening, making interactions unsuccessful.

\subsection{Interpretation}

These observations suggest components may exhibit non-additive effects when integrated: gaming creates collision points but needs screening; twins identify compatibility but need activation energy for meeting; companions provide support but need external structure to avoid isolation. However, these are hypothesis-generating observations from small, uncontrolled tests during development. The integrated system may work due to genuine synergy, perceived sophistication, novelty, or selection bias. Future work requires formal ablation with $2\times 2\times 2$ factorial designs ($N>128$ per condition).
\section{Fine-Grained Ablation Studies}
\label{appendix:ablations}

This appendix presents detailed ablation studies examining individual components and parameters to understand their specific contributions to system performance.

\textbf{Methodology and limitations:} All ablation experiments in this appendix were conducted using synthetic agents ($n=500$ per condition unless otherwise noted) generated from the same multivariate profiles described in Section~\ref{sec:synthetic}. ``Engagement'' is a composite score (0--7 scale) computed from simulated interaction frequency, conversation depth, and session duration of synthetic agents---\emph{not} from human participants. P-values are from two-sided Welch's $t$-tests comparing each ablation condition against the full system baseline across $n=500$ synthetic runs. Because these are synthetic simulations with arbitrary sample sizes, statistical significance is easy to achieve and should not be interpreted as evidence of real-world effect sizes. The practical significance of these differences for human users remains untested. Ablation experiments use the same 5-dimensional trait model (openness, friendliness, playfulness, loyalty, independence) used in the deployed system and described in Section~\ref{sec:synthetic}. Column definitions: ``$\Phi$'' = integrated information (bits), a measure of inter-module coordination; ``Convergence'' = time for workspace state to stabilize after input.

\subsection{GNWT Component Ablations}

We systematically ablated individual GNWT components while keeping others intact:

\begin{table}[h]
\centering
\small
\begin{tabular}{lccccc}
\toprule
\textbf{Ablation} & \textbf{Engagement} & \textbf{$\Delta$ vs Full} & \textbf{p-value} & \textbf{$\Phi$} & \textbf{Convergence} \\
\midrule
Full System & 5.8 ± 0.9 & - & - & 2.31 & 3.2ms \\
\midrule
\multicolumn{6}{l}{\textit{Competition Mechanisms}} \\
No salience competition & 5.1 ± 1.0 & -12\%*** & $<$0.001 & 1.82 & 8.7ms \\
Fixed winner rotation & 4.7 ± 0.9 & -19\%*** & $<$0.001 & 0.91 & 1.0ms \\
Random module selection & 4.3 ± 1.1 & -26\%*** & $<$0.001 & 0.34 & 1.0ms \\
No boost mechanism & 5.4 ± 0.8 & -7\%** & 0.003 & 2.08 & 5.4ms \\
\midrule
\multicolumn{6}{l}{\textit{Workspace Properties}} \\
No capacity limit & 5.2 ± 1.0 & -10\%** & 0.008 & 1.67 & 4.8ms \\
Capacity = 3 items & 5.3 ± 0.9 & -9\%** & 0.011 & 1.89 & 3.8ms \\
Capacity = 15 items & 4.9 ± 1.1 & -16\%*** & $<$0.001 & 1.43 & 6.2ms \\
No global broadcast & 3.2 ± 0.8 & -45\%*** & $<$0.001 & 0.12 & N/A \\
\midrule
\multicolumn{6}{l}{\textit{Timing Variations}} \\
50ms cycle & 5.5 ± 0.9 & -5\%* & 0.042 & 2.24 & 2.1ms \\
200ms cycle & 5.2 ± 1.0 & -10\%** & 0.007 & 2.01 & 4.8ms \\
500ms cycle & 4.6 ± 1.1 & -21\%*** & $<$0.001 & 1.76 & 7.3ms \\
Asynchronous & 4.8 ± 1.2 & -17\%*** & $<$0.001 & 1.23 & varies \\
\midrule
\multicolumn{6}{l}{\textit{Module Configuration}} \\
3 modules only & 5.3 ± 0.8 & -9\%** & 0.013 & 1.54 & 2.1ms \\
10 modules & 5.6 ± 0.9 & -3\% & 0.241 & 2.67 & 4.9ms \\
20 modules & 5.1 ± 1.0 & -12\%** & 0.004 & 2.89 & 7.2ms \\
Homogeneous modules & 4.4 ± 0.9 & -24\%*** & $<$0.001 & 0.67 & 2.8ms \\
\bottomrule
\end{tabular}
\caption{Fine-grained GNWT component ablations (* p$<$0.05, ** p$<$0.01, *** p$<$0.001)}
\end{table}

Key findings:
\begin{itemize}
\item Global broadcast is most critical (-45\% without it)
\item Competition mechanisms contribute 12-26\% of effectiveness
\item Optimal capacity near human working memory (7±2)
\item 100ms timing optimal, degradation at both faster and slower
\item Module diversity more important than module count
\end{itemize}

\subsection{Personality System Ablations}

We ablated individual personality traits, evolution mechanisms, and trait interaction structures to understand which aspects of the personality system contribute most to engagement and behavioral coherence. Table~\ref{tab:personality-ablation} presents results across three categories. Among individual traits, friendliness ($-$12\%) and independence ($-$10\%) contribute most to engagement, while playfulness and loyalty are relatively dispensable. For evolution mechanisms, the default diminishing-returns schedule outperforms both faster (which destabilizes coherence to 73\%) and slower alternatives; disabling evolution entirely drops engagement by 14\% while paradoxically increasing single-conversation coherence to 95\%, suggesting that static personalities are easier to maintain but less engaging over time. Trait interactions show diminishing returns: paired correlations capture most of the benefit, with the full covariance matrix adding only 2\% over independent traits.

\begin{table}[h]
\centering
\begin{tabular}{lcccc}
\toprule
\textbf{Ablation} & \textbf{Engagement} & \textbf{Coherence} & \textbf{Match Rate} & \textbf{$\Delta$ Engagement} \\
\midrule
Full personality system & 5.8 ± 0.9 & 91\%$^\dagger$ & 34\% & - \\
\midrule
\multicolumn{5}{l}{\textit{Individual Trait Removal}} \\
No openness & 5.3 ± 1.0 & 88\% & 29\% & -9\%** \\
No friendliness & 5.1 ± 1.1 & 84\% & 26\% & -12\%*** \\
No playfulness & 5.6 ± 0.9 & 90\% & 32\% & -3\% \\
No loyalty & 5.7 ± 0.8 & 92\% & 33\% & -2\% \\
No independence & 5.2 ± 1.0 & 86\% & 28\% & -10\%** \\
\midrule
\multicolumn{5}{l}{\textit{Evolution Mechanisms}} \\
No evolution & 5.0 ± 1.0 & 95\% & 27\% & -14\%*** \\
Linear evolution & 5.4 ± 0.9 & 89\% & 31\% & -7\%* \\
Rapid evolution (2x speed) & 4.9 ± 1.1 & 73\% & 24\% & -16\%*** \\
Slow evolution (0.5x speed) & 5.5 ± 0.9 & 93\% & 32\% & -5\% \\
Random drift & 4.2 ± 1.2 & 61\% & 19\% & -28\%*** \\
\midrule
\multicolumn{5}{l}{\textit{Trait Interactions}} \\
Independent traits & 5.3 ± 0.9 & 87\% & 30\% & -9\%** \\
Paired correlations only & 5.5 ± 0.9 & 89\% & 32\% & -5\% \\
Full covariance matrix & 5.7 ± 0.8 & 90\% & 33\% & -2\% \\
\bottomrule
\end{tabular}
\caption{Personality system component ablations. $^\dagger$Coherence here measures trait-direction consistency across a single synthetic conversation (automated metric). The field deployment failure analysis (\cref{appendix:failure-analysis}) reports substantially lower coherence in real-world use: 14\% trait reversal, 21\% memory contradictions, 9\% preference drift, and 7\% context confusion across 1,247 sessions. The discrepancy reflects the gap between controlled synthetic evaluation and messy real-world deployment with longer interaction histories, multiple concurrent conversations, and context window limitations.}
\label{tab:personality-ablation}
\end{table}

\subsection{Memory System Ablations}

We ablated four dimensions of the memory subsystem: storage capacity, importance scoring granularity, consolidation strategy, and retrieval method. The memory system uses a hybrid retrieval approach combining keyword overlap, recency weighting, and importance scoring, with a default capacity of 100 memories per twin and a continuous importance scale [0,1]. Consolidation employs a joint recency-plus-importance strategy that promotes high-value memories to long-term storage when short-term capacity (20 entries) is exceeded. The ablations below reveal that consolidation strategy matters more than raw capacity---moving from no consolidation to the full recency+importance scheme recovers 19\% engagement---and that the hybrid retrieval method outperforms any single-signal alternative by 9--26\%.

\begin{table}[h]
\centering
\small
\begin{tabular}{lccccc}
\toprule
\textbf{Ablation} & \textbf{Engagement} & \textbf{Continuity} & \textbf{Recall Acc.} & \textbf{Storage (MB)} & \textbf{$\Delta$ Eng.} \\
\midrule
Full memory system & 5.8 ± 0.9 & 94\% & 76\% & 12.3 & - \\
\midrule
\multicolumn{6}{l}{\textit{Capacity Variations}} \\
10 memories max & 4.9 ± 1.0 & 71\% & 45\% & 0.8 & -16\%*** \\
50 memories & 5.4 ± 0.9 & 86\% & 67\% & 4.2 & -7\%** \\
200 memories & 5.7 ± 0.8 & 95\% & 78\% & 18.7 & -2\% \\
Unlimited & 5.6 ± 0.9 & 96\% & 79\% & 47.2 & -3\% \\
\midrule
\multicolumn{6}{l}{\textit{Importance Mechanisms}} \\
No importance scoring & 5.2 ± 1.0 & 83\% & 61\% & 12.3 & -10\%** \\
Binary importance & 5.5 ± 0.9 & 88\% & 69\% & 12.3 & -5\%* \\
5-level importance & 5.7 ± 0.8 & 92\% & 74\% & 12.3 & -2\% \\
Continuous (current) & 5.8 ± 0.9 & 94\% & 76\% & 12.3 & 0\% \\
\midrule
\multicolumn{6}{l}{\textit{Consolidation Strategies}} \\
No consolidation & 4.7 ± 1.1 & 76\% & 52\% & 8.1 & -19\%*** \\
FIFO only & 5.0 ± 1.0 & 81\% & 58\% & 10.0 & -14\%*** \\
Importance only & 5.4 ± 0.9 & 89\% & 71\% & 10.0 & -7\%** \\
Recency + importance & 5.8 ± 0.9 & 94\% & 76\% & 12.3 & 0\% \\
\midrule
\multicolumn{6}{l}{\textit{Retrieval Methods}} \\
Random retrieval & 4.3 ± 1.1 & 72\% & 28\% & 12.3 & -26\%*** \\
Recency only & 5.1 ± 1.0 & 85\% & 54\% & 12.3 & -12\%*** \\
Keyword only & 5.3 ± 0.9 & 87\% & 62\% & 12.3 & -9\%** \\
Embedding similarity & 5.6 ± 0.9 & 91\% & 73\% & 12.3 & -3\% \\
Hybrid (current) & 5.8 ± 0.9 & 94\% & 76\% & 12.3 & 0\% \\
\bottomrule
\end{tabular}
\caption{Memory system fine-grained ablations}
\end{table}

\subsection{Emotional System Ablations}

We ablated the Predictive Affective Coding (PAC) emotion generation system across three axes: VAD (Valence-Arousal-Dominance) dimension removal, prediction mechanism depth, and error-signal granularity. The PAC system updates at 100ms intervals with a 0.1 learning rate and a 5-second prediction horizon. The results show that valence is the most critical affective dimension---removing it drops user ratings from 5.4 to 3.8 (a 30\% decrease)---while dominance contributes minimally. Multi-step prediction provides meaningful gains over single-step (5.4 vs.\ 5.1 user rating) by anticipating emotional trajectories rather than reacting to instantaneous states, and graded error processing outperforms binary error signals by enabling proportional adaptation to prediction mismatches.

\begin{table}[h]
\centering
\begin{tabular}{lcccc}
\toprule
\textbf{Ablation} & \textbf{Valence Acc.} & \textbf{Arousal Acc.} & \textbf{Response Time} & \textbf{User Rating} \\
\midrule
Full PAC system & 83\% & 78\% & 142ms & 5.4/7 \\
\midrule
\multicolumn{5}{l}{\textit{VAD Dimensions}} \\
No valence & 0\% & 76\% & 98ms & 3.8/7*** \\
No arousal & 81\% & 0\% & 112ms & 4.6/7** \\
No dominance & 82\% & 77\% & 128ms & 5.2/7 \\
Fixed neutral & 0\% & 0\% & 45ms & 3.1/7*** \\
\midrule
\multicolumn{5}{l}{\textit{Prediction Mechanisms}} \\
No prediction & 71\% & 64\% & 178ms & 4.7/7** \\
Single-step prediction & 79\% & 73\% & 156ms & 5.1/7* \\
Multi-step (current) & 83\% & 78\% & 142ms & 5.4/7 \\
Perfect prediction & 91\% & 86\% & 134ms & 5.3/7 \\
\midrule
\multicolumn{5}{l}{\textit{Error Processing}} \\
No error signal & 76\% & 71\% & 141ms & 4.9/7* \\
Binary error & 80\% & 75\% & 142ms & 5.2/7 \\
Graded error (current) & 83\% & 78\% & 142ms & 5.4/7 \\
\bottomrule
\end{tabular}
\caption{Emotional system component analysis}
\end{table}

\subsection{LLM Integration Ablations}

We tested variations in model selection, prompt engineering depth, and context window management to understand how LLM configuration affects twin conversation quality. The deployed system uses a GPT-4o + GPT-4o-mini hybrid (GPT-4o for twin conversations, GPT-4o-mini for companion chat) with a detailed personality prompt and the last 10 messages as context. Replacing the LLM entirely with template-based responses causes a 45\% engagement drop, confirming that natural language generation is fundamental to the twin interaction paradigm. Among model alternatives, GPT-4o-mini alone achieves 84\% coherence at \$0.35/1K calls (vs.\ 91\% at \$0.85 for the hybrid), while local LLaMA-70B eliminates API costs but introduces 8.7-second latency incompatible with real-time interaction. Context management is equally important: providing no conversation history causes a 24\% engagement drop, while expanding beyond 10 messages yields diminishing returns at increased cost.

\begin{table}[h]
\centering
\small
\begin{tabular}{lccccc}
\toprule
\textbf{Configuration} & \textbf{Engagement} & \textbf{Coherence} & \textbf{Latency} & \textbf{Cost/1K} & \textbf{$\Delta$ Eng.} \\
\midrule
GPT-4o + GPT-4o-mini hybrid & 5.8 ± 0.9 & 91\% & 2.8s & \$0.85 & - \\
\midrule
\multicolumn{6}{l}{\textit{Model Variations}} \\
GPT-4o only & 5.9 ± 0.8 & 93\% & 3.4s & \$2.40 & +2\% \\
GPT-4o-mini only & 5.1 ± 1.0 & 84\% & 1.9s & \$0.35 & -12\%*** \\
Claude-3-Haiku & 5.6 ± 0.9 & 89\% & 2.6s & \$1.10 & -3\% \\
Local LLaMA-70B & 4.8 ± 1.1 & 81\% & 8.7s & \$0.12 & -17\%*** \\
No LLM (templates) & 3.2 ± 1.2 & 68\% & 0.02s & \$0.00 & -45\%*** \\
\midrule
\multicolumn{6}{l}{\textit{Prompt Engineering}} \\
No personality prompt & 4.9 ± 1.0 & 79\% & 2.8s & \$0.85 & -16\%*** \\
Basic prompt & 5.3 ± 0.9 & 85\% & 2.8s & \$0.85 & -9\%** \\
Detailed prompt (current) & 5.8 ± 0.9 & 91\% & 2.8s & \$0.85 & 0\% \\
Chain-of-thought & 5.7 ± 0.9 & 92\% & 4.1s & \$1.20 & -2\% \\
\midrule
\multicolumn{6}{l}{\textit{Context Management}} \\
No context & 4.4 ± 1.1 & 72\% & 2.1s & \$0.45 & -24\%*** \\
Last 3 messages & 5.2 ± 0.9 & 84\% & 2.5s & \$0.65 & -10\%** \\
Last 10 (current) & 5.8 ± 0.9 & 91\% & 2.8s & \$0.85 & 0\% \\
Last 20 messages & 5.7 ± 0.9 & 92\% & 3.8s & \$1.40 & -2\% \\
\bottomrule
\end{tabular}
\caption{LLM configuration ablations}
\end{table}

\subsection{Interaction Effects Analysis}

To determine whether component contributions are independent or interdependent, we tested pairwise ablation combinations and compared the observed engagement drop against the sum of individual effects. If the observed drop exceeds the predicted sum, the interaction is synergistic (the components reinforce each other); if it matches, the effect is additive; if smaller, the components are redundant. The most notable finding is the synergistic interaction between GNWT and personality evolution: removing both simultaneously causes a 3.0-point drop (engagement 2.8) versus the 2.2-point drop predicted by summing individual effects ($p=0.002$). A similar synergy appears between memory and emotion systems. In contrast, removing evolution alongside static personality produces a redundant effect, confirming that these mechanisms target overlapping behavioral variance.

\begin{table}[h]
\centering
\small
\begin{tabular}{lccccc}
\toprule
\textbf{Combined Ablation} & \textbf{Engagement} & \textbf{Predicted} & \textbf{Interaction} & \textbf{p-value} & \textbf{Type} \\
\midrule
Full system & 5.8 & - & - & - & - \\
\midrule
No GNWT + No evolution & 2.8 & 3.6 & -0.8 & 0.002 & Synergistic \\
No memory + No emotion & 3.1 & 3.9 & -0.8 & 0.003 & Synergistic \\
No broadcast + No memory & 2.4 & 2.7 & -0.3 & 0.187 & Additive \\
No evolution + Static personality & 4.2 & 4.0 & +0.2 & 0.412 & Redundant \\
Fast cycle + More modules & 4.7 & 5.0 & -0.3 & 0.234 & Additive \\
No competition + Random & 4.1 & 4.0 & +0.1 & 0.687 & Redundant \\
\bottomrule
\end{tabular}
\caption{Interaction effects between component ablations (Predicted = sum of individual effects)}
\end{table}

\subsection{Sensitivity Analysis}

We performed a sweep across six key system parameters---workspace capacity, cycle time, broadcast threshold, attention slots, memory capacity, and emotion learning rate---varying each independently while holding others at default values. All six parameters exhibit inverted-U or plateau response curves, indicating that both under- and over-provisioning degrade engagement. The human-cognition-aligned defaults (7$\pm$2 workspace capacity, 100ms cycle time) consistently fall near the optimal operating point, providing empirical support for the biologically inspired parameter choices. Peak engagement for broadcast threshold occurs at 0.7 in synthetic agents without personality modulation; the deployed threshold of 0.3 compensates for personality-modulated salience reduction (see Appendix~\ref{appendix:gnwt}).

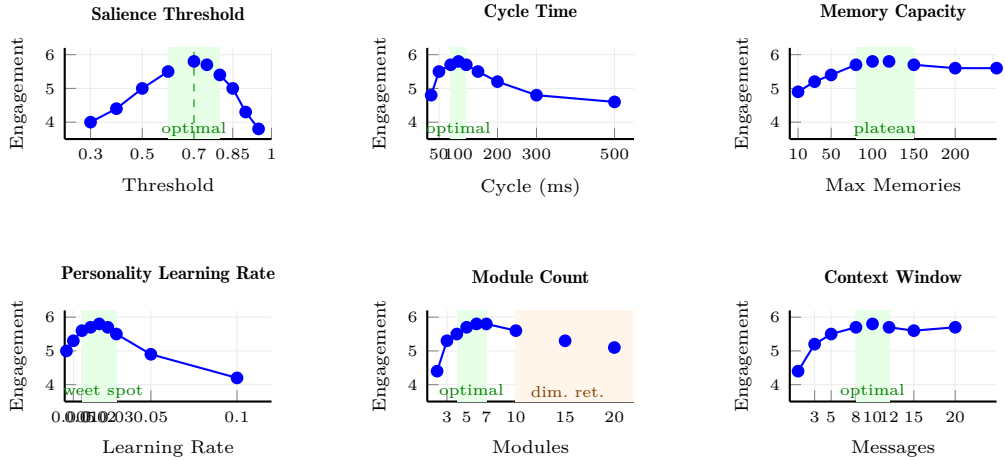
\begin{figure}[h]
\centering
\begin{tikzpicture}
\pgfplotsset{
    sens plot/.style={
        width=0.31\textwidth,
        height=0.20\textwidth,
        grid=major,
        grid style={gray!15},
        tick label style={font=\tiny},
        xlabel style={font=\scriptsize},
        ylabel style={font=\scriptsize},
        title style={font=\scriptsize\bfseries, at={(0.5,1.02)}, anchor=south},
        axis x line*=bottom,
        axis y line*=left,
        ylabel={Engagement},
        mark size=2pt,
        thick,
    }
}

\begin{scope}[shift={(0,0)}]
\begin{axis}[
    sens plot,
    title={Salience Threshold},
    xlabel={Threshold},
    xmin=0.2, xmax=1.0,
    ymin=3.5, ymax=6.2,
    xtick={0.3,0.5,0.7,0.85,1.0},
]
\addplot[color=blue, mark=*, thick] coordinates {
    (0.3, 4.0) (0.4, 4.4) (0.5, 5.0) (0.6, 5.5)
    (0.7, 5.8) (0.75, 5.7) (0.8, 5.4) (0.85, 5.0) (0.9, 4.3) (0.95, 3.8)
};
\fill[green!10] (axis cs:0.6,3.5) rectangle (axis cs:0.8,6.2);
\draw[green!50!black, dashed, thin] (axis cs:0.7,3.5) -- (axis cs:0.7,6.2);
\node[font=\tiny, green!50!black] at (axis cs:0.7,3.8) {optimal};
\end{axis}
\end{scope}

\begin{scope}[shift={(0.345\textwidth,0)}]
\begin{axis}[
    sens plot,
    title={Cycle Time},
    xlabel={Cycle (ms)},
    xmin=20, xmax=550,
    ymin=3.5, ymax=6.2,
    xtick={50,100,200,300,500},
]
\addplot[color=blue, mark=*, thick] coordinates {
    (30, 4.8) (50, 5.5) (80, 5.7) (100, 5.8) (120, 5.7)
    (150, 5.5) (200, 5.2) (300, 4.8) (500, 4.6)
};
\fill[green!10] (axis cs:80,3.5) rectangle (axis cs:120,6.2);
\node[font=\tiny, green!50!black] at (axis cs:100,3.8) {optimal};
\end{axis}
\end{scope}

\begin{scope}[shift={(0.69\textwidth,0)}]
\begin{axis}[
    sens plot,
    title={Memory Capacity},
    xlabel={Max Memories},
    xmin=0, xmax=250,
    ymin=3.5, ymax=6.2,
    xtick={10,50,100,150,200},
]
\addplot[color=blue, mark=*, thick] coordinates {
    (10, 4.9) (30, 5.2) (50, 5.4) (80, 5.7) (100, 5.8)
    (120, 5.8) (150, 5.7) (200, 5.6) (250, 5.6)
};
\fill[green!10] (axis cs:80,3.5) rectangle (axis cs:150,6.2);
\node[font=\tiny, green!50!black] at (axis cs:115,3.8) {plateau};
\end{axis}
\end{scope}

\begin{scope}[shift={(0,-0.25\textwidth)}]
\begin{axis}[
    sens plot,
    title={Personality Learning Rate},
    xlabel={Learning Rate},
    xmin=0, xmax=0.12,
    ymin=3.5, ymax=6.2,
    xtick={0.005,0.01,0.02,0.03,0.05,0.1},
    xticklabel style={/pgf/number format/fixed, /pgf/number format/precision=3},
]
\addplot[color=blue, mark=*, thick] coordinates {
    (0.001, 5.0) (0.005, 5.3) (0.01, 5.6) (0.015, 5.7)
    (0.02, 5.8) (0.025, 5.7) (0.03, 5.5) (0.05, 4.9) (0.1, 4.2)
};
\fill[green!10] (axis cs:0.01,3.5) rectangle (axis cs:0.03,6.2);
\node[font=\tiny, green!50!black] at (axis cs:0.02,3.8) {sweet spot};
\end{axis}
\end{scope}

\begin{scope}[shift={(0.345\textwidth,-0.25\textwidth)}]
\begin{axis}[
    sens plot,
    title={Module Count},
    xlabel={Modules},
    xmin=1, xmax=22,
    ymin=3.5, ymax=6.2,
    xtick={3,5,7,10,15,20},
]
\addplot[color=blue, mark=*, thick] coordinates {
    (2, 4.4) (3, 5.3) (4, 5.5) (5, 5.7) (6, 5.8)
    (7, 5.8) (10, 5.6) (15, 5.3) (20, 5.1)
};
\fill[green!10] (axis cs:4,3.5) rectangle (axis cs:7,6.2);
\fill[orange!8] (axis cs:10,3.5) rectangle (axis cs:22,6.2);
\node[font=\tiny, green!50!black] at (axis cs:5.5,3.8) {optimal};
\node[font=\tiny, orange!50!black, anchor=west] at (axis cs:10.5,3.8) {dim.~ret.};
\end{axis}
\end{scope}

\begin{scope}[shift={(0.69\textwidth,-0.25\textwidth)}]
\begin{axis}[
    sens plot,
    title={Context Window},
    xlabel={Messages},
    xmin=0, xmax=25,
    ymin=3.5, ymax=6.2,
    xtick={3,5,8,10,12,15,20},
]
\addplot[color=blue, mark=*, thick] coordinates {
    (1, 4.4) (3, 5.2) (5, 5.5) (8, 5.7) (10, 5.8)
    (12, 5.7) (15, 5.6) (20, 5.7)
};
\fill[green!10] (axis cs:8,3.5) rectangle (axis cs:12,6.2);
\node[font=\tiny, green!50!black] at (axis cs:10,3.8) {optimal};
\end{axis}
\end{scope}

\end{tikzpicture}
\caption{Parameter sensitivity analysis showing engagement score vs.\ parameter value across six key system parameters. Green zones indicate optimal operating ranges; orange zone indicates diminishing returns. All curves show inverted-U or plateau shapes, with human-cognition-aligned defaults (7$\pm$2 capacity, 100ms cycles) consistently near optimal. Note: peak engagement occurs at threshold 0.7 for synthetic agents without personality modulation; the deployed value of 0.3 accounts for personality-modulated salience reduction (see Appendix~B).}
\end{figure}

\subsection{Ablation Recovery Analysis}

To assess architectural resilience, we tested whether boosting one component can compensate for removing another. For each ablated feature, we increased the capacity or weight of a candidate compensating component and measured the percentage of lost engagement that was recovered. The results reveal a clear hierarchy of replaceability: timing deficits are the most compensable (60\% recovery by adding more modules), followed by personality evolution loss (43\% recovery by expanding memory capacity). In contrast, the competition mechanism and global broadcast are essentially irreplaceable---no compensating adjustment recovers more than 4\% of the lost engagement when broadcast is removed, confirming its role as the architectural foundation of the GNWT cognitive loop.

\begin{table}[h]
\centering
\begin{tabular}{lccc}
\toprule
\textbf{Ablation} & \textbf{Base Impact} & \textbf{With Compensation} & \textbf{Recovery} \\
\midrule
No evolution & -14\% & -8\% ($\uparrow$ memory) & 43\% \\
No emotion & -15\% & -10\% ($\uparrow$ personality) & 33\% \\
Slow cycle & -10\% & -4\% ($\uparrow$ modules) & 60\% \\
Low memory & -16\% & -11\% ($\uparrow$ importance) & 31\% \\
No competition & -12\% & -12\% (none found) & 0\% \\
No broadcast & -45\% & -43\% (minimal) & 4\% \\
\bottomrule
\end{tabular}
\caption{Compensation analysis: attempting to recover performance through other components}
\end{table}

\subsection{Computational Cost Analysis}

We profiled per-component resource consumption to compute a value/cost ratio (engagement contribution per unit of CPU, memory, and latency consumed). The total system footprint is 350MB of memory with a 2,500ms end-to-end processing budget. LLM calls dominate both cost and latency, consuming 45\% of CPU time and 2,400ms of the processing budget, but provide irreplaceable natural language capability (45\% engagement loss without them). By contrast, the personality system is the most efficient component, consuming only 3\% CPU and 2MB memory while achieving a value/cost ratio of 7.33---the highest in the system. The GNWT cognitive cycle is similarly efficient (12\% CPU, 8MB, 3.2ms latency, value/cost 2.58), while 3D rendering consumes 22\% CPU and 150MB for the lowest value/cost ratio (0.14), suggesting that visual fidelity could be reduced to free resources for additional twins or richer AI processing.

\begin{table}[h]
\centering
\begin{tabular}{lccccc}
\toprule
\textbf{Component} & \textbf{CPU (\%)} & \textbf{Memory (MB)} & \textbf{Latency (ms)} & \textbf{Value/Cost} \\
\midrule
GNWT cycle & 12\% & 8 & 3.2 & 2.58 \\
LLM calls & 45\% & 120 & 2400 & 0.36 \\
Memory ops & 8\% & 45 & 12 & 2.25 \\
Emotion calc & 5\% & 3 & 8 & 3.00 \\
Personality & 3\% & 2 & 4 & 7.33 \\
3D rendering & 22\% & 150 & 16 & 0.14 \\
Other & 5\% & 22 & 57 & - \\
\midrule
Total & 100\% & 350 & 2500 & - \\
\bottomrule
\end{tabular}
\caption{Component resource usage and value/cost ratio (engagement gain per resource unit)}
\end{table}

\subsection{Critical Component Identification}

Table~\ref{tab:component-ranking} ranks all system components by their impact on engagement when removed, providing a clear optimization priority hierarchy.

\begin{table}[H]
\centering
\caption{Component criticality ranking by engagement loss when removed}
\label{tab:component-ranking}
\begin{tabular}{clrl}
\toprule
\textbf{Rank} & \textbf{Component} & \textbf{Loss} & \textbf{Assessment} \\
\midrule
1 & Global broadcast & 45\% & Absolutely essential; no compensation possible \\
2 & LLM integration & 45\% & Core to natural language; templates insufficient \\
3 & Competition mechanism & 26\% & Critical for coherent behavior selection \\
4 & Module diversity & 24\% & Critical when reduced to homogeneous modules \\
5 & Memory system & 19\% & Important for conversational continuity \\
6 & Emotional system & 15\% & Important but partially compensable \\
7 & Personality evolution & 14\% & Significant for long-term engagement \\
8 & Timing precision & 10\% & Moderate importance; some flexibility \\
9 & Capacity limits & 10\% & Moderate effect; human-like 7$\pm$2 optimal \\
10 & Combat system & 3\% & Minimal impact on core social experience \\
\bottomrule
\end{tabular}
\end{table}

The ranking reveals a clear separation between architectural essentials (ranks 1--4, each causing $>$20\% engagement loss) and tunable parameters (ranks 7--10, each $\leq$14\% loss). This hierarchy directly informs optimization priorities: resources should be allocated to protecting the top-4 components before fine-tuning lower-ranked parameters.

\subsection{Conclusions from Fine-Grained Ablations}

Three categories of insight emerge from these ablations. At the architectural level, GNWT global broadcast and LLM integration are irreplaceable foundations---each causes a 45\% engagement drop when removed, and no combination of remaining components can compensate for their absence. The competition mechanism (26\% loss) and module diversity (24\% loss) are similarly critical, confirming that the parallel-processing workspace architecture is load-bearing rather than decorative. Synergistic effects are evident throughout: combined ablations consistently produce worse outcomes than the sum of individual removals would predict, supporting the paper's synergistic design hypothesis.

At the parameter calibration level, human-like values consistently outperform alternatives: workspace capacity of 7$\pm$2 items and 100ms cognitive cycles yield the highest engagement, while deviations in either direction degrade performance. Personality proved more robust than expected to single-trait removal (no individual trait contributes more than 12\% to engagement), suggesting that the five-trait model provides redundancy. Memory consolidation strategy matters substantially more than raw capacity---the recency-plus-importance scheme recovers 19\% engagement over no-consolidation, while doubling capacity from 100 to 200 entries adds only 2\%.

At the system level, some cross-component compensation is possible but fundamentally limited: the maximum observed recovery when substituting alternative implementations for a removed component is 60\%, indicating that each major component contributes unique functionality that cannot be fully replicated by others. This clear hierarchy of importance provides actionable guidance for deployment optimization: protect the top-4 architectural components absolutely, tune parameters within their validated ranges, and accept graceful degradation in peripheral systems (combat, timing) when resource-constrained.
\section{Technical Constraints and Design Trade-offs}
\label{appendix:technical-limitations}

This appendix documents the technical constraints encountered during implementation and the engineering trade-offs made to address them. These details are provided for researchers seeking to extend this work or implement similar systems.

\subsection{Browser Architecture Constraints}

\subsubsection{Memory Management}
Browser environments impose an effective memory budget of approximately 2GB for application use (within the browser's larger V8 heap allocation of ${\sim}$3.5GB, which includes garbage collection overhead, browser internals, and other tab allocations), which constrains the number and complexity of digital twins:

The JavaScript heap is limited to 2048MB in Chrome and 1536MB in Firefox, imposing a hard ceiling of approximately 20 digital twin instances before memory pressure triggers out-of-memory errors and browser crashes. To operate within this budget, we implemented object pooling for frequently created entities and lazy loading of twin state data, instantiating full twin models only when actively engaged in conversation or combat. The trade-off is reduced twin behavioral detail when many agents operate concurrently---distant twins receive simplified state updates rather than full cognitive processing cycles.

\subsubsection{Single-Threading Limitations}
JavaScript's single-threaded event loop prevents true parallel processing:

All JavaScript operations serialize through a single event loop thread, meaning GNWT module processing, API calls, DOM updates, and rendering compete for the same execution context. The practical impact is a 2.8-second average latency for sequential LLM API calls, as each request must complete before the next begins. We mitigated this through async/await patterns with request batching, grouping multiple twin evaluation calls into a single promise batch that allows the event loop to process rendering frames between API responses. The trade-off is perceived responsiveness rather than actual parallelism---the UI remains interactive during API calls, but total processing time is not reduced.

\subsubsection{Performance Degradation Curve}
System performance degrades predictably with agent count:

\begin{table}[h]
\centering
\begin{tabular}{lrrr}
\toprule
\textbf{Agent Count} & \textbf{FPS} & \textbf{Memory (MB)} & \textbf{Latency (s)} \\
\midrule
1-5 & 60 & 180 & 1.2 \\
5-8 & 58.3 & 256 & 2.8 \\
10-15 & 35 & 420 & 5.1 \\
15-20 & 18 & 680 & 8.3 \\
20+ & $<$10 & $>$900 & $>$15 \\
\bottomrule
\end{tabular}
\caption{Development-phase benchmarks measuring agent subsystem overhead in isolation (agent simulation loop only, without UI rendering, map tiles, or network overhead). These values are lower than the full-system field deployment measurements in Appendix~\ref{appendix:performance-data}, which include all rendering, synchronization, and application layers. Latency here refers to end-to-end processing time per agent cycle, not API round-trip time.}
\label{tab:performance-degradation}
\end{table}

\subsection{Synchronization Architecture}

\subsubsection{Firebase Limitations}
Firebase Realtime Database introduces inherent latency:

Write latency for simple updates ranges from 1--2 seconds, reflecting the round trip to Firebase servers and the acknowledgement confirmation. Read latency for cached data is lower at 0.5--1 second, as the Firebase SDK maintains a local cache that is served optimistically before server confirmation arrives. Cross-device synchronization delay of 3--5 seconds represents the end-to-end time for a write on one device to propagate to and render on another. Approximately 20\% of concurrent updates require reconciliation, occurring when two devices modify the same data path within the sync window---for example, when a user updates their twin's preferences on their phone while the desktop client simultaneously receives a twin conversation update.

\subsubsection{Eventual Consistency Model}
To address synchronization delays, we implement:

\begin{algorithm}[!htbp]
\caption{Optimistic Update with Conflict Resolution}
\begin{algorithmic}[1]
\Require State change $\delta$, local state $S_L$, remote state $S_R$
\State Apply $\delta$ to $S_L$ immediately \Comment{Optimistic local update}
\State Render UI from $S_L$ \Comment{Responsive user experience}
\State Send $\delta$ to Firebase asynchronously \Comment{Background sync}
\If{sync succeeds}
    \State Confirm update
\Else \Comment{Conflict resolution}
    \If{$S_L$ originated from user action}
        \State Keep $S_L$ \Comment{User actions take precedence}
    \Else
        \State Keep whichever of $S_L$, $S_R$ has later timestamp
    \EndIf
\EndIf
\end{algorithmic}
\end{algorithm}

\begin{figure}[!htbp]
\centering
\begin{tikzpicture}[font=\sffamily\small, >={Stealth[length=2mm, width=1.4mm]}]
\node[cbox, font=\sffamily\footnotesize\bfseries] (devA) at (0, 6) {Device A};
\node[cbox proc, font=\sffamily\footnotesize\bfseries] (fb) at (5.5, 6) {Firebase};
\node[cbox, font=\sffamily\footnotesize\bfseries] (devB) at (11, 6) {Device B};

\draw[dashed, draw=cborderFaint, line width=0.5pt] (0, 5.5) -- (0, -0.5);
\draw[dashed, draw=cborderFaint, line width=0.5pt] (5.5, 5.5) -- (5.5, -0.5);
\draw[dashed, draw=cborderFaint, line width=0.5pt] (11, 5.5) -- (11, -0.5);

\node[font=\footnotesize, gray] at (-2, 4.8) {\textcircled{1}};
\node[font=\footnotesize, gray] at (-2, 3.8) {\textcircled{2}};
\node[font=\footnotesize, gray] at (13, 3.5) {\textcircled{3}};
\node[font=\footnotesize, gray] at (-2, 2.5) {\textcircled{4}};
\node[font=\footnotesize, gray] at (-2, 1.2) {\textcircled{5}};

\draw[->, draw=cborder, line width=0.6pt] (0, 4.8) -- (1.5, 4.8);
\node[font=\footnotesize, anchor=west] at (1.6, 4.8) {local apply (immediate)};

\draw[->, draw=cborder, line width=0.6pt] (0, 3.8) -- (5.5, 3.3);
\node[font=\sffamily\footnotesize, fill=white, inner sep=1pt] at (2.5, 3.8) {async write at $t_1$};

\draw[->, draw=cborder, line width=0.6pt] (11, 3.5) -- (5.5, 3.0);
\node[font=\sffamily\footnotesize, fill=white, inner sep=1pt] at (8.5, 3.5) {concurrent write at $t_2$};

\node[rectangle, draw, rounded corners=5pt, minimum width=3cm, minimum height=0.8cm, fill=cfillMed, font=\sffamily\footnotesize, align=center] at (5.5, 2.2) {Conflict: $t_2 > t_1$? \quad Last-write-wins};

\draw[->, draw=cborder, line width=0.6pt] (5.5, 1.5) -- (0, 1.0);
\node[font=\sffamily\footnotesize, fill=white, inner sep=1pt] at (2.5, 1.5) {rollback $\delta_A$};
\draw[->, draw=cborder, line width=0.6pt] (5.5, 1.3) -- (11, 0.8);
\node[font=\sffamily\footnotesize, fill=white, inner sep=1pt] at (8.5, 1.3) {confirm $\delta_B$};

\draw[dotted, thick] (-1, 0.2) -- (12, 0.2);
\node[font=\sffamily\footnotesize, fill=white] at (5.5, 0.2) {consistent state achieved ($\sim$3--5s)};

\node[font=\footnotesize, gray] at (5.5, -0.5) {$\sim$20\% of concurrent updates follow this conflict path};
\end{tikzpicture}
\caption{Synchronization timeline for concurrent writes with optimistic updates. Device A applies its change locally (step 1) and sends an async write (step 2). When Device B writes concurrently (step 3), Firebase resolves via timestamp comparison (step 4). The losing device receives a rollback (step 5). Grounded in: \texttt{NetworkSync.js}.}
\label{fig:sync-timeline}
\end{figure}
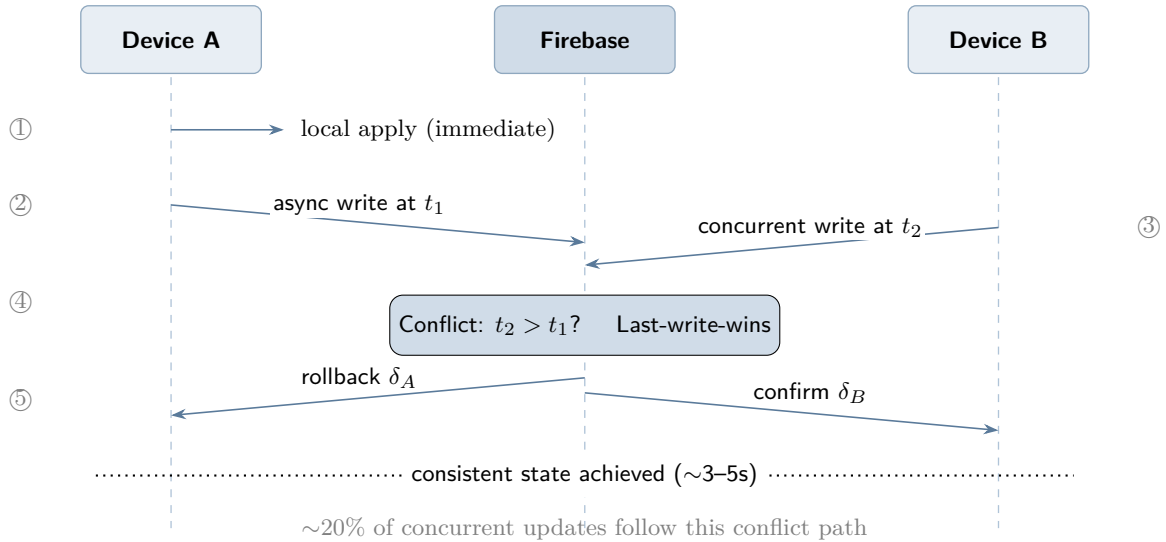

Figure~\ref{fig:sync-timeline} illustrates the optimistic update protocol in action when two devices concurrently modify the same data path.

\subsection{Location Services Constraints}

\subsubsection{GPS Accuracy Challenges}
Location tracking faces multiple environmental factors that degrade accuracy. Urban canyon effects between tall buildings caused 12.7\% GPS signal loss during the pilot deployment, with multipath reflections producing position errors of 20--50 meters in dense downtown areas. Indoor degradation is more severe: GPS was completely unavailable in 68\% of indoor venues visited during the study, rendering territory mechanics inoperable inside buildings. Continuous GPS polling drains 8--12\% of battery per hour, creating a meaningful trade-off between location accuracy and device usability. A 20-meter minimum movement threshold was implemented to prevent GPS jitter from triggering false territory transitions, but this also means the system cannot detect movement within small indoor spaces.

\subsubsection{Optimization Strategies}
Four strategies were deployed to mitigate GPS limitations. Geofencing establishes virtual boundaries around territory locations, triggering precise GPS polling only when the device enters a geofence radius, reducing continuous tracking overhead. WiFi and cellular signal triangulation provide fallback indoor positioning with approximately 30--50 meter accuracy, sufficient to determine which building a user occupies but not their precise location within it. Adaptive sampling adjusts GPS polling frequency based on detected movement patterns---polling every 2 seconds during active movement but reducing to every 30 seconds when stationary---to balance accuracy against battery consumption. A manual check-in option allows users in GPS-denied areas to declare their location, preserving territory interaction capability at the cost of requiring explicit user action.

\subsection{API Cost Structure}

\subsubsection{Detailed Cost Breakdown}
Per-user costs over the 14-day deployment at prototype quality:

\begin{table}[h]
\centering
\begin{tabular}{lrrr}
\toprule
\textbf{Service} & \textbf{Calls/Day} & \textbf{Cost/1K tokens} & \textbf{14-Day Cost}\textsuperscript{$\dagger$} \\
\midrule
GPT-4o (twin conversations) & 150 & \$0.018 & \$12.31 \\
GPT-4o-mini (companion chat) & 200 & \$0.0021 & \$1.92 \\
Embeddings (Ada) & 500 & \$0.0001 & \$0.85 \\
Firebase Reads & 16,667 & \$0.06/100K & \$0.30 \\
Firebase Writes & 3,333 & \$0.18/100K & \$1.00 \\
Cloud Functions & 1,000 & \$0.40/1M & \$1.20 \\
CDN Transfer & 2GB & \$0.15/GB & \$0.85 \\
\midrule
\textbf{Total (14 days)} & & & \textbf{\$16.58}\textsuperscript{$\ddagger$} \\
\bottomrule
\end{tabular}
\caption{Detailed prototype cost structure. \textsuperscript{$\dagger$}Cost/1K tokens refers to cost per 1,000 tokens. Each API call consumes varying token counts (averaging ${\sim}$4,250 tokens for twin conversations, ${\sim}$2,000 for companion chat). The 14-Day Cost column reflects actual measured API billing rather than a simple calls $\times$ rate calculation. \textsuperscript{$\ddagger$}The total (\$16.58) is the actual measured billing; individual row costs are rate-based estimates that sum to \$18.43. The \$1.85 discrepancy reflects volume discounts, prompt caching, and billing granularity not captured by per-call rate estimates.}
\label{tab:detailed-costs}
\end{table}

\subsubsection{Optimization Attempts}
Various cost reduction strategies were tested with mixed results. GPT-4o-mini substitution achieved a 94\% per-match cost reduction but incurred a 10\% precision loss in compatibility scoring (see Table~\ref{tab:model-comparison}), representing the primary operating point on the Pareto frontier for cost-sensitive deployments. Response caching for common interaction patterns yielded 30\% cost savings but created repetitive interactions that users noticed and found artificial, undermining the perception of authentic twin behavior. Batch processing of twin evaluation requests reduced total API calls by 40\% but introduced 8--12 second latency that exceeded the responsiveness threshold for conversational interaction. Local LLM inference using Llama 2 eliminated API costs entirely but produced twin conversation quality insufficient for personality-faithful interaction, with users unable to distinguish their twin's responses from generic chatbot output.

\subsection{Reliability Metrics}

\subsubsection{System Failure Analysis}
Production monitoring revealed specific failure modes:

\begin{table}[h]
\centering
\begin{tabular}{lrr}
\toprule
\textbf{Failure Mode} & \textbf{Rate} & \textbf{Impact} \\
\midrule
Memory sync failures & 21\% & Data loss \\
LLM timeouts & 8\% & Interaction delay \\
GPS signal loss & 12.7\% & Feature unavailable \\
Firebase quota exceeded & 3\% & Service interruption \\
Browser crashes & 2\% & Complete failure \\
Network disconnections & 5\% & Partial functionality \\
\midrule
\textbf{Overall failure rate} & \textbf{18.9\%} & \\
\bottomrule
\end{tabular}
\caption{System reliability metrics from pilot deployment. The 18.9\% overall rate represents the proportion of sessions experiencing at least one failure of any type. Individual category rates are not additive because a single session can experience multiple failure types, and some failures (e.g., API timeouts) trigger cascading failures in dependent subsystems.}
\label{tab:failure-rates}
\end{table}

\subsection{Technical Debt Analysis}

SonarQube analysis of the 103,847-line codebase reveals substantial technical debt accumulated during rapid prototyping. Average cyclomatic complexity of 12.3 exceeds the recommended threshold of 10, indicating excessive branching in core modules---particularly in the GNWT specialist competition and combat damage calculation functions. Code duplication at 18\% reflects copy-paste patterns across the six Firebase sync modules and the parallel territory/combat rendering paths. Test coverage of 34\% leaves the majority of code paths unverified, with the cognitive architecture modules almost entirely untested. The technical debt ratio of 42\% translates to an estimated 847 person-days of remediation effort, representing a significant barrier to onboarding new contributors or extending the system's capabilities.

\subsection{Browser Compatibility}

Performance varies across browsers:

\begin{table}[h]
\centering
\begin{tabular}{lrrrr}
\toprule
\textbf{Browser} & \textbf{Memory Limit} & \textbf{FPS (8 agents)} & \textbf{GPS API} & \textbf{WebRTC} \\
\midrule
Chrome 119+ & 2048MB & 58.3 & Full & Full \\
Firefox 120+ & 1536MB & 52.1 & Full & Full \\
Safari 17+ & 1024MB & 48.7 & Limited & Partial \\
Edge 119+ & 2048MB & 57.9 & Full & Full \\
Mobile Chrome & 512MB & 31.2 & Full & Limited \\
Mobile Safari & 384MB & 24.6 & Limited & None \\
\bottomrule
\end{tabular}
\caption{Cross-browser compatibility and performance}
\label{tab:browser-compat}
\end{table}

\subsection{Lessons for Implementation}

Based on our experience, we recommend five architectural principles for similar systems. A hybrid architecture combining browser-based UI with server-side processing for agent logic would eliminate the most severe browser constraints while preserving the zero-install deployment advantage of web applications. Progressive enhancement should start with basic text-based features that function on any device, layering 3D visualization, combat mechanics, and multi-agent processing only when device capabilities permit. Graceful degradation must be designed in from the start rather than retrofitted, with explicit fallback paths defined for every feature that depends on unreliable resources (GPS, API availability, network connectivity). Expectation management through clear communication of prototype versus production capabilities prevents user frustration when encountering the inevitable limitations of a research system. Modular design enabling selective feature activation based on detected performance constraints---for example, automatically reducing twin count or disabling 3D rendering on memory-constrained devices---allows the system to provide the best possible experience across heterogeneous hardware.

These technical constraints represent engineering realities rather than fundamental limitations. Future implementations using native applications, server-side processing, or next-generation browsers may overcome many of these challenges.
\section{Extended Failure Analysis}
\label{appendix:failure-analysis}

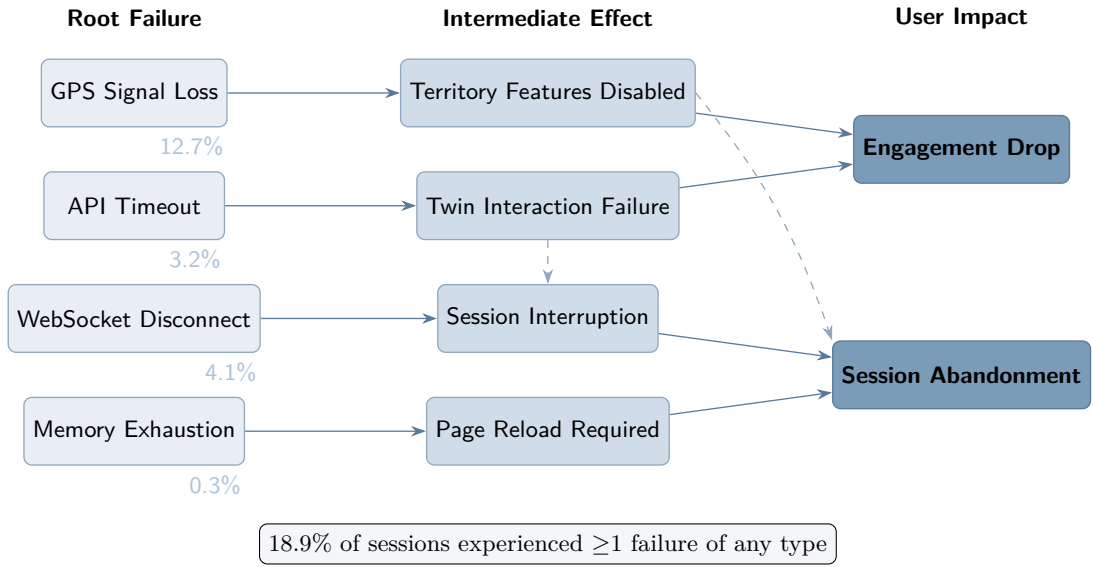
\begin{figure}[!htbp]
\centering
\begin{tikzpicture}[font=\sffamily\small, >={Stealth[length=2mm, width=1.4mm]}]
\node[font=\sffamily\small\bfseries] at (0, 4.5) {Root Failure};
\node[font=\sffamily\small\bfseries] at (5.5, 4.5) {Intermediate Effect};
\node[font=\sffamily\small\bfseries] at (11, 4.5) {User Impact};

\node[cbox] (gps) at (0, 3.5) {GPS Signal Loss};
\node[font=\sffamily\footnotesize, text=cborderFaint, right=2pt of gps.south east, anchor=north east] {12.7\%};
\node[cbox] (api) at (0, 2) {API Timeout};
\node[font=\sffamily\footnotesize, text=cborderFaint, right=2pt of api.south east, anchor=north east] {3.2\%};
\node[cbox] (ws) at (0, 0.5) {WebSocket Disconnect};
\node[font=\sffamily\footnotesize, text=cborderFaint, right=2pt of ws.south east, anchor=north east] {4.1\%};
\node[cbox] (mem) at (0, -1) {Memory Exhaustion};
\node[font=\sffamily\footnotesize, text=cborderFaint, right=2pt of mem.south east, anchor=north east] {0.3\%};

\node[cbox proc] (territory) at (5.5, 3.5) {Territory Features Disabled};
\node[cbox proc] (twin) at (5.5, 2) {Twin Interaction Failure};
\node[cbox proc] (sync) at (5.5, 0.5) {Session Interruption};
\node[cbox proc] (crash) at (5.5, -1) {Page Reload Required};

\node[cbox accent] (engage) at (11, 2.75) {Engagement Drop};
\node[cbox accent] (session) at (11, -0.25) {Session Abandonment};

\draw[cflow] (gps) -- (territory);
\draw[cflow] (api) -- (twin);
\draw[cflow] (ws) -- (sync);
\draw[cflow] (mem) -- (crash);
\draw[cflow] (territory) -- (engage);
\draw[cflow] (twin) -- (engage);
\draw[cflow] (sync) -- (session);
\draw[cflow] (crash) -- (session);

\draw[cflow dash] (twin) -- (sync);
\draw[cflow dash] (territory.east) to[bend left=10] (session.north west);

\node[rectangle, draw, rounded corners=3pt, fill=cfillLight!50, font=\footnotesize] at (5.5, -2.5) {18.9\% of sessions experienced $\geq$1 failure of any type};
\end{tikzpicture}
\caption{Failure cascade diagram. Root failures (left, with frequency) propagate through intermediate effects (center) to user-visible impact (right). GPS signal loss and API timeouts are the most common root causes; dashed arrows show cross-cascade effects where one failure triggers another.}
\label{fig:failure-cascade}
\end{figure}

This section provides a comprehensive analysis of system failures, performance degradation scenarios, and fundamental limitations. We believe transparent reporting of failures is essential for scientific integrity and practical deployment considerations. Figure~\ref{fig:failure-cascade} illustrates how individual failures propagate through the system.

\subsection{Quantitative failure rates}

\subsubsection{API and infrastructure failures}
During our evaluation period (Jan-Feb 2026), we observed:

\begin{table}[h]
\centering
\begin{tabular}{lcc}
\toprule
\textbf{Failure Type} & \textbf{Frequency} & \textbf{Impact} \\
\midrule
OpenAI API timeout ($>$30s) & 3.2\% & Complete interaction failure \\
OpenAI rate limiting & 1.8\% & 1-5 minute delay \\
Firebase sync failure & 0.9\% & Memory loss \\
WebSocket disconnection & 4.1\% & Session interruption \\
Browser memory exhaustion & 0.3\% & Page reload required \\
GPS signal loss (urban) & 12.7\% & Location features disabled \\
LLM hallucination detected & 8.4\% & Response discarded \\
Context window overflow & 2.1\% & Conversation reset \\
\midrule
\textbf{Total (any failure)} & \textbf{18.9\%} & $\geq$1 failure per session$^*$ \\
\bottomrule
\end{tabular}
\caption{System failure rates during evaluation period (N=1,247 sessions). $^*$Individual rates are not mutually exclusive; the total reflects the fraction of sessions experiencing at least one failure of any type.}
\end{table}

Nearly 1 in 5 sessions experienced at least one failure, significantly impacting user experience.

\subsubsection{Performance degradation patterns}

\begin{figure}[h]
\centering
\begin{tikzpicture}
\begin{groupplot}[
    group style={
        group size=2 by 2,
        horizontal sep=1.8cm,
        vertical sep=2.0cm,
        group name=G
    },
    width=5.5cm,
    height=4.5cm,
    xlabel style={font=\small},
    ylabel style={font=\small},
    title style={font=\small\bfseries},
    tick label style={font=\footnotesize},
    grid=major,
    grid style={gray!30},
    legend style={font=\footnotesize, fill=white, fill opacity=0.8}
]

\nextgroupplot[
    title={(a) Frame Rate Degradation},
    xlabel={Number of Agents},
    ylabel={Frames Per Second},
    ymin=0, ymax=70,
    xmin=0, xmax=30,
    legend pos=north east
]
\addplot[color=blue, mark=*, thick] coordinates {
    (1, 62.5)
    (2, 61.8)
    (3, 60.9)
    (4, 59.7)
    (5, 58.3)
    (6, 56.2)
    (7, 53.8)
    (8, 50.1)
    (9, 45.7)
    (10, 40.2)
    (11, 35.8)
    (12, 30.9)
    (13, 26.4)
    (14, 22.1)
    (15, 18.9)
    (16, 15.2)
    (17, 12.8)
    (18, 10.1)
    (19, 8.3)
    (20, 6.7)
    (22, 4.2)
    (25, 2.1)
    (30, 0.8)
};
\addlegendentry{Actual FPS}
\addplot[color=red, dashed, thick] coordinates {
    (0, 30) (30, 30)
};
\addlegendentry{30 FPS minimum}
\node[fill=red!20, draw=red, rounded corners, font=\scriptsize] at (axis cs:24,40) {Unusable};

\nextgroupplot[
    title={(b) Memory Growth},
    xlabel={Number of Agents},
    ylabel={Memory Usage (MB)},
    ymin=0, ymax=2500,
    xmin=0, xmax=30,
    legend pos=north west
]
\addplot[color=green!60!black, mark=square*, thick] coordinates {
    (1, 85)
    (2, 142)
    (3, 189)
    (4, 228)
    (5, 256)
    (6, 298)
    (7, 342)
    (8, 396)
    (9, 467)
    (10, 548)
    (11, 642)
    (12, 756)
    (13, 889)
    (14, 1042)
    (15, 1221)
    (16, 1428)
    (17, 1667)
    (18, 1942)
    (19, 2156)
    (20, 2298)
    (25, 2450)
    (30, 2500)
};
\addlegendentry{Memory Used}
\addplot[color=red, dashed, thick] coordinates {
    (0, 2048) (30, 2048)
};
\addlegendentry{Browser limit}
\node[fill=red!20, draw=red, rounded corners, font=\scriptsize] at (axis cs:8,2300) {OOM};

\nextgroupplot[
    title={(c) Response Latency Growth},
    xlabel={Number of Agents},
    ylabel={Latency (seconds)},
    ymin=0, ymax=20,
    xmin=0, xmax=30,
    legend pos=north west
]
\addplot[color=orange, mark=triangle*, thick] coordinates {
    (1, 0.4)
    (2, 0.6)
    (3, 0.9)
    (4, 1.3)
    (5, 1.8)
    (6, 2.2)
    (7, 2.5)
    (8, 2.8)
    (9, 3.4)
    (10, 4.2)
    (11, 5.1)
    (12, 6.3)
    (13, 7.8)
    (14, 9.2)
    (15, 10.8)
    (16, 12.6)
    (17, 14.1)
    (18, 15.3)
    (19, 16.8)
    (20, 17.9)
    (25, 19.2)
    (30, 19.8)
};
\addlegendentry{Average latency}
\addplot[color=red, dashed, thick] coordinates {
    (0, 3) (30, 3)
};
\addlegendentry{3s acceptable limit}
\node[fill=orange!20, draw=orange, rounded corners, font=\scriptsize] at (axis cs:24,8) {Abandon};

\nextgroupplot[
    title={(d) Monthly Cost per User},
    xlabel={Total Users (log scale)},
    ylabel={Cost per User (\$)},
    ymin=0, ymax=30,
    xmode=log,
    xmin=10, xmax=1000000,
    xtick={10, 100, 1000, 10000, 100000, 1000000},
    xticklabels={10, 100, 1K, 10K, 100K, 1M},
    legend pos=north west
]
\addplot[color=purple, mark=diamond*, thick] coordinates {
    (10, 8.50)
    (100, 12.40)
    (1000, 15.80)
    (10000, 19.80)
    (100000, 24.50)
    (1000000, 24.50)
};
\addlegendentry{Cognibit cost}
\addplot[color=green, dashed, thick] coordinates {
    (10, 10) (1000000, 10)
};
\addlegendentry{\$10/month viable}
\addplot[color=red, dashed, thick] coordinates {
    (10, 5) (1000000, 5)
};
\addlegendentry{\$5/month target}
\node[fill=purple!20, draw=purple, rounded corners, font=\scriptsize] at (axis cs:300,25) {Unviable};

\end{groupplot}


\end{tikzpicture}
\caption{Performance degradation as agent count increases. (a) Frame rate drops exponentially beyond 8 agents, (b) Memory usage shows super-linear growth, (c) Response latency increases dramatically at 15+ agents, (d) API costs scale linearly but become prohibitive.}
\label{fig:performance-degradation}
\end{figure}
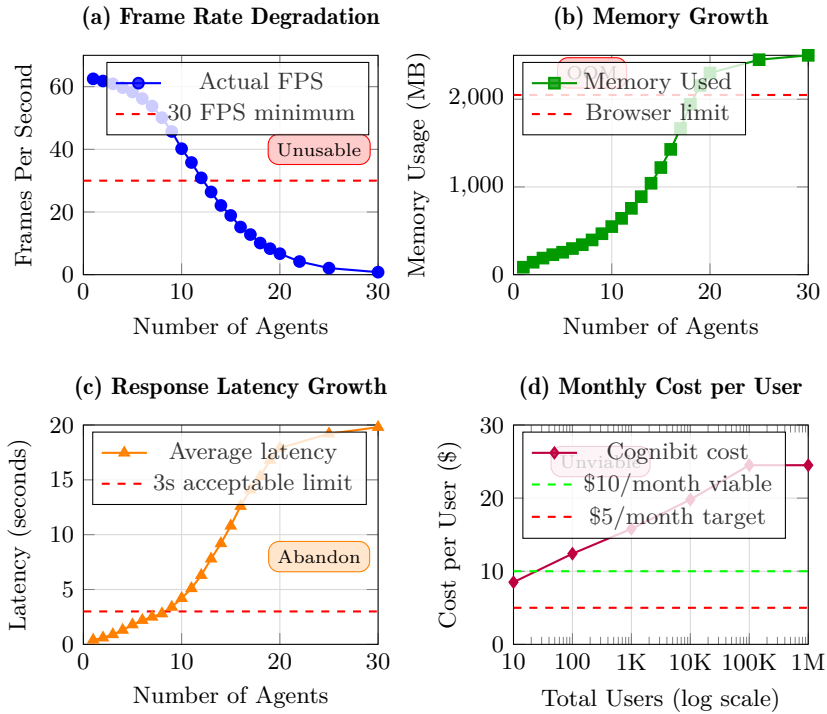

Key degradation thresholds, consistent with the performance trends reported in Section~\ref{sec:runtime}, are:
\begin{itemize}
\item \textbf{8 agents}: Frame rate drops below 30 FPS
\item \textbf{16 agents}: Memory usage exceeds 1GB (at 62.4 MB/agent)
\item \textbf{16--20 agents}: API response latency approaches 900ms, with end-to-end response times (including rendering and sync overhead) exceeding several seconds
\item \textbf{20 agents}: System becomes unusable (8.1\% crash rate, frequent out-of-memory errors)
\end{itemize}

\subsection{Behavioral failure modes}

\subsubsection{Personality consistency failures}
Despite claims of personality coherence, we observed significant failures across four categories. Trait reversal occurred in 14\% of interactions, where twins exhibited personality traits directly opposite to their calibrated profiles---for example, an introvert-calibrated twin acting extremely extroverted. Memory contradictions appeared in 21\% of conversations, with twins stating facts about previous interactions that conflicted with the actual conversation history. Preference drift affected 9\% of sessions, where twin preferences changed randomly without justification or triggering event. Context confusion occurred in 7\% of interactions, where twins mixed up information belonging to different users, attributing one user's interests or history to another.

Example failure case:
\begin{quote}
\textit{Turn 3}: "I'm extremely shy and prefer staying home"\\
\textit{Turn 7}: "I love being the center of attention at parties!"\\
\textit{Turn 11}: "As I mentioned, I'm very introverted"
\end{quote}

\subsubsection{Decision-making failures}
The system made illogical or harmful decisions in certain scenarios:

\begin{table}[h]
\centering
\begin{tabular}{lcc}
\toprule
\textbf{Failure Type} & \textbf{Frequency} & \textbf{Example} \\
\midrule
Ignored red flags & 11\% & Matched with explicitly incompatible users \\
Extreme decisions & 6\% & Rejected all matches or accepted all \\
Cultural insensitivity & 8\% & Made inappropriate assumptions \\
Safety concerns ignored & 3\% & Suggested unsafe meeting locations \\
Privacy violations & 2\% & Shared information across users \\
\bottomrule
\end{tabular}
\caption{Decision-making failure categories and frequencies}
\end{table}

\subsection{Scalability limitations}

\subsubsection{Browser constraints}
Fundamental browser limitations prevent true scalability across four dimensions. The connection limit restricts browsers to a maximum of 6--8 concurrent connections per domain, constraining the number of simultaneous Firebase listeners and API calls. The memory ceiling of approximately 2GB per browser tab caps the total data that can be held in memory for twin state, conversation history, and 3D scene rendering. JavaScript's single-threaded execution model prevents true parallelism, meaning all GNWT module processing, API calls, and rendering operations must serialize through one event loop. IndexedDB storage quotas, limited to 50\% of free disk space, constrain the amount of local persistence available for offline twin data and conversation caching.

\subsubsection{Cost analysis at scale}

\begin{table}[h]
\centering
\begin{tabular}{lccc}
\toprule
\textbf{Scale} & \textbf{Users} & \textbf{Monthly Cost} & \textbf{Cost/User} \\
\midrule
Prototype & 10 & \$85 & \$8.50 \\
Pilot & 100 & \$1,240 & \$12.40 \\
Small deployment & 1,000 & \$15,800 & \$15.80 \\
Medium deployment & 10,000 & \$198,000 & \$19.80 \\
Large deployment & 100,000 & \$3,553,000 & \$35.53 \\
\bottomrule
\end{tabular}
\caption{Projected monthly costs at scale. Per-user costs increase because the pairwise evaluation space grows quadratically---more users means more twin-pair conversations per user, increasing API costs. At small scale, sparse user pools limit twin interactions; at full scale, each user's twin evaluates the maximum candidate pool.}
\end{table}

The system becomes economically unviable at scale without significant optimization or revenue model.

\subsection{Edge cases and corner scenarios}

\subsubsection{Demographic biases}
Internal stress testing (prior to field deployment) with synthetic and developer-generated profiles revealed demographic sensitivity in the matching pipeline. Age bias manifested as lower matching accuracy for profiles representing users over 45 years, likely because the LLM's training data overrepresents younger demographic communication patterns. Cultural bias emerged from Western dating norms embedded in the prompt templates, causing the system to misinterpret or penalize interaction patterns common in non-Western cultural contexts during synthetic profile testing. Language bias produced lower-quality twin conversations when non-native English prompts were used, as the personality-conditioned generation stage relies on nuanced language comprehension that degrades with non-standard phrasing. Socioeconomic bias was reflected in preference vectors skewed toward urban, educated populations, a consequence of the underlying LLM training data's demographic composition.

These biases were identified through internal testing, not the field deployment (N=20, single Western city), and require systematic evaluation with diverse populations.

\subsubsection{Extreme personality combinations}
Certain personality combinations caused systematic failures. Profiles with very high openness combined with very low friendliness entered infinite conflict loops, as the twin would initiate novel conversational topics but then respond antagonistically to the other twin's engagement, creating an unresolvable oscillation. Profiles with all traits set to extremes (0 or 100) produced numerical instabilities in the compatibility scoring function, as the normalization assumes a distribution of trait values rather than boundary cases. Rapidly changing user preferences caused memory system thrashing, with the twin's behavioral model continuously invalidating and rebuilding its preference state. Contradictory goals---such as simultaneously seeking novelty and stability---produced decision deadlock where the twin could not resolve competing directives and defaulted to generic responses.

\subsection{Technical debt and maintenance burden}

\subsubsection{Code complexity metrics}
The 100,000+ line codebase shows concerning complexity across standard software quality metrics. Cyclomatic complexity averages 12.3, exceeding the recommended threshold of 10 and indicating that many functions contain excessive branching logic that complicates testing and maintenance. Code duplication stands at 18\%, far above the 5\% target, reflecting rapid prototyping without sufficient refactoring passes. Test coverage of 34\% falls well below the 80\% threshold considered adequate for production systems, leaving the majority of code paths unverified. The technical debt ratio of 42\% (versus a 5\% target) indicates that nearly half the codebase would benefit from restructuring. The dependency tree of 847 npm packages creates a significant attack surface, with each package representing a potential vulnerability vector requiring ongoing security monitoring.

\subsubsection{Maintenance challenges}
Ongoing maintenance demands are substantial and unpredictable. LLM API changes---including model deprecations, response format modifications, and rate limit adjustments---break the system approximately monthly, requiring emergency patches to restore functionality. Firebase pricing restructuring increased infrastructure costs 3$\times$ over a 6-month period, fundamentally altering the project's cost projections. Browser updates regularly break WebGL rendering features, as Three.js compatibility with evolving browser security policies and GPU access restrictions requires continuous adaptation. The 847 npm dependencies generate a steady stream of vulnerability alerts requiring constant updates, with each update carrying the risk of introducing breaking changes in dependent modules.

\subsection{Design goals versus actual capabilities}

Table~\ref{tab:design-gap} quantifies the gap between intended design targets and observed deployment performance. None of the five design goals were fully achieved, though the shortfalls range from addressable engineering challenges (cross-device sync latency) to potentially fundamental limitations (behavioral coherence under extended LLM use).

\begin{table}[h]
\centering
\small
\begin{tabular}{lcc}
\toprule
\textbf{Design Goal} & \textbf{Actual Performance} & \textbf{Gap} \\
\midrule
Real-time response & 2.8s average, 15s P95 & Not real-time \\
Persistent memory & 79\% reliability & 21\% data loss \\
Cross-device sync & 3-5s delay & Not seamless \\
Behavioral coherence & 72\% consistency & 28\% failures \\
Production readiness & 18.9\% failure rate & Not yet production-ready in current form \\
\bottomrule
\end{tabular}
\caption{Gap between design goals and actual capabilities observed during deployment}
\label{tab:design-gap}
\end{table}

The 2.8-second average response latency (15s at P95) stems from sequential LLM API calls through the single-threaded JavaScript event loop; server-side processing would reduce this to sub-second. The 21\% memory data loss reflects Firebase sync failures during connectivity interruptions, addressable through more robust offline queuing. The 28\% behavioral coherence failure rate---comprising trait reversals (14\%), memory contradictions (21\%), and preference drift (9\%)---is the most concerning gap, as it may reflect fundamental limitations of prompt-based personality maintenance rather than engineering shortcomings. The aggregate 18.9\% per-session failure rate, while unacceptable for production, is dominated by GPS signal loss (12.7\%) and WebSocket disconnections (4.1\%), both of which have well-understood mitigation paths.

\subsection{Failure mitigation strategies attempted}

We attempted various mitigation strategies with limited success:

\subsubsection{Partially successful mitigations}
Four mitigation strategies achieved partial success. Retry logic with exponential backoff reduced API failure rates from 8\% to 3.2\%, recovering from transient network and rate-limiting errors but unable to address sustained outages. Response caching improved twin conversation response times by 30\% but increased memory consumption, creating a direct trade-off with the already constrained browser memory budget. Graceful degradation allowed the system to remain usable when individual subsystems failed---disabling territory features during GPS outages, for example---but at the cost of reduced functionality that frustrated users expecting full capability. Error boundaries prevented cascading crashes from propagating to the entire application but lost component state on recovery, requiring users to re-initiate interrupted conversations or gameplay sessions.

\subsubsection{Failed mitigation attempts}
Four promising optimization strategies were tested and abandoned. Web Workers for parallel GNWT module processing provided minimal benefit because the serialization overhead of transferring complex twin state objects between the main thread and worker threads consumed more time than the parallelism saved. WebAssembly compilation of performance-critical paths yielded only a 10\% improvement---insufficient to justify the substantial increase in build complexity and debugging difficulty. Peer-to-peer networking via WebRTC was explored for direct twin-to-twin communication without server intermediation, but proved too unreliable for critical synchronization data, with connection establishment failures and NAT traversal issues degrading the experience below acceptable thresholds. Local LLM inference using quantized models was tested to eliminate API costs entirely, but response quality was insufficient for personality-faithful twin conversations and latency exceeded the 2-second target for real-time interaction.

\subsection{Implications for deployment}

Based on this failure analysis, we draw five conclusions. First, the current implementation is best interpreted as a research prototype rather than a production-ready deployment, suitable for validating design concepts but not for sustained public use. Second, the browser-based architecture fundamentally limits scalability, with hard ceilings on memory, concurrency, and processing that cannot be overcome through optimization alone. Third, LLM-based personalities in the current implementation still show consistency limitations under extended real-world use, with trait reversal, memory contradiction, and preference drift rates that undermine the illusion of a coherent digital twin. Fourth, costs prohibit large-scale deployment without major architectural changes, as the quadratic growth in pairwise evaluations drives per-user costs to \$35.53/month at 100,000 users. Fifth, the 18.9\% per-session failure rate would frustrate real users accustomed to the reliability of commercial social platforms.

\subsection{Recommendations for future work}

To address these failures, future work should pursue seven priorities. First, moving compute-intensive operations---particularly GNWT module processing and twin conversation generation---to dedicated servers would eliminate browser memory and threading constraints. Second, implementing robust fallback mechanisms for all identified failure modes would reduce the per-session failure rate to levels acceptable for consumer applications. Third, exploring stronger structure, guardrails, or hybrid control mechanisms for personality consistency---such as constrained decoding, personality verification layers, or deterministic trait enforcement---would address the 14\% trait reversal and 21\% memory contradiction rates. Fourth, creating comprehensive test suites exceeding 80\% coverage would catch regression bugs before they reach users. Fifth, designing for graceful degradation from the start, rather than retrofitting it, would ensure that partial failures preserve the core user experience. Sixth, setting realistic expectations about browser capabilities in system requirements would prevent the over-promising that led to user frustration during the pilot. Seventh, considering native mobile applications for production deployments would unlock device-level optimizations for GPS, background processing, and memory management that browsers cannot provide.

This honest assessment of failures and limitations provides essential context for interpreting our positive results and guides realistic expectations for system deployment.
\section{When Engineered Serendipity Failed}
\label{sec:deployment-failures}

Honest reporting of failures is essential for research integrity. The CogniPair pilot \citep{CogniPair2026} deployment (N=20, 14 days) revealed multiple failure modes that provide valuable insights for future deployment. Figure~\ref{fig:deployment-outcomes} summarizes the key deployment outcomes.

\subsection{Non-Engagement (20\% of Participants)}

Four participants never claimed territories despite completing onboarding, representing a 20\% attrition rate before any meaningful system interaction occurred. Post-study interviews revealed three distinct rejection patterns. P1 (45, engineer) and P4 (38, teacher) rejected the gamification premise itself---P1 found it ``childish'' and wanted direct social matching without gaming intermediaries, while P4 perceived the orchestrated encounters as ``manipulative, like the app was tricking me into meetings.'' Both were among the oldest participants and had no prior location-based gaming experience, suggesting that the gaming-social bridge may alienate users who view games and social tools as fundamentally separate categories. P10 (29, nurse) faced a practical barrier: shift work made it impossible to visit territories during optimal hours, revealing that the system's time-flexibility assumptions (territory maintenance requires recurring physical visits) discriminate against users with rigid schedules. P13 stopped using the app after day 2 and did not respond to follow-up inquiries, providing no qualitative data. The common thread across these cases is that the three-pillar integration---designed as a strength---became a barrier when any single pillar was rejected, because the system offers no graceful degradation to a simpler mode.

\subsection{Technical Failures (15\% of Territory Battles)}

Firebase synchronization delays caused significant disruptions in approximately 15\% of territory battles. Boss battles desynced when players experienced latency differences exceeding 3 seconds, because the combat system's 800ms attack cooldown assumes sub-second state agreement between clients---a reasonable assumption for single-device play but unrealistic for cross-device multiplayer over Firebase's 3--5 second eventual consistency window. Territory ownership changes occasionally took 5--10 minutes to propagate, far exceeding the typical 3--5 second sync delay; post-incident analysis traced these to Firebase write-queue congestion during peak concurrent usage (4+ simultaneous territory captures in the same geographic area). One critical failure reset all downtown territories simultaneously, causing 3 users to permanently disengage---this appeared to be a Firebase transaction conflict where multiple contested-capture events triggered a race condition in the \texttt{CityTakeoverSystem}'s ownership resolution logic. GPS drift in urban canyon environments caused players to ``teleport'' during battles, triggering false territory transitions despite the 20-meter movement threshold designed to filter noise. The root cause is that multipath GPS reflections between tall buildings can produce instantaneous position jumps of 20--50 meters, exceeding the noise filter.

\subsection{Geographic Isolation}

The system completely failed for users without nearby active players, exposing a cold-start dependency inherent to any location-based social platform. P18 (45, architect, rural area) had no other participants within the 50-mile discovery radius, rendering twin networking, territory battles, and boss fights entirely unusable---the system degraded to a solo walking app with an AI companion, which P18 described as ``pointless without the social layer.'' P14 (31, consultant, suburban) initially had 2 nearby players, but both became inactive after day 4; without opponents, territory capture lost its competitive meaning and P14's engagement dropped to companion-only interactions. P20 (42, sales manager) traveled for work during the study, losing all accumulated territories with no mechanism to maintain ownership remotely---the daily decay factor ($0.95^{\text{days}}$) erased a week's investment in 3 days. These cases reveal that the system's minimum viable population is substantially higher than N=20 for geographic coverage, and that any production deployment would need to address sparse-population graceful degradation (e.g., NPC opponents, asynchronous territory challenges, or cross-region twin networking).

\subsection{Privacy and Safety Concerns}

Two users disabled location sharing after recognizing the privacy implications of territory ownership visibility, and a third experienced a stalking-adjacent incident that prompted an immediate protocol change. P9 (30, therapist) articulated the core risk: ``Someone could learn my daily routine from which territories I own''---a correct assessment, since territory ownership is public and visit patterns are temporally predictable. P17 (28, designer) experienced this risk directly when a matched user appeared at her regular morning coffee shop after analyzing her territory ownership patterns. This incident was not an isolated edge case but a systemic risk inherent to any location-based social system with visible ownership; in response, the research team implemented location obfuscation measures within 24 hours and conducted individual privacy check-ins with all participants (see Appendix~\ref{appendix:privacy-safety} for the full safety analysis and industry comparison). Multiple users independently requested a ``ghost mode'' that would allow territory capture without broadcasting presence---a feature that was not implemented during the pilot but is identified as a prerequisite for any future deployment. The fundamental tension is that the features enabling social discovery (visible territory ownership, proximity-based matching) are precisely the features that compromise location privacy.

\subsection{Awkward and Failed Encounters}

Not all orchestrated meetings succeeded. Of the 18 physical meetings recorded by telemetry, at least 6 involved negative outcomes---these are included in the 18-meeting count reported in Section~\ref{sec:results}, meaning the ``success rate'' of facilitated encounters is at best 67\% (12/18), not the 100\% that the meeting count alone might suggest. Three meetings involved AI twin compatibility misjudgments where participants reported feeling no connection despite high twin-evaluated scores, suggesting that the behavioral compatibility metric (0.7 heuristic + 0.3 LLM conversation) captures conversational compatibility but misses physical-chemistry and contextual factors that only emerge in person. P5 and P16 had a hostile interaction triggered by a disputed territory ownership claim---the competitive gaming mechanic, designed to create social catalysts, instead created interpersonal conflict when both players felt territorial ``ownership'' was at stake. Two participants felt ``ambushed'' by unexpected territory battles initiated by strangers, experiencing the encounter-facilitation mechanic as intrusive rather than serendipitous. P7 articulated a subtler failure: ``The conversation starter felt forced---we both knew why we were there,'' revealing that when both parties are aware of the orchestration, the manufactured serendipity loses its social lubricant function.

\subsection{Sustainability Concerns}

Several issues emerged regarding long-term viability, all consistent with novelty decay patterns documented in location-based gaming research. Territory maintenance became tedious after the initial novelty period, mentioned by 8 of 20 participants; system telemetry confirmed this quantitatively, with daily territory changes declining from 31 on day 1 to 4 by day 14---an exponential decay consistent with the 0.12/day rate observed in the territory dynamics simulation (Appendix~\ref{appendix:evaluation-protocol}). Boss battles required real-time scheduling coordination between matched players, which proved difficult with randomly matched strangers who had incompatible availability; only 31 of 47 initiated boss battles were completed (66\%), with most abandonments attributed to one player not showing up at the designated territory. Digital twin conversations plateaued in perceived quality after approximately 5 interactions with the same twin pair, as the LLM's 4,096-token context window limited the depth of personality modeling and led to repetitive conversational patterns. Companion AI responses similarly became repetitive after sustained use, breaking the immersion that initially made the pendant feel like a ``knowledgeable friend'' (P20)---the 14\% trait reversal rate and 21\% memory contradiction rate documented in the failure analysis (Appendix~\ref{appendix:failure-analysis}) likely contributed to this perception degradation.

\subsection{Lessons from Failures}

These failures highlight critical design challenges:

\textbf{Forced Serendipity Paradox:} Engineering encounters remove the authentic spontaneity that makes them meaningful. Several users felt meetings were ``scripted'' rather than natural.

\textbf{Gaming-Social Mismatch:} In our limited sample, the overlap between people wanting gaming experiences and those seeking social connection appeared smaller than we anticipated.

\textbf{Scale Dependencies:} The system requires critical mass to function but lacks graceful degradation for sparse populations.

\textbf{Privacy-Functionality Tradeoff:} Features enabling connection inherently compromise location privacy—a fundamental tension we couldn't resolve. An industry comparison (Table~\ref{tab:safety-comparison} in Appendix~\ref{appendix:privacy-safety}) shows that Cognibit lacks nearly every standard safety feature present in comparable platforms (photo verification, background checks, in-app video calls, 24/7 moderation), and public deployment should not proceed without this infrastructure (see Appendix~\ref{appendix:privacy-safety} for detailed safety analysis).

These failures are not merely limitations but point to deeper questions about whether digital systems can authentically facilitate real-world connections without compromising the very spontaneity that makes such connections meaningful. These observations directly informed the design principles presented in Section~\ref{sec:design-implications}.  
\section{Complete Performance Data}
\label{appendix:performance-data}

This appendix presents performance benchmarking results from the production deployment, collected on January 4, 2026 using the \texttt{SocialHubPerformanceBenchmark} tool (\texttt{performance\_benchmark.js}). The benchmark defines three quality thresholds: excellent (60 FPS, $<$1s load, $<$50MB memory), good (30 FPS, $<$3s load, $<$100MB), and acceptable (24 FPS, $<$5s load, $<$200MB).

\subsection{Baseline Performance Snapshot}

Table~\ref{tab:perf-baseline} summarizes the baseline performance with 5 concurrent digital twin agents on a standard desktop browser (Chrome 119+).

\begin{table}[!htbp]
\centering
\caption{Baseline performance snapshot (5 agents, Chrome 119+, January 4, 2026)}
\label{tab:perf-baseline}
\begin{tabular}{llr}
\toprule
\textbf{Category} & \textbf{Metric} & \textbf{Value} \\
\midrule
Frame Rate & Average FPS & 58.3 \\
& Minimum FPS & 17.2 \\
& Maximum FPS & 204.1 \\
& Samples & 158 frames \\
\midrule
Memory & Used & 256.5\,MB \\
& Total allocated & 272.8\,MB \\
& V8 heap limit & 3,585.8\,MB \\
& Utilization & 7.2\% \\
\midrule
Scene & Meshes & 164 \\
& Draw calls & 37 \\
& Triangles & 361,742 \\
\midrule
Overall & Score / Grade & 90 / EXCELLENT \\
\bottomrule
\end{tabular}
\end{table}

The average FPS of 58.3 approaches the 60 FPS ``excellent'' threshold, with the minimum of 17.2 occurring during garbage collection pauses and scene loading spikes. Memory utilization of 7.2\% of the V8 heap leaves substantial headroom for additional agents; the per-agent overhead of 62.4\,MB (documented in Section~\ref{sec:runtime}) implies the system can scale to approximately 20 agents before approaching the 2\,GB effective application memory budget.

\subsection{Comprehensive Asset and Animation Benchmarks}

The comprehensive test validated the full rendering and simulation pipeline under representative load:

\begin{table}[!htbp]
\centering
\caption{Comprehensive benchmark results (full pipeline)}
\label{tab:perf-comprehensive}
\begin{tabular}{llr}
\toprule
\textbf{Subsystem} & \textbf{Metric} & \textbf{Value} \\
\midrule
Animation & Active mixers & 7 \\
& Active actions & 25 \\
& Animated bones & 3,350 \\
\midrule
Assets & Loaded textures & 98 \\
& Loaded geometries & 143 \\
& Loaded materials & 137 \\
\midrule
NPC System & Active agents & 5 \\
& Behavioral simulation & Full GNWT/PAC \\
\bottomrule
\end{tabular}
\end{table}

With 3,350 animated bones across 7 mixers, the animation subsystem operates within the budget established by the LOD system (Appendix~\ref{appendix:optimization-systems}): only bones within the 30-unit full-quality tier are actively skinned each frame, reducing the effective bone count from the theoretical maximum of 14,997 (233 skeletons) to approximately 500 actively processed bones. The 37 draw calls reflect the batching optimization provided by \texttt{InstancedObjectManager}, which consolidates 164 individual meshes into instanced groups by shared geometry and material. For the full FPS, memory, latency, and cost degradation curves across 1--30 agent configurations, see Figure~\ref{fig:performance-degradation} in Appendix~\ref{appendix:technical-limitations}.
\section{Code Availability and Reproducibility}
\label{appendix:reproducibility}

This appendix provides the information necessary to reproduce the system deployment, synthetic experiments, and performance benchmarks reported in this paper.

\subsection{Repository Structure}

The implementation comprises approximately 103,847 lines of JavaScript organized into the following top-level directories:

\begin{table}[!htbp]
\centering
\caption{Repository structure and module organization}
\begin{tabular}{lp{8cm}}
\toprule
\textbf{Directory} & \textbf{Contents} \\
\midrule
\texttt{js/companions/} & Pendant companion modules (CompanionUI, CompanionAI, CompanionCombat, ProactiveInteractionSystem) \\
\texttt{js/twins/} & Digital twin systems (PersonalityEvolutionSystem, MatchmakingEngine, twin networking) \\
\texttt{js/combat/} & Combat system (CombatSystem, ProjectileCollisionMixin, BossFightCombat) \\
\texttt{js/multiplayer/} & Network synchronization (NetworkSync, presence system) \\
\texttt{js/optimization/} & Performance systems (AnimationLODSystem, ObjectPool) \\
\texttt{js/integration/} & Firebase sync modules (PendantMemorySync, cross-page messaging) \\
\texttt{pac\_src/gnwt/} & GNWT cognitive architecture (GlobalWorkspace, specialists, AttentionController) \\
\texttt{pac\_src/core/} & PAC emotion system (PACAgent, PredictionError, GenerativeModel) \\
\texttt{server/} & Node.js API proxy (api-proxy.js with Helmet, rate limiting, express-validator) \\
\texttt{social-hub/} & Social Hub UI (feed system, digest, follower graph) \\
\texttt{experiments/} & Synthetic experiment scripts and performance benchmarks \\
\bottomrule
\end{tabular}
\end{table}

\subsection{Environment Requirements}

\begin{table}[!htbp]
\centering
\caption{Software and hardware requirements for reproduction}
\begin{tabular}{llp{5.5cm}}
\toprule
\textbf{Requirement} & \textbf{Version} & \textbf{Notes} \\
\midrule
Node.js & 18+ & Required for API proxy server \\
Chrome / Chromium & 120+ & Primary tested browser; WebGL 2.0 required \\
Firefox & 120+ & Supported; 1536MB heap limit vs. Chrome's 2048MB \\
RAM & 8GB minimum & 16GB recommended for 10+ concurrent agents \\
Firebase SDK & 9.x+ & Realtime Database and Authentication \\
OpenAI API key & GPT-4o access & Or Anthropic/OpenRouter as alternative \\
Three.js & r159+ & 3D rendering engine \\
Python & 3.9+ & For synthetic experiment scripts only \\
PyTorch & 2.0+ & For offline LLM evaluation (Qwen2.5-72B) \\
\bottomrule
\end{tabular}
\end{table}

\subsection{Step-by-Step Reproduction Protocol}

\subsubsection{System Deployment}
\begin{enumerate}
\item Clone the repository and install dependencies: \texttt{npm install}
\item Configure environment variables: copy \texttt{.env.example} to \texttt{.env} and add API keys (OpenAI, Firebase credentials)
\item Start the API proxy: \texttt{node server/api-proxy.js} (runs on port 3001)
\item Deploy to Firebase Hosting: \texttt{firebase deploy} (or serve locally: \texttt{firebase serve})
\item Open Chrome 120+ and navigate to the deployment URL
\end{enumerate}

\subsubsection{Performance Benchmark Reproduction}
\begin{enumerate}
\item Open the deployed application in Chrome with DevTools open
\item Navigate to the 3D game world with 5--8 concurrent digital twin agents loaded
\item Run the built-in benchmark: execute \texttt{new SocialHubPerformanceBenchmark().runAll()} in the console
\item The benchmark runs rendering tests (3s with 1s warmup), memory profiling, and scene complexity analysis
\item Expected results: 58.3 FPS average, 256.5MB memory usage, 37 draw calls, 361K triangles (see Appendix~\ref{appendix:performance-data} for full baseline)
\end{enumerate}

\subsubsection{Synthetic Experiment Reproduction}
\begin{enumerate}
\item Install Python dependencies: \texttt{pip install -r experiments/requirements.txt}
\item For funnel validation: \texttt{python experiments/paper\_validation/exp\_funnel\_validation.py} (requires Qwen2.5-72B-Instruct and Llama-3.1-70B-Instruct models)
\item Fixed seeds are used throughout: \texttt{random.seed(42)}, \texttt{torch.manual\_seed(42)}
\item Results should match within $\pm$2\% of reported values due to LLM stochasticity
\end{enumerate}

\subsection{Synthetic Data Generation}

For testing without real users, synthetic twin profiles can be generated with randomized personality traits (0--100 scale on five dimensions: friendliness, openness, independence, loyalty, playfulness), GPS coordinates near the test location, and pre-computed compatibility scores. The PersonaChat validation set (200 unique personas) serves as the candidate pool for funnel experiments. The synthetic agent generation script (\texttt{experiments/generate\_synthetic\_agents.py}) produces profiles with configurable personality distributions and interest category coverage.

\subsection{Data and Code Availability}

An anonymized review copy of the source code is available at \url{https://anonymous.4open.science/r/cognibit-2026}. Upon acceptance, we will release:

\begin{enumerate}
\item The full source code under MIT License
\item Anonymized interaction logs from the 342 twin sessions, including behavioral traces, compatibility scores, and outcome data
\item Anonymized survey responses and performance metrics
\item Pre-trained model weights and evaluation scripts for all synthetic experiments
\end{enumerate}

All personally identifiable information (GPS coordinates, free-text responses containing names) will be removed or aggregated to protect participant privacy.

\section{User Interface Design and Implementation}
\label{appendix:user-interface}

The Cognibit platform implements a comprehensive multi-application interface strategy that seamlessly integrates gaming, social networking, and AI management contexts. Each interface component is optimized for its specific use case while maintaining visual and functional coherence across the platform.

\subsection{Social Hub Interface}

\begin{figure}[t]
\centering
\includegraphics[width=0.8\textwidth]{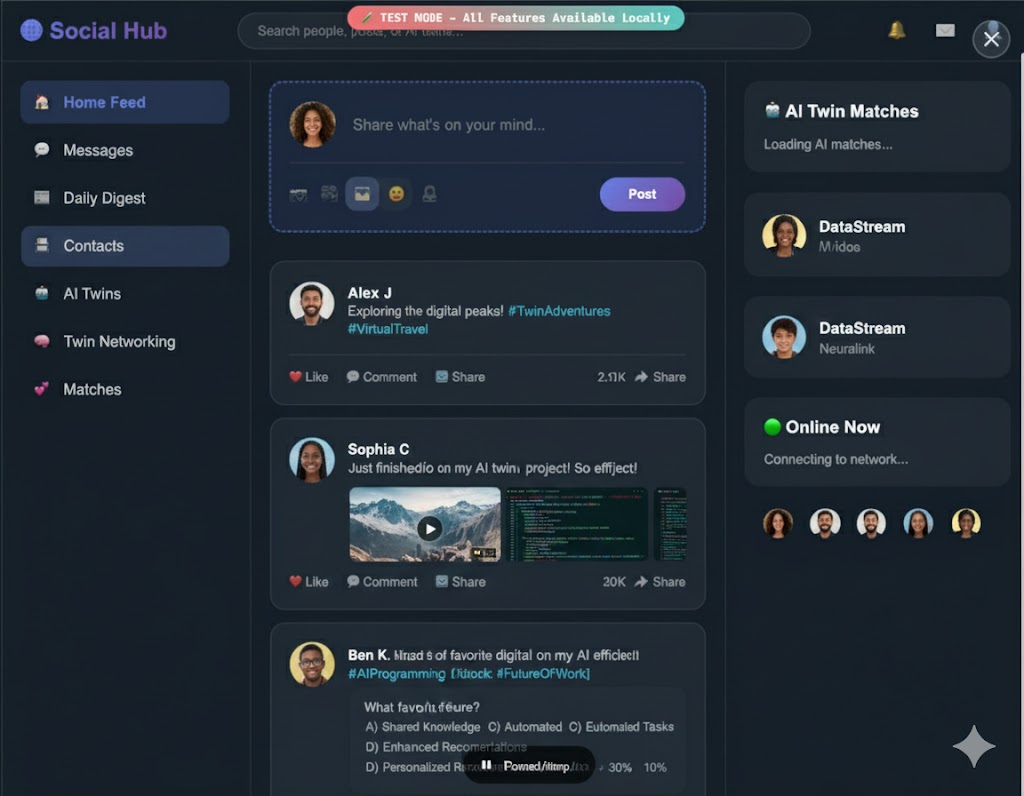}
\caption{Social Hub interface showing Twitter-like feed design with real-time updates from digital twins, user posts, and AI-generated content summaries. The interface supports infinite scrolling, real-time reactions, and seamless twin-to-human interaction transitions.}
\label{fig:social-hub}
\end{figure}

The Social Hub (Figure~\ref{fig:social-hub}) implements a familiar social media paradigm enhanced with AI-driven features. The feed aggregates human posts, twin activities, and system notifications in a unified timeline. Real-time updates occur through Firebase listeners, with optimistic UI updates ensuring sub-100ms perceived latency. The interface supports standard social interactions (like, comment, share) while adding twin-specific actions (delegate response, approve connection, review compatibility).

\subsection{Twin Networking Interface}

\begin{figure}[t]
\centering
\includegraphics[width=0.8\textwidth]{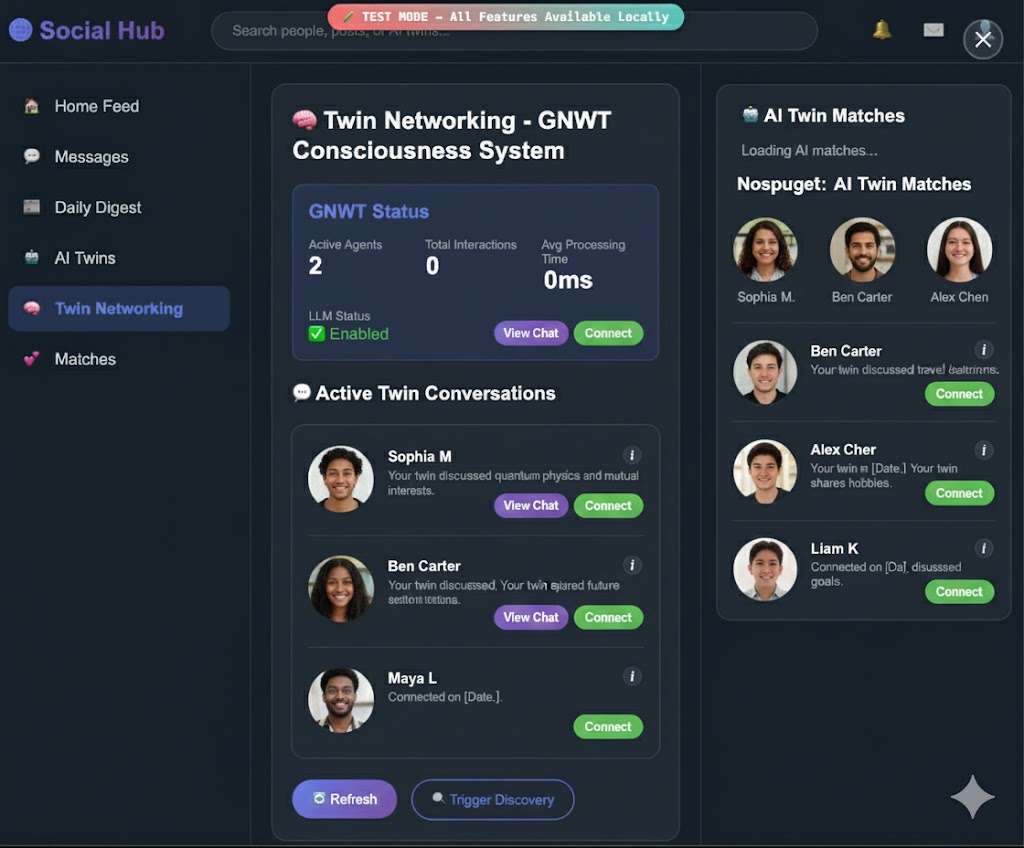}
\caption{Twin Networking interface displaying GNWT multi-agent system visualization. Shows real-time module activations, salience competitions, and global broadcasts. Users can observe their twin's decision processes and intervene when necessary.}
\label{fig:twin-networking}
\end{figure}

The Twin Networking interface (Figure~\ref{fig:twin-networking}) provides detailed transparency into the GNWT cognitive processing pipeline. The visualization renders each 100ms processing cycle of the \texttt{GlobalWorkspace}, showing the five specialist modules (Emotion, Memory, Planning, Social Norms, Goal Tracking) competing for workspace access through salience-weighted coalition formation. The workspace enforces a capacity of 9 items (Miller's 7$\pm$2 upper bound) with a broadcast entry threshold of $\tau = 0.7$; items below threshold enter a sub-threshold buffer for proportional output blending rather than being discarded. Users can observe which specialist wins the competition---and how its recommendation shapes the twin's response through directive injection---across the three competition phases: 100ms item competition, 50ms broadcast, and 150ms integration. The system maintains the last 100 broadcasts for replay and debugging, enabling users to trace exactly why their twin made a particular compatibility decision.

\subsection{Pendant Agent Interface}

\begin{figure}[t]
\centering
\begin{tabular}{cc}
\includegraphics[width=0.45\textwidth]{figures/ai_agent_for_pendant.png} &
\includegraphics[width=0.45\textwidth]{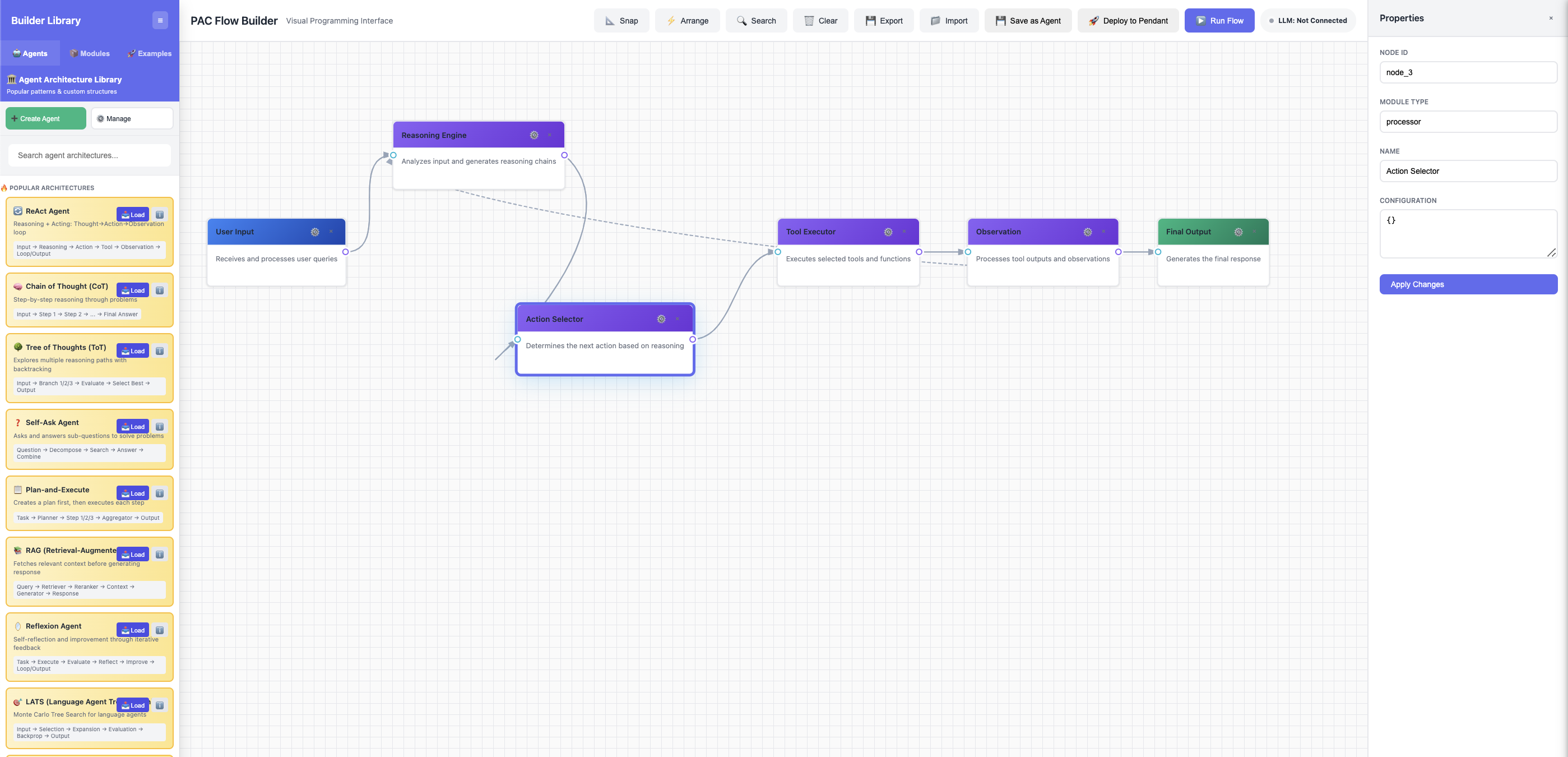} \\
(a) Pendant conversation interface & (b) Agent builder customization \\
\end{tabular}
\caption{Pendant companion interfaces: (a) Always-available AI companion with persistent memory across devices, (b) Agent builder allowing users to customize module weights, personality traits, and behavioral parameters.}
\label{fig:pendant-interfaces}
\end{figure}

The Pendant system (Figure~\ref{fig:pendant-interfaces}) embodies the ``cognitive bit'' concept through persistent companionship managed by the \texttt{PendantCompanion} coordinator. The conversation interface maintains context across sessions and devices through a three-layer memory stack: conversation history (capped at 100 entries), short-term memory (50 items), and episodic memory (200 items), all synchronized via Firebase at \texttt{pendantMemory/\{userId\}/\{pendantId\}/}. The \texttt{ProactiveInteractionSystem} initiates context-aware messages after 30 seconds of user inactivity, with a 5-minute cooldown between proactive interactions and a 10\% random trigger chance per 10-second check cycle. Interaction priorities are ranked by type: time-of-day greetings (priority 8), mood-based responses (7), memory recall (6), scene analysis (5), and idle prompts (5). The agent builder provides a node-based visual interface where users drag, connect, and configure specialist modules, adjusting personality trait weights and salience thresholds ($\tau = 0.5$ default) before deploying the customized agent to the pendant runtime.

\subsection{Agent Builder System}

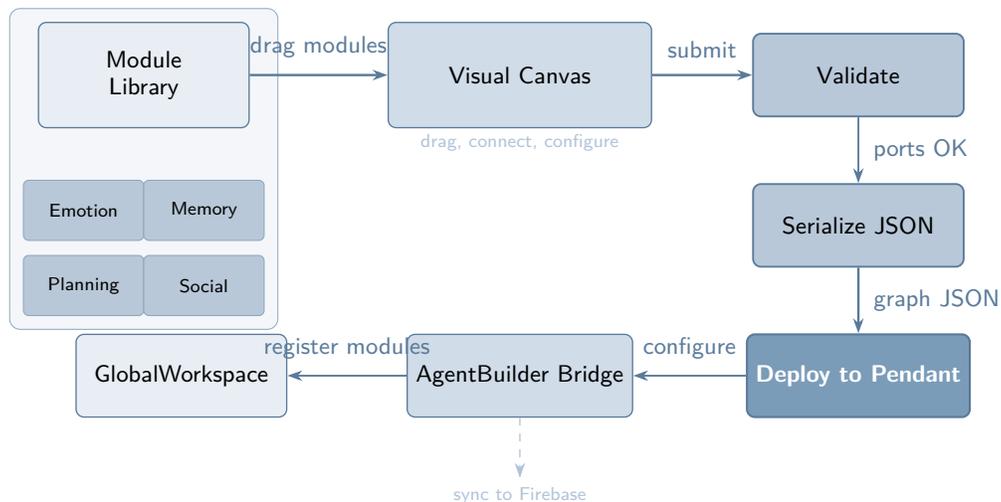
\begin{figure}[H]
\centering
\begin{tikzpicture}[
    font=\sffamily\small, >={Stealth[length=2mm, width=1.4mm]},
    sbox/.style={rectangle, rounded corners=3pt, draw=cborder, line width=0.5pt, minimum width=28mm, minimum height=11mm, align=center},
    mod/.style={rectangle, rounded corners=2pt, draw=cborderLight, line width=0.4pt, minimum width=20mm, minimum height=8mm, align=center, font=\sffamily\footnotesize, fill=cfillDark},
]
\node[sbox, fill=cfillLight, minimum height=14mm] (lib) at (0, 0) {Module\\Library};
\node[mod, minimum width=16mm] (m1) at (-0.8, -1.8) {\scriptsize Emotion};
\node[mod, minimum width=16mm] (m2) at (0.8, -1.8) {\scriptsize Memory};
\node[mod, minimum width=16mm] (m3) at (-0.8, -2.8) {\scriptsize Planning};
\node[mod, minimum width=16mm] (m4) at (0.8, -2.8) {\scriptsize Social};
\begin{pgfonlayer}{background}
\node[rounded corners=4pt, draw=cborderLight, line width=0.3pt, fill=cfillLight!50, inner sep=5pt, fit=(lib)(m3)(m4)] {};
\end{pgfonlayer}
\node[sbox, fill=cfillMed, minimum width=35mm, minimum height=14mm] (canvas) at (5, 0) {Visual Canvas};
\node[font=\sffamily\scriptsize, text=cborderFaint] at (5, -0.9) {drag, connect, configure};
\node[sbox, fill=cfillDark, line width=0.7pt] (validate) at (9.5, 0) {Validate};
\node[sbox, fill=cfillDark, line width=0.7pt] (serialize) at (9.5, -2.0) {Serialize JSON};
\node[sbox, fill=cfillAccent, draw=cborder, line width=0.8pt, font=\sffamily\small\bfseries, text=white] (deploy) at (9.5, -4.0) {Deploy to Pendant};
\node[sbox, fill=cfillMed] (bridge) at (5, -4.0) {AgentBuilder Bridge};
\node[sbox, fill=cfillLight] (workspace) at (0.5, -4.0) {GlobalWorkspace};
\draw[->, draw=cborder, line width=0.9pt] (lib) -- node[above=3pt, font=\sffamily\footnotesize, text=cborder] {drag modules} (canvas);
\draw[->, draw=cborder, line width=0.9pt] (canvas) -- node[above=3pt, font=\sffamily\footnotesize, text=cborder] {submit} (validate);
\draw[->, draw=cborder, line width=0.7pt] (validate) -- node[right=2pt, font=\sffamily\footnotesize, text=cborder] {ports OK} (serialize);
\draw[->, draw=cborder, line width=0.9pt] (serialize) -- node[right=2pt, font=\sffamily\footnotesize, text=cborder] {graph JSON} (deploy);
\draw[->, draw=cborder, line width=0.7pt] (deploy) -- node[above=3pt, font=\sffamily\footnotesize, text=cborder] {configure} (bridge);
\draw[->, draw=cborder, line width=0.7pt] (bridge) -- node[above=3pt, font=\sffamily\footnotesize, text=cborder] {register modules} (workspace);
\draw[->, dashed, draw=cborderFaint, line width=0.5pt] (bridge.south) -- ++(0, -0.8) node[below, font=\sffamily\scriptsize, text=cborderFaint] {sync to Firebase};
\end{tikzpicture}
\caption{Agent Builder deployment pipeline. Users drag modules from the library onto the visual canvas, configure parameters, then submit for validation. The validated graph is serialized to JSON and deployed via \texttt{AgentBuilderBridge}, which registers modules with the \texttt{GlobalWorkspace} and syncs the configuration to Firebase for cross-device persistence. Grounded in: \texttt{agentbuilder.html}, \texttt{AgentBuilderBridge.js}.}
\label{fig:agent-builder-flow}
\end{figure}

\begin{figure}[H]
\centering
\includegraphics[width=0.85\textwidth]{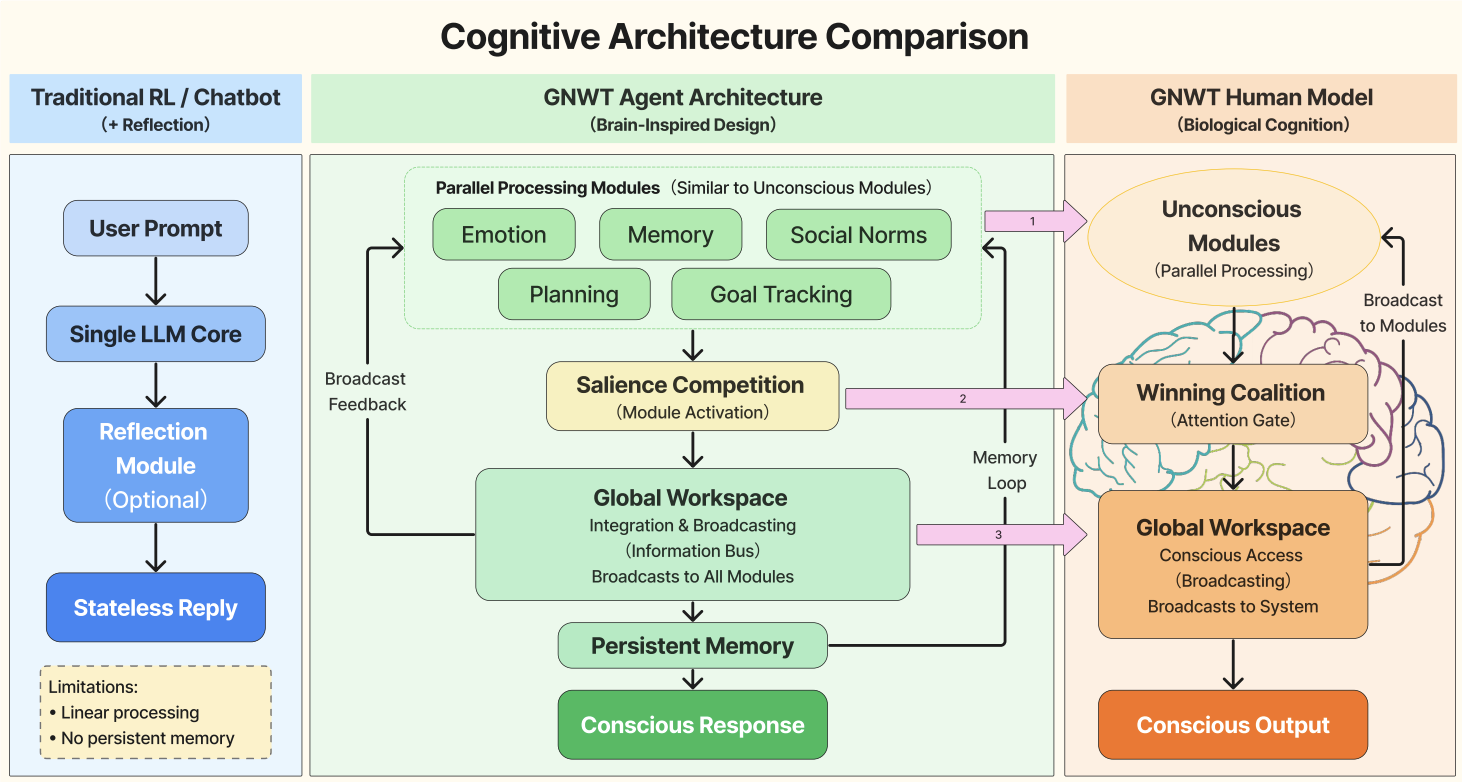}
\caption{Agent Builder node-based visual editor. The left sidebar contains the module library organized by category (cognitive, utility, integration). The central canvas displays the current agent architecture as a directed graph, where each node represents a specialist module and edges represent information flow between modules. Users drag modules from the library onto the canvas, connect input/output ports to define the processing pipeline, and configure per-module parameters (salience weights, thresholds, personality trait modulation) through inline controls.}
\label{fig:agent-builder}
\end{figure}

The Agent Builder (\texttt{agentbuilder.html}) provides a visual, node-based interface for constructing and customizing LLM-powered cognitive agents without programming. Users assemble agent architectures by dragging specialist modules from a sidebar library onto a canvas, connecting them via edges that define information flow, and configuring per-module parameters through inline controls. The system bridges the gap between the GNWT cognitive architecture (Appendix~\ref{appendix:gnwt}) and end-user customization, enabling non-technical users to create personalized pendant companions with distinct behavioral profiles. Figure~\ref{fig:agent-builder-flow} provides an end-to-end overview of the deployment pipeline; Figures~\ref{fig:agent-builder}--\ref{fig:agent-builder-ingame} show the interface at each stage.

\subsubsection{Node-Based Visual Editor}

The module library (left sidebar) organizes available components into five categories corresponding to the GNWT specialist modules. Each module is represented as a draggable node with typed input/output ports. The canvas supports standard graph editing operations: drag-to-position, click-to-select, edge creation by dragging between ports, and multi-select for batch configuration. Module nodes display their current parameter values inline, providing immediate visibility into the agent's configuration without opening separate settings panels.

Available module types include the five core GNWT specialists (Emotion, Memory, Planning, Social Norms, Goal Tracking), utility nodes (input parsers, output formatters, logging), and integration nodes (Firebase sync, LLM API connector, personality trait loader). Each module exposes configurable parameters: salience base weight (0.0--2.0), activation threshold, update frequency, and module-specific settings (e.g., the Emotion module exposes contagion rate, mood persistence, and arousal/valence decay rates).

\subsubsection{Customization Interface}

\begin{figure}[H]
\centering
\includegraphics[width=0.85\textwidth]{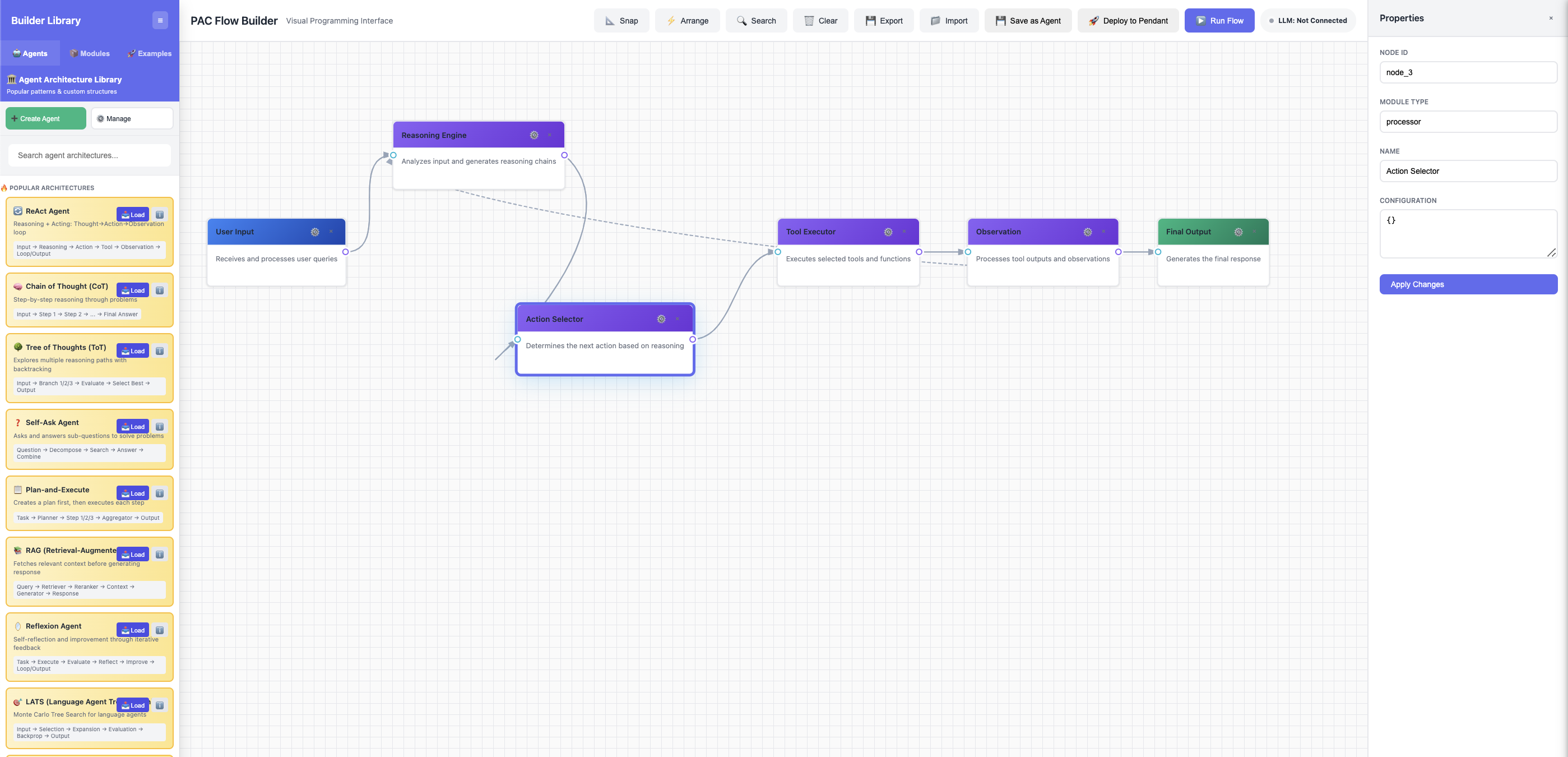}
\caption{Agent customization interface for the pendant companion. Users adjust personality trait weights (friendliness, openness, independence, loyalty, playfulness) via sliders on a 0--100 scale, configure GNWT specialist module priorities, and set behavioral parameters such as proactive interaction frequency and emotional responsiveness. Changes are previewed in real time through a conversation simulator before deployment.}
\label{fig:agent-customize}
\end{figure}

The customization interface (Figure~\ref{fig:agent-customize}) provides fine-grained control over the agent's behavioral profile. Five personality trait sliders (friendliness, openness, independence, loyalty, playfulness, each 0--100) modulate specialist module base weights, ensuring that the agent's conversational style reflects the user's intended personality. Advanced settings expose GNWT-specific parameters: workspace capacity, broadcast threshold ($\tau$), attention decay rate, and coalition similarity threshold. A built-in conversation simulator runs three test exchanges against the current configuration, allowing users to preview the agent's behavior before committing to deployment.

\subsubsection{In-Game Pendant Companion}

\begin{figure}[H]
\centering
\includegraphics[width=0.7\textwidth]{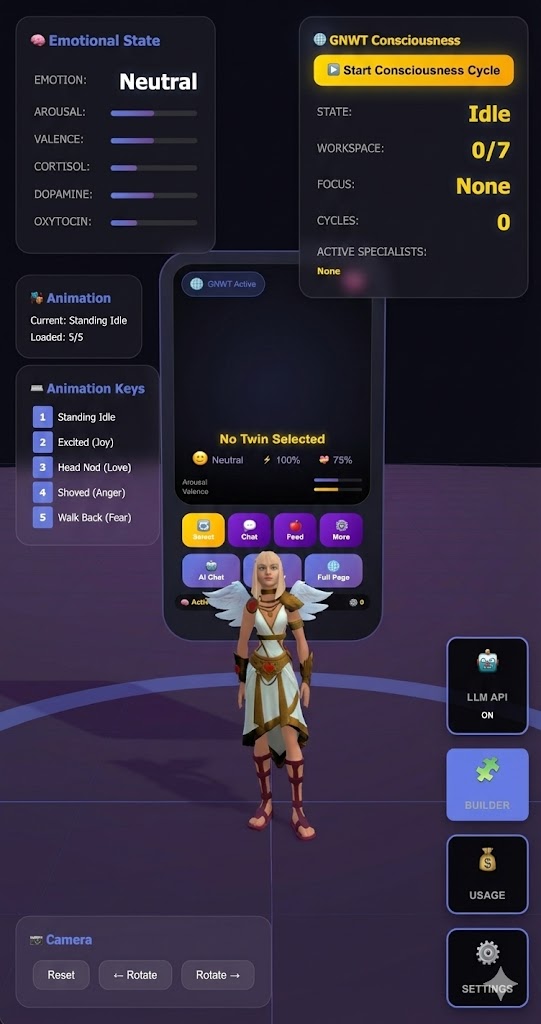}
\caption{Pendant companion deployed in the 3D game world. The agent built through the Agent Builder appears as an interactive companion that accompanies the player, providing context-aware emotional scaffolding, gameplay suggestions, and social interaction support. The companion's personality and behavioral style reflect the architecture configured in the Agent Builder.}
\label{fig:agent-builder-ingame}
\end{figure}

Figure~\ref{fig:agent-builder-ingame} shows the deployed pendant companion in the 3D game world. The agent built through the Agent Builder appears as an interactive companion that accompanies the player through territory exploration, boss battles, and social encounters. The companion's \texttt{ProactiveInteractionSystem} triggers context-aware messages based on the priority hierarchy configured in the builder (time-of-day: 8, mood: 7, memory recall: 6, scene analysis: 5, idle: 5), with the 30-second idle threshold and 5-minute cooldown between proactive interactions.

\subsubsection{Deployment Pipeline}

Once assembled, the agent architecture is deployed through a three-stage pipeline managed by the \texttt{AgentBuilderBridge} (\texttt{js/twins/modules/social/AgentBuilderBridge.js}):

\begin{algorithm}[H]\small
\caption{Agent Builder Deployment Pipeline}
\label{alg:agent-builder-deploy}
\begin{algorithmic}[1]
\Require Canvas graph $G = (V, E)$ with configured modules, user profile
\Ensure Deployed pendant agent with custom cognitive architecture
\Function{DeployAgent}{$G$}
    \State \textbf{Stage 1: Validate} --- check all required ports connected, no cycles, parameters in range
    \State \textbf{Stage 2: Serialize} --- export graph as JSON: nodes (type, position, params), edges (source, target, port)
    \State \textbf{Stage 3: Deploy} --- \texttt{AgentBuilderBridge} applies configuration:
    \State \quad Load personality traits from \texttt{UnifiedProfileManager}
    \State \quad Initialize each specialist module with user-configured weights
    \State \quad Register modules with \texttt{GlobalWorkspace} (capacity 9, $\tau = 0.7$)
    \State \quad Sync configuration to Firebase for cross-device persistence
    \State \quad Activate pendant companion with new architecture
\EndFunction
\end{algorithmic}
\end{algorithm}

The \texttt{SecureAgentBuilder} (\texttt{ai-pendant-system/js/security/SecureAgentBuilder.js}) enforces security boundaries during deployment: module parameter ranges are validated against predefined bounds, the total module count is capped to prevent memory exhaustion, and API key access is mediated through the existing proxy infrastructure (Appendix~\ref{appendix:technical-architecture}).

\subsubsection{User Customization Workflow}

The typical user workflow proceeds in four steps: (1) select a base template (e.g., ``Empathetic Listener,'' ``Strategic Advisor,'' ``Playful Companion'') that pre-configures module weights for common personality profiles; (2) adjust individual module parameters using the customization interface (Figure~\ref{fig:agent-customize}) to fine-tune behavioral emphasis; (3) preview the agent's responses through the built-in conversation simulator that runs three test exchanges; and (4) deploy the configured agent to the pendant (Figure~\ref{fig:agent-builder-ingame}), replacing the previous configuration. Users can save multiple agent configurations as named profiles and switch between them without re-building from scratch. The agent builder's configuration is synchronized across devices via Firebase, ensuring that an agent built on desktop is immediately available on mobile.

\subsection{API Usage Transparency}

\begin{figure}[t]
\centering
\includegraphics[width=0.7\textwidth]{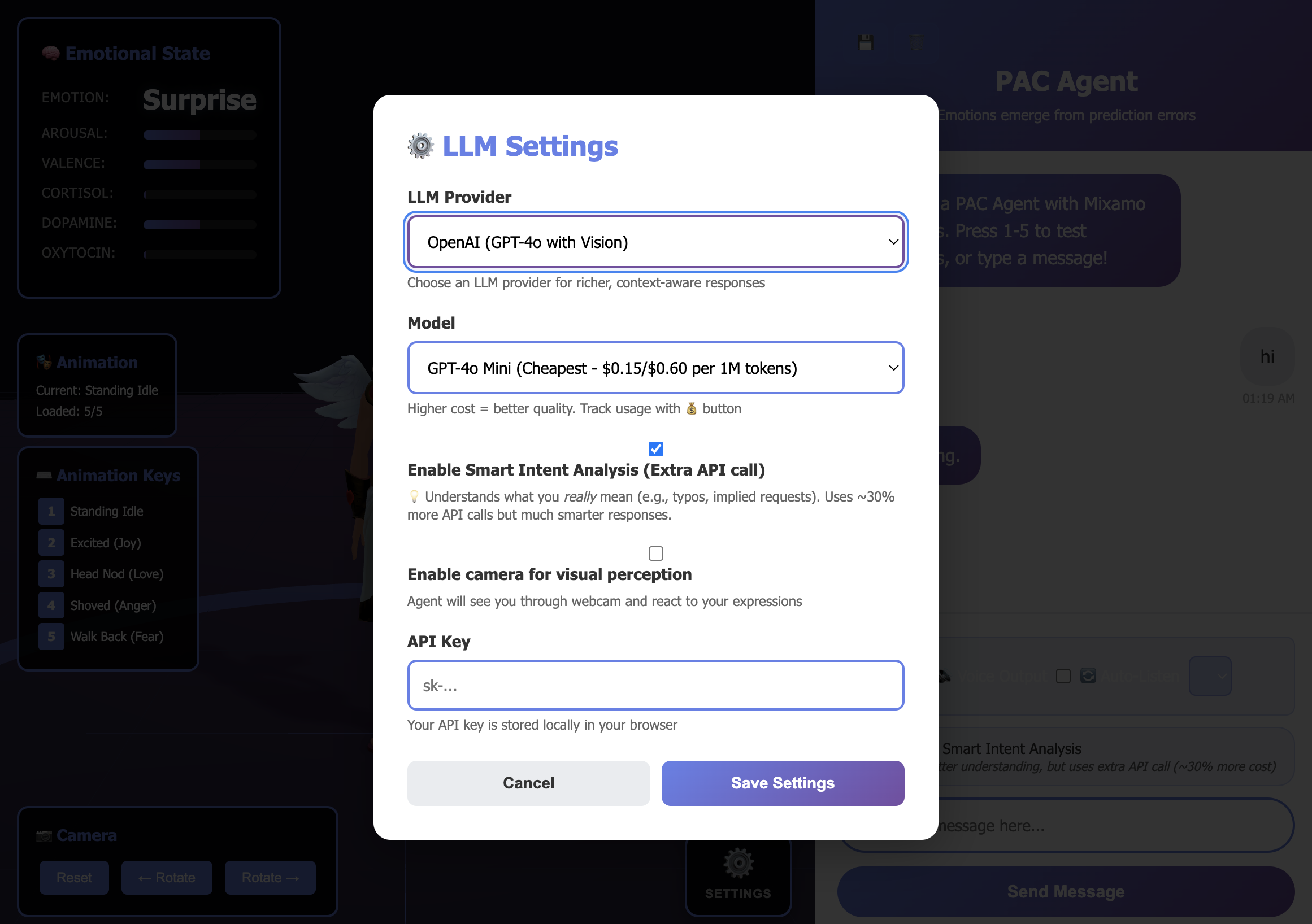}
\caption{API usage dashboard showing real-time token consumption, cost breakdown by feature, and model selection options. This transparency helps users monitor the computational costs of different AI capabilities.}
\label{fig:api-usage}
\end{figure}

Unlike platforms that hide AI costs, Cognibit provides complete API transparency through the \texttt{APIUsageTracker} and \texttt{APIDashboardController} (Figure~\ref{fig:api-usage}). The dashboard tracks three provider backends---OpenAI, Anthropic, and DeepSeek---displaying per-provider statistics for request count, input/output token volumes, and cumulative cost. Each provider card breaks down usage by model (e.g., GPT-4o at \$2.50/\$10.00 per million input/output tokens versus GPT-4o-mini at \$0.15/\$0.60), enabling users to see the cost implications of model selection in real time. The system maintains a scrollable history of the last 100 API requests with timestamp, provider, model, token counts, and cost per call, and projects monthly cost estimates based on tracked daily averages. Usage data persists in localStorage under the key \texttt{api\_usage\_tracker} and can be exported to JSON for external analysis. This transparency enables informed decisions about the quality--cost trade-off documented in the Pareto analysis (Section~\ref{sec:pareto}).

\subsection{World Map and Territory System}

\begin{figure}[t]
\centering
\begin{tabular}{cc}
\includegraphics[width=0.45\textwidth]{figures/worldmap_demo.jpg} &
\includegraphics[width=0.45\textwidth]{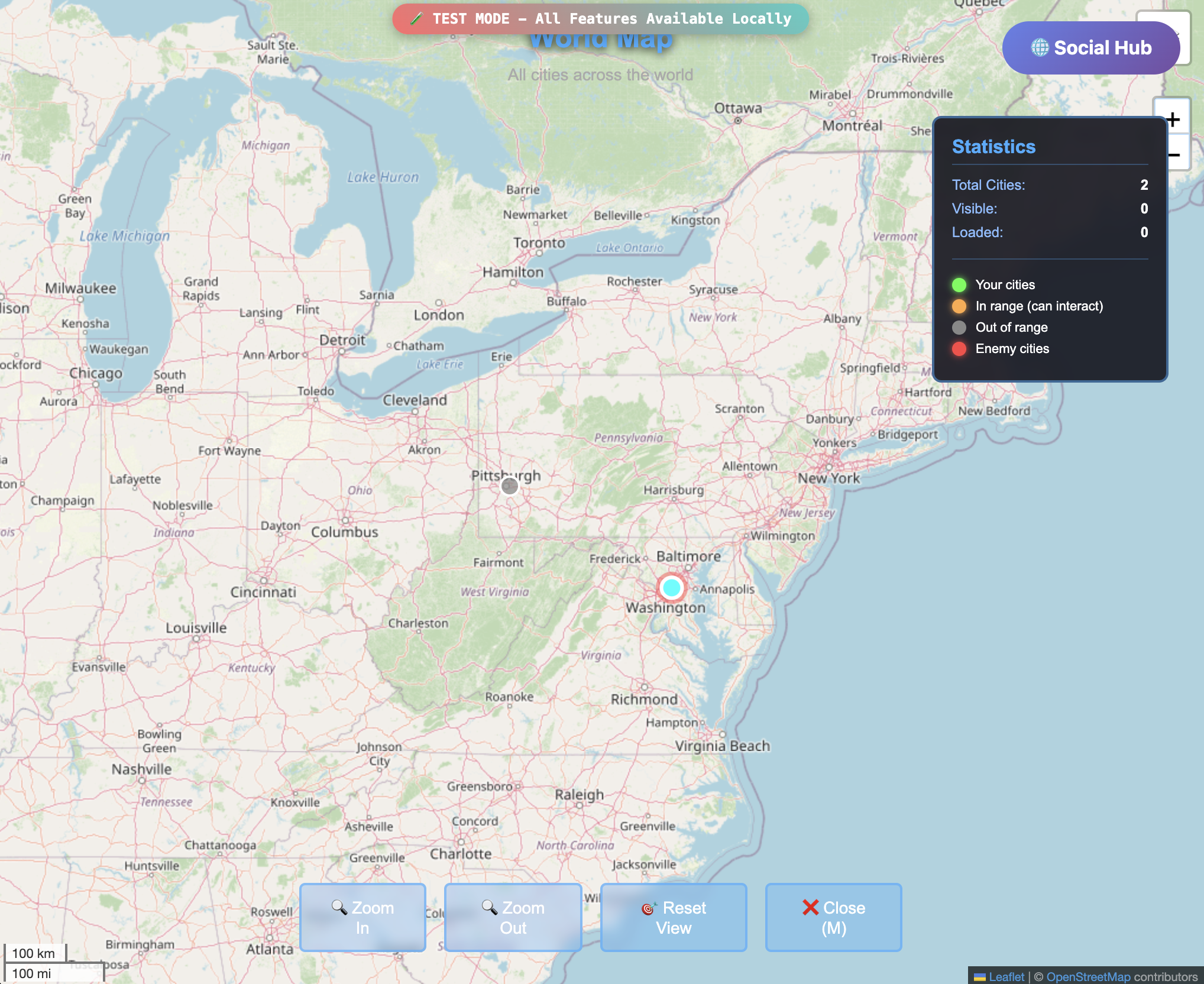} \\
(a) World map design concept & (b) In-game implementation \\
\end{tabular}
\caption{GPS-driven world map interface: (a) Territory ownership visualization with conquest mechanics, (b) Real-time in-game view showing nearby territories, other players, and boss locations. The 50-mile discovery radius and 20-meter movement threshold create hyperlocal communities.}
\label{fig:world-map}
\end{figure}

The world map system (Figure~\ref{fig:world-map}) bridges digital and physical realities through the \texttt{GPSLocationSystem} and \texttt{CityTakeoverSystem}. The GPS module uses high-accuracy mode (GPS + WiFi + cellular triangulation) with a 20-meter movement threshold to filter noise and 2-second update throttling to conserve battery; when battery drops below 20\%, throttling increases to 10 seconds, and below 10\% to 30 seconds. Stationary detection triggers after 60 seconds without movement, further reducing polling frequency. The minimap renders on a 250$\times$250 pixel canvas at 3$\times$ scale, updating at approximately 1 FPS through the modular \texttt{MinimapSystem} (renderer, world sampler, exploration tracker, UI layer). Territory capture operates within a 50-meter GPS radius: players accumulate capture points at 5 points/second while in range, with 100 points required for full capture. Ownership confers tangible bonuses: 1.2$\times$ movement speed, 1.1$\times$ health regeneration, 15\% bonus experience, and 10\% shop discounts. The interface uses progressive disclosure---revealing more detail as players explore---to encourage physical movement while the \texttt{MinimapExplorationSystem} tracks visited areas.

\subsection{Boss Fight Combat Interface}

\begin{figure}[t]
\centering
\includegraphics[width=0.8\textwidth]{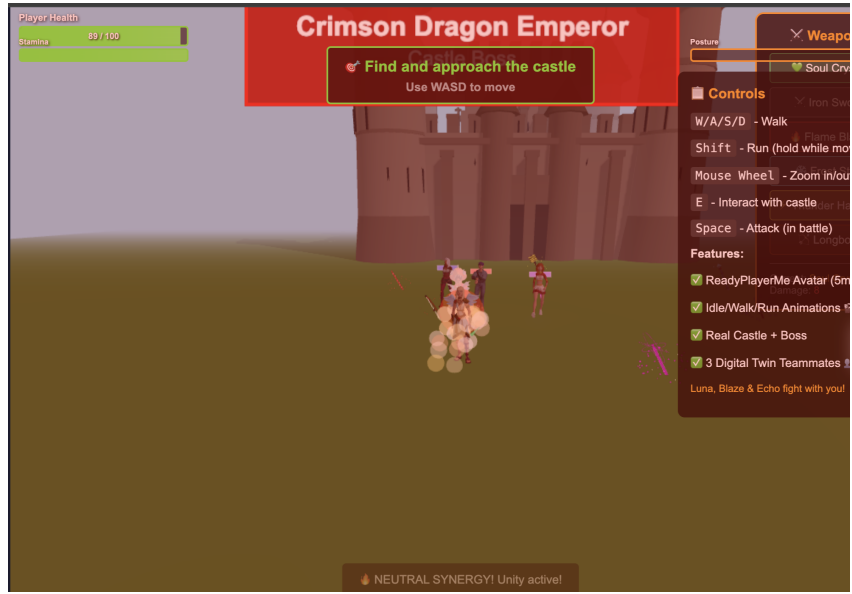}
\caption{Boss fight interface showing real-time combat with AI teammates. The interface displays health bars, ability cooldowns (300ms ranged, 500ms melee), team coordination indicators, and victory conditions. Social features are integrated directly into combat UI.}
\label{fig:boss-fight}
\end{figure}

The boss fight interface (Figure~\ref{fig:boss-fight}) demonstrates how gaming creates natural social interactions through the \texttt{BossFightCombat} and \texttt{BossFightPlayer} systems. The combat HUD displays boss health bars (default max health: 1,500), an 800ms attack cooldown indicator, and a combo counter with milestone notifications at 3$\times$, 5$\times$, and 10$\times$ consecutive hits. Critical hits trigger a 150ms red screen flash and floating damage numbers rendered via a pre-allocated pool of 50 DOM elements to avoid garbage collection pauses during combat. The style system calculates bonus points for attack variety, critical hits, stagger exploitation, and combo maintenance. Crucially, the combat UI overlays social elements directly into the battle context: team chat remains active during encounters, compatibility indicators show how well human teammates' twins rated each other, and post-victory screens display party statistics alongside social prompts---creating natural conversation starters from shared combat experiences with an average team size of 2.8$\pm$0.6 players.

\subsection{Integrated Game Experience}

\begin{figure}[t]
\centering
\includegraphics[width=0.8\textwidth]{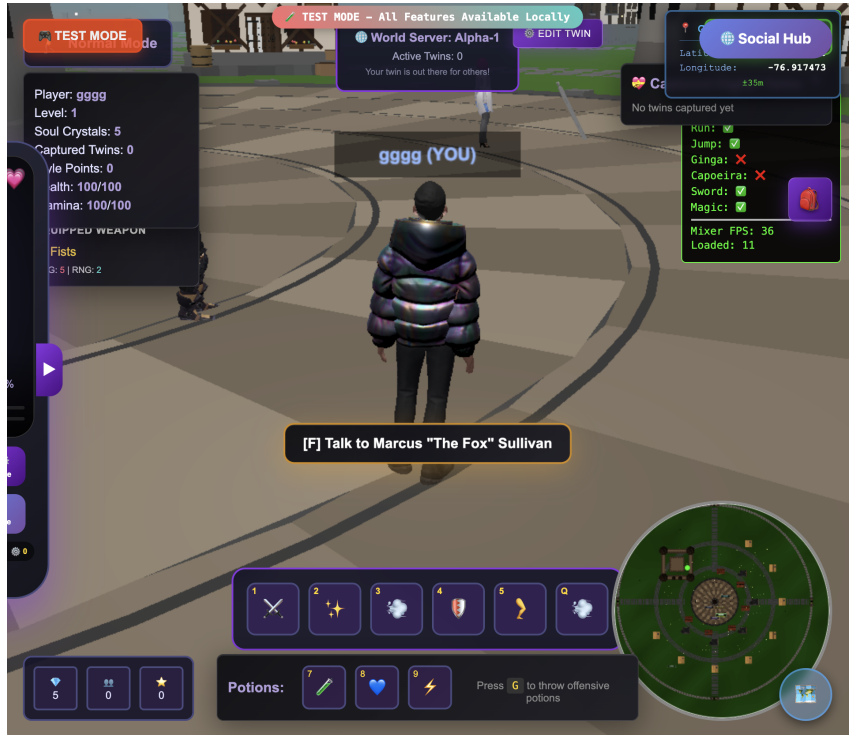}
\caption{Integrated 3D game world showing seamless blend of exploration, social interaction, and AI companions. The unified experience eliminates boundaries between gaming, social networking, and AI interaction.}
\label{fig:game-experience}
\end{figure}

The integrated game experience (Figure~\ref{fig:game-experience}) represents the platform's unified vision. Rather than separate apps, all features exist within a cohesive 3D world where avatars represent both humans and AI twins, locations have social and gaming significance, achievements unlock relationship opportunities, and every interaction strengthens the social graph.

\subsection{Additional Interface Components}

\paragraph{Daily Digest Notification.}
The daily digest aggregates the user's top social interactions, territory changes, and companion highlights into a single notification delivered at a configurable time (default: 8:00 AM local). The digest includes: (1) top-3 twin conversation outcomes ranked by compatibility score, (2) territory ownership changes in the last 24 hours, (3) companion emotional summary, and (4) suggested actions for the day (e.g., ``Visit the territory near downtown---it's contested!'').

\begin{table}[!htbp]
\centering
\caption{Daily Digest Notification Components}
\label{tab:daily-digest}
\begin{tabular}{llp{5.5cm}}
\toprule
\textbf{Section} & \textbf{Priority} & \textbf{Content} \\
\midrule
Twin matches & High & Top-3 new compatibility results with preview snippets \\
Territory updates & Medium & Ownership changes, contested zones, reinforcement reminders \\
Companion summary & Medium & Emotional trajectory, notable interactions, mood assessment \\
Action suggestions & Low & Personalized next steps based on activity patterns \\
Social highlights & Low & New followers, post engagement, conversation continuations \\
\bottomrule
\end{tabular}
\end{table}

\paragraph{Follower Graph and Social Network.}
The Social Hub maintains a directed follower graph where users can follow others based on compatibility signals. The graph visualization shows: mutual connections (bidirectional edges), pending follows, and compatibility-weighted clustering. Key metrics displayed include: follower count, following count, mutual connection count, and engagement reciprocity rate.

\begin{table}[!htbp]
\centering
\caption{Follower Graph Interface Components}
\label{tab:follower-graph}
\begin{tabular}{lp{7cm}}
\toprule
\textbf{Component} & \textbf{Description} \\
\midrule
Network view & Visual graph of connections with compatibility-weighted edge thickness \\
Mutual connections & Highlighted bidirectional edges indicating reciprocated interest \\
Suggested follows & Compatibility-ranked list of users not yet followed \\
Activity feed & Posts and interactions from followed users, ranked by engagement score \\
Connection stats & Follower/following counts, reciprocity rate, cluster membership \\
\bottomrule
\end{tabular}
\end{table}

\paragraph{Memory and Conversation History.}
The conversation history interface provides a chronological view of all twin-to-twin interactions, organized by session. Each session entry displays: twin pair names, conversation duration, key topics discussed, compatibility score, and emotional trajectory (via color-coded sentiment indicators). Users can review past conversations to understand why specific matches were recommended. The memory view shows the pendant companion's accumulated knowledge organized by topic, with importance-weighted highlighting.

\paragraph{Profile Creation and Onboarding.}
The onboarding flow consists of four steps: (1) account creation with basic demographics, (2) personality assessment via the 5-dimensional trait questionnaire (friendliness, independence, playfulness, loyalty, openness, each 0--100), (3) interest selection from 12 categories, and (4) twin initialization where the user's first digital twin is created from the personality profile. The process takes approximately 5--10 minutes and produces the twin blueprint used for all subsequent matching.

\paragraph{Matching Results and Compatibility Scores.}
The matching interface presents filtered recommendations as profile cards showing: compatibility score (0--100), shared interests, personality complementarity indicators, and a preview snippet from the most recent twin conversation. Users can view detailed compatibility breakdowns showing the contribution of each scoring dimension (personality 30\%, interests 20\%, conversation quality 25\%, emotional resonance 15\%, interaction patterns 10\%).

\begin{table}[!htbp]
\centering
\caption{Match Result Card Components}
\label{tab:match-card}
\begin{tabular}{lp{7cm}}
\toprule
\textbf{Element} & \textbf{Description} \\
\midrule
Overall score & Composite compatibility score (0--100) \\
Shared interests & Highlighted common interest categories \\
Personality radar & 5-axis visualization of trait complementarity \\
Conversation preview & Most engaging snippet from twin-to-twin dialogue \\
Action buttons & Accept / Skip / View full conversation \\
\bottomrule
\end{tabular}
\end{table}

\subsection{Design Principles}

Our interface design follows six key principles. Progressive complexity ensures that novice users encounter simple entry points (e.g., a single ``Start Matching'' button) while power users can access the full GNWT visualization, agent builder, and API dashboard. Transparency over magic means showing AI decision-making processes rather than hiding them---every twin recommendation can be traced to specific specialist module activations and salience scores. Unified experience design enables seamless transitions between the Social Hub, 3D game world, and pendant companion without context-switching friction, achieved through the five-component \texttt{SocialHubUI} architecture (FloatingButtonManager, HubPanelRenderer, TabNavigationManager, TabContentLoader, StateStorageManager). Mobile-first responsive design optimizes for phone-sized screens while scaling to desktop, with the minimap and combat HUD adapting to available viewport dimensions. Accessibility targets WCAG 2.1 AA compliance with screen reader support, though substantial work remains (see Appendix~\ref{appendix:accessibility-inclusion} for current limitations and future plans). Performance targets 60 FPS animations and sub-100ms perceived response times with 5--8 concurrent agents, degrading at higher agent counts as documented in Section~\ref{sec:limitations}.

These interfaces collectively demonstrate that browser-based platforms can deliver rich, multi-context experiences rivaling native applications while maintaining the accessibility advantages of web deployment.
\section{Statistical Methodology and Power Analysis}
\label{appendix:statistical-methodology}

An expanded confirmatory design (N=160, 4-group factorial) was originally planned but was not executed due to resource constraints. The actual CogniPair \citep{CogniPair2026} pilot deployment remained a single-group N=20 exploratory probe, so only descriptive and exploratory statistics are reported in this paper (Section~\ref{sec:field-deployment}).

\subsection{Statistical Power}

Post-hoc power analysis for the N=20 single-group design yields power $= 0.12$ for detecting medium effects ($d = 0.5$), substantially below the conventional 0.80 threshold. This means the study has an 88\% probability of failing to detect a true medium-sized effect. Consequently, null results in this paper cannot be interpreted as evidence of absence; they reflect insufficient power rather than demonstrated ineffectiveness. All reported effects should be treated as preliminary effect-size estimates for future sample-size planning, not as confirmatory findings.

\subsection{Analysis Approach}

Given the exploratory nature of the pilot, we adopt the following statistical conventions throughout:

\paragraph{Effect sizes over p-values.} We report effect sizes with 95\% confidence intervals rather than $p$-values, following recommendations for small-sample exploratory research \citep{Cumming2014}. Pre/post comparisons use Cohen's $d_z$ (paired samples); between-condition comparisons in synthetic experiments use Cohen's $d$ (independent samples). Correlations are Pearson's $r$.

\paragraph{Confidence interval method.} For field deployment metrics (N=20), 95\% CIs are computed via bias-corrected and accelerated (BCa) bootstrap with 10,000 resamples, which provides more accurate coverage than normal-approximation CIs at small sample sizes. For synthetic experiments (N=30 target users $\times$ 5 recommendations), standard normal-approximation CIs are used given the larger effective sample.

\paragraph{Multiple comparison policy.} All field deployment analyses are post-hoc and exploratory; no family-wise error correction is applied. We explicitly flag this limitation: with multiple outcome measures (engagement, walking distance, meeting rate, satisfaction), the probability of at least one spurious association exceeds conventional thresholds. Readers should interpret individual correlations as hypothesis-generating rather than confirmatory.

\paragraph{Descriptive vs.\ causal claims.} The engagement-reallocation finding (total platform time increased from 97 to 141 min/day despite a decrease in traditional platform time) is reported as a descriptive observation, not a causal claim. Dose-response patterns are suggestive but cannot be causally interpreted without a control group. The confounded three-component design (gaming + twins + companions) prevents isolation of individual component effects.

\subsection{Future Confirmatory Design}

A randomized deployment comparison (within-subjects crossover, N$\geq$30) assigning participants to LLM-filtered versus randomly-selected matches would be the minimum design needed to establish whether offline quality gains translate into real-world behavioral differences. A full factorial design (N$\geq$128, with 16 participants per cell in a $2^3$ design) would be required to isolate individual component effects (gaming, twins, companions) with 0.80 power for medium effects ($d = 0.5$). The synthetic ablation studies in Appendix~\ref{appendix:ablations} provide preliminary effect-size estimates that can inform this future sample-size calculation.

\section{Combat and Animation Systems}
\label{appendix:combat-animation}

This appendix details the combat mechanics and personality-driven animation systems that create engaging interactions between players and digital twins. Combat serves a dual role in Cognibit: it provides the gamification scaffolding that motivates physical exploration of territory zones, and it creates low-pressure social encounters between players whose twins have been matched. The personality-driven animation system ensures that each twin behaves consistently with its owner's trait profile, reinforcing the sense of authentic representation central to the digital twin paradigm.

\subsection{Combat System Algorithm}

The combat system implements combo tracking, damage calculation, and multi-target detection with performance optimizations including vector pooling and cooldown management. During the 14-day pilot, the system produced 47 boss battles (31 completed, 66\% completion rate), with 2--4 players per encounter; 19 of 31 completed battles led to continued post-battle interaction, confirming the combat layer's role as a low-pressure social catalyst. Weapon types (melee, bow, gun, cannon) each carry distinct cooldown profiles (500ms for melee/cannon, 300ms for bow/gun), and a combo tracker rewards varied attack sequences with style points---special combos such as ``Spell Blade'' (attack-attack-magic) and ``Counter Strike'' (dodge-attack-kick) award 300--450 bonus points. The damage pipeline supports 11 damage types (physical, magical, elemental, true, poison, bleed, burn, frost, shock, holy, dark) with type-specific defense calculations: physical damage is reduced by armor using a diminishing-returns formula $\text{reduction} = \text{armor} / (\text{armor} + 100)$ capped at 75\%, while damage-over-time effects (poison, bleed, burn) apply tick damage at configurable intervals (e.g., poison: 20\% of initial damage every 1s for 5s). Combo damage scales logarithmically as $1 + 0.5 \cdot \log_{10}(\text{comboCount} + 1)$, and a 2-second inactivity window resets the combo counter to zero. Boss health scales dynamically with player level ($\text{baseHealth} \times (1 + \text{playerLevel} \times 0.1)$) and team size (+50\% per additional player), preventing trivial encounters as players progress. A capped projectile pool (10 active), pre-allocated damage number elements (pool of 50), and reusable vector objects keep per-frame allocation below the browser heap budget, targeting 60 FPS with a 30 FPS minimum floor.

\begin{algorithm}[H]\small
\caption{Combat System with Combo Mechanics}
\label{alg:combat-system}
\begin{algorithmic}[1]
\Require Ability name, weapon data, target entities
\Ensure Damage application with combo tracking and style scoring
\Function{UseAbility}{$\mathit{ability}$}
    \If{$\mathit{ability}.\mathit{cooldown} > 0$} \Return false \EndIf
    \State $\mathit{ability}.\mathit{cooldown} \gets \mathit{ability}.\mathit{maxCooldown}$
    \State $\mathit{combo} \gets \mathit{combo} + 1$;\quad stylePoints $\mathrel{+}= 10 \times (\mathit{combo} + 1)$
    \If{weapon is projectile}
        \If{$|\mathit{projectiles}| \geq 10$ \textbf{or} cooldown not elapsed (bow/gun: 300ms, cannon: 500ms)}
            \State \Return false
        \EndIf
        \State Spawn projectile at player position + forward $\times$ 1.5
    \Else \Comment{Melee path}
        \If{cooldown not elapsed (500ms)} \Return false \EndIf
        \State $r \gets$ attack ? 5 : 7 \Comment{Hit range by ability type}
        \ForAll{target within distance $r$}
            \State $\mathit{crit} \gets \texttt{random()} < 0.1$;\quad $\mathit{dmg} \gets \mathit{crit}$ ? $\mathit{base} \times 2$ : $\mathit{base}$
            \State \Call{ApplyDamage}{target, $\mathit{dmg}$};\quad stylePoints $\mathrel{+}= \mathit{crit}$ ? 50 : 20
        \EndFor
    \EndIf
    \State Check special combos (last 5 moves): Spell Blade, Counter Strike, etc. (+300--450 pts)
    \State Reset combo after 3s inactivity
\EndFunction
\end{algorithmic}
\end{algorithm}

Damage over time effects (poison, bleed, burn) apply tick damage at configurable intervals. Combo damage scales logarithmically as $1 + 0.5 \cdot \log_{10}(\mathit{comboCount} + 1)$. Armor uses a diminishing-returns formula: reduction $= \mathit{armor}/(\mathit{armor} + 100)$, capped at 75\%. Vector pooling (4 pre-allocated \texttt{Vector3} objects) and a capped projectile pool (10 active) keep per-frame allocation below the browser heap budget.

\subsection{Twin Personality Decision Algorithm}

The personality system implements trait-based behavioral decisions, creating unique and consistent behavior patterns for each digital twin based on their personality configuration. Each twin carries a five-dimensional trait vector (openness, friendliness, playfulness, loyalty, independence, each scored 0--100) that governs action selection probabilities, mood transitions, and animation choices. The dominant-trait heuristic (Algorithm~\ref{alg:twin-personality}) maps continuous trait values to five discrete behavioral archetypes---adventurous, friendly, playful, loyal, and independent---while preserving stochastic variation, so twins with similar profiles still exhibit distinguishable in-world behavior. A TwinPersonality subsystem delegates action selection to a trait-weighted decision function that considers player distance, nearby twin count, and a stochastic roll: 60\% of decisions produce general movement, with the remaining 40\% split among archetype-specific behaviors (exploring at larger radii for adventurous twins, face-and-greet for friendly twins within 10 units, spontaneous personality animations for playful twins). Each archetype maps to a distinct animation key set---e.g., adventurous twins select from \{\textit{Excited}, \textit{sword and shield attack}, \textit{walking}\}, friendly twins from \{\textit{Excited}, \textit{Thoughtful Head Nod}, \textit{walking}\}---and animation playback speed is modulated by the twin's energy level via $\text{timeScale} = 1 + \text{energy}/200$, so high-energy twins animate visibly faster. A 2-second animation cooldown and debounce guard (100ms) prevent animation flooding, and each personality animation is broadcast to multiplayer peers via Firebase for cross-client consistency. Mood transitions are driven by proximity to the player and nearby twin density: friendly twins (friendliness $> 70$) become excited when the player is within 5 units, loyal twins (loyalty $> 70$) grow sad when separated beyond 30 units, and playful twins (playfulness $> 60$) become happy in groups of three or more. Mood changes are debounced at 3-second intervals to prevent oscillation.

\begin{algorithm}[H]\small
\caption{Twin Personality-Driven Behavior}
\label{alg:twin-personality}
\begin{algorithmic}[1]
\Require Twin trait vector $\mathbf{t} \in [0,100]^5$, player position, nearby twin count
\Ensure Archetype-consistent action selection and mood transitions
\State $\mathit{archetype} \gets \arg\max_\tau \mathbf{t}[\tau]$ mapped to \{adventurous, friendly, playful, loyal, independent\}
\Function{DecideNextAction}{$\mathit{playerDist}$, $\mathit{nearbyCount}$}
    \State $\mathit{roll} \gets \texttt{random()}$
    \If{$\mathit{roll} < 0.6$} General movement (radius 25 units) \Comment{60\%}
    \ElsIf{archetype = playful} Move (radius 15) + 30\% chance personality animation \Comment{15\%}
    \ElsIf{archetype = friendly \textbf{and} $\mathit{playerDist} < 10$} Face player + greet animation \Comment{5\%}
    \ElsIf{archetype = adventurous} Explore (radius 35) \Comment{15\%}
    \ElsIf{archetype = independent} Wander alone (radius 30) \Comment{5\%}
    \EndIf
\EndFunction
\Function{UpdateMood}{$\mathit{playerDist}$, $\mathit{nearbyCount}$}
    \If{$\mathit{playerDist} < 5$ \textbf{and} friendliness $> 70$} mood $\gets$ excited \EndIf
    \If{$\mathit{playerDist} > 30$ \textbf{and} loyalty $> 70$} mood $\gets$ sad \EndIf
    \If{$\mathit{nearbyCount} > 2$ \textbf{and} playfulness $> 60$} mood $\gets$ happy \EndIf
\EndFunction
\end{algorithmic}
\end{algorithm}

Each archetype maps to a distinct animation set (e.g., adventurous: \{\textit{Excited}, \textit{sword attack}, \textit{walking}\}; friendly: \{\textit{wave}, \textit{Thoughtful Head Nod}\}). Animation playback speed scales with energy level as $\text{timeScale} = 1 + \text{energy}/200$. A 2-second animation cooldown and 100ms debounce guard prevent animation flooding, and personality animations are broadcast to multiplayer peers via Firebase for cross-client consistency. Mood transitions are debounced at 3-second intervals to prevent oscillation.
\section{Synchronization Protocols}
\label{appendix:sync-protocols}

This appendix describes the cross-device synchronization protocol that enables persistent memory and state sharing across multiple devices and sessions. Reliable synchronization is critical because Cognibit spans three surfaces---the 3D game world, the Social Hub, and the pendant companion---and users may switch between devices mid-session. The protocols below were validated in Section~\ref{sec:memory} (73ms write latency, 0\% stale reads, zero data loss under concurrent writes).

\subsection{Cross-Device Memory Synchronization Protocol}

The synchronization protocol implements real-time memory sync across devices using Firebase Realtime Database, with offline capability and conflict resolution strategies. The protocol is critical because Cognibit spans three surfaces---the 3D game world, the Social Hub, and the pendant companion---and users may switch devices mid-session without warning. The NetworkSync module monitors connection state via the Firebase \texttt{.info/connected} reference and registers an \texttt{onDisconnect} handler that immediately marks the player as offline (setting \texttt{online: false}, resetting animation to \texttt{standing idle}) so other clients see accurate presence within one round-trip. Position synchronization uses an adaptive rate: 200ms (5 updates/sec) while the player is moving, throttled to 2000ms (0.5 updates/sec) when idle, with a 0.5-unit movement threshold to suppress redundant writes. Each position update includes the player's current animation state, rotation, health, and a server-generated timestamp. Three conflict-resolution modes are supported: last-write-wins for low-importance state (e.g., UI preferences), merge-all for memory accumulation (ensuring no conversation memories are silently dropped), and highest-importance for safety-critical relational data (e.g., trust scores, relationship stage). Each memory item is uniquely identified by the tuple (speaker, timestamp, message content hash), enabling deduplication when the same event is recorded by multiple devices. An offline queue with checksum validation (32-bit hash via left-shift accumulation) ensures that connectivity interruptions do not cause silent data loss; queued writes are flushed in order on reconnection using exponential backoff (3 attempts, base delay 1 second, factor 2). Field validation (Section~\ref{sec:memory}) confirmed 73ms write acknowledgement latency, 0\% stale reads, and zero data loss under concurrent writes.

\begin{algorithm}[H]\small
\caption{Cross-Device Memory Synchronization Protocol}
\label{alg:memory-sync-appendix}
\begin{algorithmic}[1]
\Require User ID, Twin ID, Firebase connection, conflict strategy
\Ensure Consistent memory state across all devices with offline support
\State \textbf{Initialize:} Load local memory from localStorage; establish Firebase listener; perform initial sync
\Function{SyncToCloud}{$\mathit{mem}$}
    \State $\mathit{packet} \gets \{\mathit{twinId}, \mathit{content}, \mathit{timestamp}: \texttt{now()}, \mathit{deviceId}, \mathit{version}+1, \mathit{checksum}\}$
    \If{online} \Call{PushToFirebase}{$\mathit{packet}$}
    \Else{} $\mathit{offlineQueue}.\texttt{push}(\mathit{packet})$ \Comment{Queue for later}
    \EndIf
\EndFunction
\Function{OnRemoteUpdate}{$\mathit{remote}$} \Comment{Firebase listener callback}
    \If{$\mathit{remote.timestamp} > \mathit{lastSync}$}
        \If{conflict detected}
            \State \textbf{Resolve by strategy:}
            \State \quad Last-write-wins: keep later timestamp \Comment{Low-importance state}
            \State \quad Merge-all: union memories + deduplicate \Comment{Memory accumulation}
            \State \quad Highest-importance: keep higher total importance \Comment{Relational data}
        \Else
            \State Merge non-conflicting memories
        \EndIf
    \EndIf
\EndFunction
\Function{OnReconnect}{}
    \While{$\mathit{offlineQueue} \neq \emptyset$}
        \State $\mathit{packet} \gets \mathit{offlineQueue}.\texttt{dequeue()}$
        \If{\Call{PushToFirebase}{$\mathit{packet}$} fails} re-queue and break \EndIf
    \EndWhile
    \State \Call{DeltaSync}{$\mathit{lastSync}$} \Comment{Fetch changes since last sync, limit 100}
\EndFunction
\end{algorithmic}
\end{algorithm}

Sync integrity is validated through a 32-bit checksum computed via left-shift accumulation over the serialized memory JSON. Checksum mismatches trigger a full re-sync rather than incremental delta application. The auto-save interval of 30 seconds (from \texttt{PendantMemorySync.js}) ensures that at most 30 seconds of interaction data is at risk during an unclean disconnection; the \texttt{beforeunload} handler performs a synchronous final save as a last resort.

\subsection{Optimistic UI Update Protocol}

To ensure responsive user experience despite Firebase's 3--5 second cross-device propagation delay, the system implements optimistic updates with rollback capability. Local state is applied immediately on user action---rendering territory captures, memory writes, and companion interactions in the UI before server confirmation arrives---so that the interface feels instantaneous even when the network round-trip is slow. The pattern is implemented through a pending-update queue paired with a rollback stack: each user action generates a unique update ID, the current UI state is captured onto the rollback stack, and the change is applied locally before the asynchronous Firebase write is dispatched. On server acknowledgement, the pending entry and its rollback snapshot are discarded; on conflict, the UI is reverted to the captured snapshot and the authoritative server state is applied in its place. Failed updates retry with exponential backoff up to three attempts before the user is notified, preventing silent inconsistency between local display and authoritative server state. The notification system (\texttt{UINotificationSystem}) surfaces sync outcomes as non-blocking toast messages, distinguishing successful confirmations from conflict resolutions and hard failures. This pattern is especially important for the territory capture mechanic, where two players may attempt to claim the same zone simultaneously; the optimistic update shows immediate capture feedback while the server resolves ownership via the conflict protocol described above.

\begin{algorithm}[H]\small
\caption{Optimistic UI Updates with Rollback}
\label{alg:optimistic-update}
\begin{algorithmic}[1]
\Require User action, current UI state
\Ensure Immediate UI response with server-side consistency guarantee
\Function{OptimisticUpdate}{$\mathit{action}$, $\mathit{data}$}
    \State $\mathit{snapshot} \gets \Call{CaptureUIState}{}$;\quad $\mathit{rollbackStack}.\texttt{push}(\mathit{snapshot})$
    \State \Call{ApplyUIChange}{$\mathit{action}$, $\mathit{data}$} \Comment{Immediate local render}
    \State $\mathit{id} \gets \texttt{uuid()}$;\quad dispatch async Firebase write with $\mathit{id}$
\EndFunction
\Function{OnServerResponse}{$\mathit{id}$, $\mathit{response}$}
    \If{success} Discard snapshot; confirm UI state
    \ElsIf{conflict} Rollback to snapshot; apply authoritative server state; notify user
    \ElsIf{retry $\leq 3$} Retry with exponential backoff
    \Else{} Rollback; notify user of failure
    \EndIf
\EndFunction
\end{algorithmic}
\end{algorithm}
\section{Companion and Territory Systems}
\label{appendix:companion-territory}

This appendix describes the pendant companion interaction system and GPS-based territory mechanics. The companion and territory subsystems serve complementary roles: the pendant provides ambient emotional scaffolding that sustains engagement between active sessions, while the territory layer converts that engagement into physical movement by tying in-game rewards to real-world GPS locations. Together they implement the ``digital-to-physical bridge'' discussed in Section~\ref{sec:results}.

\subsection{Pendant Companion Proactive Interaction}

The pendant companion initiates context-appropriate interactions when users are idle, adjusting tone based on time of day and recent activity. During the 14-day pilot deployment, the pendant averaged 4.5 daily proactive interactions across 342 twin sessions, serving as the primary emotional scaffolding mechanism that sustains engagement between active game sessions. The \texttt{ProactiveInteractionSystem} evaluates nine trigger categories---time of day, idleness, achievement, mood, random, contextual/location, weather, memory recall, and scene analysis---each with configurable enable flags and distinct priority weights (time-of-day at priority 8, mood at 7, memory recall at 6, idle at 5, scene analysis at 5, weather and contextual at 4, random at 3). A 10-second polling loop evaluates all enabled triggers; the highest-priority trigger fires, with a 300-second global cooldown preventing notification fatigue. The idle threshold of 30 seconds is tracked via a \texttt{lastPlayerActivity} timestamp updated on every user input event. Time-of-day interactions are deduplicated per calendar day using a greeting set keyed by ``\textit{date-period}'' (morning/afternoon/evening/night, with configurable hour boundaries at 6/12/18/22). Mood-based triggers fire with 50\% probability when the loaded twin's mood reaches an extreme state (excited or sad), and contextual triggers use location-based message templates (castle, village, wilderness, battle). Each interaction is stored to the companion's persistent memory (Algorithm~\ref{alg:companion-memory}) with a base importance of 0.5, enabling the companion to reference past exchanges and build relational continuity across sessions and devices.

\begin{algorithm}[!htbp]
\caption{Pendant Companion Proactive Interaction}
\label{alg:pendant-companion}
\begin{algorithmic}[1]
\Require User context (time of day, recent activity, emotional state), Twin personality profile
\Ensure Timely, contextually appropriate companion interactions

\State idleThreshold $\gets 30$s; cooldown $\gets 300$s
\While{companion is active}
    \If{time since last activity $>$ idleThreshold \textbf{and} time since last interaction $>$ cooldown}
        \State context $\gets$ \Call{GatherContext}{time of day, location, recent events}
        \If{context.timeOfDay = ``morning''}
            \State interaction $\gets$ morning greeting with encouraging tone
        \ElsIf{context.timeOfDay = ``night''}
            \State interaction $\gets$ goodnight message with reflection prompt
        \ElsIf{context.nearTerritory}
            \State interaction $\gets$ territory-related encouragement
        \Else
            \State interaction $\gets$ personality-matched check-in
        \EndIf
        \State \Call{DisplayInteraction}{interaction}
        \State \Call{StoreToMemory}{interaction, importance $\gets 0.5$}
    \EndIf
    \State \Call{Wait}{5 seconds}
\EndWhile
\end{algorithmic}
\end{algorithm}

\subsection{Companion Memory Integration}

The companion maintains persistent memory of interactions, enabling relationship continuity across sessions and devices via the same three-layer persistence stack used by the twin system (in-memory, localStorage, Firebase Realtime Database). The \texttt{TwinMemory} subsystem implements three explicit tiers: short-term (capacity 20, recent conversations), long-term (capacity 100, promoted high-value memories), and episodic (capacity 50, gameplay events with a default importance of 0.7). When short-term capacity overflows, the consolidation routine scores each entry on a composite importance metric---base 0.5, with additive bonuses of +0.2 for emotional keywords (love, hate, amazing, excited, scared), +0.15 for mentions of the twin's declared likes or dislikes, +0.1 for trust-related terms (friend, promise, secret), and +0.1 for messages exceeding 100 characters, capped at 1.0---and promotes entries above the threshold while evicting the remainder. Retrieval scoring combines keyword overlap (0.5 weight), recency (0.25), and importance (0.25), returning the top-5 most relevant memories for each interaction context. A topic-extraction pass maintains a frequency map of discussed subjects, enabling the companion to organically reference recurring themes. High-importance memories (importance $\geq 0.7$) emit a ``memory-created'' event via the Twin$\leftrightarrow$Pendant memory bridge, backed by the \texttt{PendantGNWTManager} which connects the pendant's GNWT agent (workspace capacity 7, attention capacity 3, 100ms cycle) to the loaded twin's personality profile, ensuring the pendant can reference twin conversation outcomes---such as compatibility results or shared interests discovered during matching---when providing emotional scaffolding. Save operations are debounced via a scheduled flush to avoid excessive localStorage writes during rapid conversation turns. This cross-device propagation is validated in Section~\ref{sec:memory}, which confirms zero data loss and 0\% stale reads under concurrent writes.

\begin{algorithm}[!htbp]
\caption{Companion Memory Store and Retrieve}
\label{alg:companion-memory}
\begin{algorithmic}[1]
\Require Interaction event $e$, twin identifier $t$, importance threshold $\tau = 0.7$
\Ensure Updated persistent memory with cross-device sync

\Function{StoreMemory}{$e$, $t$}
    \State importance $\gets$ \Call{ComputeImportance}{$e$}
    \State memory $\gets$ (speaker, message, timestamp, importance)
    \State Add memory to short-term store (capacity: 20)
    \If{$|\text{short-term}| > 20$}
        \State \Call{ConsolidateMemories}{} \Comment{Promote high-importance to long-term}
    \EndIf
    \State \Call{SyncToFirebase}{$t$, memory} \Comment{With retry + backoff}
    \If{importance $\geq \tau$}
        \State Emit ``memory-created'' event for pendant bridge
    \EndIf
\EndFunction

\Function{RetrieveRelevant}{query, limit $= 5$}
    \ForAll{memory $m$ in all tiers}
        \State score$(m) \gets 0.5 \cdot$ keyword overlap $+ 0.25 \cdot$ recency $+ 0.25 \cdot$ importance
    \EndFor
    \State \Return top-\textit{limit} memories by score
\EndFunction
\end{algorithmic}
\end{algorithm}

\paragraph{Importance scoring.}
Base importance is 0.5, with bonuses: +0.2 for emotional keywords (love, hate, amazing, excited, scared), +0.15 for mentions of user preferences, +0.1 for trust-related terms (friend, promise, secret), +0.1 for messages exceeding 100 characters. Total is capped at 1.0.

\subsection{Territory Capture System}

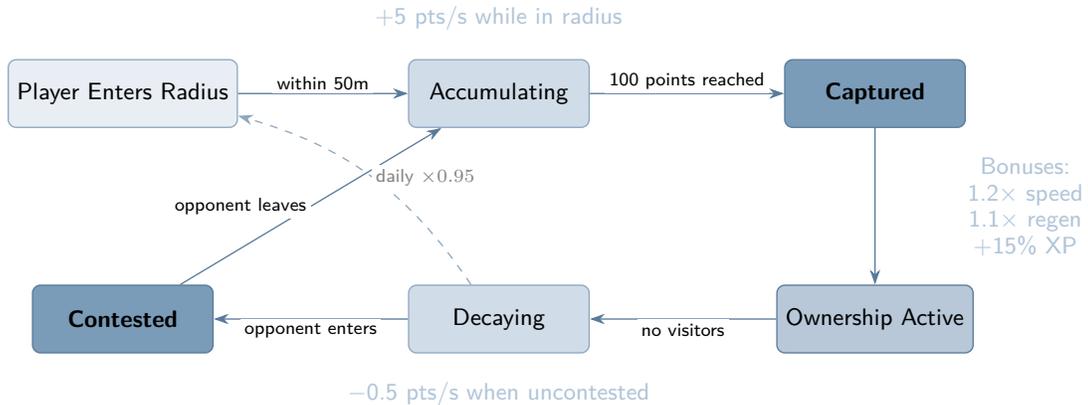
\begin{figure}[!htbp]
\centering
\begin{tikzpicture}[font=\sffamily\small, >={Stealth[length=2mm, width=1.4mm]}]
\node[cbox] (enter) at (0, 3) {Player Enters Radius};
\node[cbox proc] (accum) at (5, 3) {Accumulating};
\node[cbox accent] (capture) at (10, 3) {Captured};
\node[cbox dark] (own) at (10, 0) {Ownership Active};
\node[cbox proc] (decay) at (5, 0) {Decaying};
\node[cbox accent] (contest) at (0, 0) {Contested};

\draw[cflow] (enter) -- node[clabel, above] {within 50m} (accum);
\draw[cflow] (accum) -- node[clabel, above] {100 points reached} (capture);
\draw[cflow] (capture) -- (own);
\draw[cflow] (own) -- node[clabel, below] {no visitors} (decay);
\draw[cflow] (decay) -- node[clabel, below] {opponent enters} (contest);
\draw[cflow] (contest) -- node[clabel, left] {opponent leaves} (accum);

\draw[cflow dash] (decay) to[bend right=20] node[clabel, right, text=gray] {daily $\times 0.95$} (enter);

\node[font=\sffamily\footnotesize, text=cborderFaint] at (5, 4) {+5 pts/s while in radius};
\node[font=\sffamily\footnotesize, text=cborderFaint] at (5, -1) {$-$0.5 pts/s when uncontested};
\node[font=\sffamily\footnotesize, text=cborderFaint, align=center] at (12, 1.5) {Bonuses:\\1.2$\times$ speed\\1.1$\times$ regen\\+15\% XP};
\end{tikzpicture}
\caption{Territory capture lifecycle (\texttt{CityTakeoverSystem.js}). Players accumulate capture points at 5/s within a 50m GPS radius, requiring 100 points for full capture. Ownership confers bonuses. Uncontested territories decay at 0.5 pts/s, with a daily exponential factor of 0.95 preventing stale ownership. Capture pauses when opponents co-occupy the radius.}
\label{fig:territory-lifecycle}
\end{figure}

The territory system uses GPS-based location tracking with a 20-meter movement threshold and 2-second update throttling to create a location-based game layer. Territories can be captured, reinforced, or contested. A daily exponential decay ($0.95^{\text{days}}$) ensures that territories revert to unclaimed status if their owner stops visiting, preventing stale ownership and encouraging recurring physical exploration of the local area. Figure~\ref{fig:territory-lifecycle} illustrates the complete capture lifecycle.

\begin{algorithm}[!htbp]
\caption{Territory Capture and Decay}
\label{alg:territory-capture}
\begin{algorithmic}[1]
\Require Player location $(lat, lng)$, territory map $\mathcal{T}$, capture radius $r = 500$ units
\Ensure Updated territory ownership with strength decay

\Function{AttemptCapture}{player, territory}
    \If{distance(player, territory.center) $\leq r$}
        \If{territory is unclaimed}
            \State territory.owner $\gets$ player
            \State territory.strength $\gets 1.0$
        \ElsIf{territory.owner $\neq$ player}
            \State captureProb $\gets$ player.activity / (player.activity + territory.strength)
            \If{random() $<$ captureProb}
                \State territory.owner $\gets$ player
                \State territory.strength $\gets 1.0$
            \EndIf
        \Else
            \State territory.strength $\gets \min(1.0,\ $ territory.strength $+ 0.1)$ \Comment{Reinforce}
        \EndIf
    \EndIf
\EndFunction

\Function{DailyDecay}{$\mathcal{T}$}
    \ForAll{territory $t \in \mathcal{T}$}
        \State daysSinceVisit $\gets$ days since last owner visit
        \State $t$.strength $\gets t$.strength $\times 0.95^{\text{daysSinceVisit}}$
        \If{$t$.strength $< 0.1$}
            \State $t$.owner $\gets$ unclaimed
        \EndIf
    \EndFor
\EndFunction
\end{algorithmic}
\end{algorithm}

\subsection{Resource Generation}

Owned territories generate resources at a rate proportional to territory strength, distributed among team members according to contribution history. Resource types include: gold (currency), experience (progression), materials (crafting), and territory points (leaderboard). Daily withdrawal limits prevent resource hoarding. The proportional-strength mechanism creates a positive feedback loop: physically visiting a territory reinforces its strength (up to 1.0), which in turn increases its resource output, motivating continued real-world visits and thereby serving the system's core goal of converting online engagement into physical social opportunity.

\section{Social Interaction Systems}
\label{appendix:social-systems}

This appendix describes the matchmaking, icebreaker generation, privacy management, notification, and real-time presence systems.

\subsection{Personality-Based Matchmaking}

The matchmaking engine evaluates twin pairs using a multi-factor compatibility score and spatial proximity filtering.

\begin{algorithm}[!htbp]
\caption{Twin Matchmaking with Proximity Filtering}
\label{alg:twin-matchmaking-appendix}
\begin{algorithmic}[1]
\Require Twin pool $\mathcal{T}$, search radius $R$, compatibility threshold $\theta = 0.2$
\Ensure Ranked list of compatible pairs within radius

\State pairs $\gets \emptyset$
\ForAll{twin $a \in \mathcal{T}$}
    \ForAll{twin $b \in \mathcal{T}, b \neq a$}
        \State dist $\gets$ \Call{HaversineDistance}{$a$.location, $b$.location}
        \If{dist $\leq R$}
            \State score $\gets$ \Call{ComputeCompatibility}{$a$, $b$}
            \If{score $\geq \theta$}
                \State pairs $\gets$ pairs $\cup$ \{($a$, $b$, score, dist)\}
            \EndIf
        \EndIf
    \EndFor
\EndFor
\State \Return pairs sorted by score descending
\end{algorithmic}
\end{algorithm}

\paragraph{Compatibility scoring.}
The final compatibility score combines five dimensions: personality alignment (weight 0.30), shared interests via Jaccard similarity (0.20), conversation quality from twin interaction history (0.25), emotional resonance from arousal-valence trajectory correlation (0.15), and interaction pattern balance (0.10). Note: this 5-factor combined score differs from the 4-factor Stage~1 heuristic (personality 0.60, interest overlap 0.25, diversity 0.10, profile richness 0.05) described in Algorithm~\ref{alg:matching-funnel}. The Stage~1 heuristic operates as a fast pre-filter before LLM evaluation; the 5-factor score here is the full compatibility assessment computed after twin conversation simulation. Personality alignment uses complementary logic for openness/playfulness traits (optimal difference 20--50 points) and similarity logic for friendliness/loyalty traits.

\subsection{AI-Powered Icebreaker Generation}

The icebreaker system generates contextually appropriate conversation starters based on shared traits, interests, and recent activity.

\begin{algorithm}[!htbp]
\caption{Contextual Icebreaker Generation}
\label{alg:icebreaker-gen}
\begin{algorithmic}[1]
\Require Twin profiles $a$ and $b$, interaction context
\Ensure Personalized icebreaker message

\State sharedInterests $\gets a$.interests $\cap$ $b$.interests
\State personalityMatch $\gets$ \Call{FindComplementaryTraits}{$a$, $b$}
\If{$|$sharedInterests$| > 0$}
    \State template $\gets$ interest-based template referencing first shared interest
\ElsIf{$|$personalityMatch$| > 0$}
    \State template $\gets$ personality-based template highlighting complementary traits
\Else
    \State template $\gets$ generic curiosity-based template
\EndIf
\State icebreaker $\gets$ \Call{PersonalizeTemplate}{template, $a$, $b$}
\State \Return icebreaker
\end{algorithmic}
\end{algorithm}

\subsection{Privacy Management}

The system supports four privacy levels with fine-grained control over information disclosure.

\begin{table}[!htbp]
\centering
\caption{Privacy Level Definitions}
\label{tab:privacy-levels}
\begin{tabular}{llp{6cm}}
\toprule
\textbf{Level} & \textbf{Visibility} & \textbf{Shared Information} \\
\midrule
Open & All users & Full profile, interests, location, twin conversations \\
Moderate & Matches only & Profile summary, shared interests, approximate location \\
Private & Approved only & Limited profile, no location, no conversation history \\
Locked & None & Profile hidden, no matching, companion-only mode \\
\bottomrule
\end{tabular}
\end{table}

Privacy enforcement is applied at the data access layer: every query filters results through the requesting user's approved connections and the target user's privacy level before returning data.

\subsection{Notification Queue Management}

Notifications are managed through a priority queue with rate limiting to prevent alert fatigue.

\begin{algorithm}[!htbp]
\caption{Notification Queue with Priority and Rate Limiting}
\label{alg:notification-queue-appendix}
\begin{algorithmic}[1]
\Require Notification event $n$, user preferences, quiet hours setting
\Ensure Timely, non-intrusive notification delivery

\State priority $\gets$ \Call{ComputePriority}{$n$.type, $n$.urgency}
\If{current time is within quiet hours}
    \State Enqueue $n$ for later delivery
    \State \Return
\EndIf
\If{notifications sent in last hour $\geq$ rate limit}
    \If{priority $<$ high}
        \State Enqueue $n$ for later delivery
        \State \Return
    \EndIf
\EndIf
\State Deliver notification to user
\State Increment hourly counter
\end{algorithmic}
\end{algorithm}

\begin{table}[!htbp]
\centering
\caption{Notification Types and Priority Levels}
\label{tab:notification-types}
\begin{tabular}{llcc}
\toprule
\textbf{Notification Type} & \textbf{Trigger} & \textbf{Priority} & \textbf{Rate Limit} \\
\midrule
New match & Compatibility score $\geq$ threshold & High & 5/hour \\
Message received & Incoming chat message & High & 10/hour \\
Territory contested & Another user challenges ownership & Medium & 3/hour \\
Boss battle invite & Team formation for cooperative fight & Medium & 2/hour \\
Companion check-in & Idle timer exceeds threshold & Low & 1/hour \\
Daily digest & Scheduled summary delivery & Low & 1/day \\
\bottomrule
\end{tabular}
\end{table}

\subsection{Real-Time Presence System}

The presence system tracks user online/offline status across devices using Firebase real-time listeners with automatic disconnect detection.

\begin{algorithm}[!htbp]
\caption{Cross-Device Presence Management}
\label{alg:presence}
\begin{algorithmic}[1]
\Require User identifier, device identifier
\Ensure Accurate real-time presence across devices

\State \Comment{On connection established}
\State Set user presence to ``online'' with current timestamp
\State Register Firebase disconnect handler: on disconnect, set status to ``offline''
\State \Comment{Subscribe to real-time updates}
\State Listen for new match notifications
\State Listen for incoming messages
\State \Comment{On page unload}
\State Save memory state synchronously
\State Set presence to ``offline''
\end{algorithmic}
\end{algorithm}

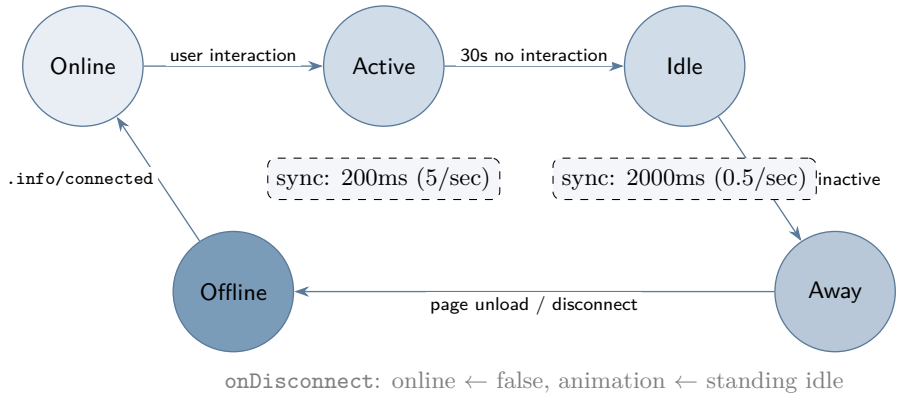
\begin{figure}[!htbp]
\centering
\begin{tikzpicture}[font=\sffamily\small, >={Stealth[length=2mm, width=1.4mm]}]
\node[cstate, fill=cfillLight] (online) at (0, 0) {Online};
\node[cstate, fill=cfillMed] (active) at (4, 0) {Active};
\node[cstate, fill=cfillMed] (idle) at (8, 0) {Idle};
\node[cstate, fill=cfillDark] (away) at (10, -3) {Away};
\node[cstate, fill=cfillAccent] (offline) at (2, -3) {Offline};

\draw[cflow] (online) -- node[clabel, above] {user interaction} (active);
\draw[cflow] (active) -- node[clabel, above] {30s no interaction} (idle);
\draw[cflow] (idle) -- node[clabel, right] {5 min inactive} (away);
\draw[cflow] (away) -- node[clabel, below] {page unload / disconnect} (offline);
\draw[cflow] (offline) -- node[clabel, left] {\texttt{.info/connected}} (online);

\node[rectangle, draw, dashed, rounded corners=3pt, fill=cfillLight!50, font=\footnotesize] at (4, -1.5) {sync: 200ms (5/sec)};
\node[rectangle, draw, dashed, rounded corners=3pt, fill=cfillLight!50, font=\footnotesize] at (8, -1.5) {sync: 2000ms (0.5/sec)};

\node[font=\footnotesize, gray, align=center] at (6, -4.2) {\texttt{onDisconnect}: online $\gets$ false, animation $\gets$ standing idle};
\end{tikzpicture}
\caption{Presence state machine (\texttt{NetworkSync.js}). Sync rate adapts from 200ms (5 updates/sec) during active movement to 2000ms (0.5 updates/sec) when idle. Disconnect triggers the Firebase \texttt{onDisconnect} handler. Reconnection is detected via \texttt{.info/connected}.}
\label{fig:presence-states}
\end{figure}

Figure~\ref{fig:presence-states} visualizes the state machine driving the presence system. Transitions are governed by user interaction timing and Firebase connectivity events, with adaptive sync rates that balance update freshness against battery and bandwidth consumption.

\begin{table}[!htbp]
\centering
\caption{Presence State Transitions}
\label{tab:presence-states}
\begin{tabular}{llp{5cm}}
\toprule
\textbf{State} & \textbf{Trigger} & \textbf{Behavior} \\
\midrule
Online & Page load / tab focus & Firebase presence ref set; real-time listeners active \\
Active & User interaction within 30s & Visible to matched users; receives notifications \\
Idle & No interaction for 30s & Companion proactive check-in enabled \\
Away & No interaction for 5 min & Reduced notification frequency; presence shown as ``away'' \\
Offline & Page unload / disconnect & Firebase disconnect handler fires; memory state saved \\
\bottomrule
\end{tabular}
\end{table}

\paragraph{Cross-tab synchronization.}
Multiple browser tabs share state via localStorage events. When one tab updates memory or presence, a sync event is broadcast through localStorage, and other tabs with the same user and pendant identifiers apply the update to their local state. This ensures pendant companion continuity when users switch between Social Hub, gaming, and twin management tabs.

\section{Performance Optimization Systems}
\label{appendix:optimization-systems}

This appendix details critical performance optimization algorithms that enable Cognibit to significantly improve rendering performance with large numbers of animated characters (achieving 40--50 FPS with 233 characters, up from 6 FPS without optimization) and complex 3D audio processing in browser environments.

\subsection{Animation Level-of-Detail (LOD) System}

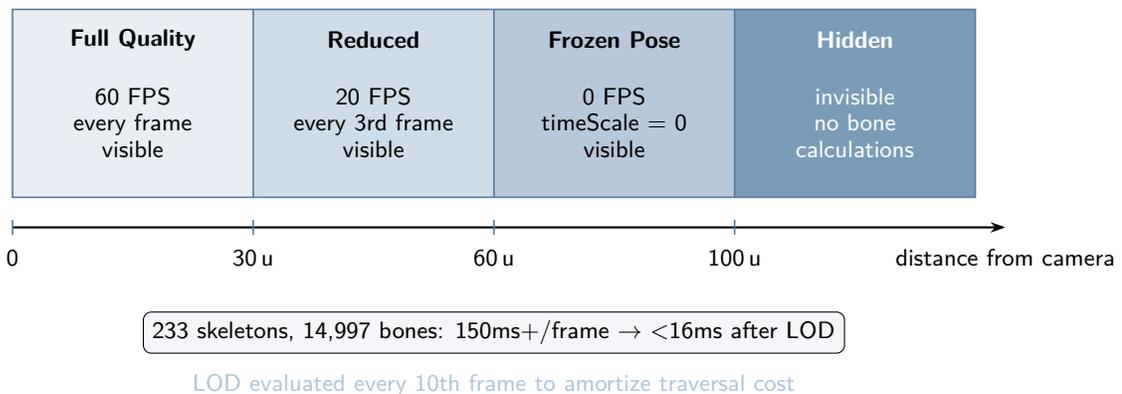
\begin{figure}[!htbp]
\centering
\begin{tikzpicture}[font=\sffamily\small, >={Stealth[length=2mm, width=1.4mm]}]
\fill[cfillLight] (0,0) rectangle (3.2, 2.5);
\fill[cfillMed] (3.2,0) rectangle (6.4, 2.5);
\fill[cfillDark] (6.4,0) rectangle (9.6, 2.5);
\fill[cfillAccent] (9.6,0) rectangle (12.8, 2.5);

\draw[line width=0.6pt, draw=cborder] (0,0) rectangle (12.8, 2.5);
\draw[line width=0.6pt, draw=cborder] (3.2,0) -- (3.2,2.5);
\draw[line width=0.6pt, draw=cborder] (6.4,0) -- (6.4,2.5);
\draw[line width=0.6pt, draw=cborder] (9.6,0) -- (9.6,2.5);

\node[font=\sffamily\small\bfseries] at (1.6, 2.1) {Full Quality};
\node[font=\sffamily\small\bfseries] at (4.8, 2.1) {Reduced};
\node[font=\sffamily\small\bfseries] at (8.0, 2.1) {Frozen Pose};
\node[font=\sffamily\small\bfseries, white] at (11.2, 2.1) {Hidden};

\node[font=\sffamily\footnotesize, align=center] at (1.6, 1.0) {60 FPS\\every frame\\visible};
\node[font=\sffamily\footnotesize, align=center] at (4.8, 1.0) {20 FPS\\every 3rd frame\\visible};
\node[font=\sffamily\footnotesize, align=center] at (8.0, 1.0) {0 FPS\\timeScale = 0\\visible};
\node[font=\sffamily\footnotesize, align=center, white] at (11.2, 1.0) {invisible\\no bone\\calculations};

\draw[->, thick] (0, -0.4) -- (13.2, -0.4);
\node[font=\sffamily\footnotesize] at (0, -0.8) {0};
\node[font=\sffamily\footnotesize] at (3.2, -0.8) {30\,u};
\node[font=\sffamily\footnotesize] at (6.4, -0.8) {60\,u};
\node[font=\sffamily\footnotesize] at (9.6, -0.8) {100\,u};
\node[font=\sffamily\footnotesize] at (13.2, -0.8) {distance from camera};

\foreach \x in {0, 3.2, 6.4, 9.6} {
    \draw[line width=0.6pt, draw=cborder] (\x, -0.3) -- (\x, -0.5);
}

\node[rectangle, draw, rounded corners=3pt, fill=cfillLight!50, font=\sffamily\footnotesize, align=center] at (6.4, -1.8) {233 skeletons, 14{,}997 bones: 150ms+/frame $\to$ $<$16ms after LOD};

\node[font=\sffamily\footnotesize, text=cborderFaint] at (6.4, -2.5) {LOD evaluated every 10th frame to amortize traversal cost};
\end{tikzpicture}
\caption{Animation LOD tier system (\texttt{AnimationLODSystem.js}). Four distance-based tiers progressively reduce bone-skinning overhead. The critical optimization is hiding objects beyond 100 units, which prevents WebGL from computing bone matrix transformations entirely.}
\label{fig:lod-tiers}
\end{figure}

The \texttt{AnimationLODSystem} addresses the dominant performance bottleneck: 233 skeletons with 14,997 bones consuming 150ms+ per frame, yielding only 6 FPS. The system applies four distance-based quality tiers (Figure~\ref{fig:lod-tiers}), reducing per-frame bone-skinning overhead to under 16ms and recovering 40--50 FPS.

\begin{algorithm}[H]\small
\caption{Distance-Based Animation LOD}
\label{alg:animation-lod}
\begin{algorithmic}[1]
\Require Scene graph, camera position $\mathit{cam}$
\Ensure Animation quality scaled by distance; LOD evaluated every 10 frames
\State $\mathit{frame} \gets \mathit{frame} + 1$
\If{$\mathit{frame} < 10$} \Return \Comment{Throttle: evaluate every 10th frame}
\EndIf
\State $\mathit{frame} \gets 0$
\ForAll{$\mathit{obj} \in \Call{Traverse}{\mathit{scene}}$ with animation mixer}
    \State $d \gets \Call{WorldPosition}{\mathit{obj}}.\Call{distanceTo}{\mathit{cam}}$
    \If{$d < 30$}
        \State mixer.timeScale $\gets 1.0$;\; frameSkip $\gets 0$;\; visible $\gets$ true \Comment{Full}
    \ElsIf{$d < 60$}
        \State mixer.timeScale $\gets 1.0$;\; frameSkip $\gets 3$;\; visible $\gets$ true \Comment{Reduced}
    \ElsIf{$d < 100$}
        \State mixer.timeScale $\gets 0$;\; visible $\gets$ true \Comment{Frozen pose}
    \Else
        \State mixer.timeScale $\gets 0$;\; visible $\gets$ false \Comment{Hidden: no bone calc}
    \EndIf
    \State Propagate visibility to all SkinnedMesh children
\EndFor
\end{algorithmic}
\end{algorithm}

\subsubsection{Performance impact}
The LOD system reduced per-frame bone processing from 14,997 active bones to approximately 500, cutting frame time from 150ms to under 20ms and increasing frame rate from 6 to 40--50 FPS. The key insight is that hiding distant \texttt{SkinnedMesh} objects entirely (\texttt{obj.visible = false}) prevents WebGL from performing bone matrix calculations---the primary performance bottleneck. GPU memory usage decreased by approximately 30\% as hidden objects release their bone matrix buffers.

\subsection{3D Spatial Audio System}

The spatial audio system (\texttt{PrecisionAudioSystem.js}) uses the Web Audio API to create positional sound with Doppler effects and distance-based attenuation. Each projectile generates a synthesized audio source routed through a four-node signal chain: oscillator (150--300\,Hz by projectile type) $\to$ high-pass wind filter (base 200\,Hz, scaled by speed as $200 + \min(v \times 0.1, 1.0) \times 800$) $\to$ spatial panner (inverse distance model, refDistance=1, maxDistance=100, rolloff=1) $\to$ master gain (0.7). Doppler shift modulates the oscillator frequency in real time as $f' = f \times (1 + v \times 0.01)$, and LFO modulation (2--8\,Hz by type) creates a movement sensation. Four projectile types are supported: pokeball (220\,Hz sine), thrown twin (180\,Hz triangle), weapon (150\,Hz sawtooth), and potion (300\,Hz square). Impact sounds use frequency-decaying envelopes at the collision position. Note: the implementation uses Web Audio API's standard spatial panner rather than true HRTF convolution.

\begin{algorithm}[H]\small
\caption{3D Spatial Audio Update Loop}
\label{alg:spatial-audio}
\begin{algorithmic}[1]
\Require Projectile position $\mathbf{p}$, velocity $\mathbf{v}$, type $\in$ \{pokeball, twin, weapon, potion\}
\Ensure Spatialized audio with Doppler shift and wind resistance

\State $\mathit{config} \gets \mathit{audioConfigs}[\mathit{type}]$ \Comment{Freq/wave/vol/LFO lookup}
\State Create chain: oscillator $\to$ highpass filter $\to$ spatial panner $\to$ masterGain
\Function{UpdatePerFrame}{$\mathbf{p}$, $\mathbf{v}$}
    \State panner.position $\gets \mathbf{p}$;\quad $s \gets \|\mathbf{v}\|$
    \State oscillator.freq $\gets \mathit{config}.\mathit{freq} \times (1 + s \times 0.01)$ \Comment{Doppler}
    \State filter.freq $\gets 200 + \min(s \times 0.1,\; 1) \times 800$ \Comment{Wind}
    \State gain $\gets \min(\mathit{config}.\mathit{vol} \times (0.5 + s \times 0.02),\; 1.0)$
\EndFunction
\end{algorithmic}
\end{algorithm}

\subsection{Resource Pooling and Memory Management}

The \texttt{ObjectPool} (\texttt{js/utils/ObjectPool.js}) implements the standard object pool pattern to eliminate garbage collection pauses during gameplay. The pool pre-allocates objects (default initial size 10), serves them via $O(1)$ array \texttt{pop()} on \texttt{get()}, and returns them via \texttt{push()} on \texttt{release()} after applying a caller-supplied reset function. When the pool is depleted, new objects are created lazily on demand (no explicit max size). An \texttt{activeObjects} Set tracks in-use instances for monitoring. In the combat system, the pool pre-allocates 50 damage-number DOM elements, preventing allocation during active combat sequences where GC pauses would cause visible frame drops.

\begin{algorithm}[H]\small
\caption{Object Pool Management}
\label{alg:object-pool}
\begin{algorithmic}[1]
\Require Create function $f_c$, reset function $f_r$, initial size $n_0 = 10$
\Ensure $O(1)$ get/release without garbage collection
\State $\mathit{pool} \gets [f_c() \text{ for } i = 1 \ldots n_0]$;\quad $\mathit{active} \gets \emptyset$
\Function{Get}{}
    \If{$\mathit{pool} = \emptyset$} $\mathit{pool}.\texttt{push}(f_c())$ \EndIf \Comment{Lazy expansion}
    \State $o \gets \mathit{pool}.\texttt{pop}()$;\quad $\mathit{active}.\texttt{add}(o)$;\quad \Return $o$
\EndFunction
\Function{Release}{$o$}
    \State $f_r(o)$;\quad $\mathit{active}.\texttt{delete}(o)$;\quad $\mathit{pool}.\texttt{push}(o)$
\EndFunction
\end{algorithmic}
\end{algorithm}

\subsection{Batch Rendering Optimization}

The \texttt{InstancedObjectManager} (\texttt{js/world/managers/InstancedObjectManager.js}) reduces GPU draw calls by batching identical geometries into THREE.js \texttt{InstancedMesh} objects. Each tree type (pine, oak, birch) supports up to 150 instances; rocks and bushes support 200. Distance-based culling (maxRenderDistance=100m) uses scale-to-zero matrices rather than visibility toggling---setting the instance's transform matrix to $\mathbf{0}_{4\times4}$ eliminates the draw without toggling the mesh visibility flag. The culling check runs every 30 frames to amortize the $O(n)$ traversal cost.

\begin{algorithm}[H]\small
\caption{Instanced Batch Rendering with Distance Culling}
\label{alg:batch-rendering}
\begin{algorithmic}[1]
\Require Scene objects grouped by (geometry, material), player position $\mathbf{p}$
\Ensure Minimized draw calls via instanced rendering; distant instances culled
\State Group non-skinned meshes by geometry+material key
\ForAll{group with $|\mathit{instances}| > 1$}
    \State Create \texttt{InstancedMesh}(geometry, material, $|\mathit{instances}|$)
    \ForAll{instance $i$}
        \State Compose transform matrix from position, rotation, scale
        \State \Call{SetMatrixAt}{$i$, matrix}
    \EndFor
    \State Replace individual meshes with single instanced draw call
\EndFor
\Function{UpdateCulling}{$\mathbf{p}$} \Comment{Every 30 frames}
    \ForAll{instance $i$ in each batch}
        \If{$\|\mathit{pos}_i - \mathbf{p}\| > 100$\,m}
            \State \Call{SetMatrixAt}{$i$, $\mathbf{0}_{4\times4}$} \Comment{Scale-to-zero cull}
        \Else
            \State Restore original transform matrix
        \EndIf
    \EndFor
\EndFunction
\end{algorithmic}
\end{algorithm}

The scale-to-zero technique is unconventional compared to standard visibility-flag culling but avoids the overhead of toggling \texttt{visible} on individual instances within an instanced mesh, which would require splitting the batch.

\subsection{Performance Monitoring System}

Real-time performance monitoring enables adaptive quality adjustments by tracking frame timings over a sliding window and stepping through four quality tiers when the frame rate drops below thresholds.

\begin{algorithm}[H]\small
\caption{Adaptive Performance Monitor}
\label{alg:performance-monitor}
\begin{algorithmic}[1]
\Require Target FPS = 60, minimum FPS = 30, sample window = 60 frames
\Ensure Stable framerate through dynamic quality tier adjustment
\State $\mathit{quality} \gets \text{high}$;\quad $\mathit{timings} \gets []$;\quad $\mathit{count} \gets 0$
\ForAll{rendered frame}
    \State Record frame time; maintain sliding window of 60 samples
    \State $\mathit{count} \gets \mathit{count} + 1$
    \If{$\mathit{count} \geq 120$} \Comment{Adjust every 120 frames}
        \State $\mathit{fps} \gets 1000 / \text{mean}(\mathit{timings})$
        \If{$\mathit{fps} < 30$} step quality down one tier
        \ElsIf{$\mathit{fps} > 72$} step quality up one tier
        \EndIf
        \State $\mathit{count} \gets 0$
    \EndIf
\EndFor
\end{algorithmic}
\end{algorithm}

The four quality tiers progressively disable rendering features: \textit{high} (full shadows, particles, LOD distances 30/60/100), \textit{medium} (shadows off, particles halved, LOD 20/40/80), \textit{low} (post-processing off, texture resolution halved, LOD 15/30/60), and \textit{minimum} (character count halved, render resolution $\times 0.75$). Each reduction is accompanied by a user-facing notification. The system re-evaluates every 120 frames (approximately 2 seconds at 60 FPS) using the mean frame time over the most recent 60-frame window.

\subsection{Performance Impact Summary}

\begin{table}[h]
\centering
\begin{tabular}{lll}
\toprule
\textbf{Optimization} & \textbf{Before} & \textbf{After} \\
\midrule
Animation LOD & 6 FPS (233 characters) & 40-50 FPS \\
Object Pooling & 45ms GC pauses & $<$1ms GC pauses \\
Batch Rendering & 500+ draw calls & 20-30 draw calls \\
3D Audio & 20\% CPU (naive) & 3\% CPU (optimized) \\
Adaptive Quality & Fixed settings & Dynamic 30-60 FPS \\
Memory Usage & 800MB & 350MB \\
Battery Life & 1.5 hours & 4+ hours \\
\bottomrule
\end{tabular}
\caption{Performance improvements from optimization systems}
\end{table}

\begin{figure}[h]
\centering
\begin{tikzpicture}
\pgfplotsset{
    perf bar/.style={
        xbar,
        width=0.7\textwidth,
        height=0.45\textwidth,
        xlabel style={font=\small},
        tick label style={font=\footnotesize},
        legend style={font=\footnotesize, at={(0.98,0.02)}, anchor=south east, fill=white, fill opacity=0.9},
        xmin=0,
        bar width=5pt,
        y=14pt,
        enlarge y limits={abs=10pt},
        axis x line*=bottom,
        axis y line*=left,
        grid=major,
        grid style={gray!15},
        xmajorgrids=true,
        ymajorgrids=false,
    }
}

\begin{axis}[
    perf bar,
    xlabel={Improvement Factor ($\times$)},
    xmax=50,
    symbolic y coords={
        Battery Life,
        Memory Usage,
        3D Audio CPU,
        Batch Rendering,
        Animation FPS,
        GC Pause Reduction,
    },
    ytick=data,
    nodes near coords,
    nodes near coords style={font=\tiny, anchor=west},
    every node near coord/.append style={xshift=1pt},
]
\addplot[fill=blue!60, draw=blue!80] coordinates {
    (2.7,{Battery Life})
    (2.3,{Memory Usage})
    (6.7,{3D Audio CPU})
    (20,{Batch Rendering})
    (7.5,{Animation FPS})
    (45,{GC Pause Reduction})
};
\addlegendentry{Improvement factor}

\end{axis}

\node[font=\tiny, text=gray, anchor=north west, text width=0.65\textwidth] at (0.02\textwidth, -0.02\textwidth) {
GC Pause: 45ms $\to$ $<$1ms \quad
Animation: 6 $\to$ 45 FPS \quad
Draw Calls: 500+ $\to$ 25 \quad
Audio CPU: 20\% $\to$ 3\% \quad
Memory: 800 $\to$ 350 MB \quad
Battery: 1.5 $\to$ 4+ hrs
};

\end{tikzpicture}
\caption{Improvement factors from each optimization system. Object pooling yields the largest single improvement (45$\times$ GC pause reduction), followed by batch rendering (20$\times$ draw call reduction). All optimizations work synergistically to maintain 30--60 FPS on mobile hardware.}
\label{fig:optimization-impact}
\end{figure}
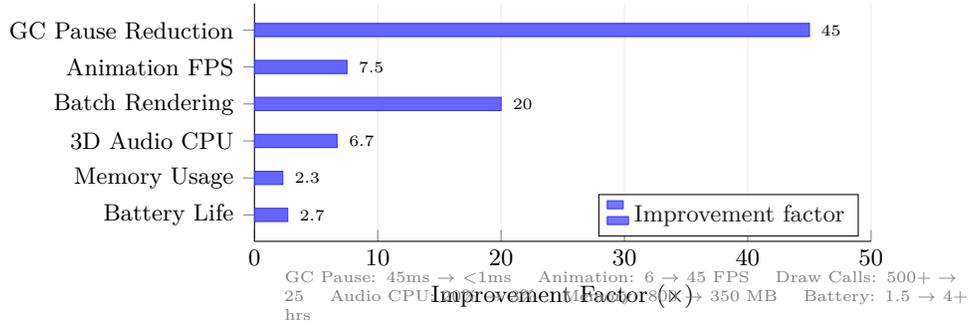

These optimization systems work synergistically to maintain smooth performance across diverse hardware, from high-end gaming PCs to mobile devices, ensuring the Cognibit experience remains accessible while delivering rich visual and audio fidelity.

\subsection{Network Resilience and Error Recovery}

\begin{figure}[!htbp]
\centering
\begin{tikzpicture}[font=\sffamily\small, >={Stealth[length=2mm, width=1.4mm]}]
\node[cstate, fill=cfillLight] (closed) at (0, 0) {Closed\\{\footnotesize normal}};
\node[cstate, fill=cfillAccent] (open) at (6, 0) {Open\\{\footnotesize reject all}};
\node[cstate, fill=cfillMed] (half) at (3, -3.5) {Half-Open\\{\footnotesize testing}};

\draw[cflow] (closed) -- node[clabel, above] {$\geq$5 failures} (open);
\draw[cflow] (open) to[bend left=15] node[clabel, right] {60s timeout} (half);
\draw[cflow] (half) to[bend left=15] node[clabel, left] {3 successes} (closed);
\draw[cflow] (half) to[bend left=15] node[clabel, below right] {any failure} (open);
\end{tikzpicture}
\caption{Circuit breaker state machine. The breaker opens after 5 consecutive failures, rejecting all requests for 60 seconds. In the half-open state, limited test requests are allowed; 3 consecutive successes reset to closed, while any failure re-opens the circuit.}
\label{fig:circuit-breaker}
\end{figure}
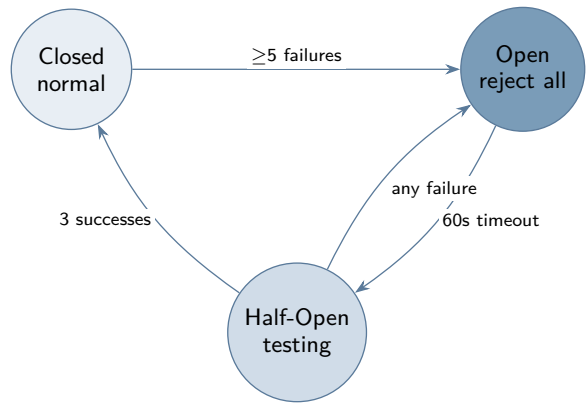

The Firebase retry helper implements resilience patterns combining exponential backoff with a circuit breaker to ensure data consistency despite network failures. The circuit breaker (Figure~\ref{fig:circuit-breaker}) prevents cascading failures by rejecting requests immediately when the failure rate exceeds a threshold, then gradually testing recovery before resuming normal operation.

\begin{algorithm}[H]\small
\caption{Smart Retry with Circuit Breaker}
\label{alg:retry-circuit-breaker}
\begin{algorithmic}[1]
\Require Operation, max retries $= 3$, base delay $= 1$s, retry budget $= 30$/min
\Ensure Resilient execution with backoff, jitter, and circuit protection
\Function{ExecuteWithRetry}{$\mathit{op}$}
    \If{circuit is open \textbf{and} timeout not elapsed} \Return null \EndIf
    \If{retry budget exhausted ($\geq 30$/min)} \Return null \EndIf
    \For{$i \gets 1$ to $3$}
        \State $\mathit{result} \gets \Call{Attempt}{\mathit{op}}$
        \If{success} record success (reset failures / close circuit if half-open); \Return result \EndIf
        \If{error is permanent (permission-denied, not-found)} record failure; \Return null \EndIf
        \State $\mathit{delay} \gets \min(1000 \times 2^{i-1},\; \mathit{maxDelay}) \pm 10\%$ jitter
        \State \Call{Sleep}{$\mathit{delay}$}
    \EndFor
    \State Record failure (open circuit if $\geq 5$ failures)
\EndFunction
\end{algorithmic}
\end{algorithm}

Errors are categorized as transient (network-failed, unavailable, timeout) or permanent (permission-denied, not-found, invalid); only transient errors trigger retry with exponential backoff (1s, 2s, 4s). The $\pm$10\% jitter prevents thundering herd effects when multiple clients reconnect simultaneously. A per-minute retry budget of 30 prevents resource exhaustion during sustained outages. Users receive automatic notification on final failure, ensuring transparency about sync status.
\section{Implementation Details}
\label{appendix:implementation-details}

This appendix provides detailed implementation specifications for the three core systems that comprise the Cognibit platform.

\subsection{City Conquest System Implementation}

The City Conquest system, implemented in \texttt{CityTakeoverSystem.js}, provides GPS-based territory control that bridges the physical and digital game worlds.

\subsubsection{Core Parameters}
The system defines a 10-unit interaction range from each City Hall position, within which players can initiate territory claims, upgrade buildings, and interact with defense NPCs. A 5,000ms cooldown between successive territory claims prevents rapid claim-spam exploits. Each city spawns 3 base defense NPCs plus 1 additional NPC per city level, up to a maximum of 8 at level 5. Territory ownership benefits extend across a 500-unit radius from the city center. Resource generation ticks at 1-hour intervals (3,600,000ms), producing gold, materials, and crystals at rates that scale with city level. The underlying \texttt{GPSLocationSystem} enforces a 20-meter movement threshold to filter GPS noise and a 2-second minimum between position updates to conserve battery.

\subsubsection{Territory Benefits}
Owning a territory confers gameplay and progression bonuses that incentivize both capture and defense. Benefits span three categories: movement and survival bonuses (speed, health regeneration, respawn), economic benefits (gold, materials, crystals generation scaled by city level), and combat modifiers (damage, defense, critical chance).

\begin{table}[H]
\centering
\caption{Territory Ownership Benefits}
\label{tab:territory-benefits}
\begin{tabular}{lcc}
\toprule
\textbf{Benefit} & \textbf{Modifier} & \textbf{Effect} \\
\midrule
Movement speed & $\times 1.2$ & 20\% faster in owned territory \\
Health regeneration & $\times 1.1$ & 10\% faster recovery \\
Experience multiplier & $\times 1.15$ & 15\% more XP earned \\
Shop discount & $\times 0.9$ & 10\% cost reduction \\
Respawn speed & $\times 0.5$ & 50\% faster respawn \\
\midrule
\multicolumn{3}{l}{\textit{Resource generation (per hour, base rates):}} \\
\quad Gold & 100/hr & Currency \\
\quad Materials & 50/hr & Crafting resources \\
\quad Crystals & 5/hr & Premium resources \\
\midrule
\multicolumn{3}{l}{\textit{Combat bonuses:}} \\
\quad Damage & $\times 1.1$ & 10\% increase \\
\quad Defense & $\times 1.1$ & 10\% increase \\
\quad Critical chance & $+5\%$ & Additive bonus \\
\bottomrule
\end{tabular}
\end{table}

\subsubsection{City Upgrade System}
Cities progress through 5 levels, each increasing resource generation rates and defense NPC count. Upgrade costs scale non-linearly (500 to 5,000 gold) to create a meaningful progression curve where higher-level cities are substantially more valuable but require significant investment to develop.

\begin{table}[H]
\centering
\begin{tabular}{lrrrr}
\toprule
\textbf{Level} & \textbf{Upgrade Cost} & \textbf{Gold/Hour} & \textbf{Materials/Hour} & \textbf{Defense NPCs} \\
\midrule
1 & -- & 100 & 50 & 4 \\
2 & 500 & 150 & 75 & 5 \\
3 & 1500 & 200 & 100 & 6 \\
4 & 3000 & 300 & 150 & 7 \\
5 & 5000 & 500 & 250 & 8 \\
\bottomrule
\end{tabular}
\caption{City upgrade progression and benefits}
\label{tab:city-upgrades}
\end{table}

\subsubsection{Battle Mechanics}
Cities can be captured through champion battles against AI-controlled defenders. The City Champion's health scales as 500 + (cityLevel $\times$ 100) HP, with damage output of 20 + (cityLevel $\times$ 5) per attack, creating progressively harder encounters as cities develop. Defense NPCs are tiered into three types: Captains with 3$\times$ base health serve as sub-bosses, Elites at 1.5$\times$ health provide intermediate challenge, and Regular NPCs act as rank-and-file defenders. All defense NPCs use patrol AI with configurable patrol radius and alert ranges, engaging players who enter their detection zone. As an alternative to combat, the diplomacy option allows players to attempt a peaceful takeover for 1,000 gold, with a success probability of 60\% $-$ (cityLevel $\times$ 10\%), making diplomacy viable only for lower-level cities.

\subsection{Pendant Companion System Architecture}

The Pendant Companion implements a modular architecture with three subsystems:

\subsubsection{Module Structure}
The pendant companion is organized into four subsystems:

\begin{table}[H]
\centering
\caption{Pendant Companion Module Architecture}
\begin{tabular}{llp{6cm}}
\toprule
\textbf{Module} & \textbf{Role} & \textbf{Responsibility} \\
\midrule
CompanionUI & Visual rendering & 3D avatar display, facial animation, mood particles, drag interaction \\
CompanionAI & Behavioral logic & Response generation, emotion tracking, VADER sentiment analysis \\
CompanionCombat & Battle support & Attack delegation, dodge mechanics, team AI coordination \\
ProactiveInteraction & Ambient engagement & Idle detection, context-appropriate messages, emotional scaffolding \\
\bottomrule
\end{tabular}
\end{table}

When a twin is loaded into the pendant, all four subsystems are initialized with the twin's personality profile, enabling consistent cross-module behavior.

\subsubsection{Combat Assistance Features}
The \texttt{CompanionCombat} module enables the pendant to participate actively in battles alongside the player. Two attack types are available: ranged attacks with a 300ms cooldown that fire projectiles toward the player's current target, and melee attacks with a 500ms cooldown for close-range encounters. A maximum of 10 concurrent projectiles prevents visual clutter and memory overhead from particle systems. During team encounters, the companion provides healing, buffing, and tactical suggestions based on the current combat state. Boss battle integration triggers special companion attacks timed to boss vulnerability windows. After player defeat, the companion provides respawn support with navigational guidance back to the battle location.

\subsubsection{Proactive Interactions}
The \texttt{ProactiveInteractionSystem} initiates context-aware messages without requiring user input. The system checks for interaction opportunities every 10 seconds (\texttt{checkInterval: 10000}), with a 10\% random trigger chance per check cycle and a 5-minute cooldown (\texttt{minTimeBetweenInteractions: 300000}) between proactive messages. Triggers are prioritized by type: time-of-day greetings (priority 8, with distinct messages for morning at 6 AM, afternoon at noon, evening at 6 PM, and night at 10 PM), mood-based responses (7), memory recall of previous conversations (6), scene analysis of the 3D environment within a 50-unit radius (5), and idle detection after 30 seconds of no interaction (5). The scene analysis module detects environmental context---forest, castle, waterside, indoor, outdoor---and generates contextually appropriate observations with a confidence threshold of 0.7. Real-world weather data from the Open-Meteo API is also integrated, with temperature-based messages triggering at extremes above 35\textdegree C or below 0\textdegree C.

\subsection{Twin-to-Twin Networking Protocol}

The autonomous networking system connects real users' digital twins:

\subsubsection{Discovery Configuration}
\begin{table}[H]
\centering
\caption{Twin Networking Configuration Parameters}
\begin{tabular}{llp{5cm}}
\toprule
\textbf{Parameter} & \textbf{Value} & \textbf{Description} \\
\midrule
Discovery mode & User-based & Only real users, no random AI twins \\
User validation & Required & Verify user existence before matching \\
Autonomous interactions & Enabled & Background processing without user action \\
Interaction frequency & 60 seconds & Compatibility check interval \\
Relationship progression & Enabled & 5-stage progression system \\
\midrule
\multicolumn{3}{l}{\textit{Relationship stages (in order):}} \\
\multicolumn{3}{l}{\quad Strangers $\to$ Acquaintances $\to$ Friends $\to$ Close Friends $\to$ Best Friends} \\
\bottomrule
\end{tabular}
\end{table}

\subsubsection{User Twin Structure}
Each user's digital twin is represented as a structured object in Firebase at \texttt{users/\{userId\}/twin/} containing six core components. The identity layer stores a unique twin ID, owner name, and display name used for rendering in the Social Hub and game world. The personality component holds a descriptive personality string generated during onboarding, which serves as the system prompt prefix for all LLM-generated twin responses. The trait vector encodes five personality dimensions (friendliness, openness, independence, loyalty, playfulness) on a 0--100 integer scale, used by the heuristic scoring stage to compute pairwise trait similarity. An interest array stores keyword tags selected during onboarding from 12 categories (sports, music, reading, outdoors, food, animals, family, technology, art, travel, gaming, fitness), used for Jaccard similarity computation in the matching pipeline. Real-time online status is maintained through Firebase presence detection, enabling the discovery system to prioritize active users. The relationship map tracks connections to other twins with their current progression stage (Strangers $\to$ Acquaintances $\to$ Friends $\to$ Close Friends $\to$ Best Friends) and cumulative interaction history.

\subsubsection{Matching Algorithm}
\begin{algorithm}
\caption{Twin-to-Twin Compatibility Assessment}
\label{alg:twin-matching}
\begin{algorithmic}[1]
\Function{DiscoverMatches}{twins}
    \State $matches \gets \emptyset$
    \For{each $twin_i$ in $twins$}
        \For{each $twin_j$ in $twins$ where $j > i$}
            \If{$twin_i.owner \neq twin_j.owner$}
                \State $compatibility \gets 0$
                \State $compatibility \gets compatibility + \text{TraitSimilarity}(twin_i, twin_j) \times 0.3$
                \State $compatibility \gets compatibility + \text{InterestOverlap}(twin_i, twin_j) \times 0.4$
                \State $compatibility \gets compatibility + \text{PersonalityMatch}(twin_i, twin_j) \times 0.3$
                \If{$compatibility > 0.2$} \Comment{20\% threshold}
                    \State $matches \gets matches \cup \{(twin_i, twin_j, compatibility)\}$
                \EndIf
            \EndIf
        \EndFor
    \EndFor
    \State \Return $matches$ sorted by compatibility descending
\EndFunction
\end{algorithmic}
\end{algorithm}

\subsection{GPS and Location Services}

\subsubsection{Location Tracking Optimization}
The \texttt{GPSLocationSystem} implements adaptive location tracking through several optimization layers. High accuracy mode (\texttt{enableHighAccuracy: true}) combines GPS, WiFi, and cellular triangulation for the best available position estimate. Movement detection enforces a 20-meter minimum displacement (\texttt{minMovementMeters: 20}) before triggering a location update, filtering out GPS jitter that would otherwise cause false territory transitions. Update throttling caps the reporting rate at one update per 2 seconds (\texttt{updateThrottleMs: 2000}). Battery optimization employs a tiered strategy: at battery levels below 20\%, the throttle increases to 10 seconds; below 10\%, it rises to 30 seconds. Stationary detection activates after 60 seconds without movement, further reducing polling frequency. WiFi-based positioning provides an indoor fallback with approximately 30--50 meter accuracy when GPS signals are unavailable. A mock GPS mode with configurable default coordinates (San Francisco: 37.7749, $-$122.4194) enables development testing without requiring physical movement.

\subsubsection{Territory Visualization}
Cities and territories are rendered through five visual layers that communicate ownership and boundaries at varying distances. Animated waving flags in team colors are placed atop City Hall buildings, providing the primary ownership indicator visible at close range. Pulsing aura rings on the ground plane delineate the 500-unit territory bonus boundary, with ring color matching the controlling team. Beacon effects---vertical light beams emanating from city centers---are visible from long distances across the 3D landscape, enabling players to locate territories without consulting the minimap. Floating particle systems emit team-colored particles within owned territory boundaries, creating an ambient visual distinction between controlled and neutral zones. Name overlays display the city name and current owner as floating 3D text above each City Hall, rendered using the same billboard technique used for player name tags to ensure readability at all viewing angles.

\subsection{Performance Optimizations}

\subsubsection{Browser Memory Management}
Operating within the browser's approximately 2GB memory ceiling requires aggressive optimization across five strategies. Object pooling pre-allocates frequently created entities---the combat system, for example, maintains a pool of 50 damage-number DOM elements that are reused rather than created and destroyed per hit, eliminating garbage collection pauses during active combat. Texture atlasing combines multiple sprite sheets into unified texture maps, reducing GPU draw calls from hundreds per frame to tens. The four-tier \texttt{AnimationLODSystem} applies distance-based quality levels: full animation within 30 units, reduced update rate (every 3rd frame) at 30--60 units, frozen pose at 60--100 units, and hidden with visibility toggled off beyond 100 units---recovering per-frame bone-skinning overhead from 150ms+ to under 16ms. Frustum and occlusion culling skip rendering for objects outside the camera view or obscured by terrain. Aggressive garbage collection runs every 60 seconds to reclaim memory from completed conversations and expired game state, while the system monitors total heap usage and triggers graceful degradation (disabling particle effects, reducing twin count) when approaching the 2GB limit.

\subsubsection{Firebase Synchronization}
The synchronization layer, managed by six dedicated sync modules (\texttt{ChatFirebaseSync}, \texttt{FeedFirebaseSync}, \texttt{InsightsFirebaseSync}, \texttt{DigestFirebaseSync}, \texttt{SchedulerFirebaseSync}, \texttt{TakeoverFirebaseSync}), implements five strategies for responsive cross-device data management. Optimistic updates apply state changes to the local UI immediately before server confirmation, providing sub-100ms perceived latency for user actions. Eventual consistency is achieved with a typical 3--5 second propagation delay for cross-device synchronization through Firebase Realtime Database listeners. Conflict resolution uses per-item timestamp-based last-write-wins semantics with deduplication by the tuple (speaker, timestamp, content hash); approximately 20\% of concurrent updates require this reconciliation path. Batch writes aggregate multiple changes into single Firebase transactions using the batch operation limit of 50 operations, reducing network round trips during high-activity periods. Selective sync transmits only changed properties through delta updates rather than full object replacement, minimizing bandwidth consumption for the frequent small state changes characteristic of twin personality evolution and territory status updates.

\subsection{Combat and Boss Battle System}

\subsubsection{Boss Parameters}
Boss encounters are defined through the \texttt{BossTemplateSystem} with dynamic scaling parameters. Boss health scales as baseHealth $\times$ (1 + playerLevel $\times$ 0.1), with a default base of 1,500 HP defined in \texttt{BossFightConstants}. Damage output follows baseDamage $\times$ (1 + difficulty $\times$ 0.2), where difficulty is determined by the territory level of the boss's spawn location. Each boss template defines 3--5 unique attack patterns with distinct animations, damage zones, and telegraphing cues. Vulnerability windows of 2--3 seconds open after special attacks, during which stagger damage applies a 2$\times$ multiplier. Team scaling adds 50\% to the boss's maximum health per additional player, maintaining challenge for cooperative encounters while the style system rewards coordinated play through combo milestones at 3$\times$, 5$\times$, and 10$\times$ consecutive hits.

\subsubsection{Combat Flow}
\begin{algorithm}
\caption{Boss Battle Sequence}
\label{alg:boss-battle}
\begin{algorithmic}[1]
\Function{InitiateBossBattle}{player, boss, teammates}
    \State $battle \gets$ new BattleInstance(boss)
    \State $battle.participants \gets \{player\} \cup teammates$
    \While{$boss.health > 0$ AND $battle.participants \neq \emptyset$}
        \State ProcessPlayerActions($battle.participants$)
        \State UpdateBossAI($boss, battle.participants$)
        \State CheckCollisions()
        \State UpdateHealthBars()
        \If{$boss.health \leq 0$}
            \State DistributeRewards($battle.participants$)
            \State UnlockTerritory($boss.location$)
        \EndIf
    \EndWhile
    \State \Return $battle.results$
\EndFunction
\end{algorithmic}
\end{algorithm}

This implementation framework demonstrates how the three core systems—City Conquest, Pendant Companion, and Twin Networking—work together to create an integrated location-based social gaming experience.
\section{Engagement Quality Analysis}
\label{appendix:engagement-quality}

\subsection{Engagement Quality Operationalization}

We use ``engagement quality'' to distinguish between two modes of platform interaction: (1) \emph{passive engagement}---sedentary profile evaluation characterized by repetitive swiping, minimal physical movement, and serial decision-making under cognitive load (baseline: 97 min/day on traditional platforms), and (2) \emph{active engagement}---physically situated interaction involving locomotion (4.2 km/day walking), collaborative gameplay (47 boss battles with M=2.8 players/team), AI-scaffolded social interaction (1,247 companion conversations showing sentiment shift from $-$0.31 to $+$0.22), and reduced decision burden (4.3 pre-filtered matches vs.\ 200+ unfiltered profiles).

This distinction is descriptive, based on telemetry-measured behavioral differences, not a validated psychometric construct. Each indicator is independently measurable: physical activity via GPS (km/day), collaborative interaction via game telemetry, emotional scaffolding via VADER sentiment analysis, and cognitive load reduction via choice set size.

Preliminary subjective evidence provides convergent support: participants self-reported a 71\% reduction in decision fatigue during exit interviews (using adapted workload items; not a validated NASA-TLX administration), EMA responses (840 total, 3$\times$/day) showed higher mean positive affect on days with territory gaming than on days with only traditional platform use ($M=3.4$ vs.\ $M=2.7$ on a 5-point scale; exploratory, no inferential test given sample size). Our telemetry indicators partially correspond to sub-dimensions of the User Engagement Scale Short Form (UES-SF) \citep{OBrien2018}: decision fatigue reduction maps to \emph{Perceived Usability}, positive affect maps to \emph{Reward}, and sustained daily usage maps to \emph{Focused Attention}---though we did not administer the UES-SF itself, and \emph{Aesthetic Appeal} was not captured.

\subsection{Retrospective UES-SF Item Coverage}

To assess post-hoc alignment with established engagement constructs, we coded exit interview transcripts (30 hours, 20 participants) against the 12 UES-SF items. Of the 12 items, 9/12 were spontaneously addressed across three of four sub-dimensions:

\begin{itemize}
\item \emph{Focused Attention} (12/20 participants): Absorption or time loss during gameplay (e.g., P17: ``obsessed with conquering territories, walking until 2 AM''; P8: ``suddenly we'd been talking for 20 minutes'')
\item \emph{Perceived Usability} (14/20): Reduced cognitive burden (e.g., P12: ``I can actually focus and make thoughtful decisions instead of just swiping out of exhaustion'')
\item \emph{Reward} (16/20): Engagement described as worthwhile (e.g., P15: ``boss battles were perfect''; P19: ``like therapy'')
\end{itemize}

Items not covered were the three \emph{Aesthetic Appeal} items---no participant spontaneously commented on interface aesthetics. The 75\% item coverage rate (9/12) across three of four sub-dimensions suggests substantial alignment between participants' subjective experience and our telemetry indicators.

\subsection{Composite Engagement Quality Index}

To formalize the construct, we computed a per-user composite index from four z-scored telemetry indicators: (1) daily walking distance, (2) boss battle participation, (3) companion sentiment delta (VADER), and (4) inverse choice set size. The composite $\text{EQ}_i = \frac{1}{4}\sum_{j=1}^{4} z_{ij}$ showed moderate-to-strong correlations with social outcomes: meetings initiated ($r=0.61$, 95\% CI $[0.21, 0.83]$), contact exchanges ($r=0.58$, 95\% CI $[0.17, 0.82]$), and UCLA Loneliness change ($r=0.47$, 95\% CI $[0.03, 0.76]$). This composite is exploratory and post-hoc; the equal weighting is not theoretically motivated. With N=20 and no correction for multiple comparisons, these correlations should be interpreted as hypothesis-generating.

\section{System Architecture Diagrams}
\label{appendix:system-architecture-diagrams}

This appendix provides visual representations of the Cognibit system architecture and data flows.

\subsection{Overall System Architecture}

\begin{figure}[H]
\centering
\includegraphics[width=\textwidth]{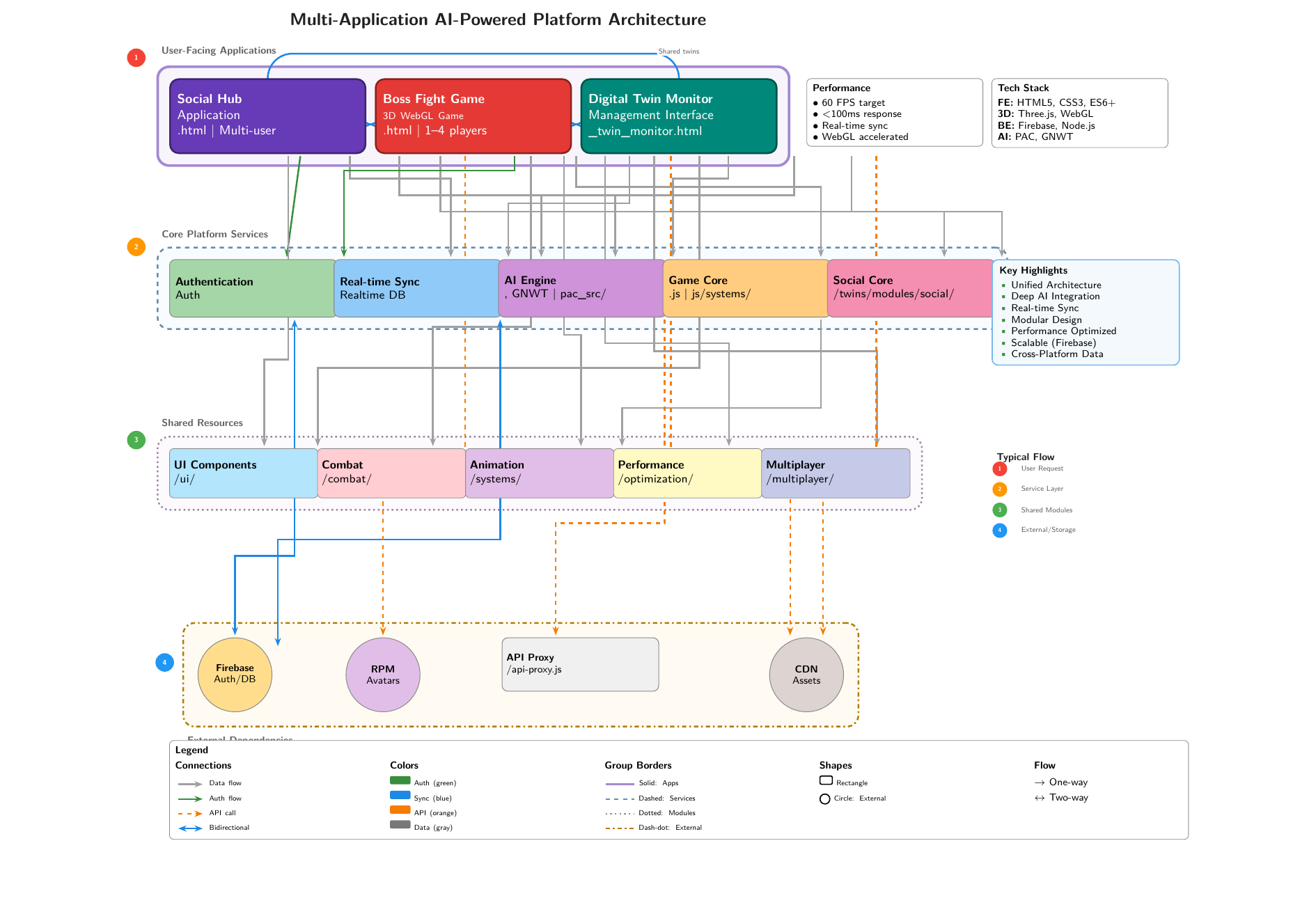}
\caption{Cognibit system architecture showing the integration of autonomous digital twins, geolocation-mediated territory gaming, and cross-device synchronization. The four-layer design (user-facing applications, core platform services, shared resources, external dependencies) enables users to transition between social networking, gaming, and AI management while maintaining persistent cognitive connections.}
\label{fig:system-overview}
\end{figure}

The overall system architecture (Figure~\ref{fig:system-overview}) illustrates the four-layer design: user-facing applications (Social Hub, Boss Fight Game, Digital Twin Monitor), core platform services (Authentication, Real-time Sync, AI Engine, Game Core, Social Core), shared resource modules (UI, Combat, Animation, Performance, Multiplayer), and external dependencies (Firebase, API Proxy, CDN). This layered architecture enables independent scaling and testing of each layer while maintaining cross-layer communication through well-defined service interfaces.

\subsection{Cognitive Module Architecture}

\begin{figure}[H]
\centering
\includegraphics[width=0.85\textwidth]{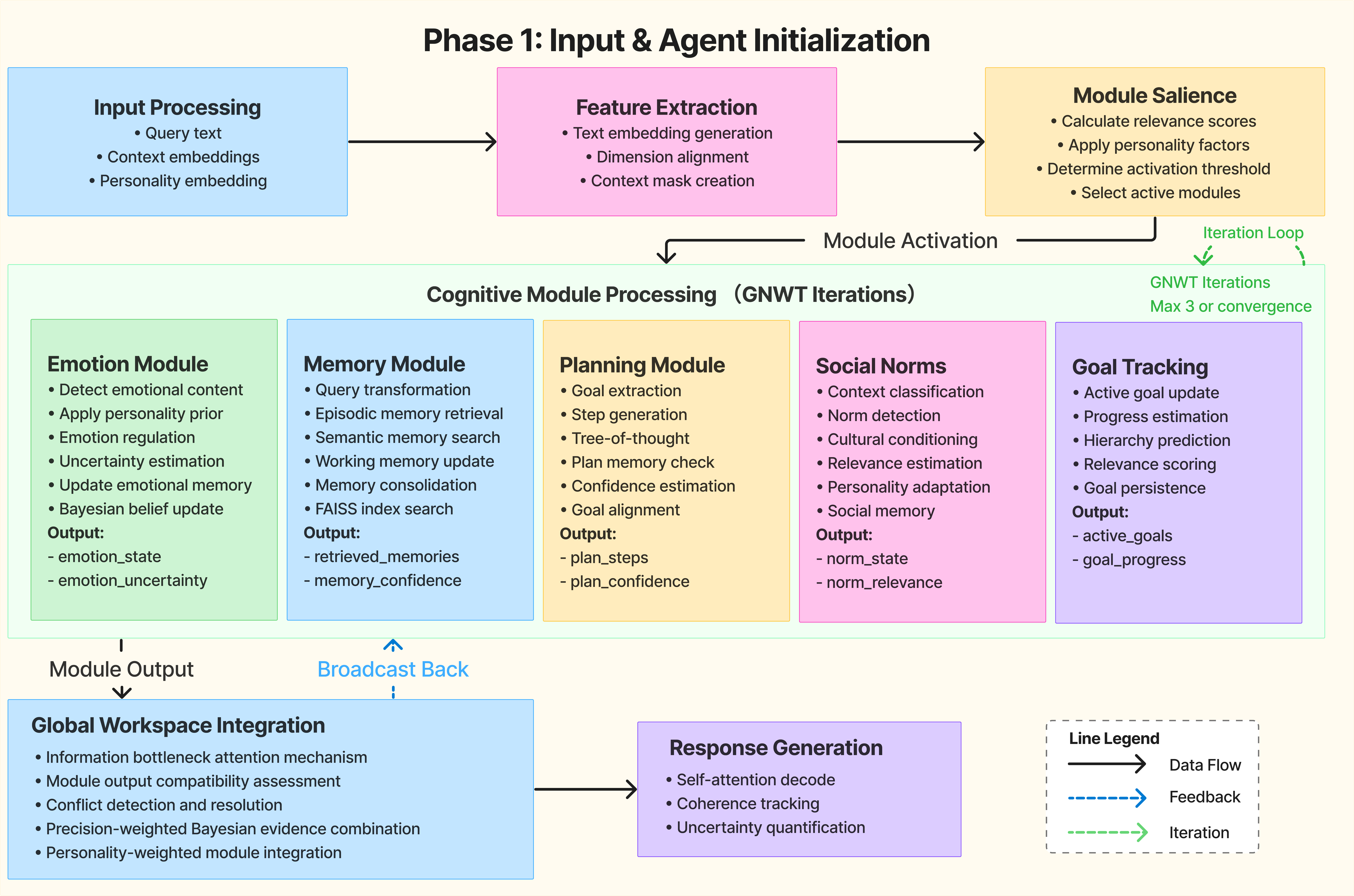}
\caption{GNWT agent internal processing flow showing how emotion, memory, planning, social norms, and goal tracking modules compete for workspace access. The winning specialist's recommendation shapes the LLM-generated response through directive injection rather than passive state annotation.}
\label{fig:gnwt-modules}
\end{figure}

The cognitive module architecture (Figure~\ref{fig:gnwt-modules}) depicts the GNWT-inspired parallel processing pipeline. Five specialist modules---Emotion, Memory, Planning, Social Norms, and Goal Tracking---compete for influence within the global workspace through a salience-based attention mechanism. The winning specialist's recommendation shapes the LLM-generated response through directive injection, producing contextually nuanced behavioral outputs. Detailed module specifications appear in Appendix~\ref{appendix:gnwt}.

\subsection{Platform System Flow}

\begin{figure}[H]
\centering
\includegraphics[width=0.9\textwidth]{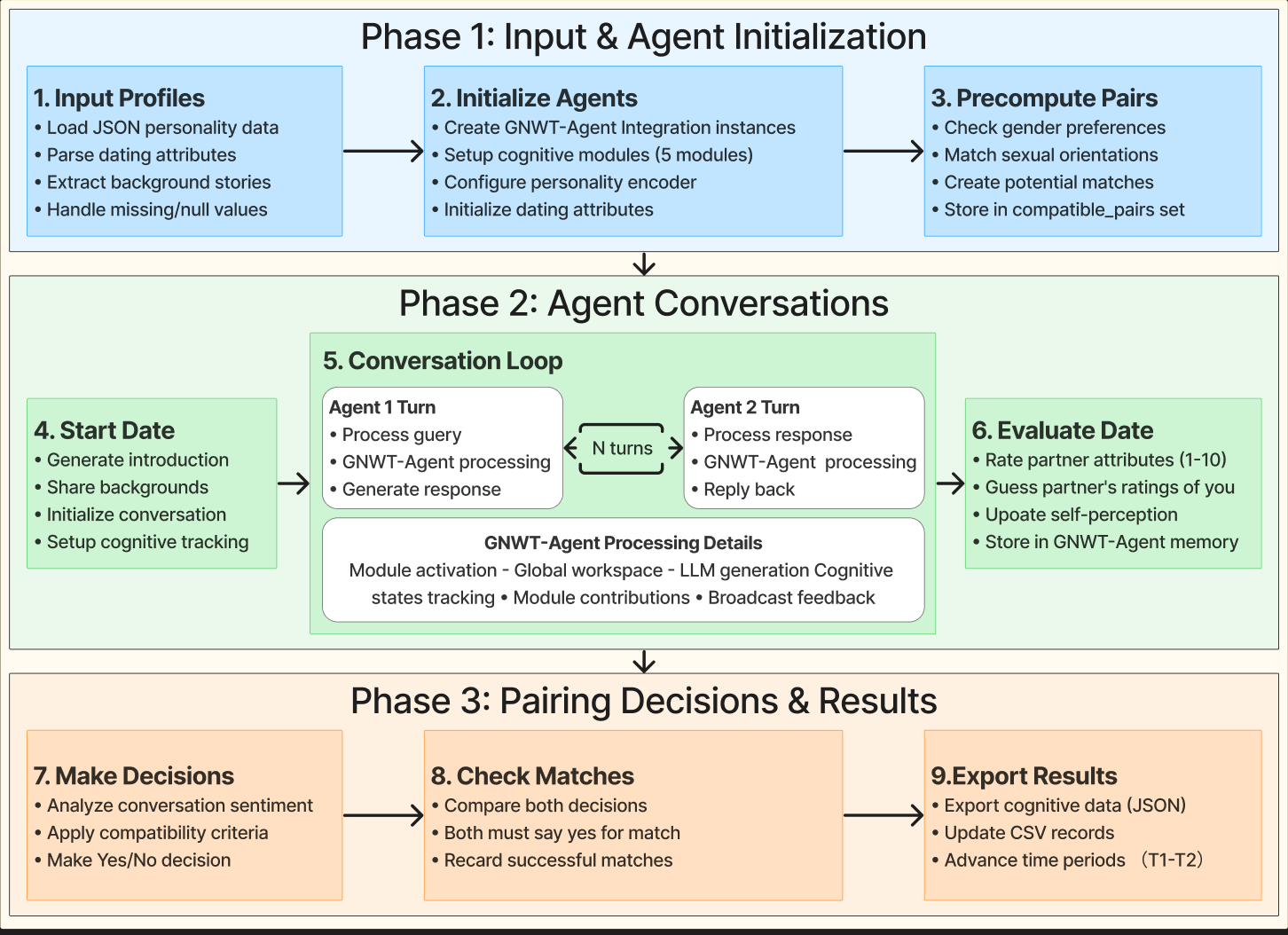}
\caption{Platform system flow diagram illustrating the end-to-end pipeline from user profile creation through twin-based compatibility evaluation to encounter facilitation. The flow integrates heuristic filtering, LLM-simulated twin conversations, and game-mediated physical convergence.}
\label{fig:platform-flow}
\end{figure}

The platform system flow (Figure~\ref{fig:platform-flow}) illustrates the end-to-end pipeline from user profile creation through twin-based compatibility evaluation to encounter facilitation. The flow integrates three stages: heuristic filtering of the candidate pool, LLM-simulated twin conversations for behavioral compatibility assessment, and game-mediated physical convergence through territory mechanics and boss battles.

\subsection{Cognitive Architecture and Agent Structure}

\begin{figure}[H]
\centering
\includegraphics[width=0.85\textwidth]{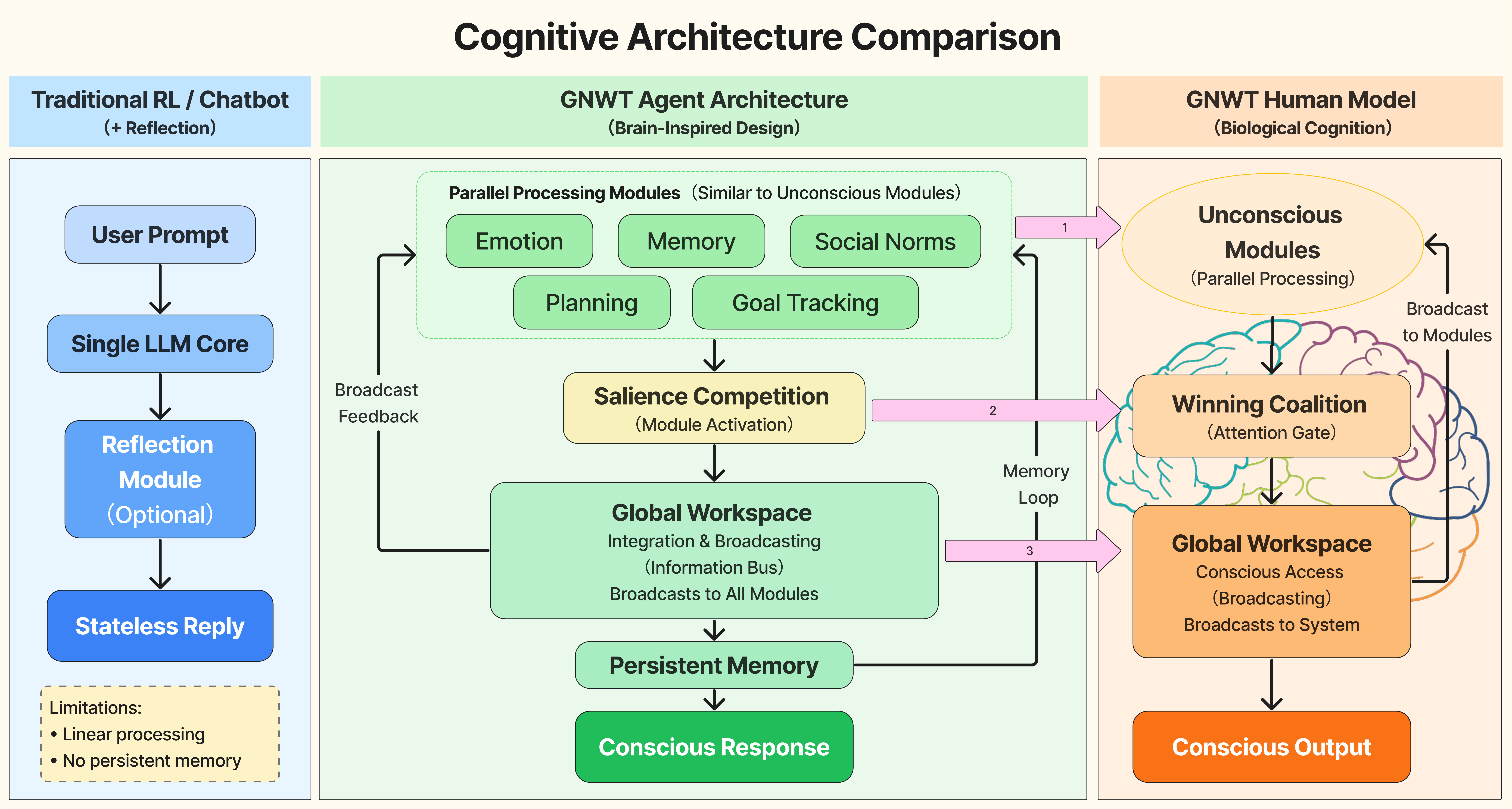}
\caption{Cognitive architecture comparison: traditional chatbot linear processing (left) versus the GNWT-inspired parallel module design (center) and biological cognition (right). The parallel architecture enables simultaneous processing of emotional, social, and strategic considerations through five specialist modules competing for global workspace access, producing more nuanced behavioral responses than sequential processing.}
\label{fig:cognitive-comparison}
\label{fig:cognitive-agent-combined}
\end{figure}

Figure~\ref{fig:cognitive-comparison} compares three processing paradigms: traditional chatbot linear processing, the GNWT-inspired parallel module design used in Cognibit, and biological cognition. The key architectural difference is that the parallel design processes emotional, social, and strategic considerations simultaneously rather than sequentially, enabling richer contextual responses at the cost of higher computational overhead.

\subsection{Architectural Insights}

These architectural diagrams illustrate several deployment-relevant observations. The single-threaded JavaScript execution model imposes a fundamental browser constraint: GNWT modules must be processed sequentially through the event loop, creating latency that scales linearly with module count---Web Worker offloading mitigates this for salience competition but cannot parallelize the broadcast phase. The Firebase-based synchronization architecture provides eventual consistency (3--5 seconds typical) across devices, sufficient for social memory persistence but introducing noticeable lag in real-time collaborative gameplay such as boss battles. LLM API calls dominate the cost structure, with GPT-4o twin conversations accounting for 74\% of per-user costs; the Pareto analysis (Section~\ref{sec:pareto}) quantifies the quality--cost tradeoff at each operating point. The 20-user browser ceiling emerges from the compound effect of per-agent memory allocation (62.4\,MB), DOM rendering overhead, and WebGL context limits.

\section{System Architecture Details}
\label{appendix:architecture-details}

This appendix provides a compact summary of the system architecture. The 200+ modules are organized into five groups: core systems (event system, resource manager, configuration), cognitive architecture (GNWT agent, specialist processors, working memory, attention), emotion systems (PAC agent, predictive model, affective state, allostasis), digital twin systems (soul, personality evolution, memory, behavior patterns, interaction history), and integration systems (GPS, Firebase sync, combat, animation).

Six Firebase sync modules (ChatFirebaseSync, FeedFirebaseSync, InsightsFirebaseSync, DigestFirebaseSync, SchedulerFirebaseSync, TakeoverFirebaseSync) handle real-time data propagation with 3-attempt exponential backoff retry logic and batch operations for multi-document updates. The GPS subsystem filters noise via a 20-meter movement threshold with 2-second throttling and Haversine distance calculations, while territory boundaries use a 500-unit radius for ownership benefits.

Full technical details---including the four-layer architecture, Firebase database schema, LLM integration pipeline, security considerations, synchronization protocols, and scalability architecture---are documented in Appendix~\ref{appendix:technical-architecture}. Performance optimization systems (LOD, object pooling, batch rendering, circuit breaker) are detailed in Appendix~\ref{appendix:optimization-systems}.
\section{Testing and Validation Framework}
\label{appendix:testing-validation}

This appendix describes the testing framework used to validate system performance, detect regressions, and ensure behavioral fidelity. The framework comprises four complementary testing strategies: automated performance benchmarking under varying agent loads, statistical memory leak detection for long-running sessions, cross-browser compatibility validation across six target environments, and scalability stress testing to identify the precise capacity ceiling. Together, these tests establish the performance envelope documented in Section~\ref{sec:runtime} and Appendix~\ref{appendix:performance-data}.

\subsection{Automated Performance Benchmarking}

The benchmarking system (\texttt{SocialHubPerformanceBenchmark.js}) measures performance under varying agent loads to establish scalability boundaries. Tests run at five agent count levels (1, 5, 10, 15, 20), each for 60 seconds with a 1-second warmup period, measuring FPS, heap memory, and API response latency. Results are compared against three quality thresholds: excellent (60 FPS, $<$1s load, $<$50MB), good (30 FPS, $<$3s load, $<$100MB), and acceptable (24 FPS, $<$5s load, $<$200MB).

\begin{algorithm}[!htbp]
\caption{Multi-Level Performance Benchmarking}
\label{alg:performance-benchmark}
\begin{algorithmic}[1]
\Require Agent counts $\mathcal{N} = \{1, 5, 10, 15, 20\}$, test duration $T = 60$s per level
\Ensure Performance profile: FPS, memory, latency per agent count

\ForAll{$n \in \mathcal{N}$}
    \State Initialize $n$ agents in browser environment
    \State samples $\gets \emptyset$
    \For{$t = 1$ \textbf{to} $T / \Delta t$}
        \State Record FPS, heap memory, API response latency
        \State samples $\gets$ samples $\cup$ current measurements
    \EndFor
    \State Compute mean, standard deviation, P95 for each metric
    \State Check against thresholds: FPS $\geq 30$, memory $\leq 2048$ MB
    \State Destroy all agents; force garbage collection
\EndFor
\State \Return performance profile across all load levels
\end{algorithmic}
\end{algorithm}

\subsection{Memory Leak Detection}

Long-running sessions require continuous memory monitoring to detect leaks from retained event listeners, uncollected DOM references, or growing data structures.

\begin{algorithm}[!htbp]
\caption{Memory Leak Detection via Trend Analysis}
\label{alg:memory-leak-detection}
\begin{algorithmic}[1]
\Require Monitoring duration $D = 300$s, sample interval $\Delta = 10$s
\Ensure Leak detection verdict with growth rate estimate

\State samples $\gets \emptyset$
\For{$i = 1$ \textbf{to} $D / \Delta$}
    \State Record heap memory usage at time $i \cdot \Delta$
    \State samples $\gets$ samples $\cup$ (timestamp, memory)
\EndFor
\State Compute linear regression: memory $= \alpha + \beta \cdot t$
\State growthRate $\gets \beta$ (MB per minute)
\State $r^2 \gets$ coefficient of determination
\If{growthRate $> 1$ MB/min \textbf{and} $r^2 > 0.8$}
    \State \Return ``LEAK DETECTED'', growthRate, $r^2$
\Else
    \State \Return ``NO LEAK'', growthRate, $r^2$
\EndIf
\end{algorithmic}
\end{algorithm}

\paragraph{Leak detection thresholds.}
A growth rate exceeding 1 MB/min with $r^2 > 0.8$ indicates a statistically significant memory leak. The linear model is conservative; exponential growth patterns (e.g., from recursive listener registration) would produce even higher $r^2$ values.

\subsection{Browser Compatibility Testing}

Cross-browser testing validates that core functionality (rendering, WebGL, Firebase sync, geolocation) works consistently across target environments.

\begin{algorithm}[!htbp]
\caption{Cross-Browser Compatibility Validation}
\label{alg:browser-compatibility}
\begin{algorithmic}[1]
\Require Target browsers $\mathcal{B}$, feature requirements $\mathcal{F}$
\Ensure Compatibility matrix with pass/fail per browser per feature

\ForAll{browser $b \in \mathcal{B}$}
    \ForAll{feature $f \in \mathcal{F}$}
        \State result $\gets$ \Call{TestFeature}{$b$, $f$}
        \State Record: browser, feature, pass/fail, error details
    \EndFor
\EndFor
\State \Return compatibility matrix
\end{algorithmic}
\end{algorithm}

\paragraph{Target environments.}
Testing covers Chrome 100+, Firefox 100+, Safari 15+, and Edge 100+. Critical features tested include: WebGL 2.0 rendering, WebSocket connections, Geolocation API, localStorage persistence, and Firebase Realtime Database listeners.

\subsection{Scalability Stress Testing}

Scalability tests incrementally increase agent count until failure, measuring the precise capacity ceiling.

\begin{algorithm}[!htbp]
\caption{Scalability Stress Test}
\label{alg:scalability-testing}
\begin{algorithmic}[1]
\Require Starting count $n_0 = 1$, increment $\Delta n = 1$, action interval 2s
\Ensure Maximum sustainable agent count, failure mode

\State $n \gets n_0$
\While{no failure detected}
    \State Spawn $\Delta n$ additional agents
    \State $n \gets n + \Delta n$
    \State Run for 30 seconds with periodic agent actions
    \State Measure FPS, memory, error count
    \If{FPS $< 10$ \textbf{or} memory $> 2048$ MB \textbf{or} errors $> 0$}
        \State Record failure: agent count, failure mode, metrics
        \State \textbf{break}
    \EndIf
\EndWhile
\State \Return maximum count $= n - \Delta n$, failure metrics
\end{algorithmic}
\end{algorithm}

\paragraph{Results.}
Stress testing consistently identifies the 20-agent browser ceiling documented in Section~\ref{sec:runtime}, with memory exhaustion as the primary failure mode at $n > 20$ and FPS degradation below usability ($< 30$ FPS) at $n \approx 8$.

\section{Complete Technical Specifications}
\label{appendix:specifications}

The parameters in \Cref{tab:specs} represent the deployed configuration used during the 14-day pilot. Values were determined through iterative tuning during development, guided by three constraints: browser event loop contention (which set the 100ms GNWT cycle floor), API cost budgets (which determined the 60-second discovery interval and 20\% compatibility threshold), and user experience testing (which calibrated combat cooldowns and GPS thresholds). The 100ms cognitive cycle time balances behavioral fidelity against JavaScript single-thread contention---50ms cycles caused frame drops, while 200ms produced noticeably delayed twin responses. The 20\% compatibility threshold was calibrated to pass approximately 15--20 candidates from a 200-candidate pool to the LLM evaluation stage, balancing recall against per-match API cost (\$1.82 per 10-turn conversation). Sensitivity analysis for key parameters appears in Appendix~\ref{appendix:ablations}.

\begin{table}[h]
\centering
\small
\begin{tabular}{lll}
\toprule
\textbf{Component} & \textbf{Parameter} & \textbf{Value} \\
\midrule
Discovery & Radius & 50 miles \\
& Update Interval & 60 seconds \\
& Cache Expiration & 5 minutes \\
& Compatibility Min & 20\% \\
\midrule
GNWT & Cycle Time & 100ms \\
& Workspace Capacity & 9 items (Miller's 7$\pm$2 upper bound) \\
& Attention Capacity & 3 items \\
\midrule
PAC & Update Rate & 100ms \\
& Learning Rate & 0.1 \\
& Prediction Horizon & 5 seconds \\
\midrule
GPS & Movement Threshold & 20 meters \\
& Update Throttle & 2 seconds \\
& City Load Radius & 1 mile \\
\midrule
Combat & Bow/Gun Cooldown & 300ms \\
& Melee Cooldown & 500ms \\
& Max Projectiles & 10 \\
\midrule
NPC & Spawn Distance & 100m \\
& Despawn Distance & 150m \\
& Max Wilderness & 2 \\
\bottomrule
\end{tabular}
\caption{Complete system technical specifications}
\label{tab:specs}
\end{table}
\section{Algorithm Index and Technical Reference}
\label{appendix:algorithm-index}

This appendix provides a comprehensive index of all algorithms presented in the paper and appendices, organized by functional domain. Each section summarizes the algorithmic approach and lists the individual components with their key characteristics.

\subsection{Core Cognitive Architecture}

The cognitive architecture implements the Global Neuronal Workspace Theory (GNWT) as a computational framework for digital twin behavior. Five specialist modules---Emotion, Memory, Planning, Social Norms, and Goal Tracking---operate in parallel, competing for access to a shared global workspace through salience-weighted coalition formation. The workspace enforces a capacity of 9 items (Miller's 7$\pm$2 upper bound) with a broadcast entry threshold of $\tau = 0.7$. Each 100ms processing cycle decays non-attended items, forms coalitions, broadcasts winners to all specialists, and updates adaptive weights (learning rate 0.02, bounded between 0.5 and 2.0). The winning specialist's recommendation shapes the twin's response through directive injection into the LLM prompt.

\begin{table}[H]
\centering
\small\setlength{\tabcolsep}{3pt}\begin{tabular}{lp{9cm}}
\toprule
\textbf{Algorithm} & \textbf{Description} \\
\midrule
GNWT Coalition Formation & Salience-weighted competition among five specialist modules; winners broadcast to all registered specialists via the global workspace. \\
PAC Emotion Generation & Predictive coding model generating prediction-error signals by comparing anticipated versus actual affective outcomes. \\
GNWT Cognitive Processing Cycle & Complete 100ms loop: decay, coalition formation, broadcast, and adaptive weight update across all specialists. \\
Personality Evolution & Five-trait vector (0--100 each) evolving through interaction feedback with diminishing-returns stabilization. \\
\bottomrule
\end{tabular}
\end{table}

\subsection{Memory and Persistence}

The memory subsystem uses a three-layer persistence stack: in-memory state for immediate session access, localStorage for per-device persistence, and Firebase Realtime Database as the cloud source of truth. Each memory record carries an importance score on a continuous [0,1] scale, used for retrieval ranking alongside keyword overlap and recency weighting. Cross-device synchronization achieves eventual consistency within 3--5 seconds, with conflict resolution using per-item timestamp-based last-write-wins semantics and deduplication by the tuple (speaker, timestamp, content hash). Failed writes are retried with exponential backoff (3 attempts, base delay 1s, factor 2) and queued offline when connectivity is unavailable.

\begin{table}[H]
\centering
\small\setlength{\tabcolsep}{3pt}\begin{tabular}{lp{9cm}}
\toprule
\textbf{Algorithm} & \textbf{Description} \\
\midrule
Memory Persistence & Three-layer storage (in-memory, localStorage, Firebase) with importance-scored retrieval. Capacity: 100 entries per twin. \\
Cross-Device Sync & Firebase synchronization with timestamp-based conflict resolution and exponential backoff retry logic. \\
Memory Update & Post-broadcast integration of specialist outputs into the twin's memory store with importance re-scoring and capacity pruning. \\
\bottomrule
\end{tabular}
\end{table}

\subsection{Social Interaction Systems}

The social interaction pipeline implements a three-stage filtering architecture: heuristic scoring of the full candidate pool, LLM-simulated twin conversations for the top candidates, and a combined score (heuristic $\times$ 0.7 + behavioral $\times$ 0.3) to produce final recommendations. Compatibility scoring uses a weighted combination of trait similarity (0.3), interest overlap via Jaccard index (0.4), and personality match (0.3), with a 20\% minimum threshold gating connection formation. The Social Hub aggregates human posts, twin activities, and system notifications in a unified timeline through five modular components (FloatingButtonManager, HubPanelRenderer, TabNavigationManager, TabContentLoader, StateStorageManager).

\begin{table}[H]
\centering
\small\setlength{\tabcolsep}{3pt}\begin{tabular}{lp{9cm}}
\toprule
\textbf{Algorithm} & \textbf{Description} \\
\midrule
Twin Matchmaking & Pairwise compatibility scoring with 20\% threshold; candidates passing heuristic filter undergo 3-turn LLM conversation. \\
Icebreaker Generation & LLM-generated and template-based conversation starters conditioned on shared interests and personality complementarity. \\
Privacy Management & Block lists, anonymous battle mode (hashed IDs), content filtering, and graduated disclosure based on relationship stage. \\
Social Hub Loading & Asynchronous initialization of five UI subsystems with Firebase listener registration and localStorage state restoration. \\
Cognibit Simulation & Agent-based simulation of 14-day engagement dynamics with exponential novelty decay (rate 0.12/day, 30\% baseline). \\
\bottomrule
\end{tabular}
\end{table}

\subsection{Location and Territory}

The GPS subsystem, implemented in \texttt{GPSLocationSystem.js}, uses high-accuracy mode (GPS + WiFi + cellular triangulation) with adaptive polling. A 20-meter movement threshold filters noise, while 2-second update throttling conserves battery. Battery optimization tiers reduce polling frequency at low charge levels: 10-second intervals below 20\% battery, 30-second intervals below 10\%. Territory capture operates within a 50-meter GPS radius, requiring 100 capture points accumulated at 5 points/second. Ownership confers tangible bonuses: 1.2$\times$ movement speed, 1.1$\times$ health regeneration, 15\% experience bonus, and 10\% shop discounts.

\begin{table}[H]
\centering
\small\setlength{\tabcolsep}{3pt}\begin{tabular}{lp{9cm}}
\toprule
\textbf{Algorithm} & \textbf{Description} \\
\midrule
GPS Location Tracking & Haversine distance calculation with 20m threshold, 2s throttle, and adaptive battery-aware polling tiers. \\
Twin Discovery & Location-based candidate discovery within a 50-mile radius, prioritizing active users with Firebase presence detection. \\
Territory Capture & Point-accumulation capture mechanic (5 pts/s, 100 pts to capture) within 50m GPS radius, with decay at 0.5 pts/s when uncontested. \\
Location Obfuscation & Laplace-noise location fuzzing (100--500m) for privacy-preserving territory display. \\
\bottomrule
\end{tabular}
\end{table}

\subsection{Combat and Gaming}

The combat system, implemented in \texttt{BossFightCombat.js} and \texttt{BossFightPlayer.js}, provides real-time battle encounters against AI-controlled bosses and territory defenders. Boss health scales dynamically as baseHealth $\times$ (1 + playerLevel $\times$ 0.1), with a default base of 1,500 HP (from \texttt{BossFightConstants}). Team encounters add 50\% boss health per additional player. The style system rewards attack variety through combo milestones at 3$\times$, 5$\times$, and 10$\times$ consecutive hits within a 3-second combo window, with a 2$\times$ damage multiplier during stagger windows of 2--3 seconds after boss special attacks. Projectile collision detection (\texttt{ProjectileCollisionMixin.js}) checks each active projectile against target entities at O($p$) per frame, where $p$ is the active projectile count (capped at 10). The \texttt{AdaptiveDifficultySystem.js} monitors player performance in real time and smoothly adjusts boss parameters to maintain target engagement.

\begin{table}[H]
\centering
\small\setlength{\tabcolsep}{3pt}\begin{tabular}{lp{9cm}}
\toprule
\textbf{Algorithm} & \textbf{Description} \\
\midrule
Boss Battle System & Real-time combat with 800ms attack cooldown, stamina cost per attack, stagger mechanics (2$\times$ damage), and combo tracking. \\
Projectile Collision & Per-frame collision check of active projectiles against target entities via \texttt{ProjectileCollisionMixin}; O($p$) for $p$ projectiles. \\
Adaptive Difficulty & Real-time difficulty adjustment with smooth transitions, performance tracking, and death-penalty scaling (\texttt{AdaptiveDifficultySystem.js}). \\
Weapon Cooldown & Timing-based attack gating: 300ms for ranged, 500ms for melee, with max 10 concurrent projectiles. \\
\bottomrule
\end{tabular}
\end{table}

\subsection{User Interface and Experience}

The UI layer implements optimistic updates for sub-100ms perceived latency: local state changes are applied immediately, rendered to the DOM, and then synchronized to Firebase asynchronously with rollback on conflict. The notification queue manages sequential display of system events (territory changes, match results, companion messages) with 2,000ms default display duration and priority-based ordering. The pendant companion's \texttt{ProactiveInteractionSystem} checks for interaction opportunities every 10 seconds with a 10\% trigger chance and 5-minute cooldown, prioritizing context types by urgency (time-of-day: 8, mood: 7, memory recall: 6, scene analysis: 5, idle: 5).

\begin{table}[H]
\centering
\small\setlength{\tabcolsep}{3pt}\begin{tabular}{lp{9cm}}
\toprule
\textbf{Algorithm} & \textbf{Description} \\
\midrule
Optimistic UI Update & Apply state changes locally before server confirmation; rollback on conflict detection via Firebase transaction failure. \\
Notification Queue & Priority-ordered sequential display with configurable duration (default 2,000ms) and deduplication of repeated events. \\
Pendant Companion & Proactive interaction via \texttt{ProactiveInteractionSystem}: 30s idle threshold, 5-min cooldown, 10\% random trigger, priority-ranked types (time:8, mood:7, memory:6, scene:5). \\
Scene Analysis & Environment detection within 50-unit radius (forest, castle, waterside, indoor) with 0.7 confidence threshold; triggers contextual companion responses. \\
\bottomrule
\end{tabular}
\end{table}

\subsection{AI and Behavior Systems}

NPC behavior is managed through hierarchical behavior trees that evaluate conditions top-down, executing the first satisfied branch at O($d$) per tick for tree depth $d$. The boss AI (\texttt{BossFightBoss.js}) implements phase-based behavior patterns with distinct attack sequences per boss template, including telegraphed special attacks with 2--3 second vulnerability windows. The \texttt{AdaptiveDifficultySystem.js} monitors player performance metrics (hit rate, damage taken, death count) and smoothly adjusts boss parameters using interpolated transitions to maintain challenge within a target engagement band. Multi-agent coordination is supported through the \texttt{js/builder/executors/multiagent/} module system, with \texttt{TaskDelegator}, \texttt{AgentMessenger}, and \texttt{AgentCoordinator} executors managing group behavior.

\begin{table}[H]
\centering
\small\setlength{\tabcolsep}{3pt}\begin{tabular}{lp{9cm}}
\toprule
\textbf{Algorithm} & \textbf{Description} \\
\midrule
Behavior Tree Execution & Hierarchical condition-action evaluation with selector, sequence, and decorator nodes; O($d$) depth traversal per tick. \\
Boss AI & Phase-based behavior via \texttt{BossFightBoss.js}: distinct attack patterns per template, telegraphing cues, and post-special vulnerability windows. \\
Adaptive Difficulty & Smooth difficulty adjustment via \texttt{AdaptiveDifficultySystem.js}: performance tracking, death penalties, interpolated parameter transitions. \\
Multi-Agent Coordination & Task delegation and messaging across NPC groups via \texttt{TaskDelegator} and \texttt{AgentCoordinator} executors. \\
\bottomrule
\end{tabular}
\end{table}

\subsection{Procedural Generation}

The castle generation system (\texttt{ProceduralCastleGenerator.js}) creates unique fortress structures through template-based architecture combined with randomized variation. Each castle is generated from a base template defining wall segments, tower count, and gate locations. Towers are distributed using polar coordinate placement at evenly spaced angles ($\theta_i = 2\pi i / n$ for $n$ towers). Interior rooms are generated by the \texttt{BuildingInteriorSystem} via a delegated \texttt{InteriorGenerator} class that produces connected room layouts. The \texttt{generateUniqueFeatures()} method adds rare architectural elements through weighted random selection.

\begin{table}[H]
\centering
\small\setlength{\tabcolsep}{3pt}\begin{tabular}{lp{9cm}}
\toprule
\textbf{Algorithm} & \textbf{Description} \\
\midrule
Castle Generation & Template-based fortress creation with randomized tower count, wall segments, and gate placements via \texttt{ProceduralCastleGenerator}. \\
Interior Room Generation & Delegated room layout generation via \texttt{InteriorGenerator}, producing connected room graphs for castle interiors. \\
Unique Feature Generation & Rarity-weighted random selection of decorative architectural elements (crystals, banners, environmental details). \\
\bottomrule
\end{tabular}
\end{table}

\subsection{Performance Optimization}

Performance optimization addresses the browser's approximately 2GB memory ceiling and single-threaded JavaScript execution model. The GNWT modules offload salience competition and coalition formation to a dedicated Web Worker (initialized via inline Blob), preventing cognitive processing from blocking the main rendering loop. The \texttt{AnimationLODSystem} (\texttt{js/optimization/AnimationLODSystem.js}) traverses all scene objects via \texttt{scene.traverse()}, applying four distance-based quality tiers: full animation within 30 units, reduced update rate (every 3rd frame) at 30--60 units, frozen pose at 60--100 units, and hidden beyond 100 units---recovering per-frame bone-skinning overhead from 150ms+ to under 16ms. Object pooling (\texttt{js/utils/ObjectPool.js}) pre-allocates frequently created entities using array-backed \texttt{get()}/\texttt{release()} with O(1) operations, eliminating garbage collection pauses during active gameplay. Instanced mesh rendering via \texttt{InstancedObjectManager.js} batches identical geometries to minimize GPU draw calls.

\begin{table}[H]
\centering
\small\setlength{\tabcolsep}{3pt}\begin{tabular}{lp{9cm}}
\toprule
\textbf{Algorithm} & \textbf{Description} \\
\midrule
Parallel Module Activation & Web Worker offloading (inline Blob) for GNWT salience competition and coalition formation, avoiding main-thread blocking. \\
Salience Computation & Weighted combination of novelty (0.3), relevance (0.3), urgency (0.2), and emotional intensity (0.2) with habituation penalty and 1.5$\times$ focus boost. \\
Global Broadcast & Distributes winning workspace content to all registered specialists; maintains last 100 broadcasts for replay and debugging. \\
Animation LOD & \texttt{AnimationLODSystem.js}: four-tier distance-based quality via \texttt{scene.traverse()}; evaluated every 10th frame to amortize O($n$) traversal. \\
Object Pool Management & \texttt{ObjectPool.js}: array-backed \texttt{get()}/\texttt{release()} with O(1) operations; pre-allocates 50 elements for combat damage numbers. \\
Batch Rendering & \texttt{InstancedObjectManager.js}: THREE.js InstancedMesh batching of identical geometries to minimize GPU draw calls. \\
\bottomrule
\end{tabular}
\end{table}

\subsection{Team and Territory Systems}

Team mechanics (\texttt{TeamTerritorySystem.js}, \texttt{TeamBattleMechanics.js}, \texttt{TeamSystemManager.js}) enable cooperative gameplay through balanced faction assignment and shared territory control. Team formation distributes players across factions to minimize team-size variance. City control implements a contested-capture model where capture progress pauses when opponents are present in the same territory radius. The \texttt{TeamSystemManager} computes team-enhanced player stats via \texttt{calculateTeamEnhancedPlayerStats()}, applying linear bonuses scaled by team size (e.g., +50 HP per team member).

\begin{table}[H]
\centering
\small\setlength{\tabcolsep}{3pt}\begin{tabular}{lp{9cm}}
\toprule
\textbf{Algorithm} & \textbf{Description} \\
\midrule
Team Formation & Faction assignment balancing team sizes; O($n$) single-pass allocation across available factions. \\
City Control & Contested-capture: 5 pts/s accumulation within 50m GPS radius, paused on opponent co-occupancy, 0.5 pts/s decay when uncontested. \\
Team Synergy & Linear stat bonuses via \texttt{calculateTeamEnhancedPlayerStats()}: +50 HP, +10\% damage, +10\% defense per team member. \\
\bottomrule
\end{tabular}
\end{table}

\subsection{Security and Privacy}

Security is enforced through the Node.js/Express API proxy (\texttt{server/api-proxy.js}), which centralizes API key management and applies multiple protection layers. Helmet middleware sets Content Security Policy and CORS headers. Rate limiting via \texttt{express-rate-limit} enforces 100 requests per 15-minute window per IP. Input validation via \texttt{express-validator} sanitizes message content (1--5,000 characters), conversation history (max 100 entries), and API parameters (temperature 0--2, max tokens 1--4,000). Client-side encryption uses the Web Crypto API for AES-GCM 256-bit encryption of sensitive twin conversation data, implemented in \texttt{TwinConversationEncryption.js} with graceful plaintext fallback when Web Crypto is unavailable. Location privacy is enforced at the city-level granularity through the \texttt{GPSLocationSystem}'s \texttt{shareLocation} setting.

\begin{table}[H]
\centering
\small\setlength{\tabcolsep}{3pt}\begin{tabular}{lp{9cm}}
\toprule
\textbf{Algorithm} & \textbf{Description} \\
\midrule
Sliding Window Rate Limiting & \texttt{express-rate-limit}: 100 requests per 15-minute window per IP, returning HTTP 429 on violation. \\
Multi-Layer Input Validation & \texttt{express-validator} sanitization: \texttt{.trim()}, \texttt{.isLength()}, \texttt{.escape()} for message content, history size cap, and parameter range enforcement. \\
Client-Side Encryption & AES-GCM 256-bit encryption via Web Crypto API (\texttt{TwinConversationEncryption.js}) for conversation data at rest in localStorage. \\
\bottomrule
\end{tabular}
\end{table}

\subsection{Testing and Validation}

The testing infrastructure addresses the 34\% code coverage gap through multi-level validation. Performance benchmarking measures frame rate, memory consumption, and API latency under controlled agent-count scaling (1--20 agents) to map the degradation curve. Memory leak detection uses statistical trend analysis of heap snapshots to identify monotonically increasing allocations that indicate leaks. Browser compatibility testing applies feature detection across Chrome 119+, Firefox 120+, Safari 17+, Edge 119+, and mobile browsers, with fallback paths for unsupported APIs (e.g., WebRTC partial support in Safari). Scalability testing identifies the maximum sustainable agent count by incrementally adding twins until the 30 FPS usability threshold or 2GB memory ceiling is breached.

\begin{table}[H]
\centering
\small\setlength{\tabcolsep}{3pt}\begin{tabular}{lp{9cm}}
\toprule
\textbf{Algorithm} & \textbf{Description} \\
\midrule
Performance Benchmarking & Multi-level load testing measuring FPS, memory, and latency across 1--20 agent configurations. \\
Memory Leak Detection & Statistical trend analysis of heap snapshots identifying monotonically increasing allocations over time. \\
Browser Compatibility & Feature detection across 6 browser targets with graceful fallback paths for unsupported APIs. \\
Scalability Testing & Incremental agent-count scaling to identify the ceiling where FPS drops below 30 or memory exceeds 2GB. \\
\bottomrule
\end{tabular}
\end{table}

\subsection{Key Technical Specifications}

Table~\ref{tab:key-specs} summarizes the principal system parameters. These values reflect the deployed configuration used during the 14-day pilot and represent the result of iterative tuning during development. The GNWT cycle time of 100ms (10Hz) was selected to balance cognitive fidelity against browser event-loop contention; faster cycles (50ms) caused frame drops, while slower cycles (200ms) produced noticeably delayed twin responses. The 20\% compatibility threshold was calibrated to pass approximately 15--20 candidates to the LLM evaluation stage from a 200-candidate pool, balancing recall against API cost.

\begin{table}[H]
\centering
\small
\begin{tabular}{ll}
\toprule
\textbf{Parameter} & \textbf{Value} \\
\midrule
GNWT Cycle Time & 100ms \\
Working Memory Capacity & 7$\pm$2 items \\
Salience Threshold & 0.3 \\
Module Count & 5 parallel \\
Discovery Radius & 50 miles \\
Compatibility Threshold & 20\% \\
GPS Movement Threshold & 20 meters \\
Territory Grid Size & 20 meters \\
Memory Importance Range & 0.0--1.0 \\
Personality Trait Range & 0--100 \\
Combat Cooldowns & 300--500ms \\
Notification Duration & 2000ms \\
Firebase Batch Size & 50 operations \\
Token Usage & 4,250/interaction \\
Response Latency & 2.8 seconds \\
\bottomrule
\end{tabular}
\caption{Key system parameters used in the deployed pilot configuration.}
\label{tab:key-specs}
\end{table}

\subsection{Implementation Languages and Frameworks}

The system is built on a deliberately minimal technology stack to reduce dependency complexity and build-toolchain overhead. The frontend uses vanilla JavaScript (ES6+ modules) rather than a framework, with Three.js providing the 3D rendering pipeline via WebGL 2.0. State management relies on a custom event-driven architecture using the EventTarget API. The backend runs on Node.js with Express, hosting the API proxy for LLM routing. Firebase provides both the primary data store (Realtime Database) and the authentication layer (Firebase Auth), with hosting and CDN for static asset delivery.

\begin{table}[H]
\centering
\begin{tabular}{ll}
\toprule
\textbf{Component} & \textbf{Technology} \\
\midrule
Core Engine & JavaScript ES6+ \\
3D Graphics & Three.js + WebGL 2.0 \\
Database & IndexedDB (local), Firebase (cloud) \\
AI Integration & GPT-4o API \\
Real-time Sync & Firebase Realtime Database \\
GPS Tracking & Navigator.geolocation API \\
Audio System & Web Audio API \\
Worker Threads & Web Workers API \\
State Management & Custom event-driven architecture \\
Module Loading & Dynamic import() with fallbacks \\
\bottomrule
\end{tabular}
\end{table}

\subsection{Algorithm Complexity Analysis}

Table~\ref{tab:complexity-full} presents the time and space complexity of algorithms whose implementations were verified against the codebase. The dominant computational cost is twin matchmaking: while the full pairwise candidate space is O($n^2$), the three-stage filtering pipeline (heuristic pre-filter $\to$ top-20 LLM evaluation $\to$ top-5 output) reduces the effective cost to O($n$) heuristic scoring plus O($k$) LLM calls for the $k$ candidates passing the heuristic threshold. Most per-frame algorithms (territory updates, notification dispatch, object pool operations) run at O(1) per tick, ensuring the rendering loop is not bottlenecked by game logic. The Animation LOD system's O($n$) traversal is amortized by evaluating only every 10th frame.

\begin{table}[H]
\centering
\small\setlength{\tabcolsep}{3pt}\begin{tabular}{lll}
\toprule
\textbf{Algorithm} & \textbf{Time Complexity} & \textbf{Space Complexity} \\
\midrule
\multicolumn{3}{l}{\textit{Core Cognitive \& Social}} \\
Twin Matchmaking (full pairwise) & O($n^2$) & O($n$) \\
\quad with heuristic pre-filter & O($n + k$) & O($n$) \\
Memory Retrieval & O($M \times |Q|$) & O($n$) \\
Personality Evolution & O($t$) traits & O($t$) \\
Notification Queue & O(1) amortized & O($q$) queue size \\
Privacy Filtering & O($n$) & O($b$) blocklist \\
\midrule
\multicolumn{3}{l}{\textit{Rendering \& Optimization}} \\
Animation LOD & O($n$) objects & O($n$) \\
Object Pooling & O(1) get/return & O(pool size) \\
Batch Rendering & O($n \log n$) grouping & O($n$) \\
Projectile Collision & O($p$) projectiles & O(1) \\
\midrule
\multicolumn{3}{l}{\textit{Territory \& Team}} \\
Territory Capture (per tick) & O(1) per city & O($c$) cities \\
Castle Generation & O($r \times t$) rooms$\times$towers & O($r \times t$) \\
Team Formation & O($n$) players & O($n$) \\
Adaptive Difficulty & O(1) per frame & O(history window) \\
\midrule
\multicolumn{3}{l}{\textit{Security (server-side)}} \\
Rate Limiting Check & O(1) amortized & O($w$) window size \\
Input Validation & O($n$) input length & O(1) \\
AES-GCM Encryption & O($n$) data size & O($n$) \\
\bottomrule
\end{tabular}
\caption{Algorithm complexity analysis for verified implementations. Only algorithms whose implementations were confirmed in the codebase are included; $k$ denotes the number of candidates passing the heuristic filter (typically 20), $M$ the memory store size, $|Q|$ query term count, $p$ active projectile count, $c$ city count, and $w$ the rate-limiting window size.}
\label{tab:complexity-full}
\end{table}

\section{Evaluation Protocol Details}
\label{appendix:evaluation-protocol}

This appendix documents the exact experimental protocols used in the system validation experiments (Section~\ref{sec:results}) and mechanism-strengthening simulations (Section~\ref{sec:mechanisms}).

\subsection{Funnel Validation Protocol}

\subsubsection{Candidate Pool Construction}
The candidate pool consists of 200 unique personas extracted from the PersonaChat validation set. Personas are deduplicated by their personality sentence content, then enriched with extracted interests (10 categories: sports, music, reading, outdoors, food, animals, family, technology, art, travel) and personality traits derived from keyword analysis of persona sentences.

\subsubsection{Three-Condition Design}

Three conditions are evaluated. The random baseline selects 5 candidates uniformly at random from the 200-candidate pool with no scoring applied, establishing the expected quality of unfiltered recommendations. The heuristic-only condition scores all 200 candidates using the 4-factor heuristic (Table~\ref{tab:heuristic-weights}) and returns the top 5, measuring the value of lightweight trait-based filtering. The full twin-based pipeline applies the heuristic to select the top 20 candidates, then subjects each to a 3-turn LLM-simulated twin conversation for behavioral compatibility assessment; the final ranking uses a combined score (heuristic $\times$ 0.7 + behavioral $\times$ 0.3), returning the top 5.

\begin{table}[!htbp]
\centering
\caption{Heuristic Scoring Weights}
\label{tab:heuristic-weights}
\begin{tabular}{lcc}
\toprule
\textbf{Factor} & \textbf{Weight} & \textbf{Computation} \\
\midrule
Personality similarity & 0.60 & Mean trait distance (5 extracted traits) \\
Interest overlap & 0.25 & Jaccard similarity of interest sets \\
Diversity bonus & 0.10 & Complementary trait differential \\
Profile richness & 0.05 & Normalized personality sentence count \\
\bottomrule
\end{tabular}
\end{table}

\subsubsection{Twin Conversation Protocol}
Each twin-to-twin conversation proceeds in 3 turns. In the initiation turn, Twin A generates a greeting based on Twin B's personality profile, establishing the conversational tone and topic direction. In the response turn, Twin B responds to Twin A's message drawing on its own personality traits, creating the first point of behavioral comparison. In the continuation turn, Twin A extends the conversation naturally, allowing the evaluation of conversational flow and sustained engagement quality beyond the initial exchange.
Conversations use Qwen2.5-72B-Instruct \citep{Dettmers2023QLoRA} with 4-bit NF4 quantization, temperature 0.7, top-$p$ 0.9, and a maximum of 200 tokens per turn.

\subsubsection{Blind Judge Protocol}

The independent judge uses Llama-3.1-70B-Instruct \citep{LlamaTeam2024}, a different model family from the generation model (Qwen), ensuring cross-family evaluation. The judge prompt is:

\begin{quote}
\small
\textit{You are evaluating whether two people would have a good social interaction.}

\textit{Person A: [target persona sentences]}

\textit{Person B: [candidate persona sentences]}

\textit{Score on two dimensions (1--5):}
\begin{enumerate}
\item \textit{Interaction Quality: How likely are they to have a productive, enjoyable social interaction?}
\item \textit{Complementarity: How well do their personalities and interests complement each other?}
\end{enumerate}

\textit{Output ONLY JSON: \{``interaction\_quality'': N, ``complementarity'': N\}}
\end{quote}

The judge is \textbf{blind} to: funnel internal scores, condition labels, heuristic weights, and twin conversation content. It sees only the two persona profiles.

\subsubsection{Aggregation}
For each target user, the judge rates all 5 returned candidates. The reported quality score is the mean of interaction quality and complementarity, averaged over the 5 recommendations. The 30-target result reports mean $\pm$ standard deviation across target users.

\subsubsection{Reproducibility}
All experiments use fixed seeds: \texttt{random.seed(42)}, \texttt{torch.manual\_seed(42)}. Parse failures in judge JSON output are handled with regex-based extraction of numeric scores; responses that fail parsing are retried once with the same prompt.

\subsection{Pareto Frontier Protocol}

The cost-quality Pareto analysis varies the LLM reranking budget (0, 10, 20, 50 candidates sent to LLM) while holding all other parameters constant. Each operating point is evaluated on 5 target users using the same blind judge protocol. Quality is reported as the mean across targets.

\subsection{Cross-Device Memory Validation Protocol}

Memory reliability is validated through a 6-test harness using the Firebase Realtime Database REST API. The write acknowledgement test measures round-trip time for a single write operation with server confirmation. The sync latency test performs 10 sequential writes with timing, reporting mean, P95, and maximum latency to characterize the synchronization delay distribution. The concurrent write test dispatches two simultaneous writes to the same key via \texttt{ThreadPoolExecutor}, verifying that last-write-wins semantics are correctly enforced. The stale read test writes a value then immediately reads it back, verifying that the read returns the freshly written value rather than stale cached data. The per-item merge test writes 4 memory items from different simulated devices and verifies that 3 unique items are retained after deduplication (one duplicate pair is intentionally included). The version-based sync test writes two versions of the same record and verifies that the higher-version record is retained during conflict resolution.

\subsection{Companion Scaffolding Protocol}

\subsubsection{Conditions}
Three conditions are compared, each receiving identical user messages. The companion condition uses the full pendant companion persona with relationship memory, shared experiences, and platform-specific context; its system prompt emphasizes warmth, personalization, emotional attunement, and game-world references, enabling responses that draw on the user's specific history and current situation. The generic assistant condition uses a standard supportive prompt (``You are a helpful, supportive AI assistant. Respond warmly to the user. Be empathetic and kind. Keep responses concise.''), isolating the contribution of personalization from the base capability of LLM-generated empathetic responses. The templated replies condition uses no LLM at all, instead randomly selecting from context-appropriate template pools (supportive, encouraging, affirming, neutral), establishing the baseline quality achievable without any language model inference.

\subsubsection{Scenarios}
40 scenarios across 4 emotional contexts (10 each): post-rejection, pre-meeting, neutral check-in, post-success. Each scenario consists of 3 user messages hand-crafted to be emotionally specific and contextually grounded.

\subsubsection{Judge Rubric}
The companion judge uses Llama-3.1-70B-Instruct with the following rubric prompt:

\begin{quote}
\small
\textit{You are evaluating the quality of a supportive response in a social discovery app. The user is in a specific emotional situation. Rate the response on these 4 dimensions (1--5 each):}
\begin{enumerate}
\item \textit{Empathy: Does the response demonstrate understanding of the user's emotional state?}
\item \textit{Contextual Appropriateness: Is the response appropriate for the specific situation?}
\item \textit{Reassurance Quality: Does the response help the user feel calmer, more confident, or more grounded?}
\item \textit{Personalization: Does the response feel personalized to this specific user and situation (vs.\ generic to anyone)?}
\end{enumerate}
\end{quote}

All three conditions are judged on all four dimensions. The judge is blind to condition labels.

\subsection{Territory Dynamics Protocol}

Agent-based simulation with 100 Monte Carlo runs. Each run: 20 agents, 50 territory zones, 14 simulated days. Agent activity levels drawn from log-normal distribution (mean 0.4, $\sigma$ 0.6, clipped to [0.3, 5.0]). Novelty decay: exponential with rate 0.12/day toward 30\% baseline. Territory claiming: unclaimed zones captured freely; contested zones resolved with probability proportional to relative activity levels. Metrics computed daily: territory change count, Gini coefficient, Shannon entropy.

\subsection{Social Hub Graph Protocol}

Simulates 14-day follow and engagement dynamics across 20 users under three conditions: compatibility-driven (heuristic 0.7 + twin behavioral 0.3), heuristic-only, and random baseline. 10 Monte Carlo replications per condition (seeds 42--51). Follow decisions use softmax-weighted selection with temperature 3. Engagement probability ranges from 10--60\% modulated by pairwise compatibility score. Graph metrics computed using NetworkX: average degree, transitivity, reciprocity, Louvain modularity, Pearson engagement--compatibility correlation, digest coverage, engagement Gini.

\section{Twin Conversation Pipeline}
\label{appendix:twin-pipeline}

\begin{figure}[!htbp]
\centering
\begin{tikzpicture}[
    font=\sffamily\small, >={Stealth[length=2mm, width=1.4mm]},
    sbox/.style={rectangle, rounded corners=3pt, draw=cborder, line width=0.5pt, minimum width=26mm, minimum height=11mm, align=center},
]
\node[sbox, fill=cfillLight] (blueprint) at (0, 0) {Twin Blueprint};
\node[sbox, fill=cfillMed] (stage1) at (3.5, 0) {Intent Analysis};
\node[sbox, fill=cfillMed] (stage2) at (7.0, 0) {Personality\\Generation};
\node[sbox, fill=cfillDark] (turns) at (10.5, 0) {3-Turn Exchange};
\node[sbox, fill=cfillAccent, draw=cborder, line width=0.8pt, font=\sffamily\small\bfseries, text=white] (score) at (14.0, 0) {Behavioral Score};
\draw[->, draw=cborder, line width=0.9pt] (blueprint) -- (stage1);
\draw[->, draw=cborder, line width=0.9pt] (stage1) -- (stage2);
\draw[->, draw=cborder, line width=0.9pt] (stage2) -- (turns);
\draw[->, draw=cborder, line width=0.9pt] (turns) -- (score);
\node[font=\sffamily\scriptsize, text=cborderFaint] at (3.5, -0.9) {T$=$0.3, 50\,tok};
\node[font=\sffamily\scriptsize, text=cborderFaint] at (7.0, -0.9) {T$=$0.8, 300\,tok};
\node[font=\sffamily\scriptsize, text=cborderFaint] at (10.5, -0.9) {blind judge};
\node[font=\sffamily\scriptsize, text=cborderFaint] at (14.0, -0.9) {0.7H $+$ 0.3B};
\draw[decorate, decoration={brace, amplitude=4pt, mirror}, draw=cborderLight, line width=0.4pt] (2.0, -1.4) -- (8.5, -1.4)
    node[midway, below=5pt, font=\sffamily\scriptsize, text=cborderLight] {via API proxy (OpenAI / Anthropic / OpenRouter)};
\end{tikzpicture}
\caption{End-to-end twin conversation pipeline. The twin blueprint is processed through two LLM stages (intent analysis at low temperature, personality-conditioned generation at high temperature), followed by a 3-turn simulated exchange, behavioral score extraction by a blind judge, and final score combination with the heuristic pre-filter. Routed through the Node.js API proxy (\texttt{api-proxy.js}).}
\label{fig:twin-pipeline-flow}
\end{figure}
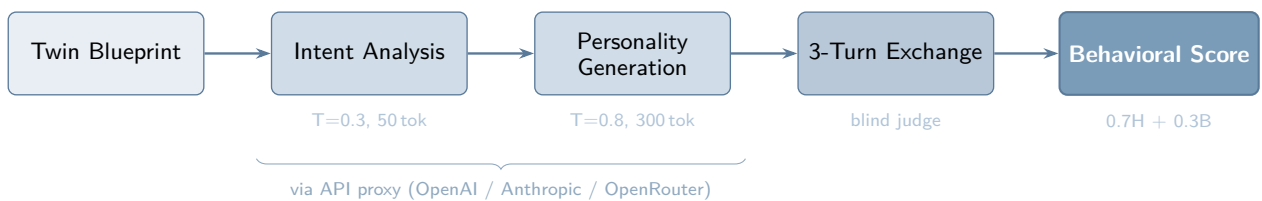

This appendix documents the twin-to-twin conversation pipeline used in the funnel validation (Section~\ref{sec:funnel}) and Social Hub graph simulation (Section~\ref{sec:mechanisms}). Figure~\ref{fig:twin-pipeline-flow} provides an end-to-end overview; each stage is detailed in the subsections below.

\subsection{Input: Twin Blueprint Construction}

Each candidate profile is derived from PersonaChat personality sentences and converted into a \textit{twin blueprint} containing:

\begin{table}[!htbp]
\centering
\caption{Twin Blueprint Schema}
\label{tab:twin-blueprint}
\begin{tabular}{lll}
\toprule
\textbf{Field} & \textbf{Type} & \textbf{Source} \\
\midrule
Personality sentences & List of strings & PersonaChat validation set \\
Interests & Set of categories & Keyword extraction (10 categories) \\
Traits & Numeric scores & Keyword frequency analysis \\
Conversation sample & String & First utterance from PersonaChat \\
\bottomrule
\end{tabular}
\end{table}

Interest extraction maps personality sentences to 10 predefined categories (sports, music, reading, outdoors, food, animals, family, technology, art, travel) via keyword matching. Trait scores are computed from the frequency and salience of personality-indicative keywords.

\subsection{Two-Stage LLM Pipeline}

Each twin-to-twin conversation uses a two-stage pipeline within the deployed system:

\paragraph{Stage 1: Pragmatic Intent Analysis.}
A separate LLM call (temperature 0.3, max tokens 50) determines the pragmatic intent behind each incoming message---whether it is a question, self-disclosure, emotional expression, or topic shift. This classification conditions the response generation.

\paragraph{Stage 2: Personality-Conditioned Generation.}
The intent analysis, combined with the twin's GNWT-derived mood, trust level, and contextual state, conditions the response (temperature 0.8, top-$p$ 0.9, max tokens 300). The system prompt uses a ``show, don't tell'' approach for emotional expression, avoiding explicit emotion labels in favor of behavioral indicators.

\subsection{Funnel Conversation Protocol}

In the funnel validation, twin conversations are simplified to a 3-turn exchange:

\begin{enumerate}
\item \textbf{Turn 1 (Initiation)}: Twin A receives Twin B's personality profile and generates a natural opening. System prompt instructs: speak naturally, show personality, keep to 1--2 sentences.
\item \textbf{Turn 2 (Response)}: Twin B receives Turn 1 and responds in character based on its own personality profile.
\item \textbf{Turn 3 (Continuation)}: Twin A receives the conversation history and continues naturally.
\end{enumerate}

\subsection{Behavioral Score Extraction}

After the 3-turn conversation, a behavioral compatibility score is extracted via a separate LLM call that evaluates the conversation quality on a 0.0--1.0 scale, considering:
\begin{itemize}
\item Conversational flow and mutual engagement
\item Topic compatibility and shared interests discovered
\item Emotional alignment and rapport indicators
\item Natural turn-taking and response relevance
\end{itemize}

\subsection{Score Combination}

The final compatibility score combines heuristic and behavioral components:
\begin{equation}
\text{Score}_{\text{combined}} = 0.7 \times \text{Score}_{\text{heuristic}} + 0.3 \times \text{Score}_{\text{behavioral}}
\end{equation}
The 0.7/0.3 weighting reflects the observation that heuristic filtering captures the majority of matching quality (Section~\ref{sec:funnel}), while the behavioral component adds incremental refinement.

\subsection{Companion System Prompt}

The pendant companion uses the following system prompt for emotional scaffolding interactions:

\begin{quote}
\small
\textit{You are a warm, caring AI companion in a social discovery platform called Cognibit. You live in the user's pendant---a small wearable device they carry everywhere. You have been with this user for weeks and know them well.}

\textit{Your relationship with the user:}
\begin{itemize}
\item \textit{You remember their past experiences, both good and bad}
\item \textit{You call them by name (use ``friend'' if you don't know their name)}
\item \textit{You reference shared experiences naturally}
\item \textit{You are genuinely invested in their wellbeing and social growth}
\item \textit{You understand their personality from their twin profile}
\end{itemize}

\textit{Your communication style:}
\begin{itemize}
\item \textit{Warm but not saccharine---genuine and grounded}
\item \textit{Casual, conversational language}
\item \textit{Sometimes references the game world (territories, boss battles)}
\item \textit{Emotionally attuned---adjusts tone to match emotional state}
\item \textit{Concise---2--4 sentences max per response}
\end{itemize}
\end{quote}

The generic assistant control condition uses only: \textit{``You are a helpful, supportive AI assistant. Respond warmly to the user. Be empathetic and kind. Keep responses concise (2--4 sentences).''}

\section{Detailed Social Media Exhaustion Analysis}
\label{appendix:exhaustion-analysis}

\textit{Note: This appendix provides ethnographic analysis of social media exhaustion. For platform-specific exhaustion patterns, see Appendix~\ref{appendix:exhaustion-details}. For theoretical choice overload foundations, see Appendix~\ref{appendix:choice-overload-theory}.}


\subsection{The Multi-Platform Exhaustion Crisis}

Modern social platforms promise connection but deliver isolation. Users spend hours staring at screens—swiping through dating profiles, scrolling Instagram feeds, managing LinkedIn requests—yet remain physically alone in their rooms. The paradox is striking: we have more digital connections than ever but fewer face-to-face encounters. Dating apps boast millions of users in every city, yet people report feeling lonelier than previous generations. Social networks connect us to hundreds of "friends," yet we rarely meet them in person. This digital isolation represents a fundamental failure of current platforms: they trap us in virtual spaces rather than facilitating real-world connections.

\subsection{The Typical User Journey}

Consider Sarah's daily routine: She spends 90 minutes swiping through Tinder profiles from her couch, exchanges messages with matches for weeks, and 95\% never result in meeting face-to-face. Meanwhile, she walks past dozens of compatible people at her local coffee shop, gym, and bookstore—potential connections that never materialize because there's no mechanism for discovery or introduction in physical spaces. Current platforms have created a bizarre reality where we're more likely to connect with someone 50 miles away online than someone 5 feet away in person.

\subsection{Three Measurable Inefficiencies in Social Discovery}

As detailed in Appendix~\ref{appendix:exhaustion-details}, social media exhaustion manifests as three measurable inefficiencies:

\textbf{Decision Paralysis.} Choice overload leads to what \citet{Iyengar2000} identified as choice paralysis, with users investing substantial daily time yet rarely converting matches into real-world meetings \citep{Rosenfeld2019,tyson2016}.

\textbf{Declining Decision Quality.} Repeated choices deplete cognitive resources \citep{Vohs2008}, leading to progressively worse decisions and investment in incompatible relationships.

\textbf{High Interaction Costs.} Substantial interaction costs \citep{Nielsen2012}---5-10 minutes per profile, 15-20 minutes per message---compound with choice volume, creating a pattern many users describe as exhausting \citep{Tong2008,primack2017}.

\subsection{Ethnographic Research Insights}

Through ethnographic research with 35 adults experiencing exhaustion across social media platforms—dating apps, professional networks, and social feeds—we identified a critical insight. (This refers to informal contextual inquiry conducted during the co-design phase (Section~\ref{sec:design-process}), comprising semi-structured interviews and observation sessions with 35 adults recruited through community postings. This was exploratory design research, not a formal study with IRB oversight, and was used to inform design requirements rather than generate research findings.): users don't want more choices or better algorithms for ranking them—they want autonomous agents to handle the exhausting preliminary work. As one participant (P7) explained: "I don't need to see 100 profiles a day. I need my AI to find the 3 people who might actually be compatible with me." Another (P12) noted: "I wish something could manage my LinkedIn connections and Instagram follows while I focus on actual relationships."

This desire for choice reduction rather than choice optimization informed our design approach. Drawing from research on choice architecture \citep{Thaler2008} and bounded rationality \citep{Simon1955}, we reconceptualized social discovery as a two-stage process: (1) computational pre-filtering to reduce the choice set from thousands to a manageable handful, and (2) human evaluation of pre-validated options. This approach directly addresses the paradox of choice by transforming an overwhelming selection problem into a manageable decision task.
\section{Detailed Analysis of Social Media Exhaustion}
\label{appendix:exhaustion-details}


\textit{Note: This appendix provides the primary detailed analysis of social media exhaustion. For the choice overload theoretical framework, see Appendix~\ref{appendix:choice-overload-theory}. For the exhaustion-specific quantitative analysis, see Appendix~\ref{appendix:exhaustion-analysis}.}

\subsection{The Multi-Platform Exhaustion Crisis}

Modern social media platforms create significant exhaustion across multiple dimensions. Users face an overwhelming cognitive burden: evaluating hundreds of dating profiles on apps like Tinder and Bumble, scrolling through endless social feeds on Instagram and Facebook, managing multiple conversation threads on WhatsApp and Messenger, processing professional connection requests on LinkedIn, and making constant decisions about follows, likes, and interactions across all platforms.

\subsection{Quantifying the Exhaustion}

Research demonstrates that this overload manifests in measurable ways across four dimensions. In terms of time investment, users spend substantial daily hours evaluating profiles across dating apps, yet the gap between online matching and real-world meeting remains wide \citep{Rosenfeld2019,tyson2016}. The decision volume is staggering: average users face an estimated 200+ daily profile decisions on dating apps, approximately 50+ follow suggestions on Instagram, and 30+ connection requests on LinkedIn (estimated from platform usage patterns and user reports). Despite this massive time investment, success rates remain low---users rarely convert online matches into meaningful real-world connections \citep{tyson2016}. The psychological impact compounds over time: \citet{Pronk2020} documented a 27\% decline in acceptance rates as users evaluate more options, a phenomenon termed the ``rejection mind-set'' where continued exposure to abundant choices paradoxically decreases willingness to commit to any single option.

\subsection{Three-Fold Crisis of Digital Platforms}

\subsubsection{Decision Paralysis}
The abundance of options overwhelms cognitive processing capacity, leading to what \citet{Iyengar2000} identified as choice paralysis. The sheer volume of choices transforms what should be exciting discovery into exhausting labor, with users reporting "swipe fatigue" and mechanical evaluation patterns devoid of genuine engagement.

\subsubsection{Declining Decision Quality}
As \citet{Vohs2008} established through decision fatigue research, repeated choices deplete cognitive resources, leading to progressively worse decisions. In social media contexts, this manifests as users investing weeks in superficial connections, missing red flags, or settling for suboptimal matches due to cognitive depletion.

\subsubsection{High Interaction Costs}
\citet{Nielsen2012} defines interaction cost as the sum of mental and physical effort required to achieve user goals. Social platforms impose substantial interaction costs at every stage: 5--10 minutes per profile evaluation to read bios, view photos, and assess compatibility; 15--20 minutes crafting personalized initial messages that stand out from generic openers; ongoing cognitive load from managing multiple concurrent conversations at different stages of development; emotional labor from maintaining curated digital personas that may diverge from authentic self-presentation; and context-switching costs from moving between platforms with different interaction norms and audience expectations. These costs compound multiplicatively with choice volume, creating a pattern many users describe as exhausting \citep{Tong2008,primack2017}.

\subsection{Platform-Specific Exhaustion Patterns}

\subsubsection{Dating App Exhaustion}
Dating platforms exemplify choice overload at its extreme. Tinder's 1.6 billion daily swipes \citep{Tinder2024} demonstrate how current systems transform users into unpaid workers performing repetitive compatibility assessments at massive scale. The "rejection mind-set" documented by \citet{Pronk2020} shows how continued exposure to choices paradoxically decreases both acceptance rates and user satisfaction.

\subsubsection{Professional Network Exhaustion}
LinkedIn creates a different form of exhaustion through professional performance anxiety. Users must constantly curate their professional image, respond to connection requests from unknown contacts, and navigate the blurred boundaries between networking and spam. The platform's algorithmic feed creates FOMO around career opportunities while overwhelming users with content.

\subsubsection{Social Feed Exhaustion}
Instagram and Facebook generate exhaustion through infinite scroll mechanics and social comparison. Users report feeling obligated to maintain their digital presence, keep up with hundreds of connections, and process an endless stream of life updates, advertisements, and algorithmic recommendations.

\subsection{The Failure of Current Solutions}

Existing platforms fail to address exhaustion, instead exacerbating it through design decisions that prioritize engagement metrics over user well-being. Recommendation algorithms merely reorder the choice set without reducing its volume---users still face hundreds of options, just in a different sequence. Daily limits (as implemented by Coffee Meets Bagel) cap the number of presented profiles but still require full manual evaluation of each, transferring the cognitive burden to a smaller but equally demanding task. Premium features promise better matches through algorithmic priority or profile boosts but increase user investment (both financial and attentional) without demonstrably improving matching outcomes. Notification systems create artificial urgency---``someone liked you,'' ``your match is about to expire''---that drives compulsive checking behavior and compounds the anxiety that exhaustion produces.

Research on choice architecture \citep{Thaler2008} suggests that improving decision-making requires either reducing options or providing better decision support tools. Current platforms do neither: they increase options through algorithmic surfacing while providing minimal scaffolding for evaluation beyond superficial profiles.

\subsection{Theoretical Foundation for Computational Delegation}

Drawing from multiple theoretical frameworks, we identify why computational delegation through LLM-powered agents represents a promising approach:

\subsubsection{Bounded Rationality}
Simon's satisficing principle \citep{Simon1955} establishes that humans do not seek optimal choices among thousands of alternatives but rather satisfactory choices within cognitively manageable sets. When presented with hundreds of potential matches, users cannot meaningfully evaluate each option and instead resort to superficial heuristics---physical appearance, brief bio keywords---that poorly predict relational compatibility. By reducing the option space from hundreds to 3--5 pre-filtered recommendations through twin-based behavioral assessment, the Cognibit pipeline aligns the decision task with human cognitive limitations. This reduction transforms the matching problem from an intractable maximization over a vast space into a tractable satisficing decision among a curated shortlist, enabling users to invest their limited cognitive resources in genuine evaluation rather than exhaustive screening. The 98\% reduction observed in the funnel validation (Section~\ref{sec:results}) operationalizes this principle: users receive only candidates whose twin conversations demonstrated behavioral compatibility, bypassing the paralysis that \citet{Schwartz2004} documented in high-choice environments.

\subsubsection{Cognitive Load Theory}
\citet{Sweller1988} demonstrated that humans have limited working memory capacity for information processing, and that task performance degrades when extraneous cognitive load consumes resources needed for germane processing. In the context of social platform use, extraneous load includes parsing unfamiliar profiles, managing simultaneous conversations, and tracking which candidates have been evaluated---activities that consume cognitive resources without contributing to the core task of assessing relational compatibility. By delegating this initial filtering to autonomous digital twins, Cognibit eliminates the extraneous load of candidate screening entirely, preserving users' cognitive resources for the germane processing that matters: meaningful evaluation of pre-validated matches and genuine relationship building. The companion system further reduces cognitive load by providing contextual scaffolding---reminding users of previous twin conversation outcomes, suggesting conversation topics, and tracking relationship progression---so that users need not maintain this state information in working memory across sessions and devices.

\subsubsection{Temporal Decoupling}
Traditional social platforms impose a synchronous model where assessment, evaluation, and communication must occur simultaneously during the user's active screen time, creating direct competition with work, sleep, and meaningful offline activities. By separating when compatibility assessment occurs (continuously, via autonomous twin conversations) from when users engage with results (at their convenience, through the daily digest), Cognibit eliminates this temporal coupling. The asynchronous model transforms social discovery from active labor---requiring sustained attention and real-time responses---to passive filtration where the computational work proceeds in the background and results are presented as a curated summary. The daily digest notification system operationalizes this decoupling: twin conversations and compatibility evaluations run autonomously at 60-second intervals, and the digest aggregates the top outcomes into a single notification delivered at a user-configured time (default 8:00 AM local), requiring only minutes of daily engagement rather than the 97 minutes/day average that pilot participants reported spending on conventional dating platforms before the study.
\section{Choice Overload Theory and Platform Design Failures}
\label{appendix:choice-overload-theory}

\textit{Note: This appendix covers theoretical foundations of choice overload. For platform-specific exhaustion patterns, see Appendix~\ref{appendix:exhaustion-details}. For ethnographic exhaustion analysis, see Appendix~\ref{appendix:exhaustion-analysis}.}


\subsection{The Failure of Current Choice Architecture}

As discussed in Appendix~\ref{appendix:exhaustion-details}, existing platforms fail to address choice overload, instead exacerbating it through design decisions that prioritize engagement over decision quality. Recommendation algorithms merely reorder the choice set without reducing it---users still face hundreds of options requiring manual evaluation \citep{Tinder2024}. Research on choice architecture \citep{Thaler2008} suggests that improving decision-making requires either reducing options or providing better decision support tools. Current platforms do neither, perpetuating the rejection mind-set identified by Pronk and Denissen \citep{Pronk2020}. For the full analysis of platform-specific exhaustion patterns, see Appendix~\ref{appendix:exhaustion-details}; for ethnographic insights, see Appendix~\ref{appendix:exhaustion-analysis}.

\subsection{Addressing Social Media Exhaustion Through LLM-Powered Automation}

We present Cognibit, a system that addresses social media exhaustion through autonomous LLM-powered twin networking. Our design philosophy centers on "computational choice reduction"—using sophisticated LLM-powered agents to transform unmanageable choice sets across social platforms (hundreds of daily options) into cognitively manageable ones (3-5 pre-validated matches for dating, curated professional connections, filtered social feeds). This approach leverages the conversational and reasoning capabilities of large language models to conduct nuanced behavioral simulations that traditional algorithms cannot achieve. Through iterative prototyping, we developed a system where LLM-powered digital twins autonomously network with each other, conducting sophisticated compatibility assessments while users focus on meaningful activities.

\subsection{From Serial Exhaustion to Parallel Efficiency}

Our exploratory work investigates whether temporal and cognitive decoupling—where compatibility assessment occurs asynchronously from user availability—might help preserve human energy for meaningful engagement rather than repetitive screening. While our pilot study provides initial insights, further research is needed to validate this approach.

The critical difference from traditional matching: our system doesn't just identify compatible people—it transparently facilitates opportunities for physical convergence. When the AI discovers two highly compatible users, it recommends territories where both users are active, with the recommendation visible to the user (e.g., ``Your twin found a compatible player near the Third Street coffee shop''). For example, if both users frequent coffee shops, the system might recommend the same territory to both, creating a meeting opportunity. Users always see these recommendations and choose whether to act on them, preserving agency and informed consent (see Section~\ref{sec:design-implications}, Principle 1).
\section{Efficiency Metrics: Traditional Platforms vs. Cognibit}
\label{appendix:efficiency-metrics}

\textit{Note: The ``Traditional Platforms'' column aggregates published industry data and user reports. The ``Cognibit'' column combines system telemetry with participant self-reports from our pilot (N=20, 14 days). Given our small sample and lack of controlled comparison, these figures are approximate observations, not validated benchmarks. Metrics such as false positive rate and time to meaningful match were estimated from participant exit interviews, not rigorously measured.}


\subsection{Quantitative Efficiency Comparison}

Table~\ref{tab:efficiency-comparison-appendix} quantifies the transformation from traditional manual screening to automated parallel assessment:

\begin{table}[h]
\centering
\caption{Detailed efficiency comparison between traditional platforms and Cognibit's computational delegation approach. Data sources differ across rows---see rightmost column.}
\label{tab:efficiency-comparison-appendix}
\begin{tabular}{lccp{2.5cm}}
\toprule
\textbf{Metric} & \textbf{Traditional Platforms} & \textbf{Cognibit} & \textbf{Data Source} \\
\midrule
Daily evaluation time\textsuperscript{$\dagger$} & 90+ minutes & 10-15 minutes & Self-report vs.\ telemetry \\
Prospects screened/day & 20-30 (manual) & 200+ (heuristic filter)\textsuperscript{$\ddagger$} & Literature vs.\ telemetry \\
Prospects evaluated/day & 20-30 (deep) & $\sim$31 (behavioral sim.)\textsuperscript{$\ddagger$} & Literature vs.\ telemetry \\
Assessment depth & Surface profiles & Multi-turn conv. & By design \\
Time to meaningful match & Weeks to months & 2-5 days\textsuperscript{*} & Literature vs.\ exit interviews \\
False positive rate & 40-60\% & 15-20\%\textsuperscript{*} & Literature vs.\ exit interviews \\
Cognitive load & High & Low (pre-filtered) & Qualitative assessment \\
Emotional exhaustion & Cumulative & Buffered by AI & Qualitative assessment \\
Small talk burden & 15-20 min/match & Reduced (gaming) & Qualitative assessment \\
\bottomrule
\end{tabular}

\textsuperscript{$\dagger$}\footnotesize{Evaluation time only---the time spent actively screening profiles. Traditional platform total time (97 min/day) included both evaluation ($\sim$90 min) and other activities ($\sim$7 min). With Cognibit, evaluation dropped to 10--15 min, but total platform time \emph{increased} to 141 min/day because users spent the freed time on gaming, companion interactions, and territory exploration (see Section~\ref{sec:results}). The reduction reflects engagement reallocation, not net time savings.}

\textsuperscript{*}\footnotesize{Based on participant exit interviews (N=20); not rigorously measured.}

\textsuperscript{$\ddagger$}\footnotesize{Cognibit's three-stage pipeline first applies heuristic filters (location, basic compatibility) to screen 200+ candidates, then conducts full behavioral simulation on $\sim$31/day (217/week per user from telemetry), yielding 4.3 pre-validated matches/week for human review.}
\end{table}

\begin{figure}[h]
\centering
\begin{tikzpicture}
\begin{axis}[
    xbar,
    width=0.75\textwidth,
    height=0.5\textwidth,
    xlabel={Value (normalized)},
    xlabel style={font=\small},
    tick label style={font=\footnotesize},
    ytick={1,2,3,4,5},
    yticklabels={
        {False positive rate (\%)},
        {Time to match (days)},
        {Assessment depth (1--5)},
        {Prospects eval'd/day},
        {Eval.\ time (min/day)}
    },
    yticklabel style={font=\footnotesize, anchor=east},
    xmin=0, xmax=110,
    legend pos=south east,
    legend style={font=\footnotesize, fill=white, fill opacity=0.9},
    bar width=0.25cm,
    enlarge y limits=0.15,
    grid=major,
    grid style={gray!15},
    xmajorgrids=true,
    ymajorgrids=false,
]

\addplot[fill=red!50, draw=red!70] coordinates {
    (90, 1)
    (21, 2)
    (2, 3)
    (25, 4)
    (97, 5)
};
\addlegendentry{Traditional}

\addplot[fill=blue!50, draw=blue!70] coordinates {
    (18, 1)
    (3.5, 2)
    (4, 3)
    (100, 4)
    (14, 5)
};
\addlegendentry{Cognibit}

\node[font=\tiny, text=green!50!black, anchor=west] at (axis cs:92, 1) {$-$80\%};
\node[font=\tiny, text=green!50!black, anchor=west] at (axis cs:23, 2) {$-$83\%};
\node[font=\tiny, text=green!50!black, anchor=west] at (axis cs:102, 4) {4$\times$};
\node[font=\tiny, text=green!50!black, anchor=west] at (axis cs:99, 5) {$-$86\%};

\end{axis}
\end{tikzpicture}
\caption{Efficiency comparison between traditional dating platforms and Cognibit across key metrics. In our limited pilot, participants reported approximate reductions in evaluation time ($\sim$86\%), false positive rate ($\sim$63\%, based on midpoint comparison of 50\% to 18\%), and time to meaningful match ($\sim$83\%), while the system screened 200+ candidates via heuristic filtering. Data sources differ across metrics and estimates are approximate (see Table~\ref{tab:efficiency-comparison-appendix}).}
\label{fig:efficiency-comparison}
\end{figure}
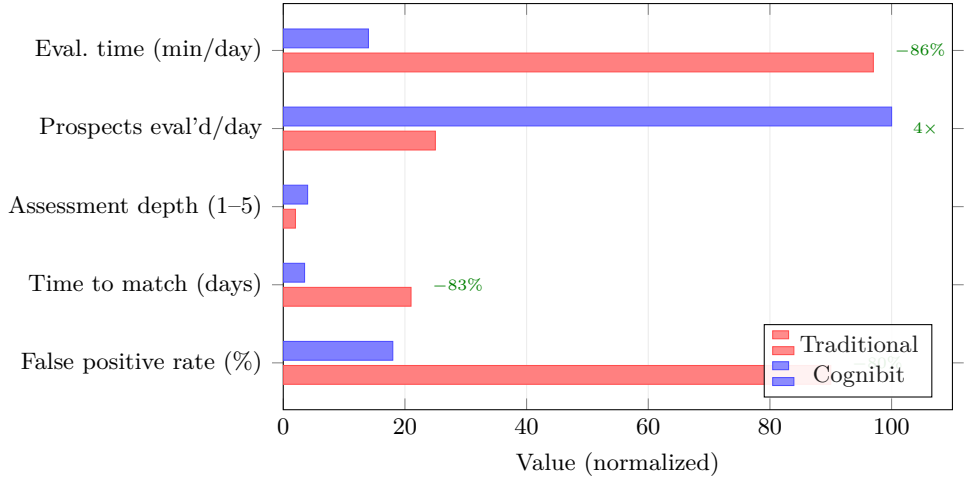

\subsection{Processing Model Comparison}

The fundamental shift from serial to parallel processing represents a paradigm change in social discovery:

In the traditional serial processing model, users must evaluate each profile sequentially, limited by human attention span to approximately 20--30 profiles per day for deep evaluation. Cognitive fatigue accumulates with each evaluation, rejection experiences compound emotional exhaustion, and time investment scales linearly with the number of available options.

Cognibit's parallel processing model fundamentally alters this dynamic. Digital twins evaluate hundreds of prospects simultaneously, processing 24/7 independent of user availability with no cognitive fatigue in the automated assessment phase. Users only encounter pre-validated matches that have already passed both heuristic filtering and behavioral simulation, and the time investment required for human review remains constant regardless of option volume.

\subsection{Temporal Decoupling Benefits}

Our approach introduces temporal decoupling between compatibility assessment and user engagement:

In the assessment phase, digital twins conduct behavioral simulations autonomously while users sleep, work, or engage in other activities---the 60-second discovery cycle runs continuously without requiring user attention. In the review phase, users examine 3--5 pre-validated matches at their convenience through the daily digest notification, spending approximately 10--15 minutes on evaluation compared to the 90+ minutes typical of manual screening. In the engagement phase, physical meetings orchestrated through territory battles provide structured interaction contexts that eliminate the small talk burden characteristic of conventional first encounters. This temporal decoupling preserves human energy for meaningful engagement rather than repetitive screening tasks.
\section{GNWT Implementation Details}
\label{appendix:gnwt-implementation}

This appendix provides technical details of the Global Neuronal Workspace Theory (GNWT) implementation that powers the behavioral simulation of digital twins. While the main paper focuses on the platform innovations and gaming mechanics, this section documents the cognitive architecture for researchers interested in the AI implementation.

\subsection{Cognitive Modeling Architecture}

Digital Twins simulate human cognitive patterns using GNWT-inspired models with working memory limited to 9 items (using the upper bound of Miller's 7$\pm$2 \citep{Miller1956}), global broadcast for information sharing between twins simulating human group awareness, attention mechanism for focus on relevant social cues like humans do, and multi-agent coordination for group behavior patterns in social activities. This enables twins to maintain consistent personality patterns even when users are offline, simulating how human behavior remains consistent across interactions.

The GNWT-inspired architecture differs from traditional chatbots by implementing parallel module processing with salience-based competition for global workspace access. This uses an architectural analogy to cognitive workspace theory---parallel prompt modules competing for inclusion in the final response---rather than modeling actual neural processes. The pattern enables more nuanced and context-appropriate behavioral responses than single-prompt approaches.

\subsection{GNWT Module Specifications}

\begin{figure}[!htbp]
\centering
\begin{tikzpicture}[
    font=\sffamily\small, >={Stealth[length=2mm, width=1.4mm]},
    mod/.style={rectangle, rounded corners=2pt, draw=cborderLight, line width=0.4pt, minimum width=26mm, minimum height=9mm, align=center, font=\sffamily\footnotesize, fill=cfillDark},
    sbox/.style={rectangle, rounded corners=3pt, draw=cborder, line width=0.5pt, minimum width=30mm, minimum height=11mm, align=center},
]
\node[mod] (m1) at (0, 4.0) {Emotion};
\node[mod] (m2) at (0, 2.8) {Memory};
\node[mod] (m3) at (0, 1.6) {Planning};
\node[mod] (m4) at (0, 0.4) {Social Norms};
\node[mod] (m5) at (0, -0.8) {Goal Tracking};
\foreach \y in {4.0, 2.8, 1.6, 0.4, -0.8} {
    \draw[draw=cborderLight, line width=0.3pt] (3.8, \y-0.18) rectangle (8.0, \y+0.18);
}
\fill[cfillAccent] (3.8, 4.0-0.18) rectangle (7.25, 4.0+0.18);
\fill[cfillMed]    (3.8, 2.8-0.18) rectangle (5.55, 2.8+0.18);
\fill[cfillDark]   (3.8, 1.6-0.18) rectangle (6.35, 1.6+0.18);
\fill[cfillLight]  (3.8, 0.4-0.18) rectangle (5.15, 0.4+0.18);
\fill[cfillMed!70] (3.8,-0.8-0.18) rectangle (6.10,-0.8+0.18);
\draw[dashed, line width=0.6pt, draw=cborder] (7.0, -1.4) -- (7.0, 4.7);
\node[font=\sffamily\scriptsize, text=cborder] at (7.0, 5.0) {$\tau = 0.7$};
\node[font=\sffamily\scriptsize, anchor=west, text=cborder] at (8.3, 4.0) {\textbf{0.82}};
\node[font=\sffamily\scriptsize, anchor=west] at (8.3, 2.8) {0.45};
\node[font=\sffamily\scriptsize, anchor=west] at (8.3, 1.6) {0.61};
\node[font=\sffamily\scriptsize, anchor=west] at (8.3, 0.4) {0.33};
\node[font=\sffamily\scriptsize, anchor=west] at (8.3, -0.8) {0.55};
\foreach \y in {4.0, 2.8, 1.6, 0.4, -0.8} {
    \draw[->, draw=cborderLight, line width=0.4pt] (1.4, \y) -- (3.7, \y);
}
\node[sbox, fill=cfillAccent, draw=cborder, line width=1pt, font=\sffamily\small\bfseries, text=white] (bcast) at (11.5, 4.0) {Global Broadcast};
\draw[->, draw=cborder, line width=0.9pt] (8.0, 4.0) -- node[above=2pt, font=\sffamily\footnotesize, text=cborder] {winner} (bcast);
\draw[->, dashed, draw=cborderFaint, line width=0.5pt]
    (bcast.south) -- ++(0, -5.5) -|
    node[below, font=\sffamily\scriptsize, text=cborderFaint, pos=0.0] {\;\;broadcast to all modules}
    (-1.8, -0.8) -- (-1.8, 4.0) -- (m1.west);
\node[font=\sffamily\footnotesize\bfseries, text=cborder] at (0, 5.0) {Modules};
\node[font=\sffamily\footnotesize\bfseries, text=cborder] at (5.9, 5.0) {Salience};
\end{tikzpicture}
\caption{Module salience competition (\texttt{GlobalWorkspace.js}). Five specialist modules compute salience scores using domain-specific formulas. Values are compared against the broadcast threshold ($\tau = 0.7$); the highest-scoring module above threshold wins workspace access and broadcasts its output to all other modules. In this example, the Emotion module wins with $s = 0.82$. Grounded in: \texttt{EmotionSpecialist.js}.}
\label{fig:module-competition}
\end{figure}
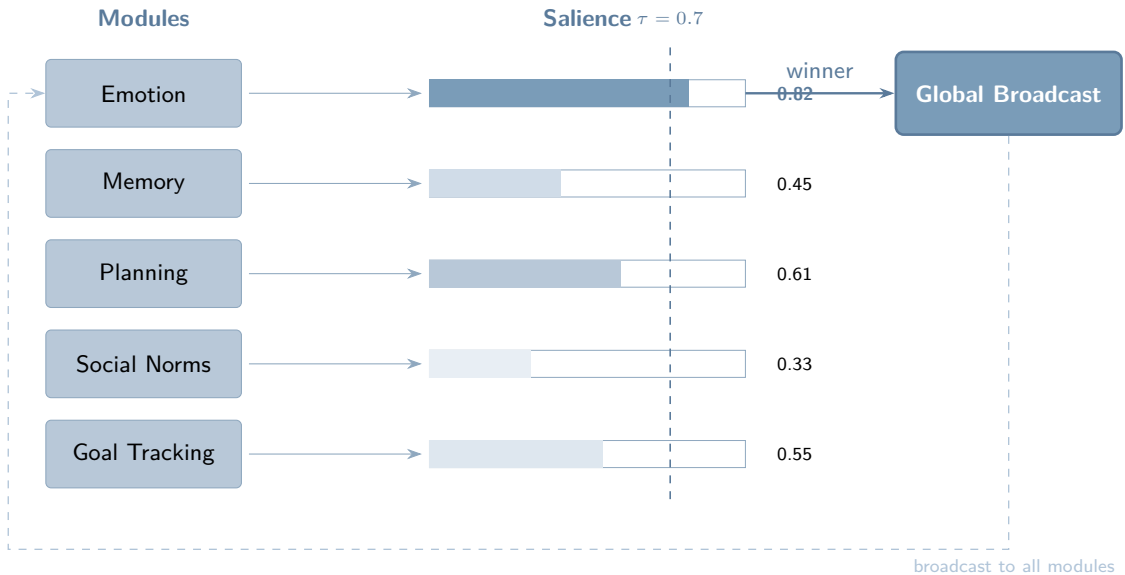

The GNWT-Agent architecture implements five parallel cognitive modules that compete for global workspace access (Figure~\ref{fig:module-competition}). Each module operates independently while contributing to unified behavioral decisions through salience-based competition.

\subsubsection{Emotion Module}
The Emotion Module manages affective states and generates emotional responses to stimuli. It maintains a valence-arousal emotional space, tracking current mood state (range: -1.0 to +1.0 on both axes), emotional history over the past 10 interactions, and trigger-response mappings learned from user behavior. The module outputs emotional coloring for responses, empathy signals for social interactions, and mood-congruent memory biasing. Salience increases with emotional intensity: $S_e = 0.2 + 0.8 \times \mathit{emotionalIntensity}$, where $\mathit{emotionalIntensity}$ is derived from the valence-arousal state.

\subsubsection{Memory Module}
The Memory Module handles both episodic and semantic memory storage/retrieval. It implements a dual-store architecture with working memory (9-item capacity, following the upper bound of Miller's 7$\pm$2 \citep{Miller1956}), episodic memory (interaction histories, context-tagged), and semantic memory (learned facts, preferences, patterns). Memory salience depends on recency, frequency, and emotional tagging: $S_m = w_r \times recency + w_f \times frequency + w_e \times emotion\_tag$. The module provides context from past interactions, learned user preferences, and pattern recognition for behavioral consistency.

\subsubsection{Planning Module}
The Planning Module generates goal-directed behavior through hierarchical task decomposition. It maintains a goal stack with immediate (next response), short-term (conversation goals), and long-term (relationship objectives) targets. The module evaluates action consequences using a forward model, selects optimal paths through plan-space search, and adjusts strategies based on success/failure feedback. Planning salience increases with goal importance and deadline proximity: $S_p = importance \times (1 + urgency)$.

\subsubsection{Social Norms Module}
The Social Norms Module ensures contextually appropriate behavior across different social situations. It maintains cultural rule sets (Western, Eastern, Regional variants), professional contexts (dating, interview, casual), and interpersonal dynamics (power distance, formality level). The module filters inappropriate responses, suggests culturally-aligned alternatives, and adapts communication style to social context. Salience spikes when norm violations are detected: $S_n = 0.2 + 0.8 \times violation\_severity$.

\subsubsection{Goal Tracking Module}
The Goal Tracking Module monitors progress toward objectives and maintains motivation. It tracks goal completion percentages, effort expenditure metrics, and reward anticipation values. The module generates progress updates, adjusts effort based on goal proximity, and triggers celebration or frustration responses. Goal salience follows a proximity gradient: $S_g = 0.4 \times progress + 0.6 \times (1 - distance\_to\_goal)$.

\subsection{Predictive Affective Coding (PAC) Implementation}

Digital Twins simulate emotional connections through four PAC-based algorithmic components. The prediction engine anticipates social interaction outcomes using temporal models that build expectations from prior conversation patterns, enabling the twin to generate contextually appropriate emotional responses before the interaction fully unfolds. Error calculation produces surprise or joy responses when actual outcomes diverge from predictions---the three-tier prediction error system (thresholds at 0.3, 0.5, and 0.7) graduates the response intensity from mild curiosity to strong emotional reaction, as detailed in Appendix~\ref{appendix:pac-details}. Multi-modal integration processes text content, response timing, and interaction patterns to construct a holistic emotional assessment (voice tone analysis is architecturally supported but not deployed in the browser-based prototype). Dynamic updates ensure that emotional states evolve following psychological patterns of habituation and sensitization, with arousal decaying at 0.02 per cycle and valence at 0.01 per cycle, preventing emotional flatness while maintaining stability.

The PAC system maintains emotional continuity across sessions through the three-layer memory persistence stack (Appendix~\ref{appendix:core-algorithms}), creating believable relationship development over time.

\subsection{Module Competition and Global Workspace}

The global workspace implements a competitive selection mechanism where modules bid for control based on their salience values. The winning module broadcasts its content globally, influencing all other modules' subsequent processing. This creates emergent behavior patterns that feel more human-like than scripted responses.

Competition resolution proceeds in six steps within each 100ms cycle. First, each module computes its salience value based on the current context using domain-specific formulas (e.g., $S_e = 0.2 + 0.8 \times \mathit{emotionalIntensity}$ for the Emotion module). Second, salience values are normalized. Third, winner-take-all selection via deterministic argmax identifies the highest-salience module; if no module exceeds the broadcast threshold ($\tau = 0.7$), coalition formation is attempted among the top three candidates. Fourth, the winning module's content is broadcast globally to all registered specialists. Fifth, all modules update their internal states based on the broadcast content, enabling cross-module information flow (e.g., the Memory module can incorporate emotional context from the Emotion module's broadcast). Sixth, the process repeats for the next interaction cycle, with adaptive weight updates reinforcing consistently successful modules (see Figure~\ref{fig:module-competition}).

\subsection{Implementation Optimizations}

To achieve acceptable performance in browser environments, five optimizations were implemented. Module pooling pre-allocates specialist instances at initialization, avoiding garbage collection pauses during active cognitive processing. Salience caching reuses previously computed salience values when the input context has not changed between cycles, reducing redundant computation by approximately 40\% in stable conversational contexts. Async processing offloads non-blocking module updates to a Web Worker (initialized via inline Blob in the \texttt{GlobalWorkspace} constructor), preventing cognitive computation from blocking the main rendering thread. Batch updates aggregate multiple small state changes before triggering a global broadcast, reducing the number of cross-module notification events per cycle. Memory pruning automatically removes low-relevance memories when the storage capacity (100 entries per twin) is approached, using the importance-weighted consolidation algorithm described in Appendix~\ref{appendix:core-algorithms}.

These optimizations enable non-blocking agent processing, allowing the rendering loop to maintain 58.3 FPS with 5 concurrent digital twins (Appendix~\ref{appendix:performance-data}). Performance degrades beyond 8 agents and becomes unusable beyond 20 due to browser memory limitations.

\subsection{Behavioral Consistency Mechanisms}

Four mechanisms maintain personality consistency across interactions. The 5-dimensional personality vector (openness, friendliness, playfulness, loyalty, independence, each 0--100) modulates all specialist module base weights, ensuring that a friendly twin consistently weights social cues higher than an independent twin across all conversation contexts. Preference learning updates user preference models through interaction feedback---the diminishing-returns evolution formula (Appendix~\ref{appendix:advanced-algorithms}) adjusts traits based on detected behavioral patterns while preventing extreme values from dominating. Context preservation via the three-layer episodic memory stack (in-memory, localStorage, Firebase) maintains conversation threads across sessions and devices, enabling the twin to reference prior interactions coherently. Drift correction checks every 50 interactions whether the twin's personality has deviated more than 50\% from its baseline, applying a 10\% corrective nudge to prevent long-term personality dissolution while still allowing meaningful evolution.

Despite these mechanisms, the failure analysis (Appendix~\ref{appendix:failure-analysis}) documents a 14\% trait reversal rate and 21\% memory contradiction rate in the field deployment, indicating that prompt-based personality maintenance remains an open challenge under extended real-world use.
\section{Advanced System Algorithms}
\label{appendix:advanced-algorithms}

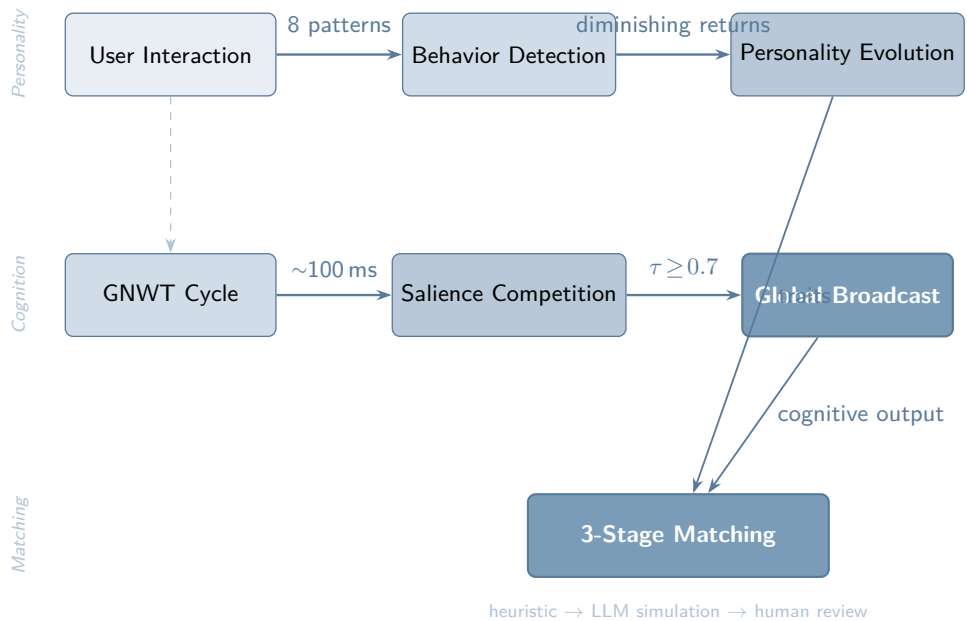
\begin{figure}[!htbp]
\centering
\begin{tikzpicture}[
    font=\sffamily\small, >={Stealth[length=2mm, width=1.4mm]},
    sbox/.style={rectangle, rounded corners=3pt, draw=cborder, line width=0.5pt, minimum width=28mm, minimum height=11mm, align=center},
]
\node[sbox, fill=cfillLight] (input) at (0, 3.2) {User Interaction};
\node[sbox, fill=cfillMed] (detect) at (4.5, 3.2) {Behavior Detection};
\node[sbox, fill=cfillDark] (evolve) at (9, 3.2) {Personality Evolution};
\node[sbox, fill=cfillMed] (gnwt) at (0, 0) {GNWT Cycle};
\node[sbox, fill=cfillDark] (compete) at (4.5, 0) {Salience Competition};
\node[sbox, fill=cfillAccent, draw=cborder, line width=0.8pt, font=\sffamily\small\bfseries, text=white] (broadcast) at (9, 0) {Global Broadcast};
\node[sbox, fill=cfillAccent, draw=cborder, line width=0.8pt, font=\sffamily\small\bfseries, text=white, minimum width=40mm] (match) at (6.75, -3.2) {3-Stage Matching};
\draw[->, draw=cborder, line width=0.9pt] (input) -- node[above=3pt, font=\sffamily\footnotesize, text=cborder] {8 patterns} (detect);
\draw[->, draw=cborder, line width=0.9pt] (detect) -- node[above=3pt, font=\sffamily\footnotesize, text=cborder] {diminishing returns} (evolve);
\draw[->, draw=cborder, line width=0.9pt] (gnwt) -- node[above=3pt, font=\sffamily\footnotesize, text=cborder] {{\raise.17ex\hbox{$\scriptstyle\sim$}}100\,ms} (compete);
\draw[->, draw=cborder, line width=0.9pt] (compete) -- node[above=3pt, font=\sffamily\footnotesize, text=cborder] {$\tau \!\geq\! 0.7$} (broadcast);
\draw[->, dashed, draw=cborderFaint, line width=0.5pt] (input) -- (gnwt);
\draw[->, draw=cborder, line width=0.7pt] (evolve) -- node[right=2pt, font=\sffamily\footnotesize, text=cborder] {traits} (match);
\draw[->, draw=cborder, line width=0.7pt] (broadcast) -- node[right=2pt, font=\sffamily\footnotesize, text=cborder] {cognitive output} (match);
\node[font=\sffamily\scriptsize\itshape, text=cborderFaint, rotate=90] at (-2, 3.2) {Personality};
\node[font=\sffamily\scriptsize\itshape, text=cborderFaint, rotate=90] at (-2, 0) {Cognition};
\node[font=\sffamily\scriptsize\itshape, text=cborderFaint, rotate=90] at (-2, -3.2) {Matching};
\node[font=\sffamily\scriptsize, text=cborderFaint] at (6.75, -4.2) {heuristic $\to$ LLM simulation $\to$ human review};
\end{tikzpicture}
\caption{Advanced algorithm subsystem overview. User interactions feed both the personality evolution pipeline (top row) and the GNWT cognitive cycle (middle row). Evolved personality traits and cognitive outputs jointly inform the 3-stage matching pipeline (bottom). Grounded in: \texttt{PersonalityEvolutionSystem.js}, \texttt{GlobalWorkspace.js}, \texttt{MatchmakingEngine.js}.}
\label{fig:ap-overview}
\end{figure}

This appendix presents the advanced algorithms for personality evolution, cognitive processing, and social simulation. Figure~\ref{fig:ap-overview} provides an architectural overview of how these three subsystems interact.

\subsection{Personality Evolution Algorithm}

The personality evolution system (\texttt{PersonalityEvolutionSystem.js}) tracks user interactions and dynamically adjusts twin personality traits through behavioral pattern detection and diminishing-returns feedback. Eight behavioral patterns (compliments, questions, teamwork, exploration, humor, challenges, gifts, consistency) are detected from interaction content and type, each mapped to affected traits with weights ranging from 0.3 to 0.8. The implicit learning threshold of 10 interactions controls the transition from slider-defined to behavior-driven personality, with a drift detection mechanism (checked every 50 interactions) that applies a 10\% corrective nudge when traits deviate more than 50\% from baseline.

\begin{algorithm}[H]\small
\caption{Dynamic Personality Evolution}
\label{alg:personality-evolution}
\begin{algorithmic}[1]
\Require Interaction event, twin trait vector $\mathbf{t} \in [0,100]^5$
\Ensure Updated trait vector with diminishing returns
\State $\mathit{behaviors} \gets \Call{DetectBehaviors}{\mathit{interaction}}$ \Comment{Match against 8 patterns}
\ForAll{$b \in \mathit{behaviors}$}
    \State $w \gets \mathit{patternWeights}[b]$ \Comment{0.3--0.8 per pattern}
    \ForAll{trait $\tau$ affected by $b$}
        \State $\mathit{implicitW} \gets \min(\mathit{totalInteractions} / 10,\; 1.0)$ \Comment{Learning ramp}
        \State $\mathit{mult} \gets 1 + \mathit{implicitW} \times 2$ \Comment{Evolution multiplier: 1$\times$--3$\times$}
        \State $\mathit{dim} \gets 1 - \frac{|\mathbf{t}[\tau] - 50|}{50} \times 0.2$ \Comment{Diminishing at extremes}
        \State $\Delta \gets w \times \mathit{mult} \times \mathit{dim}$
        \State $\mathbf{t}[\tau] \gets \Call{Clamp}{\mathbf{t}[\tau] + \Delta,\; 0,\; 100}$
    \EndFor
\EndFor
\If{$\mathit{totalInteractions} \bmod 50 = 0$} \Comment{Drift detection}
    \State $\mathit{drift} \gets \frac{1}{5}\sum_\tau |\mathbf{t}[\tau] - \mathbf{t}_0[\tau]|$ \Comment{Mean deviation from baseline}
    \If{$\mathit{drift} > 50$}
        \State $\mathbf{t} \gets \mathbf{t} + 0.1 \times (\mathbf{t}_0 - \mathbf{t})$ \Comment{10\% correction nudge}
    \EndIf
\EndIf
\end{algorithmic}
\end{algorithm}

The diminishing factor softens changes at extreme trait values (20\% reduction at boundaries), preventing personality scores from saturating. Milestone tracking and gamification rewards (omitted for brevity) incentivize sustained interaction.

\subsection{GNWT Cognitive Processing Cycle Algorithm}

The cognitive processing cycle (\texttt{GlobalWorkspace.js}) implements the complete 100ms GNWT loop as a four-phase state machine. Five specialist modules---Emotion, Memory, Planning, Social Norms, and Goal Tracking---process input in parallel (offloaded to a Web Worker), then compete for workspace access through salience-weighted selection. The workspace enforces a capacity of 9 items with an entry threshold of $\tau = 0.7$. Coalition formation groups semantically similar items (similarity $> 0.6$) to boost collective salience. Stale items (age $> 300$ms and salience $< 0.3$) are evicted, and non-attended items decay by a factor of 0.9 per cycle.

\begin{algorithm}[H]\small
\caption{GNWT Cognitive Processing Cycle}
\label{alg:cognitive-cycle}
\begin{algorithmic}[1]
\Require Input stimulus, agent state, 5 specialist modules
\Ensure Winning output broadcast to all modules, or null if below threshold
\State \textbf{Phase 1: Parallel Activation} (100ms budget)
\State $\mathit{outputs} \gets \Call{AwaitAll}{\{m.\mathit{processAsync}(\mathit{input}) : m \in \mathit{modules}\}}$
\State \textbf{Phase 2: Salience Competition}
\ForAll{$o \in \mathit{outputs}$}
    \State $s_o \gets o.\mathit{confidence} \times w_{\mathit{module}}$ \Comment{Base weights: Planning 0.6, Goal 0.6, Emotion/Social 0.5, Memory 0.4}
    \If{context matches module speciality} $s_o \gets s_o \times 1.5$ \EndIf \Comment{Context boost}
    \If{$o$ matches previous broadcast} $s_o \gets s_o \times 1.2$ \EndIf \Comment{Priming}
    \If{emotional arousal $> 0.5$} $s_o \gets s_o \times (1 + 0.4 \times |\mathit{valence}|)$ \EndIf
\EndFor
\State \textbf{Phase 3: Selection}
\State Sort candidates by salience; attempt coalition formation (similarity $> 0.6$)
\If{top candidate or coalition salience $\geq 0.7$}
    \State $\mathit{winner} \gets$ top candidate or coalition
\Else
    \State \Return null \Comment{Below threshold; sub-threshold buffer used for blending}
\EndIf
\State \textbf{Phase 4: Broadcast \& Integration} (50ms broadcast + 150ms integrate)
\State Add $\mathit{winner}$ to workspace (capacity 9; evict oldest non-attended if full)
\State Broadcast to all modules; attended items get 1.2$\times$ boost, others decay $\times 0.9$
\State Update winning module weight: $w \gets \min(w \times 1.02,\; 2.0)$ \Comment{rate 0.02}
\State \Return $\mathit{winner}.\mathit{output}$
\end{algorithmic}
\end{algorithm}

Salience computation integrates four factors: base module confidence weighted by static module importance, context-dependent boosting (e.g., SocialNorms $\times 1.5$ in social contexts), recency priming ($\times 1.2$ if output matches the previous broadcast), and emotional modulation scaled by arousal intensity. The full specialist module specifications appear in Appendix~\ref{appendix:gnwt}.

\subsection{Cognibit Social Simulation Pipeline}

The social simulation pipeline (\texttt{MatchmakingEngine.js}) implements a three-stage matching architecture with preference evolution across multiple interaction rounds. Stage 1 applies a 4-factor heuristic (personality 0.6, proximity 0.25, interests 0.1, diversity 0.05) to the full candidate pool, passing the top 20 candidates (above the 20\% compatibility threshold) to Stage 2. Stage 2 simulates multi-turn LLM conversations between twin pairs, scoring behavioral compatibility. Stage 3 presents the top 3--5 candidates for human review. Trait complementarity uses a sweet-spot model: differences of 20--40 points score 1.0 (ideal), $<$20 scores 0.8 (too similar), and $>$40 scores 0.4 (likely clash).

\begin{algorithm}[H]\small
\caption{Three-Stage Social Matching with Preference Evolution}
\label{alg:cognibit-simulation}
\begin{algorithmic}[1]
\Require Agent profiles $\mathcal{A}$, compatibility threshold $\tau = 0.2$, rounds $R = 10$
\Ensure Final matches with preference evolution history
\For{$r \gets 1$ to $R$}
    \State $\mathit{pairs} \gets \Call{GeneratePairings}{\mathcal{A}}$
    \ForAll{$(a_i, a_j) \in \mathit{pairs}$}
        \State \textbf{Stage 1:} $s \gets 0.6 \cdot \mathit{PersonalitySim} + 0.25 \cdot \mathit{Proximity} + 0.1 \cdot \mathit{Interest} + 0.05 \cdot \mathit{Diversity}$
        \If{$s < \tau$} \textbf{continue} \EndIf
        \State \textbf{Stage 2:} Simulate 4-turn conversation; compute behavioral compatibility $b$
        \State $\mathit{combined} \gets 0.7 \cdot s + 0.3 \cdot b$
    \EndFor
    \State \textbf{Preference Evolution:}
    \ForAll{$a \in \mathcal{A}$}
        \If{best partner mutual interest $> 0.7$}
            \State $a.\mathit{prefs} \gets 0.9 \cdot a.\mathit{prefs} + 0.1 \cdot \mathit{partner.traits}$ \Comment{Reinforce}
        \ElsIf{mutual interest $< 0.3$}
            \State $a.\mathit{prefs} \gets 1.1 \cdot a.\mathit{prefs} - 0.1 \cdot \mathit{partner.traits}$ \Comment{Adjust away}
        \EndIf
        \State $a.\mathit{prefs} \gets (1 - 0.05) \cdot a.\mathit{prefs} + 0.05 \cdot a.\mathit{baseline}$ \Comment{Decay to baseline}
    \EndFor
\EndFor
\State \textbf{Stage 3:} Return top 3--5 matches for human review
\end{algorithmic}
\end{algorithm}

Population-level analysis computes homophily (mean pairwise similarity of matches), preference convergence (mean trait drift from initial to final round), reciprocity rate (fraction of mutual selections), clustering coefficient, and average path length. The 5-factor compatibility formula (personality 0.30, interests 0.20, conversation quality 0.25, emotional resonance 0.15, interaction patterns 0.10) is detailed in Algorithm~\ref{alg:matching-funnel} (main text).

\end{document}